\documentstyle[12pt,epsfig]{article}
\begin{document}

%\twocolumn[\hsize\textwidth\columnwidth\hsize\csname@twocolumnfalse\endcsname

\vskip 1cm

\title{\bf Dynamics of glassy systems}
\vskip 20pt
\author{
{\bf  Leticia F. Cugliandolo}
\\
 Laboratoire de Physique Th{\'e}orique, Ecole Normale Sup{\'e}rieure,
\\
24 rue Lhomond 75231 Paris Cedex 05 France and 
\\
Laboratoire de Physique Th{\'e}orique  et Hautes {\'E}nergies, 
Jussieu, \\
1er {\'e}tage,  Tour 16, 4 Place Jussieu, 
75252 Paris Cedex 05, France
}

\date{October 12, 2002}
\maketitle
\begin{abstract}
These lecture notes can be read in two ways. The first two Sections
contain a review of the phenomenology of several physical systems with 
slow nonequilibrium dynamics. In the Conclusions we summarize the 
scenario for this temporal evolution derived from 
the solution to some solvable models 
($p$ spin and the like) that are intimately
connected to the mode coupling approach (and similar ones) 
to super-cooled liquids. At the end we list a number of open 
problems of great relevance in this context.
These Sections can be read independently 
of the body of the paper where we present
some of the basic analytic 
techniques used to study  the out of equilibrium dynamics of
classical and quantum models with and without disorder.
We start the technical part 
by briefly discussing the role played by the environment 
and by introducing and comparing its representation in 
the equilibrium and dynamic treatment of classical and quantum systems. 
We next explain the role played by explicit quenched disorder 
in both approaches. 
Later on we focus on analytical techniques;
we expand on the dynamic functional methods, and the 
diagrammatic expansions and resummations used to 
derive macroscopic equations from the microscopic dynamics. We show why
the macroscopic 
dynamic equations for disordered models and those resulting from 
self-consistent approximations to non-disordered ones coincide.
We review some generic properties of dynamic systems evolving out 
of equilibrium like the modifications of the fluctuation-dissipation 
theorem, generic scaling forms of the correlation functions, etc. 
Finally we solve a family of mean-field models.
The connection between the dynamic 
treatment and the analysis of the free-energy landscape of these models
is also presented. We use pedagogical examples all along these lectures 
to illustrate the properties and results. 

\end{abstract}

\newpage

\tableofcontents 

\newpage
\section{Introduction}
\label{introduction}
\setcounter{equation}{0}
\renewcommand{\theequation}{\thesection.\arabic{equation}}

Graduate and undergraduate courses on statistical mechanics and 
thermodynamics are usually devoted to the theory of 
macroscopic systems in thermal equilibrium.
In many experimental realizations, actually some of the more interesting
ones at present, the  situation is, however, very different. 
The systems are in contact with equilibrated environments 
but, for one reason or another, they do not manage to equilibrate 
with them. The systems evolve in time in an out of equilibrium manner.

The list of systems evolving out of equilibrium  is very long.
The reasons for not reaching equilibrium with the environment are 
also varied. The most common cases are those in which the time needed
to equilibrate the sample falls beyond the experimental time-window.
We discuss them in the context of domain growth, phase 
separation and classical and quantum glassy systems. 
Another important cause for lack of equilibration is the 
action of external forces that 
drive the samples out of equilibrium.
In this context we discuss the rheological properties of glass forming
liquids and glassy materials, that 
are closely related to the relaxation of the same systems. The driven 
dynamics of granular matter is another example of this kind. Finally,
we briefly touch another type of problem that has received much attention 
in recent years: the relaxation and 
weakly driven dynamics of elastic manifolds in random 
potentials that model
magnetic domain walls in 
disordered materials, superconductors, Wigner crystals, etc.

A common feature among the relaxing and weakly driven 
examples cited above is that 
they evolve {\it very slowly}. Thus, they belong  to a particular
class of the full set of non-equilibrium systems. Exploiting the fact that 
their dynamics is slow, and other more subtle properties that we shall
discuss along these notes,  
we can hope to develop a common theoretical description for all of them. 

In these lectures we focus on the study of a family of simple models
that can be adapted to mimic the above mentioned physical 
systems. Typically, these models are fully connected interacting 
spin systems or models of interacting particles in infinite dimensions.
They can be seen as the equivalent of the 
fully connected Ising model for ferromagnetism that 
correctly predicts the existence of a thermodynamic transition and 
the nature of the two phases but fails to describe the dependence on
dimensionality or the precise critical behavior.
Similarly, the schematic models  
do not include a notion of distance inside the system. This crude 
approximation
allows one to solve the dynamics explicitly, paying the price of 
loosing information about the behavior in real space.
Models of finite manifolds embedded
in infinite dimensional spaces and under the effect of 
random potentials are generalizations of the schematic models that 
capture partial spatial information. They are also solvable analytically.
Interestingly enough, one finds that these models realize several 
phenomenological approaches to 
glassy physics that have been known for long. Moreover, 
their dynamic macroscopic equations coincide with those arising from
the self-consistent approximations used to analyze more realistic models
as, for instance, the mode-coupling approach to super-cooled liquids.
Having an exact solution is very important in 
many respects. Firstly, it establishes the phenomenological approaches
on firmer bases. Secondly, it allows one to set clear limits of 
validity of the self-consistent approximations to realistic models.
Thirdly, being defined 
by interacting potentials their free-energy density is accessible 
to analytical studies, from which one obtains the organization of
metastable states and relates it to the dynamic properties.
Fourthly, many important and common features of systems evolving 
slowly out of equilibrium have been discovered in the analytic solution 
to these models. Fifthly, one is able to identify some of the missing
ingredients needed for a more accurate description of real systems. Even if
their treatment has been too tough to implement correctly yet, it is important
to know in which direction one could try improving the analytical study.
In short, they constitute a very useful ``laboratory'' where
one identifies general trends 
that can be later tested numerically and experimentally 
in more realistic models and real materials. 

Although the models on which we concentrate are simple in the 
sense discussed above one needs to master many analytical methods 
to extract all the richness of their behavior. These methods are 
not completely standard and are not comprehensibly described
in textbooks or lecture notes. For this 
reason, we try to present a rather complete and detailed 
introduction  to them.
We also discuss
the scenario for glassy dynamics that stems out of this analysis. 
Finally, we mention several lines for future work that are currently
being explored by several groups trying to go beyond the fully-connected 
and infinite dimensional models. 

The lecture notes are organized as follows. In the next Section we 
introduce the phenomenology of the physical systems we are interested
in paying special attention to the dynamic
properties that we later describe analytically. The rest of the 
lecture notes are more technical. We start by reviewing very briefly 
several theoretical approaches to the glassy problem in 
Section~\ref{section:theoretical}.
This summary is certainly not exhaustive but it may serve as a source of 
references. Since in the rest of the notes we shall develop classical and 
quantum systems in parallel we devote 
Section~\ref{systemsincontactwithreservoirs}
to discuss how should one model the coupling between a system and 
its environment in both cases. In Section~\ref{section:observables} 
we set the notation 
and we define several useful observables for spin and particle systems.
The subject of Section~\ref{sec:prob} is a 
brief discussion of Fokker-Planck and Kramers 
processes and how they describe the 
approach to equilibrium. In the following Section
we introduce the fluctuation-dissipation theorem, valid for systems 
evolving in equilibrium. Once these generic properties
are established it  will become simpler to discuss how they are modified in 
a system that evolves far from equilibrium. In 
Section~\ref{generating-functionals} we
explain the functional formalism that allows one to derive a generic
dynamic effective action and the dynamic equations that 
we present in Section~\ref{dynamicequations}. We also 
introduce the very useful
super-symmetric formulation of classical stochastic processes. 
In Section~\ref{diagrammatic-techniques} we discuss an 
alternative method to obtain 
approximate dynamic equations. We focus on the mode-coupling
({\sc mc}) approximation to show how averaging over disorder in 
some random models eliminates the same family of diagrams that the 
{\sc mc} procedure neglects for non-disordered ones. The rest 
of the lectures are dedicated to the solution to the dynamic
equations of the schematic models. We present it in as much generality as
possible in Section~\ref{sec:lowT}. 
We discuss three generic results that were obtained when 
solving these equations without assuming equilibrium: the definition
of correlation scales (Section~\ref{time-scales}), the modifications of the 
fluctuation-dissipation theorem (Section~\ref{Modifs}) and the definition
of effective temperatures (Section~\ref{temp_intro}). We also discuss 
how these properties appear in variations of the models that mimic
the physical systems introduced in Section~\ref{interestingproblems} and
in finite $d$ toy models, simulations and experiments.
Finally, in Section~\ref{section:tap}
we relate the dynamic results to the organization of metastable states
via the static and dynamic approach of Thouless, Anderson and Palmer
({\sc tap}). We briefly discuss the connection between it
and the studies of the potential energy landscape (Edwards measure and 
inherent structure approach).
Finally, in Section~\ref{conclusions} we summarize the 
scenario for the glass transition and glassy dynamics derived from 
solvable mean-field models and 
present some of the lines for future research in this area.

\section{Some physical systems out of equilibrium}
\label{interestingproblems}
\setcounter{equation}{0}
\renewcommand{\theequation}{\thesection.\arabic{equation}}

In this Section we summarize the phenomenology of a number of systems
with slow dynamics. We especially signal the features that we expect
to capture with an analytical approach.

\subsection{Domain growth}
\label{subsec:domain}

Out of equilibrium relaxational dynamics occurs, for instance, when one 
suddenly quenches a system with ferromagnetic interactions 
from its high temperature phase into
its low temperature phase. When the system is in contact 
with a thermal bath at temperature $T> T_c$ ($T_c$ is the Curie critical 
temperature) the system is disordered and the instantaneous 
averaged magnetization 
vanishes at all times. (The average refers here to a coarse graining  
over a region of linear size, $\ell$, with $\xi \ll \ell$ and $\xi$ the 
correlation length.) 
All the one-time properties, such as the instantaneous averaged magnetization
or the static magnetic susceptibility, can be computed using the 
Gibbs-Boltzmann distribution, $P_{\sc gb}$. 
The system evolves in time but in a very simple manner
controlled by $P_{\sc gb}$.
All properties of the equilibrium dynamics hold and
any two-time 
correlation function is invariant under translations of time. 
Instead, if one externally and suddenly changes $T$ 
to set it below the Curie temperature, $T<T_c$, 
the system evolves from the
very disordered initial condition via the growth of domains
of up and down magnetic order. With simple arguments one shows that 
the typical linear size of the domains grows as a power of the time spent in 
the low temperature phase, ${\cal R}(t_w)= \Upsilon(T) \, 
t_w^{1/z}$~\cite{Bray}.
We call waiting-time, $t_w$, the time spent since the 
entrance in the low-$T$ phase. The dynamic exponent $z$ 
depends on
the kind of microscopic dynamics considered, {\it e.g.} for 
non-conserved order parameter $z=2$ while for conserved order 
parameter $z=3$.
All $T$-dependence is concentrated in the prefactor $\Upsilon(T)$.   
After the initial quench 
the system is a superposition 
of up and down domains and the magnetization averaged over the full 
system ($\ell = L$) vanishes. During coarsening domains
grow. This  is a {\it scaling regime} in which the system is 
statistically invariant under rescaling by the typical length
${\cal R}(t)$. Typically, 
at a time $t_{\sc req}\approx L^z$ the 
system orders and the averaged magnetization, $m$, 
equals the magnetization of the conquering domain, say $m>0$.
{\it Restricted ergodicity or equilibrium} within one ergodic component
holds in the sense that time and ensemble averages
can be exchanged {\it if and only if} 
the time average is taken over a time window
$t_{\sc req} < t < t_{\sc erg}$ and the ensemble average is 
restricted to the configurations with positive magnetization.   
Since a rare fluctuation might lead the system to reverse 
from $m>0$  to $m<0$ another, still longer, characteristic time, 
$t_{\sc erg}$, appears. 
This time is also a function of the size of the system $L$ and it is
such that for times that are much  longer than it
complete ergodicity is restored. One can estimate it to be of 
Arrhenius type $t_{\sc erg} \sim \exp[cL^{d-1}/(k_BT)]$, with $c$ a
constant, $d$ the dimension of space and $cL^{d-1}$ the free-energy barrier
to be surmounted to go from one ergodic component to the other.

\vspace{0.5cm}
\begin{figure}[h]
\centerline{
\psfig{file=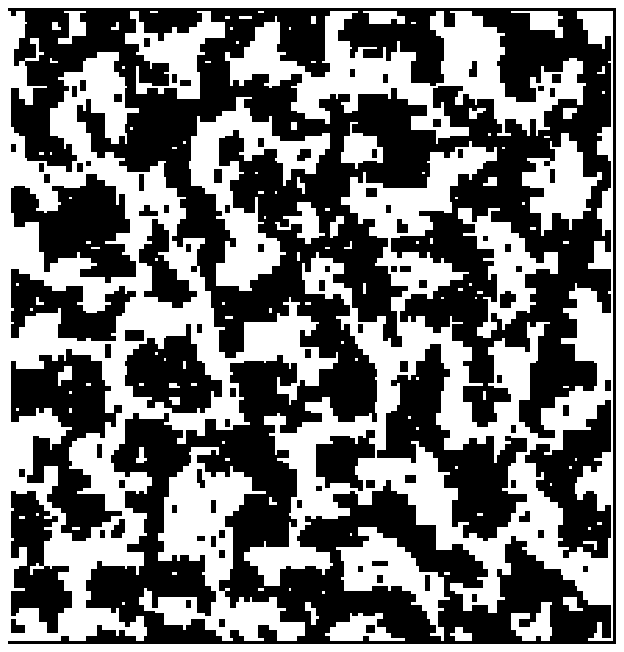,width=4cm}
\hspace{0.5cm}
\psfig{file=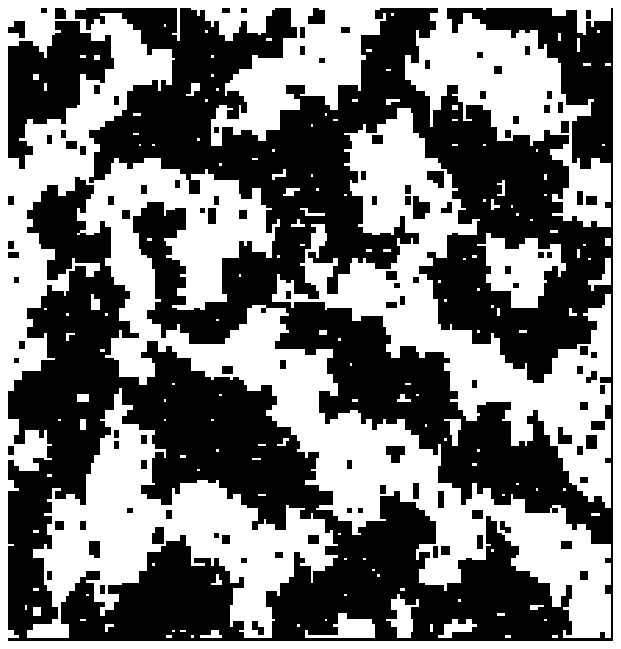,width=4cm}
\hspace{0.5cm}
\psfig{file=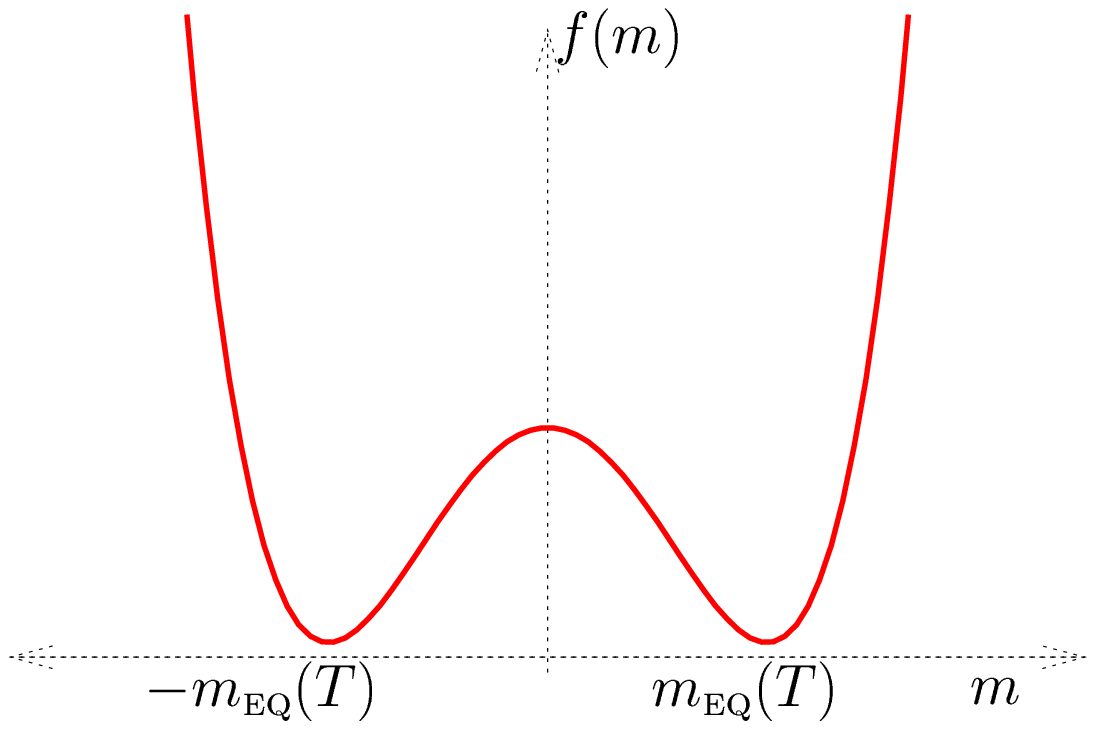,width=6cm}
}
\vspace{0.25cm}
\caption{Two snapshots of a $2d$ cut of a $3d$ lattice undergoing 
ferromagnetic domain growth with non-conserved order parameter
after a quench from $T\to\infty$ to $T<T_c$ at time $t=0$. 
On the left, $t_w=10^3$ Montecarlo steps (MCs).
On the right, $t_w=10^5$ MCs. Sketch of the Landau free-energy density
$f(m)$ in the low-$T$ phase. The evolution of the system is partially
described by the evolution of a point in this free-energy landscape.
Just after the quench $m=0$ and coarsening is visualized
in this plot as a static point on top of the barrier. After $t_{\sc req}$
the point falls into the well around the magnetization of the 
conquering domain. After $t_{\sc erg}$ ergodicity is restored and the point 
jumps the barrier via thermal activation~\cite{clarify}.
}
\label{aging}
\end{figure}

The low temperature phase can also be reached with an annealing, {\it e.g.}
by modifying the external temperature in steps of length 
$\Delta t$ and magnitude $\Delta T<0$ until reaching the working temperature
$T$. Since the entrance in the low-$T$  phase the system
coarsens.
If the prefactor $\Upsilon(T)$ increases with $T$,
after a time $t_w$ since crossing $T_c$
a system prepared with a slow cooling rate will have much larger 
structures than one of the same age that has been 
quenched into the ordered phase.
The dynamics in isothermal conditions is 
basically identical in both cases but the starting configuration at the 
final temperature $T$ is 
however very different, the annealed system looking older than the 
quenched one. The presence or absence of 
cooling rate dependences as well as the effect of temperature on the 
low-$T$ dynamics allow one to distinguish among different
glassy system. We shall come back to this issue when discussing 
structural and spin glasses.

On the right panel of  Fig.~\ref{aging} we plot the 
two-welled Landau free-energy density $f(m)$ against $m$. 
Transverse to the 
$m$ direction there are $2^{L^d}-1$ other directions that complete the 
phase space of the spin model. Note that when $L\to\infty$ phase space
is infinite dimensional even if real space is finite 
dimensional, $d<\infty$. 
If one wishes to view the dynamics as the wandering of a point,
that represents the instantaneous configuration of the system,
in the free-energy landscape, the one-dimensional
plot in Fig.~\ref{aging} might be useful only if used carefully. 
The initial configuration after a quench corresponds to the 
top of the barrier ($m=0$). This unstable point hides 
the $2^{L^d}-1\gg 1$, for $L\gg a$,  transverse directions
($a$ is the lattice spacing). The domain growth 
process takes place while the representative point sits on the 
top of the barrier. Falling into one well corresponds to the 
growth of the conquering domain. Finally, jumping over the 
barrier is the activated process of reversing the full system. As long
as $t< t_{\sc req}$ the dynamics is highly non-trivial while viewed
on this plot it looks trivial, with the representative point 
simply sitting on the border between the two 
basins of attraction of the equilibrium states 
$m=\pm m_{\sc eq}$~\cite{Kula}. 

Importantly enough, both $t_{\sc req}$ and $t_{\sc erg}$ grow with $L$ 
and, if the thermodynamic limit $L\to\infty$ is taken at the outset,
diverging times with $L$ cannot be reached and physical times are always
smaller  than $t_{\sc req}$. The system cannot equilibrate with its 
environment and the non-equilibrium domain growth process goes on for ever.

\subsection{Glasses}
\label{subsec:glassy}

The domain growth example is very useful to visualize a 
non-equilibrium evolution. The mechanism behind the dynamics 
is clear and the growth of order can be easily 
identified. In other systems that undergo a non-equilibrium 
evolution whether there is a growing 
order controlling the evolution is still an open question. 
Glassy systems are one such example.

\begin{figure}[h]
\centerline{
\psfig{file=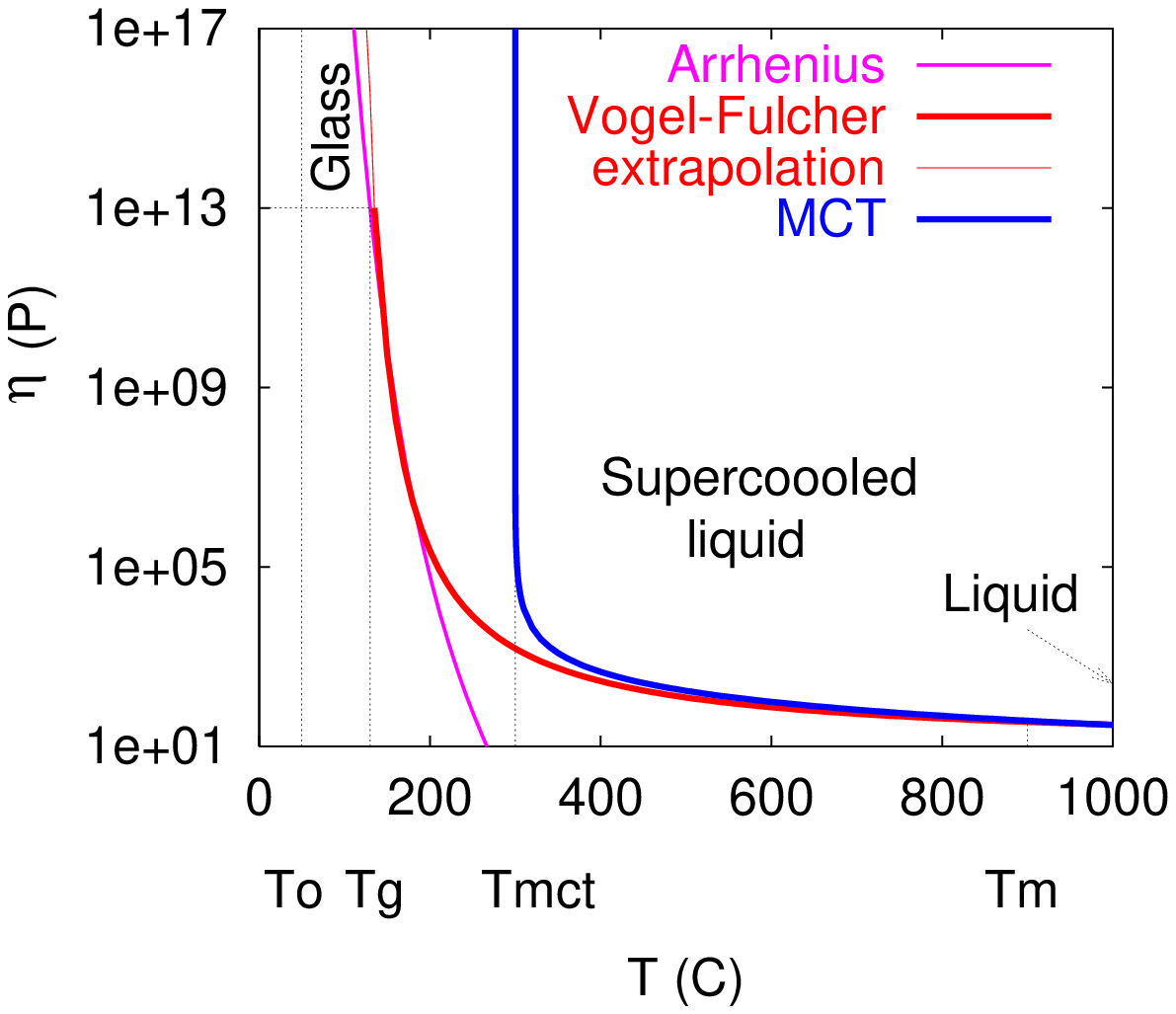,width=8cm}
\psfig{file=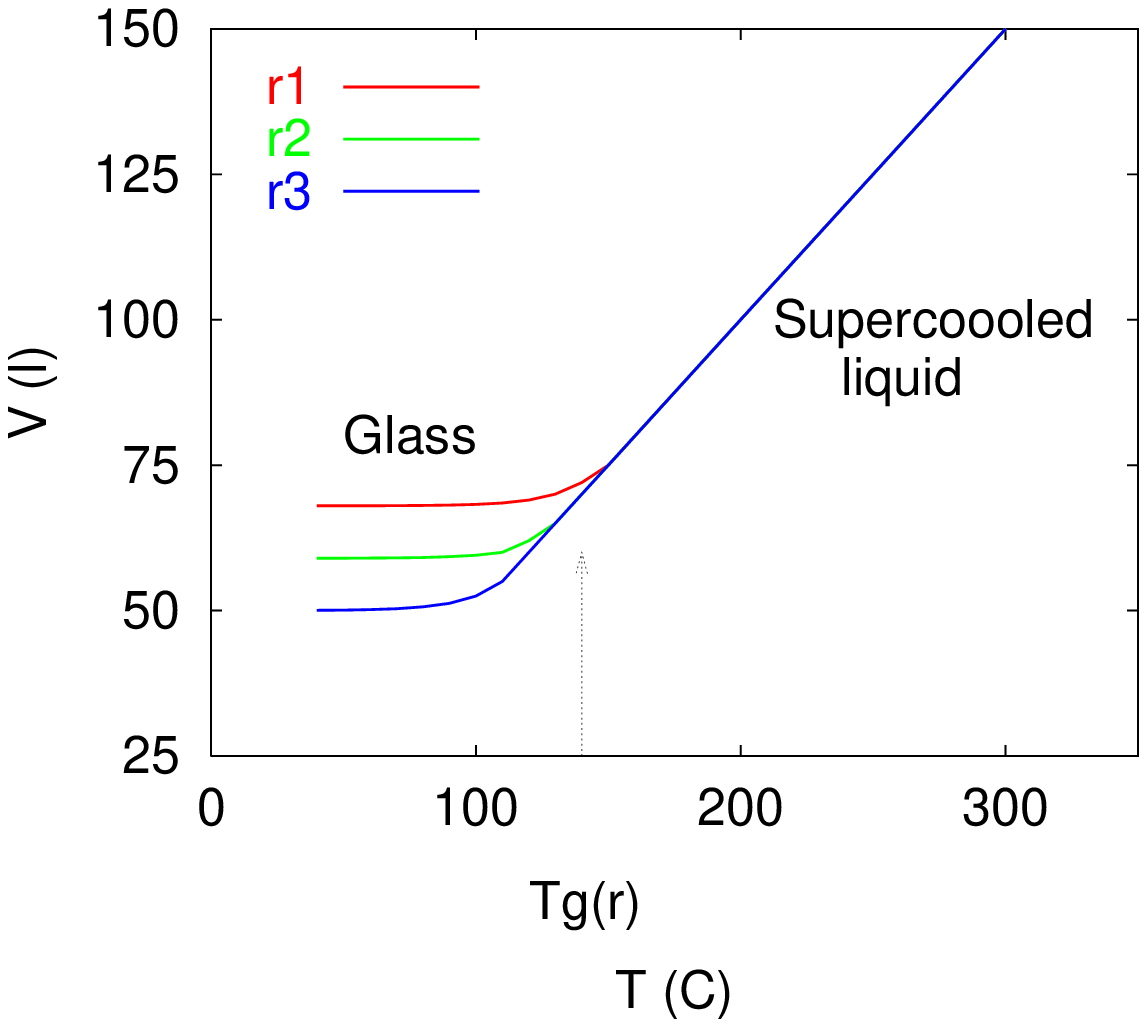,width=8cm}
}
\caption{Left: sketch of the viscosity against temperature approaching $T_g$. 
Rough comparison between different scaling forms. (Arrhenius:
$A=3900$C, $\eta_0=10^{-6}$P. Vogel-Fulcher: $A=500$C, $T_0=100.$C,
$\eta_0=10$P. {\sc mct}: $\eta=\eta_0/(T-T_d)^\gamma$ with 
$T_d=300$C, $\gamma=0.7$, $\eta_0=1700$P.) 
Right: cooling rate dependence of the volume,
$r1>r2>r3$.}
\label{viscosity}
\end{figure}

Understanding the glass transition and glassy dynamics 
is one of the greatest challenges in theoretical physics.
The glassy problem can be summarized as follows~\cite{review-glasses}. 
Take a 
liquid at high temperature $T_i$ 
and quench it at a constant rate, $r\equiv -\Delta T/\Delta t$. 
On each temperature step, the viscosity relaxes rather quickly 
and with a simple analysis one estimates its asymptotic value to trace 
a curve $\eta(T)$ that is 
sketched in the left panel of Fig.~\ref{viscosity} (red curve). 
This curve has several remarkable features. At very high temperature 
$\eta(T)$ very slowly grows with decreasing $T$. 
Decreasing the external 
temperature still the system approaches the crystallization (or melting) 
transition $T_m$.
If the cooling rate $r$ is sufficiently fast, 
this transition is avoided and the system enters a metastable 
{\it super-cooled liquid} phase where the viscosity grows
very quickly with decreasing $T$. Indeed, one typically observes 
that when $T$ changes by, say, $100^oC$, the viscosity 
jumps by approximately $10$ orders of magnitude. In consequence, the 
dynamics of the liquid slows down enormously. For several liquids
the form of the $\eta(T)$ curve can be described with a Vogel-Fulcher
law, 
\begin{equation}
\eta(T) = \eta_0 \exp\left( \frac{A}{T-T_o}\right)
\; ,
\label{vogel}
\end{equation} 
that predicts a divergence of $\eta(T)$ when $T\to T_o$. These 
liquids are conventionally called {\it fragile}. (Within the 
experimental precision $T_o$ coincides with the Kauzmann temperature 
$T_K$ where the extrapolation  of the difference between the 
entropy of the liquid and the crystal vanish.)
However,
this fitting procedure can be criticized for several reasons. 
In many cases 
the form of the fitting function strongly depends on the temperature
window chosen for the fit. Even more important is the 
fact that the dynamics becomes so slow when $T$ decreases that 
the time needed to equilibrate the sample goes beyond the minimal 
cooling rate, or the maximum 
time reachable in the experience. Below a temperature $T_g(>T_o)$ 
one can no longer equilibrate the sample. 
One of the most clear signatures of the absence of equilibrium 
below $T_g$ is the fact that the measurement of, {\it e.g.}, the volume 
as a function of temperature, depends on the cooling rate, $r$, 
used to reach $T$. Moreover, $T_g$ decreases with decreasing cooling rate.
Hence, the so-called ``glass transition'' is not a true thermodynamic
transition but 
a dynamic crossover from the super-cooled liquid phase, where the 
dynamics is slow but occurs as in equilibrium~\cite{metastable}, 
to the glass phase where
the system does not manage to equilibrate with its environment.
These features are schematically shown in the right panel of 
Fig.~\ref{viscosity}.
Going back to the 
interpretation of Eq.~(\ref{vogel}), some authors extrapolate it below 
$T_g$, where viscosity measurements are not possible, and interpret the 
divergence at $T_o$ as the signature of a true thermodynamic transition.
This means that even in the limit of vanishing cooling rate one would
observe a transition from the super-cooled liquid to an ideal glass 
phase (still metastable with respect to the crystal) at $T_o$. Others 
prefer a scenario in which Eq.~(\ref{vogel}) has no meaning below $T_g$.

In this sense, it is important to stress that the viscosity of 
many glass forming liquids, as the
silica materials that give rise to window glass, 
is well fitted by an Arrhenius law 
\begin{equation}
\eta(T) = \eta_0 \exp\left(\frac{A}{T}\right)
\end{equation} 
that diverges at $T\to 0$. These are the so-called {\it strong} 
glasses. It has been suggested that actually all glasses are 
strong since one can always fit $\eta(T)$ with an Arrhenius law 
if the temperature window used for the fit is close enough 
to $T_g$ (as suggested in Fig.~\ref{viscosity}). 
This is the reason why some authors prefer a 
scenario in which the super-cooled liquid phase extends all the way up to 
$T=0$ when an infinitely slow cooling rate is used.  
The curve labeled {\sc mct} in the same figure
shows the prediction of the models we shall discuss in these notes
($p$ spin models and mode-coupling theory).
In short, this approach predicts a 
dynamic transition at $T_d$ that is typically 
higher than $T_g$, with a power law divergence of $\eta$. 
Albeit this and other defects, this approach is
successful in many respects since it yields a satisfactory {\it qualitative} 
description of the phenomenology of super-cooled liquids and glasses.

\begin{figure}[h]
\centerline{
\psfig{file=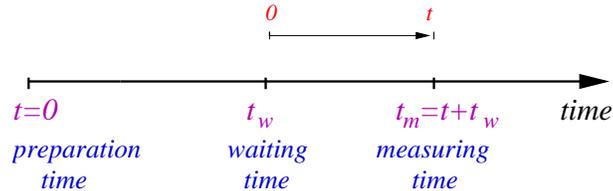,width=8cm}
}
\caption{Characteristic times. The waiting and measuring times are
{\it experimental times} $t_{\sc exp}$. The equilibration time
$t_{\sc eq}$ can be shorter or longer than them leading to equilibrium
or non-equilibrium dynamics, respectively.}
\label{characteristic-times}
\end{figure}

For the purpose of our discussion the important point to 
stress is the fact that the liquid falls out of equilibrium 
at the (cooling rate dependent) temperature $T_g$. Indeed, even the 
``state'' reached by the system below $T_g$ depends on the cooling 
rate as seen, for instance, in measurements of volume, entropy, 
etc. as functions of $T$, see the right panel in Fig.~\ref{viscosity}. 
The slower the cooling rate, the deeper one penetrates below the 
{\it threshold} level corresponding to $r\to\infty$.   
The dynamics below $T_g$ occurs out of equilibrium since 
$t_{\sc eq} > t_{\sc exp}$, 
see the sketch in Fig.~\ref{characteristic-times}. 
The properties of the 
system cannot be described with the use of $P_{\sc gb}$
and a more sophisticated analysis has to be developed.

\begin{figure}[h]
\centerline{
\psfig{file=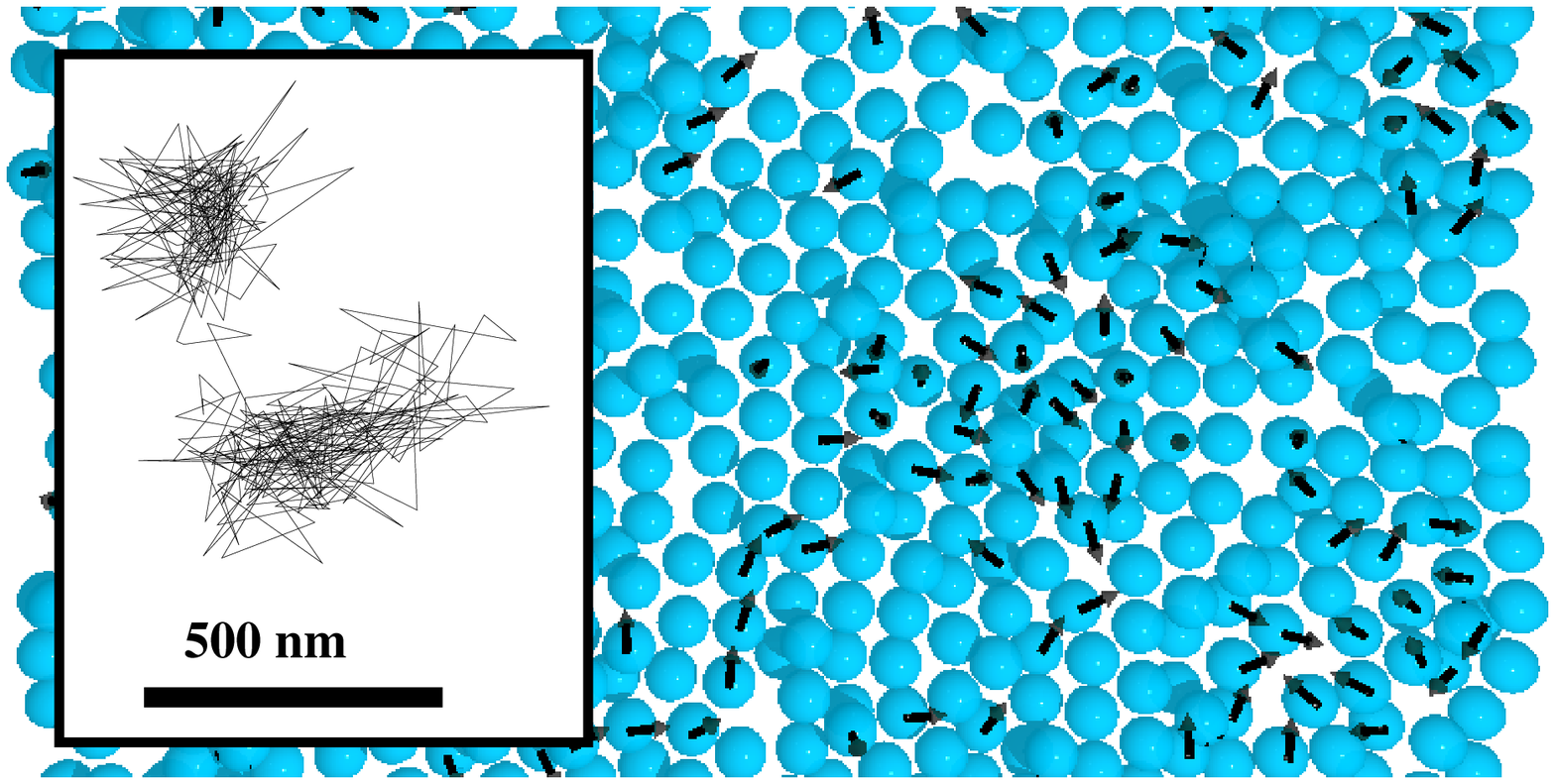,width=6cm}
\hspace{1cm}
\psfig{file=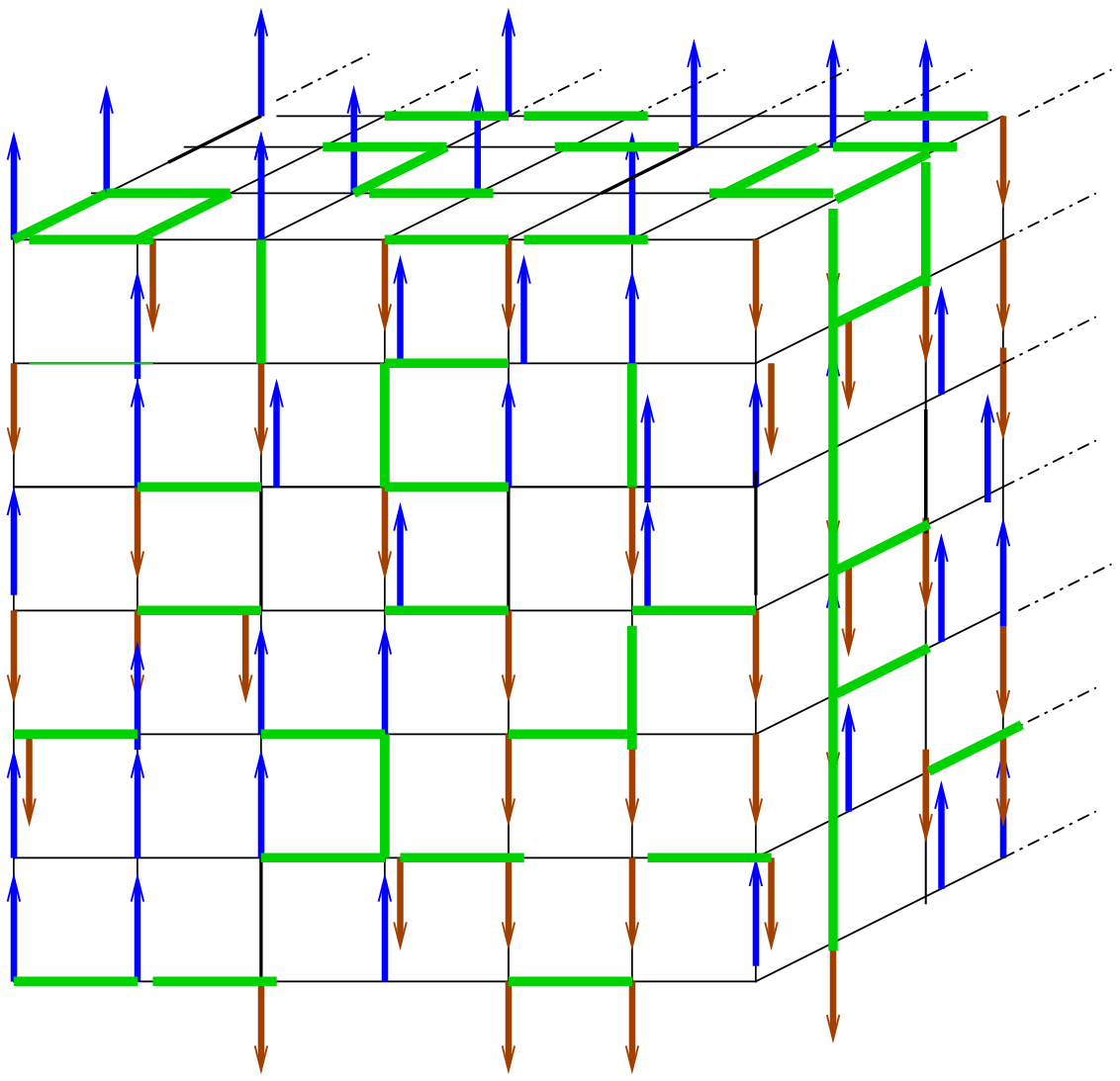,width=4cm}
}
\caption{Different types of glasses: on the left, a colloidal system
(image taken from \cite{Weeks}); on the right a representation of the 
$3d$ {\sc ea} spin-glass model.}
\label{different-glasses}
\end{figure}

The above 
paragraphs were devoted to the discussion of the liquid to glass transition 
and it implicitly assumed that the system was made 
of molecules in interaction. In the case of simple liquids, 
in which the constituents have no structure,  one can describe the 
relevant interactions with a two body potential
\begin{equation}
E=\sum_{i\neq j} V(|{\vec r}_i - {\vec r}_j|)
\; . 
\label{energy1}
\end{equation}
Hard spheres are the simplest example of this kind 
where the only interaction are hard core ones
that forbid the penetration of one particle by another. Adding 
polymers in the solution that hosts the spheres one can tune 
a repulsive interaction between the particles. (See 
Fig.~\ref{different-glasses}-left below for a snapshot of one such 
experimental system~\cite{Weeks}.) Favorite potentials used in numerical
simulations are the Lennard-Jones and soft-sphere ones
\begin{equation}
V(|{\vec r}_i - {\vec r}_j|) = 
4 \epsilon_{ij} 
\left[ \left( \frac{\sigma_{ij}}{|{\vec r}_i - {\vec r}_j|} \right)^{12}
       - \left( \frac{\sigma_{ij}}{|{\vec r}_i - {\vec r}_j|} \right)^6
\right]
\; .
\label{lennard-jones}
\end{equation}
To avoid crystallization one usually uses a binary system with 
$N_A$ particles with mass $m_A$ 
and $N_B$ particles with mass $m_B$. The parameter 
$\sigma_{AA}$ fixes the length scale, 
$\epsilon_{AA}$ the energy scale and $m_A$ the mass scale.
The time scale is then given by 
$t_0\equiv\sqrt{m_A \sigma_{AA}^2/\epsilon_{AA}}$. 
For a given density, {\it e.g.} $\rho=1.2$,
an adequate choice of the remaining parameters yields
the expected properties of a liquid or a glass 
with a numerical transition at a temperature $T_g$ where the 
relaxation time goes beyond the time accessible with the 
simulation. In the soft-sphere model one only keeps the 
repulsive term in the potential.
Many other types of glasses are known in Nature. 
For instance, plastics as PVC whose mesoscopic constituents are polymers
 also undergo a glass transition
very similar to the one described previously. 
Several types of interactions between the monomers that 
form the macromolecules
are also of two-body type and they are repertoriated
in the literature. The dynamics of particle systems is given by 
Newton's equations.

\subsection{Spin-glasses}
\label{subsec:spin-glasses}

More exotic types of glasses have been studied 
for long. Spin-glasses have attracted the 
attention of experimentalists and theoreticians as a prototypical 
system with 
{\it quenched disorder}~\cite{spin-glasses-old,Levetal,Mepavi,Kawamura}. 
These systems   
are magnetic alloys in which magnetic impurities are replaced in 
a magnetically inert host. The impurities occupy random positions
and are not displaced within the sample in experimental times.
The interactions between the impurities
depend on the distance between them. 
Since the latter are random, the interactions themselves 
take random values that change in sign very quickly. A number of 
experimental realizations exist. 
As in other glassy systems, 
spin-glasses fall out of equilibrium at a transition temperature $T_g$
when usual cooling rate procedures are used. From 
experimental and numerical results
near $T_g$ complemented with standard
critical analysis, it is rather 
generally believed that the transition between the paramagnetic and 
the spin-glass phase is, in this case, a true thermodynamic
transition. This is at variance with what occurs in structural and 
polymeric glasses. Another important difference
with structural glasses and systems undergoing simple coarsening as 
ferromagnets is that the magnitude of cooling rate dependences is 
quite negligible suggesting that for spin-glasses 
the preferred configurations at
one temperature are totally different from the ones at any other 
temperature.
Still, slightly farther away from the 
transition one can no longer equilibrate the spin-glasses and 
observes typical non-equilibrium effects. 

Edwards and Anderson proposed a simplified model for spin-glasses
in which one represents the magnetic impurities with Ising spins
placed on the vertices of a three dimensional cubic lattice. 
The random nature of the interactions are mimicked with
first neighbors random interactions between the spins 
taken from a Gaussian (or bimodal) probability distribution with 
zero mean and variance $[J_{ij}^2]={\tilde J}^2/(2z)$ 
where $z$ is the connectivity  of the lattice. (Hereafter we  
denote with square brackets the average over disorder.) 
The Hamiltonian is 
\begin{equation}
H_J[{\vec S}] = - \sum_{\langle ij\rangle} J_{ij} s_i s_j  
\; .
\label{3DEA-hamil}
\end{equation}
where the vector $\vec S$ encodes the full set of spins in the 
sample $\vec S=(s_1,s_2,\dots,s_N)$ and 
${\langle ij\rangle}$ represents nearest neighbors on the lattice.
In the fully connected limit in which each spin interacts with all 
others the sum runs over all pair of spins and the 
unusual normalization of the $J_{ij}$s, $[J_{ij}^2]={\tilde J}^2/(2N)$,
ensures a correct thermodynamic limit. This is the 
Sherrington - Kirkpatrick ({\sc sk}) model

For many years it was common lore that the presence of explicit 
quenched disorder made spin-glasses intrinsically different from 
glasses where no quenched random forces have been identified. 
This belief was in part motivated by the analytical treatment 
used to study the {\it equilibrium properties}
of the spin-glass phase in {\sc sk}, namely, the replica trick. 
More recently, after a series of seminal papers by Kirkpatrick,
Thirumalai and Wolynes~\cite{Kithwo}, 
and the later solution to 
the non equilibrium dynamics of several glassy 
models~\cite{Cuku1,review}, 
it has been realized that the presence of quenched disorder 
is not that relevant. All glasses can be analyzed on the same footing.
The schematic mean-field model for glasses, the $p$ spin model, 
has quenched Gaussian interactions and its Hamiltonian is
\begin{equation}
H_J[{\vec S}] = - \sum_{\langle i_1 i_2 \dots i_p\rangle} 
J_{i_1 i_2 \dots i_p} s_{i_1} s_{i_2} \dots s_{i_p}  
\; .
\label{pspsin-hamil}
\end{equation}
($[J_{i_1\dots i_p}]=0$ and 
$[J_{i_1\dots i_p}^2]={\tilde J}^2 p!/(2N^{p-1})$, we henceforth 
set $\tilde J=1$.) Glauber's rule for Ising
spins or Langevin equations for soft spins define the 
microscopic evolution.

\subsection{Quantum fluctuations}
\label{subsec:quantum}

Different driving  dynamics slightly modify the picture
just presented.
Quantum glassy phases, where quantum fluctuations are at least 
as important as thermal activation, 
 have been identified in a number of materials.
Two such examples are the spin glass compound
and the amorphous insulator studied in  
\cite{Aeppli} and \cite{Osheroff}, respectively. 
Another interesting realization is the so-called
Coulomb glass in which localized electrons interact via Coulomb 
two-body potentials and hop between localization centers~\cite{Zvi}. 
In all these  systems the dynamics is extremely slow and  
strong history dependence as 
well as other glassy features have been 
observed. 

Models for magnetic compounds are constructed with 
$SU(2)$ spins while models for particle 
systems are quantized with the usual commutation relations between 
coordinate and momenta. In both cases the dynamics is fixed by 
Heisenberg equations.

\subsection{Rheology and granular matter}
\label{subsec:driven}

\begin{figure}[h]
\centerline{
\psfig{file=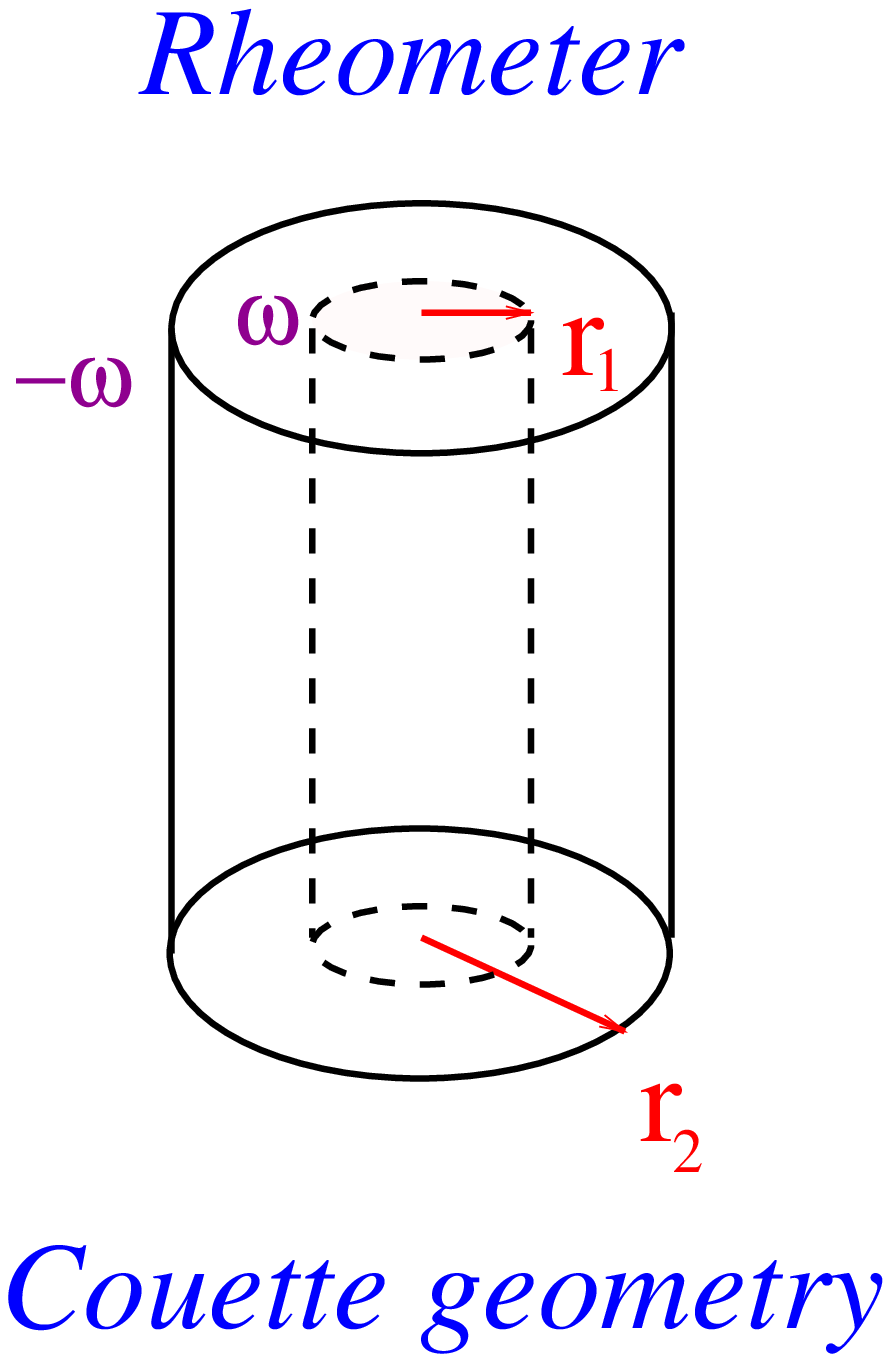,width=3.cm}
\hspace{1cm}
\psfig{file=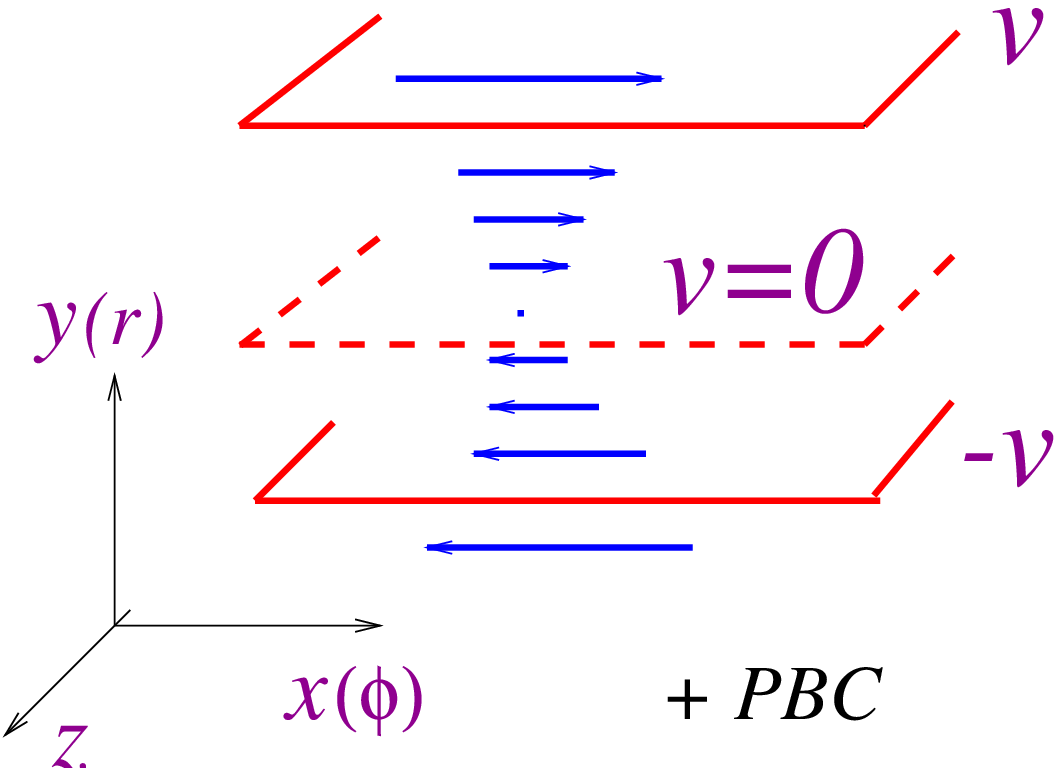,width=4.5cm}
}
\caption{Left: A Couette cell used to shear a liquid.
The internal and external walls turn with opposite angular velocities
and the fluid is included in between them.
Right: a cut of the Couette cell. ({\sc pbc}: periodic boundary conditions.)}
\label{shear-fig}
\end{figure}

The above examples concern systems that are not able to reach 
equilibrium with their  environments in a reasonable time but that, let
evolve on astronomical time-scales, will eventually equilibrate.
Other ways of establishing non-equilibrium states with slow dynamics 
are possible if one externally drives the samples~\cite{Cates-Adjari}. 

A dense liquid can be driven to a slow 
out of equilibrium stationary regime by a weak shear. A shear
is an example of a  
force that does not derive from a potential, {\it i.e.} it cannot 
be written as $\vec f=-\partial V/\partial \vec r$. The simplest way to 
apply a shear on a liquid is by means of a rheometer. In  
Fig.~\ref{shear-fig} we show one with a Couette geometry.
The shear modifies the dynamic equations for the fluid by adding 
an advection ${\vec v} {\vec \nabla} \rho({\vec r},t)$. In the planar limit
${\vec v}= \dot\gamma y \hat e_x$ where $\vec v$ is the 
velocity of the fluid, $\rho$ its density and $\dot \gamma$ the
shearing rate. 
One mimics such a force in a spin system with a non-symmetric 
force, {\it i.e.} the force exerted by the spin $i$ on 
the spin $j$ is not equal to the force exerted by the spin $j$ on 
the spin $i$~\cite{Cukulepe,Ludo}, {\it e.g.}
\begin{equation}
f_i =\alpha \sum_{j(\neq i)} J_{ij} s_j 
\;\; \mbox{with} \;\; J_{ij} \neq J_{ji} 
\; .
\label{non-symmetric-force}
\end{equation}
(The motivation to define such a force comes from neural nets where the 
synapsis have a direction.)
It is clear that this force 
cannot be written as the variation of a potential 
energy with respect to $s_i$, it violates detail balance, 
and an equilibrium measure cannot describe its effects~\cite{Gr}. 

A weak shear has a spectacular effect on the relaxation of 
liquids~\cite{review-rheology}. 
Usually, the viscosity as a function of the shear rate, $\dot\gamma$,
has a Newtonian plateau at small $\dot\gamma$ that crosses over to 
a decreasing function that is approximately given by $\eta \approx 
{\dot\gamma}^{-2/3}$. Hence, by shearing the liquid 
one facilitates its flow and the relaxation time decreases
with increasing $\dot\gamma$. Moreover, 
one introduces a shear-dependent time scale $t_{\sc sh}$
that plays an important role in aging experiments as discussed 
in Section~\ref{subsec:aging}.
 
Another family of materials that have captured the attention of 
experimentalists and theoreticians in recent  years
is granular matter~\cite{review_sid}. 
Since the potential energy needed to displace a macroscopic
grain by a distance equal to 
its diameter, $m g d$,  
is much larger than the characteristic thermal energy, $k_BT$,
thermal activation is totally irrelevant for systems made of macroscopic
grains. Therefore, in the absence of external driving granular matter
is blocked in metastable states and 
there exists no statistical mechanics approach capable 
of describing its static behavior. 
Instead, when energy is pumped in in the form of 
shearing, vibration or tapping, transitions between the otherwise 
metastable states occur and granular matter slowly relaxes towards 
configurations with higher densities. 
When trying to model these systems it is also important 
to keep in mind that dissipation is not given by the 
usual Ohmic form proportional to the velocity of the grains, 
$-\gamma v$, but it is much more cumbersome.  
Glassy features such as hysteresis
as a function of the amount of energy injected, 
slow dynamics~\cite{review_sid}, and 
non stationary correlations~\cite{Josserand,aging-granular,Barrat-granular} 
have been exhibited. 

The effect of the external drive can be described by 
applying time-dependent oscillatory forces, {\it e.g.} 
$f_i(\omega,t) = A \sin(\omega t)$, to each spin variable
in model~(\ref{pspsin-hamil})~\cite{Becuig}. One could also 
include complicated sources of dissipation by modifying the 
noise kernels obtained in 
Section~\ref{systemsincontactwithreservoirs} for a usual 
equilibrated bath.

\subsection{Elastic manifolds in random potentials}
\label{subsec:elastic}

The motion of a $d$ dimensional directed elastic manifold 
embedded in an $N$ dimensional space in the presence of 
quenched random disorder has a bearing in several areas
of physics~\cite{manifolds,Giamarchi}. (The total dimension 
of space is $d+N$.)
The case $d=0$ represents a particle in a
random potential. With $d=1$, $N=2$ and an attractive 
punctual disorder one models, for 
instance, a single vortex in a dirty superconductor. 
When $d=2$ and $N=1$ one describes the dynamics of
a directed interface in three dimensions.  The standard
model is
\begin{equation}
H=\int d^d x \left[ (\vec\nabla \vec\Phi(\vec x))^2 + 
V(\vec \Phi,\vec x) \right]
\label{manifold-hamil}
\end{equation}
where $\vec \Phi(\vec x)=(\Phi_1,\dots,\Phi_N)(\vec x)$
and $\vec x = (x_1,\dots,x_d)$
represents the transverse position of the point $\vec x$ on the 
manifold and $V(\vec\Phi,\vec x)$ is  a Gaussian random 
potential, with zero 
mean and correlations $[V(\vec\Phi,\vec x) V(\vec\Phi',\vec x')]=
-N {\cal V}\left[(\vec \Phi-\vec\Phi')^2/N\right] \, 
\delta^d(\vec x-\vec x')$. 

The study of these 
problems has been boosted by the advent of high-$T_c$ superconductivity.
Other physical systems that are modeled with similar Hamiltonians are
Wigner crystals, vortex lattices, charge density waves, etc.
in the presence of disorder. All these problems have an
underlying periodic structure that is modified by elastic 
distortions, topological defects and external quenched disorder. 
They can be set into motion with an external force and the velocity-force
characteristics has several interesting features that have been 
much studied. One observes a depinning transition at $T=0$, creep 
dynamics when the applied force is weak and $T>0$, hysteresis, 
etc. (The driven motion is achieved, for instance, 
by applying an external current in the case of the vortex systems
and the current-voltage characteristics is monitored.) 
The relaxational and driven 
dynamics of these systems show similarities but also 
marked differences with that of glasses and spin-glasses
that are possibly due to the existence of an ordered underlying structure.

\subsection{Aging}
\label{subsec:aging}

Aging means that older systems relax in a slower manner than 
younger ones~\cite{Struick}. One defines the age of a system as the time
spent in the phase under study. For instance, the age of a system that 
is suddenly quenched from high-$T$ to low-$T$ 
is simply $t_w$. The aging properties are studied 
by monitoring the time evolution of correlation and 
response functions. In the former experiments one lets the system 
evolve and compares its configuration at the waiting-time $t_w$ 
with the one reached at the subsequent time $\tau+t_w$. 
In the latter one perturbs the 
system at $t_w$ with, {\it e.g.} a dc or an ac  field,  and 
follows the evolution of the linear response to the perturbation.
In the glassy phase both correlations and responses depend on $t_w$ 
in an aging manner and, within the experimentally accessible time-window,
this trend does not show any tendency to stop. At temperatures that are close
but above $T_g$  
one observes ``interrupted aging'', that is, a dependence on the 
age of the system until it reaches the equilibration time ($t_w > t_{\sc eq}$)
where the dynamics crosses over to an equilibrium one. In equilibrium 
correlation and response measurements are related in a system independent
manner by the fluctuation-dissipation theorem (see Section~\ref{section:fdt}).
Out of equilibrium this general relation does not hold and, as we
shall explain in Sections~\ref{Modifs} and \ref{temp_intro}, 
important information can be 
extracted from its modifications.

Aging has an easy interpretation within coarsening systems. 
While the averaged domain size ${\cal R}(t_w)$
grows with $t_w$, its rate of increase $d_{t_w}{\cal R}(t_w)$ decreases, 
{\it e.g.} $d_{t_w}{\cal R}(t_w)\sim t_w^{1/z-1}$ in the example
explained in Section~\ref{subsec:domain}. The motion of interfaces 
slows down as time elapses. Comparing the configuration
at $t_w$ and at a later time $\tau+t_w$, one finds a clear separation 
of time-scales depending on the relative value of $\tau$ with respect to 
$t_w$. If $\tau\ll \tau_0(t_w) \equiv 1/d_{t_w} \ln {\cal R}(t_w)$, 
one has ${\cal R}(\tau+t_w) \sim {\cal R}(t_w) +  
d_{t_w}{\cal R}(t_w) \tau \sim {\cal R}(t_w)$, and the 
domain walls do not move. 
The overlap between the configurations at $t_w$ and $\tau+t_w$
is a sum of overlaps between domains of one and another type. 
(This holds when the ratio between the 
number of spin in the surface of the domains and in the 
bulk vanishes in the thermodynamic limit. Some non-standard
systems have fractal scaling of the interfaces and the volume of the 
domains might also have a fractal 
dimensionality~\cite{spin-glasses-fractal}.)
Due to 
thermal fluctuations within the domains, the correlation
decays as in equilibrium 
from $1$ at equal times to $m^2_{\sc eq}(T)$ when $\tau$ increases
while still satisfying the constraint $\tau\ll \tau_0(t_w)$. For $\tau$'s 
beyond
this limit, the correlations decay below  $m^2_{\sc eq}(T)$
since the interfaces move and
one compares configurations with very different domain structures
as shown, for instance, in the two snapshots in Fig.~\ref{aging}. 

In structural glasses, a pictorial explanation of aging is also possible
imagining that each particle sees a cage made of its 
neighbors. When $\tau$ is short compared to a 
characteristic time $\tau_0(t_w)$ each  particle rapidly rattles within its
cage and the decorrelation is only characterized by thermal fluctuations.
The correlations decay in a stationary manner from its value at equal times
to a value called $q_{\sc ea}$ that will be 
defined precisely in Section~\ref{section:solution} 
[in the domain growth example 
$q_{\sc ea}=m^2_{\sc eq}(T)$]. When $\tau$ increases, the motion of the 
particles destroys the original cages and one sees the structural relaxation.
The waiting-time dependence implies that the cages are stiffer when time
evolves. The motion of a tagged particle is depicted in the left panel of 
Fig.~\ref{different-glasses}~\cite{Weeks}. 
One sees how it first rattles within a cage
to later make a long displacement and start rattling within another 
cage.

\begin{figure}[h]
\centerline{
\psfig{file=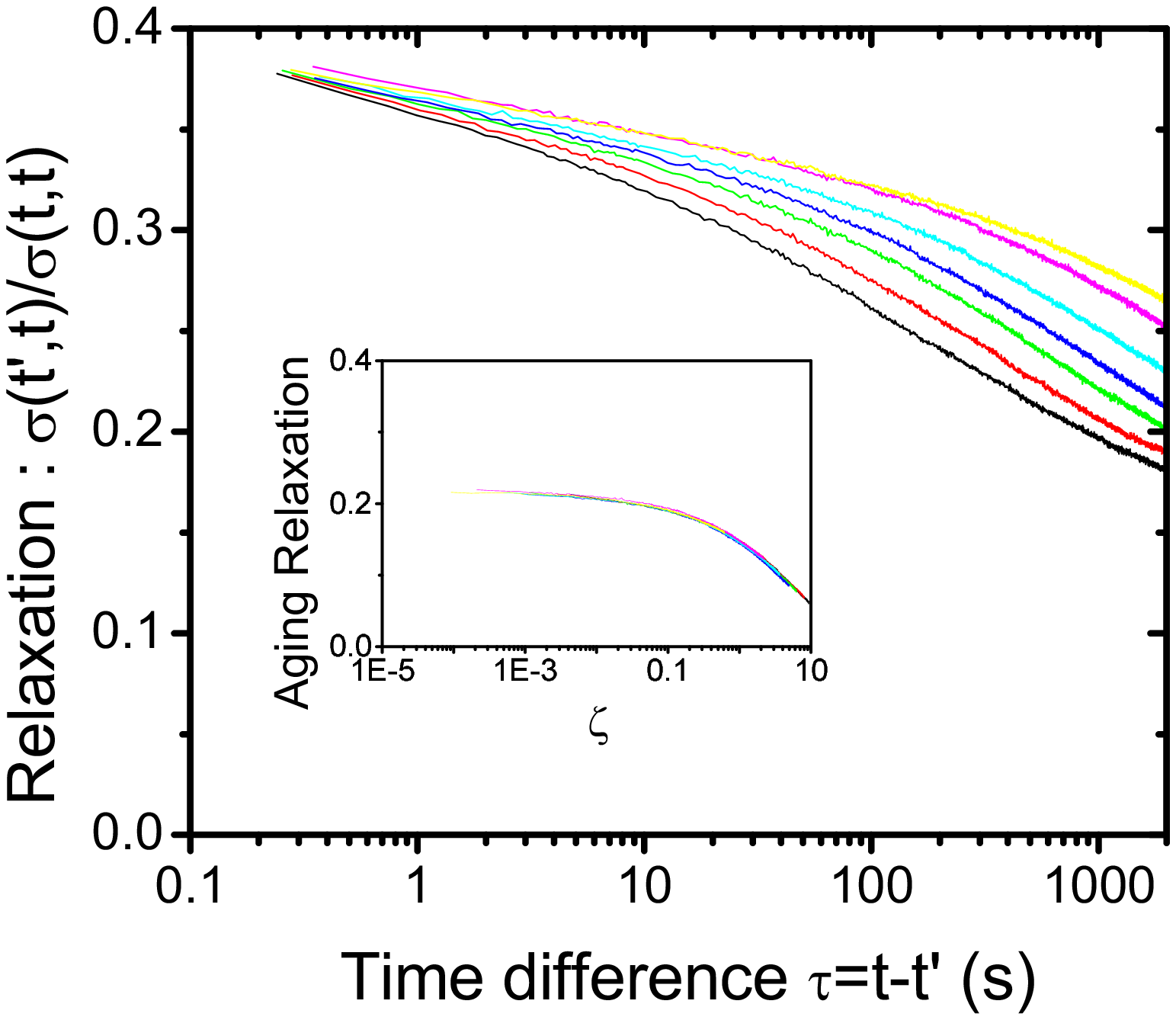,width=5cm}
\hspace{1cm}
\psfig{file=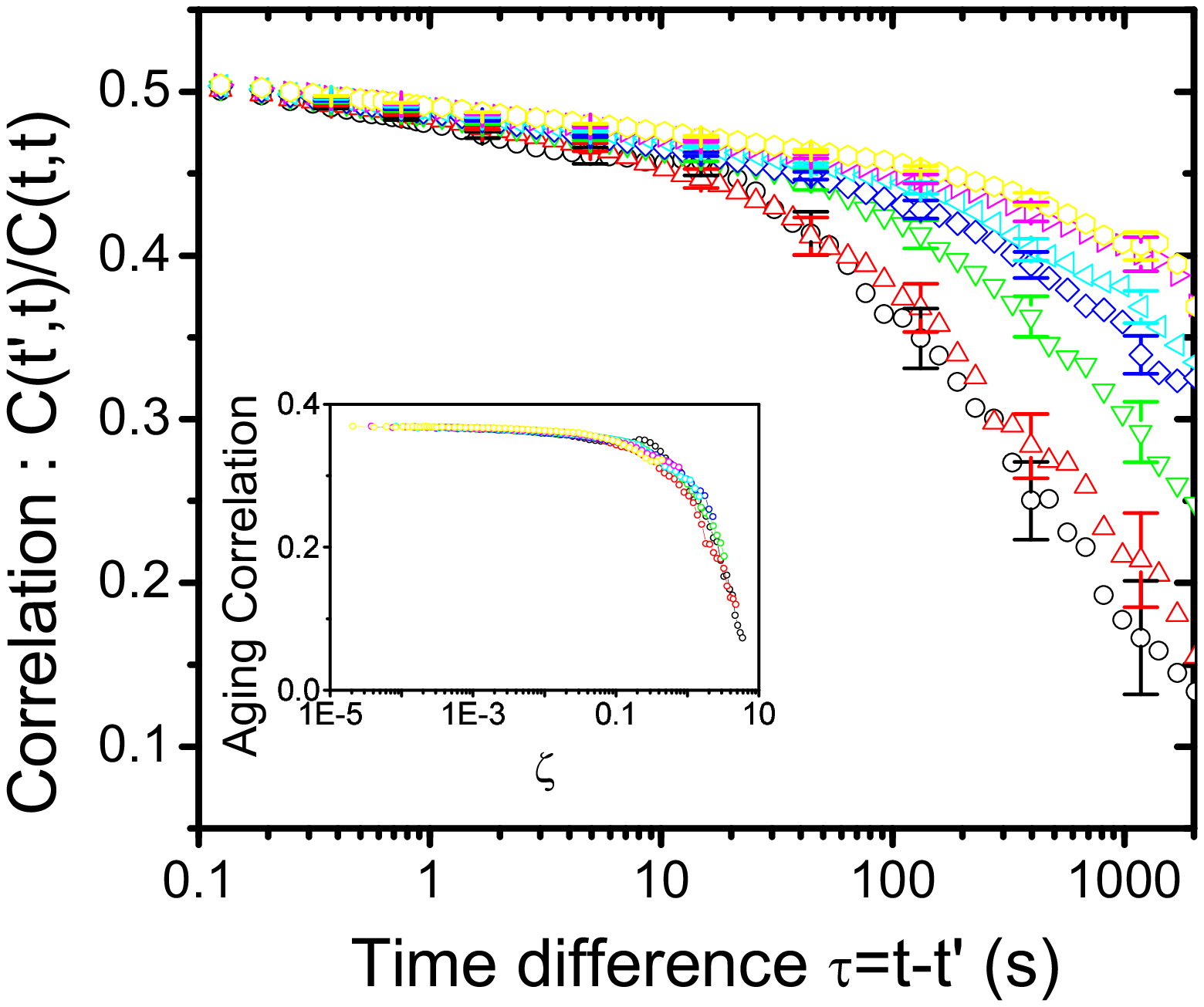,width=5cm}
}
\caption{Aging in the thiospinel insulator spin-glass. 
Decay of the thermoremanent magnetization (left) and
correlations between magnetic 
fluctuations (right). From left to right curves for increasing waiting-times.
Inset: scaling. (Curves taken from~\cite{Didier}, see~\cite{experiments}
and \cite{Didier} for details.)}
\label{aging-exp}
\end{figure}

In spin-glasses no consensus as to which is the origin of aging 
has been reached. Still, the qualitative behavior of correlations
and responses is rather close to the one in domain growth and structural 
glasses. In Fig.~\ref{aging-exp} we show the decay of the 
thermoremanent magnetization
(an integrated linear response) and the correlations between the 
fluctuations of the magnetization in a 
spin-glass~\cite{experiments,Didier}. 
 
\begin{figure}[h]
\centerline{
\psfig{file=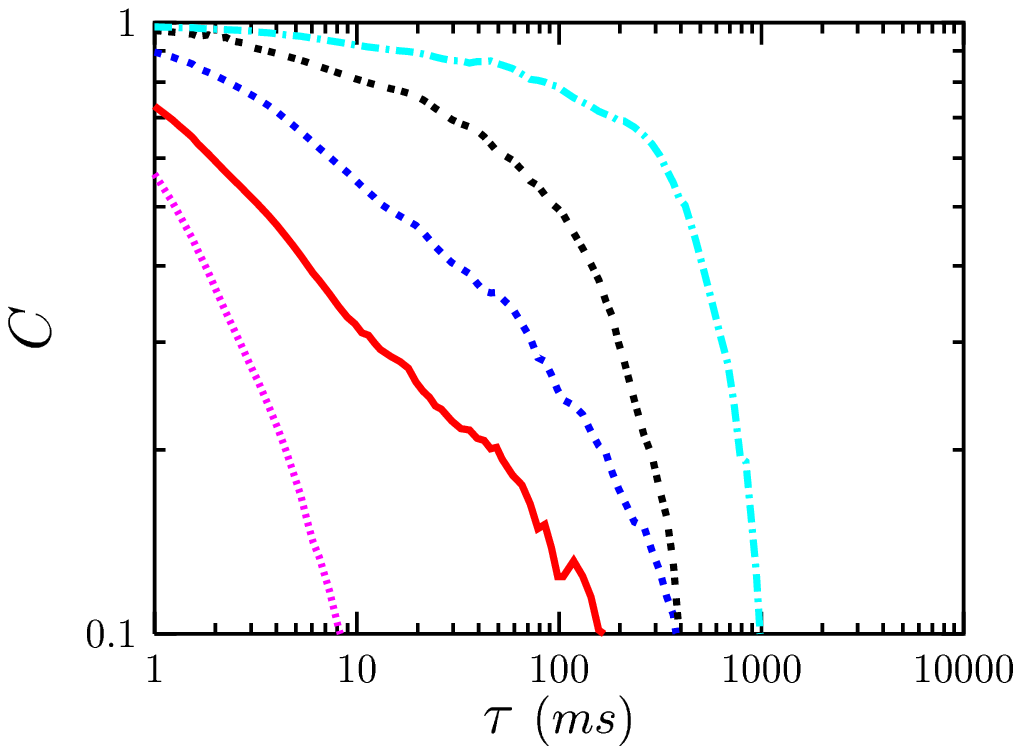,width=6.5cm}
\hspace{-1cm}
\psfig{file=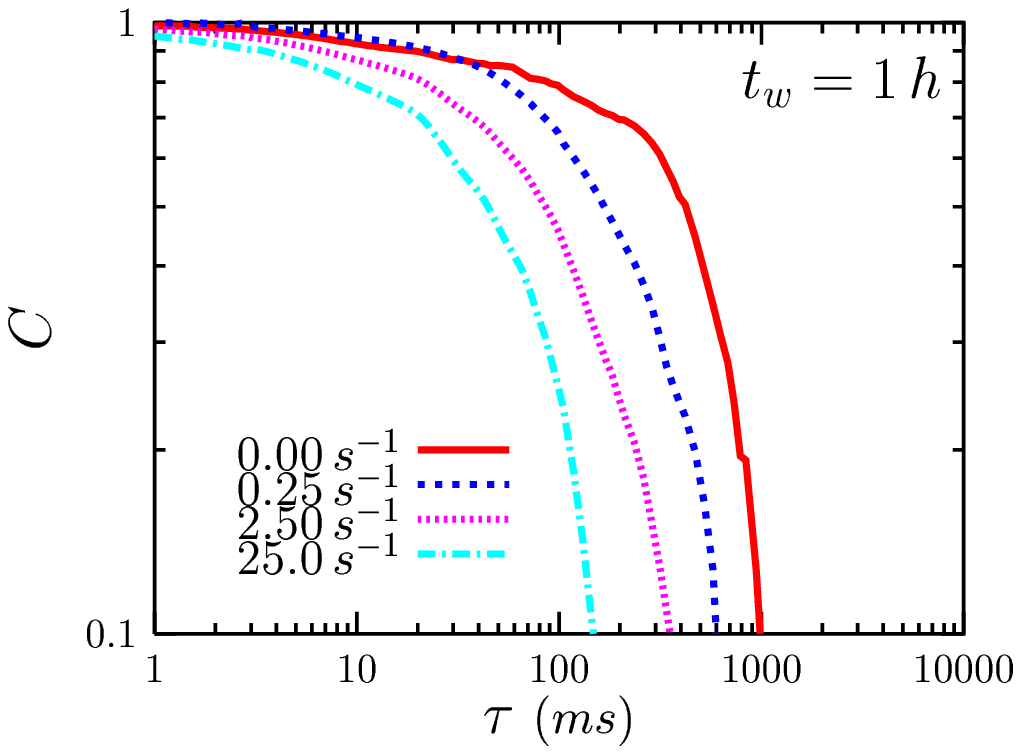,width=6.5cm}
}
\caption{
Aging and interrupted aging in laponite. Decay of the 
correlation function in a relaxing (left) and a sheared (right) sample.
In the aging case different curves correspond to increasing
waiting-times from left to right, $t_w=10,20,30,40,50$ min.
(note that the decay occurs for $\tau \ll t_w$
since  the waiting-times are rather short and the sample is still very far
from equilibrium; {\it cfr.} with the results in 
A. Knaebel {\it et al} in \cite{softglassy}). 
In the sheared case, the sample has been let wait for $t_w=1\,h$,
four shear rates, $\dot\gamma$, with values indicated in the key were applied
at this instant and the decay of the stationary correlation was 
recorded~\cite{Daniel-Bonn}.}
\label{agingandrheology}
\end{figure}

Shearing may have a very strong effect on an aging system. In some 
cases it introduces a 
characteristic time $t_{\sc sh}$ that yields the longest relaxation 
time. Thus, aging is interrupted 
for waiting-times that are longer than $t_{\sc sh}$
(see Fig.~\ref{agingandrheology}). 
This effect has been known for long 
experimentally~\cite{review-rheology} and 
it has been found and explored recently within the theoretical framework
that we review~\cite{jorge,rheology_theor}. 
Experiments in other soft glassy materials with aging and 
aging interrupted by shear can be found in ~\cite{softglassy}.
Some examples 
where the effect of shearing is not as spectacular are also known.
For instance, in a phase separating mixture sheared in one direction
the domains stop growing in the transverse direction
while they continue to grow longitudinally
(see~\cite{Ludo} and references therein). This is a subject of 
intensive research.

Granular matter is usually driven with periodic forces or tapping. 
These perturbations pump energy into the system that is evacuated 
via friction and introduce a time-scale $t_{\sc osc}$ that is simply
the period of the oscillation. How these forces influence 
the aging properties is much less known and it is the subject of 
current investigations. For the moment, most of the work in this 
direction has been numerical~\cite{aging-granular,Barrat-granular} and also 
analytical within the kind of models we shall solve~\cite{Becuig}
(see~\cite{Josserand} for some experimental studies).

The aging properties of relaxing manifolds in random potentials
have been found analytically~\cite{Cukule} and 
numerically~\cite{yoshino}. The experiment of Portier 
{\it et al}~\cite{Portier} shows aging features in a high-$T_c$ 
superconductor~\cite{Vabe}
that are also observed numerically~\cite{Exartier,Exartier-these}. However, 
other experimental studies using different protocols
have not exhibited these properties~\cite{Simon}. More 
experiments are needed to get a better 
understanding of the non-equilibrium relaxation of these systems.
An external electric current drives a vortex system 
in its transverse direction due to Lorentz forces. How the longitudinal 
and transverse aging properties of the system are modified in the 
moving phases is another problem that deserves further 
study~\cite{Koltonetal}. The comparison between the dynamics 
of the moving vortex lattice,
the tapped dynamics of granular matter and the sheared dynamics of 
dense liquids is also interesting and begins to be done.

\vspace{-1cm}

\subsection{Summary}

In this Section we introduced several examples of macroscopic 
systems that evolve with a slow non-equilibrium macroscopic dynamics. 
The microscopic dynamics governing the evolution of the constituents
of each system is very different. For 
systems undergoing domain growth and classical glasses, $T$ 
is the external control parameter that generates fluctuations. 
For glasses at very low $T$ thermal fluctuations are  
almost completely irrelevant while quantum fluctuations become 
important
and drive the dynamics.
For granular matter $T$ is again irrelevant and the systems'
relaxation is due to 
the external drive. In the elastic systems an underlying ordered structure 
is visible and may cause important differences in the macroscopic dynamics
with respect to other glassy systems where no quasi order has been identified.
An ubiquitous phenomenon in these slowly evolving systems out of equilibrium
is (sometimes interrupted) aging or the breakdown of stationarity.
Even if all these systems
seem to be totally different, a common formalism to study their macroscopic 
dynamics is now being used and a common picture starts to develop.

In the rest of these lectures we shall explain the main technical 
tools needed to study simplified models for these real systems. 
We shall describe the main features of the out-coming
scenario making contact with the phenomenology introduced in this 
Section.

\section{Theoretical approach}
\label{section:theoretical}
\setcounter{equation}{0}
\renewcommand{\theequation}{\thesection.\arabic{equation}}

Besides many phenomenological descriptions of the glass transition and 
glassy dynamics proposed long ago, recently, several theoretical
approaches have been developed. In the following we briefly
describe some of the main ones.

\vspace{0.25cm}
\noindent
{\it Dynamics in the phase space.} 
\vspace{0.25cm}

The instantaneous configuration of 
the full system is a point in phase space that moves as time goes on
(see the right panel in Fig.~\ref{aging}). In 
a whole family of models one assumes a free-energy landscape on 
phase space with wells and barriers and proposes 
that the point evolution is given by some dynamic prescription on this 
space. Choosing a convenient distribution and organization
of wells and barriers, aging effects
are captured by such models.
(An average over different systems or over different parts of the same
system is implicitly assumed in order to obtain smooth results for the 
observables.) In ``trap models'', for instance, each well has an 
associated trapping-time such that once the system falls in the trap
it has to wait this trapping-time to escape from it. A useful choice 
is a L\'evy  distribution of trapping-times that is not 
bounded (this is related to assuming that the depth of wells is not
bounded from below). 
The dynamics is such that whenever the point leaves a
trap a complete ``renewal'' takes place in the sense that in the next 
time step it chooses any trap from the ensemble 
with equal probability. For simple probability
reasons one can prove that at a time $t_w$ after the starting time 
the system will be trapped in a well with life-time $\sim t_w$. 
This model is due to J-P Bouchaud and it leads to 
aging~\cite{trap-model,Bertin,Sane,Fielding} (for other trap models
in the glassy context see~\cite{traps-old}).

\vspace{0.25cm}
\noindent
{\it Domain growh.} 
\vspace{0.25cm}

A different approach most commonly used to describe the 
dynamics of spin-glasses assumes that the evolution is driven by the 
growth of domains of two (or a few) competing phases. The 
non-equilibrium dynamics is then very similar to ferromagnetic coarsening
slowed down by the existence of disorder and/or competing interactions.
To obtain concrete predictions one either uses scaling arguments that lead
to the ``droplet model'' for spin-glasses~\cite{droplets} and its extension 
to other glassy systems or one solves exactly simple models in low 
dimensions, typically $d=1$~\cite{kinetic,Stinchcome}. 
A drawback of this approach is that it 
has been hard to identify the growing structures in glassy systems.

\vspace{0.25cm}
\noindent {\it Kinetic models.} 
\vspace{0.25cm}

A third approach used to model the dynamics of 
glasses consists in proposing purely kinetic models with no underlying
energy function. The dynamic rules are chosen so as to slow 
down the dynamics and lead to a dynamic arrest at some critical
temperature or density.  Interestingly enough some 
tunning the nature of the constraints these models may behave as 
strong, fragile or may even show a fragile-to-strong crossover and 
they can have non-trivial long-lasting nonequilibrium dynamics
without having an underlying static transition. 
Several such models exist, see~\cite{kinetic}
for a collection of  articles on this subject and \cite{kinetic2}.

\vspace{0.25cm}
\noindent {\it Mean-field models.}
\vspace{0.25cm}

Many of the recent  developments in the understanding of the similarities 
in the behavior of {\it a priori} so different systems  
as the ones described in Section~\ref{interestingproblems}
are based on the analysis of mean-field models.
These models are defined by Hamiltonians with 
long range interactions [{\it e.g.} sum over all pairs of spins in 
Eq.~(\ref{3DEA-hamil})] 
or in infinite dimensions [{\it e.g.} take $N\to\infty$ in 
Eq.~(\ref{manifold-hamil})]. Sometimes
it is convenient to include quenched disordered interactions
though this is not necessary. 

The static properties of these models are 
accessible by the  usual statistical mechanics analysis
complemented by the less standard replica trick
when disorder exists~\cite{Mepavi}. The organization of 
metastable states or, in other words, the full structure of the 
relevant Landau-type free-energy landscape, is also accessible 
with time-independent calculations called {\sc tap} approach in the 
context of spin-glass theory~\cite{tap}.

As for dynamics we first have to define how the microscopic 
variables evolve in time.
In  classical particle systems 
Newton equations determine their evolution.
In quantum systems Heisenberg equations do the same.
Since in realistic situations
the systems of interest are in contact with their environments 
we present the modelization
of the coupled systems in Section~\ref{systemsincontactwithreservoirs}. 
We show how the effect of the environment translates into noise in both 
cases. For classical models the elimination of the bath variables leads to 
Langevin equations. Non-potential forces
and vibrations in classical systems are easily included by adding 
terms to the Langevin equations.
Once the microscopic dynamics
is given the techniques described in the rest of the notes allow
us to solve these schematic models. 

These models capture many 
properties of real systems. Nevertheless, their mean-field character 
implies many drawbacks. For example, one finds  that 
they have sharp dynamic (with no thermodynamic anomaly) 
and static transitions at temperatures $T_d$ and 
$T_s$, ($T_d>T_s$).  The relationship between the dynamic 
solution and the organization of metastable states in the relevant 
free-energy landscape can be made precise and it allows us to
understand the existence of these two separate transitions. 
In real systems, however, there is no sharp dynamic transition 
since $T_g$ is actually a crossover, 
while $T_s$ might not exist (recall the discussion on $T_o$ in 
Section~\ref{subsec:glassy}). In spite of this and other defects, one 
of the interests in these models and their 
solutions is that they have a great power of prediction
of so far unknown effects and that they act as a source of 
inspiration for searching new features in the numerical and experimental
study of more realistic models and systems.
The rather accurate comparison to numerical 
simulations~\cite{suecos,numerics,Picco}, 
 and, to the extent of 
their availability, 
experiments~\cite{experiments,Gris,Ciliberto,Teff_exp,Didier}, 
supports the proposal that the  mechanism in these models is similar 
to the one 
responsible for the glass transition, and the glassy dynamics,
in real materials. Indeed, there is growing consensus in that 
they provide an exaggerated realization of the actual glassy phenomenon.
It is worth mentioning too that some of the features found in the 
mean-field models that we shall explain below have also been analysed
within the other approaches mentioned above. 

In the following we present the asymptotic 
analytic solution to this family of models.
Some of the ingredients missing
in their full analytic solution, that would 
render their description of real materials more accurate,
have been identified (analysis of the dynamics in time-scales 
that diverge with $N$, description of dynamic heterogeneities, etc.). 
For the moment, their complete analytical treatment has proven too difficult
but some recent articles report partial success and suggest interesting
ways to follow up. We shall come back to these issues in the Conclusions
and Perspectives.

\section{Systems in contact with environments}
\label{systemsincontactwithreservoirs}
\setcounter{equation}{0}
\renewcommand{\theequation}{\thesection.\arabic{equation}}

In the typical experimental protocols discussed in 
Section~\ref{introduction} one quenches 
the sample and subsequently follows its 
evolution in time.
Once arrived at the final point in parameter
space, if no external forces are applied,
the system relaxes and its energy density 
decreases towards an asymptotic value. Hence, the system is 
not isolated, but in contact with an environment that 
acts as a source of dissipation.  The system plus environment is 
``closed'' while the system alone is ``open''.  The first question 
to answer is how to model the coupled classical and quantum system.

\subsection{Modeling the coupled system}

The nature of the 
environment, {\it e.g.} whether it can be modeled by a  classical 
or a quantum ensemble, depends on the problem under study. 
The choice of the coupling between system and environment is determined
by the symmetry properties of the system and by physical intuition.
Weiss' textbook~\cite{Weiss} has a very complete description of this
problem. We here explain the main ingredients of the modellization.
The generic problem we want to study is
\begin{equation}
H_{\sc tot} = H_{\sc syst} + H_{\sc env} + H_{\sc int}+H_{\sc counter}
\; .
\label{Htot}
\end{equation}
Until otherwise stated, 
we focus on a single particle coupled to an 
environment. 
$H_{\sc syst}$ is the  Hamiltonian of the isolated 
particle, $H_{\sc syst}= p^2/(2M) + V(q)$, 
with $p$ and $q$ its momentum and position. 
$H_{\sc env}$ is the Hamiltonian of a thermal bath that, again for simplicity,
we take to be an ensemble of $N$ independent harmonic oscillators
\begin{equation} 
H_{\sc env}= \sum_{a=1}^{N_b} \frac{\pi_a^2}{2m_a} + 
\frac{m_a \omega_a^2}{2} x_a^2 
\; .
\label{osc-bath}
\end{equation}
This is indeed a very usual choice since, for example, 
it may represent phonons.
$H_{\sc int}$ is the coupling between system and environment.
We restrict the following discussion to an interaction that is 
 linear in the oscillator and  
particle coordinates, 
$H_{\sc int}=q\sum_{a=1}^{N_b} c_a x_a$, with $c_a$ the coupling constants.
The calculations can be easily generalized to an interaction with a  
more complicated dependence on the system's coordinate, $F(q)$,  
that may be dictated by the symmetries of the system. We discuss 
the last term, $H_{\sc counter}$, below.

\subsubsection{Statics}
\label{staticproperties}

\vspace{0.25cm}
\noindent {\it Classical problems} 
\vspace{0.25cm}

Let us first show how the static properties of a 
classical system are modified by its interaction with a classical 
thermal bath. If the coupled system 
is in equilibrium,
it is described by a partition function given by a sum over the 
combined phase space of the system and environment.
Having chosen a bath of harmonic oscillators, the integration over the 
bath variables can be readily performed; this calculation 
yields the {\it reduced}
partition function that is written as an integration over the phase 
space of the system only. One can easily prove that
the mass of the system gets (negatively) 
renormalized due to the coupling to the 
environment~\cite{comment-renorm}. 
Therefore one introduces the counter-term 
\begin{equation}
H_{\sc counter} =
\frac{1}{2}\sum_{a=1}^{N_b} 
\frac{c_a^2}{m_a \omega_a^2} \; q^2 
\end{equation}
in such a way to eliminate 
the mass renormalization and to recover the partition function of 
the isolated system. 

\vspace{0.25cm}
\noindent {\it Quantum problems} 
\vspace{0.25cm}

If one includes quantum fluctuations to describe 
the system and environment, the situation is slightly more 
complex. The relevant quantity to study is 
the density matrix of the full system that, for instance, can be represented 
as a path integral on imaginary time~\cite{Zinn}. 
The contribution of the environment 
to the effective action is quadratic and its variables can be 
integrated away to yield a reduced density matrix. As opposed to the 
classical case, the interaction with the reservoir not only 
induces a (negative) mass renormalization but it also generates a 
{\it retarded quadratic interaction}
\begin{equation}
\int_0^{\beta\hbar} d\tau \int_0^\tau d\tau' \, x(\tau) \,  K(\tau-\tau')
\, x(\tau')
\label{retarded}
\end{equation} 
controlled by the kernel
\begin{equation}
K(\tau) = 
\frac{2}{\pi\hbar\beta} 
\sum_{n=-\infty}^\infty 
\int_0^\infty d\omega \; \frac{I(\omega)}{\omega} \;  
\frac{\nu_n^2}{\nu_n^2+\omega^2} \,
\exp(i\nu_n\tau)
\; ,
\end{equation}
with $\nu_n$ the Matsubara frequencies, 
$\nu_n=2\pi n /\hbar\beta$, $n$ an integer in $(-\infty,\infty)$
and $I(\omega)$ the spectral density of the bath,
\begin{equation}
I(\omega) =\frac{\pi}{2} \sum_{a=1}^{N_b} \frac{c_a^2}{m_a\omega_a}
\delta(\omega-\omega_a) 
\; ,
\label{spectral}
\end{equation}
that is a smooth function of $\omega$ usually taken to be 
$
I(\omega) = \gamma \omega \left(\omega/\omega_s\right)^{s-1} 
\exp(-\omega/\Lambda)
\; ,
$
with $\gamma$ the ``friction coefficient'', $\omega_s$ a constant,
$\Lambda$ a high frequency cut-off and $s$ a parameter that characterizes
different baths: $s=1$ is Ohmic (and leads to the usual white noise when
$\Lambda\to\infty$), $s>1$ is super-Ohmic and $s<1$ is sub-Ohmic.
As in the classical case, one includes a  
counter-term to cancel the mass renormalization 
but the retarded interaction (\ref{retarded}) cannot be eliminated.

\subsubsection{Dynamics}

The distinction between the effect of a 
reservoir on the statistic properties of a   
classical and a quantum system is absent from a full dynamic
treatment where  
the coupling to the environment always 
leads to a {\it retarded} interaction. 
In classical problems one generally argues that the retarded
interaction can be simply replaced by a local one, {\it i.e.}
one uses white noises, if long enough time-scales are explored. 
In quantum problems the same simplification 
is not justified in general.

\vspace{0.25cm}
\noindent
{\it Classical problems}
\vspace{0.25cm}

The dissipative dynamics of a classical system in contact with an 
environment is usually described by a phenomenological 
Langevin equation. If the system
is simply given by a particle of mass $M$, whose position is denoted by  
$q$, this equation reads 
\begin{eqnarray}
M \ddot q(t) + \int_0^t dt' \; \gamma(t-t') \dot q(t') &=& -V'(q(t)) + \xi(t)
\label{lang1}
\\
\langle \xi(t)\xi(t') \rangle &=& k_B T \gamma(t-t')
\label{lang2}
\end{eqnarray}
where $\gamma(t-t')$ is a retarded friction and $\xi(t)$ is a time-dependent 
Gaussian random force with zero mean and 
correlation given by Eq.~(\ref{lang2}). We adopt angular brackets 
to denote averages over the noise. 

The Langevin equation was first introduced in the context of Brownian
motion and later used in a variety of problems with dissipation.
It can be derived from the coupled system defined in 
Eq.~(\ref{Htot})~\cite{Zwanzig,Weiss} (see 
Appendix~\ref{sec:langevin_general}).
Indeed, if one assumes that the initial coordinates and 
momenta of the oscillators have a canonical distribution at an inverse
temperature $\beta$ [shifted by the coupling to the particle $x_a(0)\to
x_a(0)-c_a/(m_a \omega_a^2) q(0)$~], 
Newton's equation for the evolution of the particle becomes 
Eq.~(\ref{lang1}). The random nature of the force $\xi$ is due to the 
randomicity in the initial values of the position and momenta
of the oscillators.  
The use of an  equilibrium measure for the 
distribution of the oscillators implies the invariance under time 
translations of the friction kernel, $\gamma(t-t')$, and its 
relation to the noise-noise correlation 
in Eq.~(\ref{lang2}). The latter  
is a fluctuation-dissipation 
theorem that holds for the bath variables (see 
Appendix~\ref{sec:langevin_general} and 
Section~\ref{section:fdt}). 
Different forms of $\gamma(t-t')$ can be generated 
by different choices of the
ensemble of oscillators. Typically, the 
decay of $\gamma(t-t')$ occurs in a finite relaxation time, 
$\tau_\xi$.
When the minimum 
observation time is of the order of or shorter than 
$\tau_\xi$ one has a ``colored
noise''. Instead, when it is much longer than $\tau_\xi$, any 
time difference satisfies $t-t'\gg \tau_\xi$ and the kernel 
can be approximated with a delta function 
$\gamma(t-t') \sim 2 \gamma \delta(t-t')$, which corresponds to  
a white noise. For the classical problems we shall deal with 
the white-noise approximation is adequate.

Once the Langevin equation has been established on a firmer ground, 
it can be used as a starting point to study the dynamics of 
more complicated classical systems in contact with classical 
environments. The description of the dynamics of a 
macroscopic system with dissipation
is then given by $N$ coupled Langevin equations
with $N$ the number of dynamic degrees
of freedom. (In this case one usually 
couples an independent set of oscillators to each microscopic variable of
the system, other choices lead to more complicated equations.)
Time-dependent, $f_i(t)$,  and constant non-potential forces, $f_i^{\sc np}$, 
as the ones
applied to granular matter and in rheological measurements are simply 
included as part of the deterministic force. In the white noise 
limit
\begin{eqnarray}
M \ddot q_i(t) + \gamma \dot q_i(t) 
&=& -\frac{\delta V(\vec q)}{\delta q_i(t)} + f_i(t) + f_i^{\sc np} + \xi_i(t)
\; ,
\label{lang3}
\\
\langle \xi_i(t)\xi_j(t') \rangle &=& 2 \gamma k_B T \, \delta_{ij} 
\delta(t-t')
\; .
\label{lang4}
\end{eqnarray} 

A continuous Langevin equation for classical spins is usually written 
replacing the hard Ising constraint $s_i=\pm 1$ by a soft one
implemented with a potential term of the form $V(s_i) = u (s_i^2-1)^2$
with $u$ a coupling strength that one eventually takes to infinity.
The soft spins are continuous unbounded variables $s_i\in (-\infty,\infty)$
but the potential energy favors the configurations with $s_i$ close to 
$\pm 1$. Even simpler models are constructed with spherical spins,
that are also continuous variables globally constrained to satisfy 
$\sum_{i=1}^N s_i^2=N$.

\vspace{0.25cm}
\noindent
{\it Quantum problems}
\vspace{0.25cm}

Even if several attempts to write down quantum versions of the 
Langevin equation appeared in the literature, these methods
remain very much model dependent and difficult to generalize~\cite{Weiss}. 
A more convenient way to analyze the dynamics of 
a coupled system and environment with
quantum fluctuations is the  functional 
Schwinger-Keldysh formalism. We postpone the discussion of the 
effect of the quantum reservoir on the quantum system to 
Section~\ref{generating-functionals} where we introduce this formalism.
In short, the effect of the environment is to introduce retarded terms
in the dynamic action that are similar to the ones in Eq.~(\ref{retarded}) 
but in real time. The white approximation is not acceptable
and one is forced to keep the non-local in time kernels.

\section{Observables and averages}
\label{section:observables}
\setcounter{equation}{0}
\renewcommand{\theequation}{\thesection.\arabic{equation}}

As usual in statistical and quantum mechanics 
meaningful quantities are averaged observables. For an 
equilibrated system, due to ergodicity, one can either take an ensemble 
average or an average over a sufficiently long time-window. Out of equilibrium 
these do not coincide in general. In this 
Section we define averaging procedures for classical and quantum
problems out of equilibrium
and we set the notation used in the rest of the notes. 

\subsection{Classical systems} 
\label{average-thermal}

The interaction with the environment induces fluctuations and 
the Langevin equation is solved in a probabilistic sense, 
\begin{equation}
q^{sol}_{\xi_k}(t) = {\cal F}[(\xi_k),q_0,t] 
\; .
\label{sol_stoch}
\end{equation}
The index $k$ labels different realizations of the thermal history, 
{\it i.e.} different realizations of the noise at each instant. 
$q_0$ is the initial condition $q_0=q(0)$ and $(\xi_k)$ encodes all
noise values in the interval $[0,t]$.  
We discretize time, $t_a= a \delta$ with $\delta$ the time spacing
and $a=0,1.\dots$ The 
total time is $t=\delta {\cal T}$. We shall be interested in the 
limit ${\cal T}\to\infty$ and $\delta\to 0$ with $t$ fixed.
Equation (\ref{sol_stoch}) means that there is a different solution
for each noise history.

 Any {\it one-time} functional of $q$, 
$A[q](t)$, must be averaged over all histories of the thermal 
noise to obtain a deterministic result
\begin{equation}
\langle  A[q](t) \rangle =
\lim_{{\cal N}\to \infty} 
\sum_{k=1}^{{\cal N}} A[q^{sol}_{\xi_k}](t) P(\xi_k) 
= 
\int {\cal D}\xi \, P(\xi) \, A[q^{sol}_{\xi}](t)
\; .
\label{calG}
\end{equation}
${\cal N}$ is the number of noise realizations.
$P(\xi_k)$ is the probability distribution of the $k$-th thermal history.
For a Gaussian noise
\begin{equation}
P(\xi_k) \propto \exp\left[-\frac{1}{2k_BT} \sum_{ab} \xi_k(t_a) 
\gamma^{-1}(t_a-t_b) \xi_k(t_b) \right]
\; .
\label{gaussian-noise}
\end{equation}
The measure of the functional integral is just 
${\cal D}\xi \equiv \prod_{a} d\xi(t_a)$.

{\it Two-time functions} characterize the out of 
equilibrium dynamics in a more detailed way and they are 
defined as 
\begin{equation}
C_{AB}(t,t') \equiv 
\langle A[q](t) B[q](t') \rangle
= \int {\cal D}\xi \, P(\xi) \, A[q^{sol}_{\xi}](t) B[q^{sol}_{\xi}](t')
\; .
\end{equation}
The observable $B[q]$ is measured
at time $t'$, the observable $A[q]$ is measured at time $t$
for each noise realization and the average is taken 
afterwards. 

%\vspace{-1cm}
\begin{figure}[ht]
\centerline{
\psfig{file=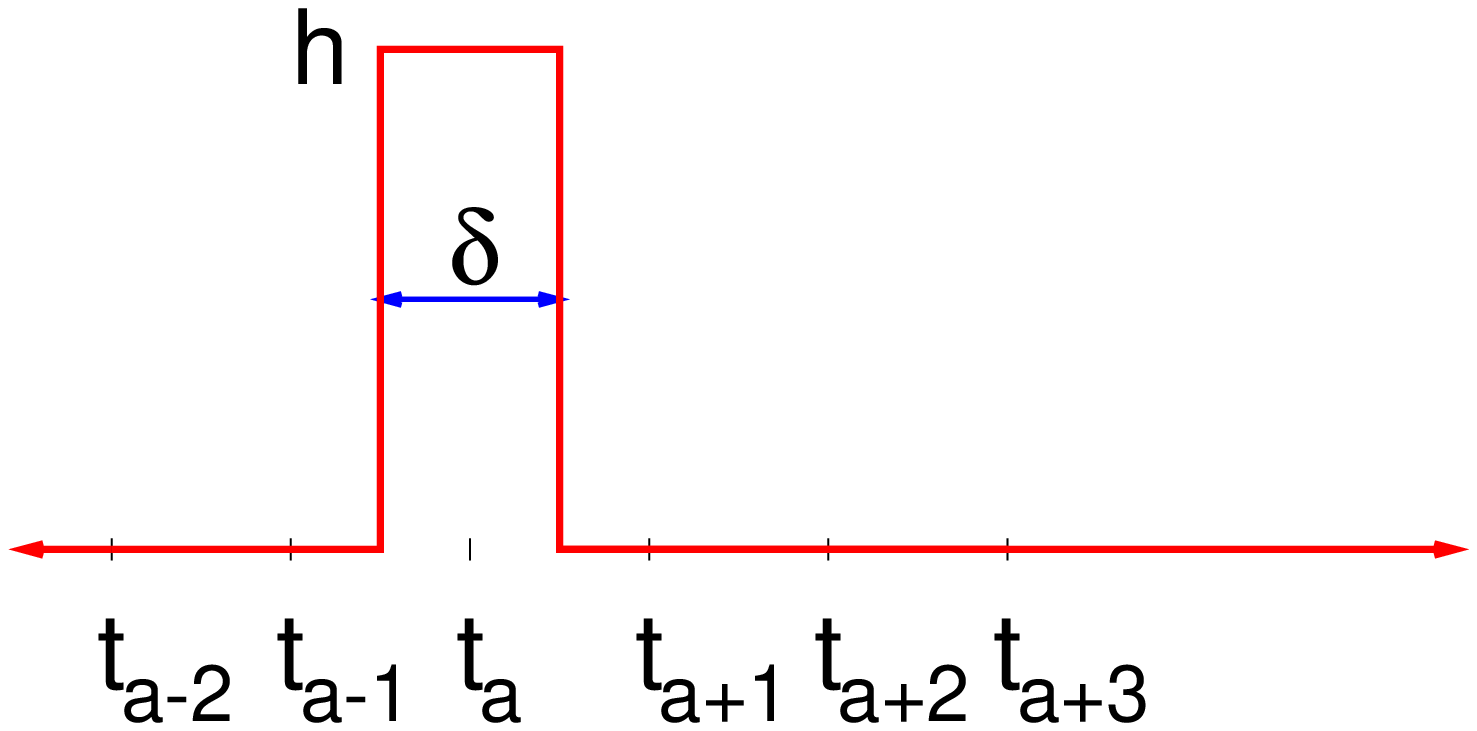,width=4.5cm}
\hspace{2.5cm}
\psfig{file=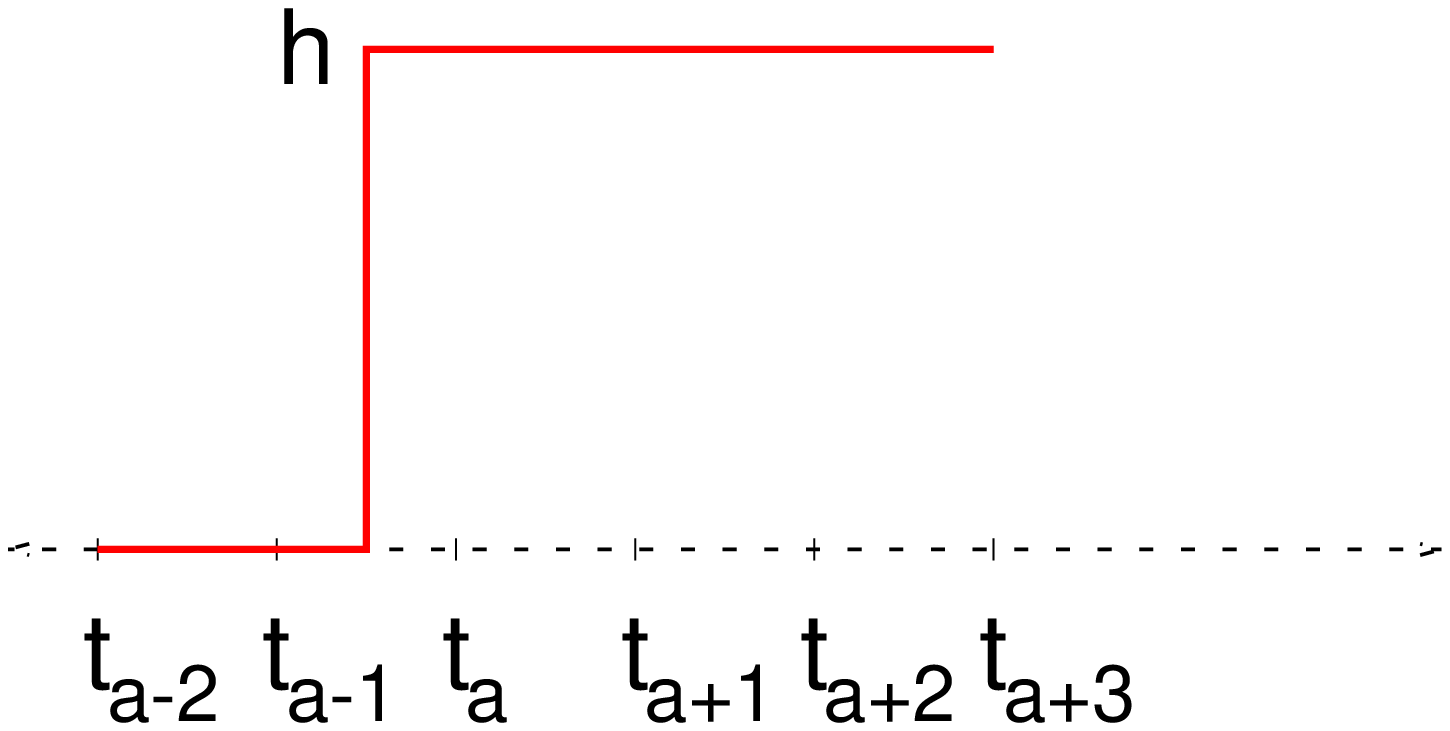,width=4.5cm}
}
\vspace{0.25cm}
\caption{Left: an instantaneous perturbation applied at $t_a$.
Right: a step perturbation applied at $t_a$ and held constant for
all subsequent times.}
\label{pert-resp}
\end{figure}

The instantaneous linear response is also a two-time function.
Imagine that $q$ represents the position of a Brownian 
particle that one {\it kicks}
with a 
weak perturbing force at time $t'=t_a$ (see Fig.~\ref{pert-resp}). 
The subsequent position
of the particle is modified by the perturbation. The linear
response is given  by the comparison of the
perturbed dynamics with the unperturbed one, in which no force has been 
applied, up to linear order in the perturbation:
\begin{equation}
R_{AB}(t,t') \equiv \left. \frac{\delta \langle A[q]\rangle(t)}
{\delta h_B(t')} \right|_{h_B=0}
\; . 
\label{linear-response}
\end{equation}
The subindex $B$ indicates that the perturbation applied at $t'$ 
is conjugated to 
the observable $B[q]$ when
the Hamiltonian is modified as
%\begin{equation}
$H \to H - h_B B[q]$. 
%\; . 
%\end{equation}
The subindex $A$ indicates that we examine how the observable $A[q]$ 
at time $t$ 
reacts to the perturbation. At the end of the calculation 
we set $h_B=0$ to extract the linear response. Keeping $h_B\neq 0$ 
yields information about the nonlinear terms in the response function.
For {\it causal} systems the response function vanishes if $t'>t$.

In future Sections we shall be interested in the integrated linear response
rather than the instantaneous one. This quantity 
represents the linear response of the system to a {\it step}-like 
perturbation of duration $t-t'$ that starts at $t'$, as represented on 
the right panel of Fig.~\ref{pert-resp}:
\begin{equation}
\chi_{AB}(t,t') \equiv \int_{t'}^t dt'' \; R_{AB}(t,t'')
\; .
\label{int-linear-response}
\end{equation}

Rather often results are 
presented in the frequency domain. One defines the Fourier 
transform and its inverse
\begin{eqnarray}
\tilde A(\omega) = \int_{-\infty}^\infty  dt \, 
\exp\left(-i\omega t\right) \, A(t)
\; ,
\label{fourier}
\;\;\;\;\;\;\;\;
%\\
A(t) = \int_{-\infty}^\infty  \frac{d\omega}{2\pi} \, 
\exp\left(i\omega t\right) \, \tilde A(\omega)
\; .
\label{inv-fourier}
\end{eqnarray}
For a stationary process,  
the linear susceptibility, $\tilde\chi(\omega)$,
is simply given by
the Fourier transform of the linear response (\ref{linear-response}). 
However
$R_{AB}(t,t')$ is not necessarily stationary out of equilibrium.
Hence, we define two generalized linear susceptibilities,
\begin{eqnarray}
\tilde \chi^{(1)}_{AB}(\omega,t') &\equiv& \int_0^\infty d\tau \; 
\exp\left(-i\omega\tau \right) R_{AB}(t'+\tau,t')
\; ,
\label{chi1}
\\
\tilde \chi^{(2)}_{AB}(\omega,t) &\equiv& \int_0^\infty d\tau \; 
\exp\left(-i\omega\tau \right) R_{AB}(t,t-\tau)
\; ,
\label{chi2}
\end{eqnarray}
that reduce to the  well-known expression for $\tilde\chi_{AB}(\omega)$
in a stationary system. Note that in the first line we kept the
shorter time ($t'$) fixed while in the second line we kept the 
longer time ($t$) fixed.
These expressions have a real and an imaginary part that yield
the in-phase ($\tilde\chi'$) 
and the out-of-phase ($\tilde\chi''$) susceptibilities, respectively. 
The integrations run over positive values of 
$\tau$ only due to causality. 

%{\it Exercise: Compute the average position, the two-time correlation 
%and the linear response for the biased one dimensional random walk in $d=1$
%in the limit in which inertia, $M \ddot q(t)$, can be neglected and for a
%white noise. Relate these two-time functions. Repeat this calculation for 
%a particle in a harmonic potential $V(q)=kq^2/2$. Study the long 
%times limit for $k$ and $f$ fixed, $k\to 0 $ and $f$ fixed, etc. Find when
%the correlation and response become stationary and when there is a simple
%relation between them ({\sc fdt}) as discussed in Section~\ref{section:fdt}.} 

Up to this point we have discussed the simple case in which  observables
only depend on time. More generally, one is interested in extending
the above definitions to field theories. In the context of 
liquids, glasses, etc.  the generic  observables 
$A$ and $B$ depend on the positions and momenta of the particles. 
A key quantity is the density
%\begin{equation}
$\rho(\vec r, t) = 
\sum_{\alpha=1}^N \delta(\vec r -\vec r_\alpha)
$, 
%\end{equation}  
where $\vec r$ is a $d$ dimensional vector in real space and 
$\vec r_\alpha$ are the positions of the $N$ particles in the 
system.  From the density-density correlator 
%\begin{equation}
$
N^{-1} G(\vec r, t; \vec r', t') \equiv \langle 
\rho(\vec r' + \vec r, t) \rho(\vec r',t') \rangle
$ 
%\label{density-density}
%\end{equation}
one defines the more useful van Hove correlator
%\begin{equation}
$
G_{\sc vh}(\vec r; t, t') \equiv 
\int d^dr' \; G(\vec r, t; \vec r', t')
$,
%\; ,
%\end{equation}
that measures the probability of finding a particle $i$ within a 
volume $d^d r$ around the point $\vec r$ at time $t$ given that there
was a particle $j$ at the origin at time $t'$. The normalization 
factor fixes the 
number of particles,  $\int d^d r G_{\sc vh}(\vec r,t) =N$. 
The density-density and van Hove correlators can be naturally separated
in two contributions, a {\it self} and a {\it distinct} part. In the 
former, one adds over equal particles only while in the latter one 
adds over distinct particles. 
The {\it two-time intermediate scattering function} is constructed 
with the components of the real-space Fourier transform 
of the density at different times:
%\begin{equation}
$
N^{-1} F(\vec k;t,t') 
\equiv \langle \rho(\vec k,t)  \rho^*(\vec k,t') \rangle
= \langle \rho(\vec k,t)  \rho(-\vec k,t') \rangle
$.
%\; .
%\end{equation}
When times are equal one recovers the {\it structure factor}, 
$S(\vec k;t)=F(\vec k;t,t)$ that at long times approaches 
a limit $S(\vec k; t) \to S(\vec k)$ even for glassy systems. 

The two-time intermediate scattering function is measurable via 
neutron scattering. Indeed, one can easily extend the proof described 
{\it e.g.} in \cite{Hansen} to the non-equilibrium case, to show that 
the cross-section per nucleus is related to the intermediate 
scattering function. This relation suggests to call the self
and distinct correlators, {\it incoherent} and {\it coherent} ones,
respectively. Many times, one defines the correlators of the 
local density fluctuations, $\delta\rho(\vec r,t) \equiv 
\rho(\vec r, t)-\langle \rho(\vec r,t)\rangle$. The modification of the 
correlations defined above follow straightforwardly.
A detailed description of the properties of these correlations 
in an equilibrated liquid can be found in \cite{Hansen}. 

Up to now  we have only discussed one-point 
and two-point functions. In general problems, higher order
functions are not trivially related to the previous ones and 
bear richer information. These are
{\it four-point functions}, 
$\langle A(t) B(t') C(t'') D(t'') \rangle$, 
or any other form of a more general type. In most
of the solvable models we shall discuss below, and in most of the 
approximations used to analyze realistic models, higher-order 
functions do not appear. The reasons for their disappearance
are manifold. In simplified models one can 
simply prove that higher order functions are exactly given 
in terms of one and two-point functions. In more realistic cases, 
higher order functions are 
approximated with expressions that depend on one and 
two-point functions only. This is done, for instance, in 
Gaussian approximations and mode-coupling theory.
However, a complete solution to a finite dimensional model should 
be able to predict the behavior of such higher order correlations. 

\subsection{Quantum problems}
\label{subsec:quantum-observables}

A quantum mechanical operator, $\hat A$,  in 
the Heisenberg representation evolves according to 
\begin{equation}
\hat A(t) = \exp\left(\frac{i H t}{\hbar} \right) \hat A(0) 
\exp\left(-\frac{i H t}{\hbar}\right)
\; .
\label{Heisenberg}
\end{equation}
Ensemble averages are defined  as
%\begin{equation}
$
\langle \hat A(t) \rangle \equiv 
\mbox{Tr} \left( \hat A(t) \hat \rho(0) \right)/
\mbox{Tr} \, \hat \rho(0)  
$,
%\end{equation}
where $\hat \rho(0)$ is the initial density operator
and the trace is defined in the usual way, 
$\mbox{Tr} [ \, \bullet \, ] \equiv \sum_\alpha \langle \psi_\alpha | 
\bullet | 
\psi_\alpha \rangle$, 
with $\{\psi_\alpha\}$ an orthonormal basis in Fock space.
The normalization factor is the partition function 
$Z\equiv {\mbox{Tr} \, \hat \rho(0)} $.
Two-time dependent correlation functions are introduced as
%\begin{equation}
$
\langle \hat A(t) \hat B(t') \rangle \equiv 
Z^{-1} \mbox{Tr} \left( \hat A(t)  B(t') \hat \rho(0) \right)
$.
%\end{equation}
Clearly $\langle \hat A(t) \hat B(t') \rangle \neq \langle 
\hat B(t') \hat A(t) \rangle$
and one can define symmetrized and anti - symmetrized correlations:
\begin{eqnarray}
C_{\{A,B\}}(t,t') &=& \frac12 \, \langle \hat A(t) \hat B(t') + 
\hat B(t') \hat A(t) \rangle
\; ,
\label{symm}\\
C_{[A,B]}(t,t') &=&  \frac12 \,  \langle \hat A(t) \hat B(t') - 
\hat B(t') \hat A(t) \rangle
\; ,
\label{antisymm}
\end{eqnarray}
respectively.
The linear response and the integrated linear response 
are defined just as in the classical case, see
Eqs.~(\ref{linear-response}) and~(\ref{int-linear-response}).
%\begin{equation}
%R_{AB}(t,t') \equiv \left. \frac{\delta \hat A(t)}{\delta h_B(t')} 
%\right|_{h_B=0}
%\; . 
%\label{resp-quantum}
%\end{equation}
%The subindex $B$ indicates that the perturbation is conjugated to 
%the observable $B$ and it has been applied at time $t'$. This means 
%that at time $t'$ the Hamiltonian is modified according to 
%\begin{equation}
%\hat H \to \hat H - h_B \hat B 
%\; . 
%\end{equation}
In linear response theory, in and out of equilibrium, 
$R_{AB}(t,t')$ and the anti-symmetrized correlation $C_{[A,B]}(t,t')$
are related by the Kubo formula~\cite{Parisi-book}
(see Appendix~\ref{app:Kubo-formula})
\begin{equation}
R_{AB}(t,t') = \frac{i}{\hbar} \; \theta(t-t') \, 
\langle \, [ \hat A(t), \hat B(t') ] \, \rangle = 
\frac{2 i}{\hbar} \; \theta(t-t') \, 
C_{[A,B]}(t,t')
\label{FDTasym}
\; .
\end{equation}

\subsection{Average over disorder}

Time independent {\it quenched} random forces 
and interactions 
exist in some of the models and systems that we study.
We shall be mostly interested in quantities 
averaged over the distribution of disorder that 
we  denote with square brackets
$[\langle A\rangle]$. 

Averaging over disorder is a 
delicate matter when one wishes to compute {\it static} 
properties. For instance, one 
has to resort to the sometimes contested replica 
trick~\cite{Parisi}. We shall see in Section~\ref{generating-functionals}
that in a full dynamic treatment with no special initial conditions 
there is no need to introduce 
replicas and the formalism is totally free from ambiguities.

%{\it Exercise: Compute the same quantities already calculated
%for the biased random walk when the time-independent force $f$ 
%applied on the Brownian particle is random, with $[f]=0$ and $[f^2]=\Delta$.}

\section{Time dependent probability distributions}
\label{sec:prob}
\setcounter{equation}{0}
\renewcommand{\theequation}{\thesection.\arabic{equation}}

In Section~\ref{systemsincontactwithreservoirs} we 
derived a Langevin equation with white noise 
as the microscopic dynamic equation controlling the evolution 
of the classical problems we shall study.
In this Section  we prove some properties of the equilibrium 
dynamics of classical models with dynamics given by these equations.
One can simply modify this proof for a generic Markov process, 
a classical problem with colored noise and a quantum model.
%In Appendix~\ref{app:master} we recall the definition of Markov processes
%and we derive the properties of equilibrium dynamics for them.

\subsection{The Fokker -- Planck and Kramers equations}
\label{subsection:fokker}

The Fokker-Planck and Kramers equations are particular
master equations that hold 
%approximately for Markov 
%processes in which $W(q_0\to q_1)$ is very peaked around $q_0 \sim q_1$ 
%and $T(q_0 \to q_1)$ varies slowly with $q_1$~\cite{vanKampen} 
%and 
exactly for a Langevin process with white noise. 
%We concentrate
%on the latter case.
The probability distribution of the thermal noise, $P(\xi)$,
induces a time-dependent probability distribution of the dynamic variables
$q$ and $v$ (or $p=v/M$):
%that we call $P(q,v,t)$. 
%This probability distribution is defined
%as 
\begin{equation}
P(q,v,t) = \int {\cal D} \xi \, P(\xi) \, 
\delta\left(q - q_\xi^{sol}(t) \right) 
\delta\left(v - v_\xi^{sol}(t) \right) 
%= 
%\langle 
%\delta\left(q - q_\xi^{sol}(t) \right) 
%\delta\left(v - v_\xi^{sol}(t) \right) 
%\rangle  
\; .
\label{Pq}
\end{equation}
that satisfies the Kramers equation
\begin{equation}
\gamma \frac{\partial P(q,v,t)}{\partial t} = 
-\frac{\partial}{\partial q} \left( v P(q,v,t) \right) 
+
\frac{\partial }{\partial v}
\left[ 
\left( 
v + \frac{V'(q)}{M} + \frac{\gamma k_B T}{M} \frac{\partial}{\partial v}
\right)
P(q,v,t)
\right]
\; .
\end{equation}
For colored noises one cannot derive a simple differential 
equation for $P(q,v,t)$; indeed, it is clear that in these 
cases the stochastic process is not Markovian.
The averages over thermal histories can be expressed in terms
of  $P(q,v,t)$,
%\begin{equation}
$
\langle  A[q,v]\rangle (t)= 
\int {\cal D}q {\cal D}v \, P(q,v,t) \, A[q,v]
$.
%\end{equation}
%If the dynamics of $q$ and $v$ is governed by a Langevin equation of the 
%type (\ref{lang1})-(\ref{lang2}) with a white noise
%one can prove that 

When the inertial term in the Langevin equation can be dropped, 
$P(q,v,t)$ is replaced by an exclusive function of $q$,  
$P(q,t)$, defined as 
\begin{equation}
P(q,t) = 
\int {\cal D} \xi \, P(\xi) \, 
\delta\left(q - q_\xi^{sol}(t) \right) 
\label{delta-fp}
\end{equation} 
and determined by the Fokker-Planck equation,
\begin{equation}
\gamma \frac{\partial P(q,t)}{\partial t} = 
\frac{\partial}{\partial q} 
\left[ P(q,t) \, V'(q) \right] +
k_B T \frac{\partial^2}{\partial q^2} P(q,t)
\; .
\label{fokker-planck}
\end{equation}
It is very important to note that the balancing of factors on the 
right-hand-side ({\sc rhs}) of the Fokker-Planck and the Kramers equations
is a direct consequence of the 
equilibration of the noise (see 
Appendix~\ref{sec:langevin_general}).
This relation is known under the name of Einstein relation or
fluctuation -- dissipation 
theorem of the second kind (according to Kubo). We shall see its
implications below.

%{\it Exercise: derive the Fokker-Planck equation from a Langevin 
%process. Hint: write the Langevin equation using a discretized 
%time grid and expand the delta function in Eq.~(\ref{delta-fp}).}   

\subsection{Approach to equilibrium}
\label{approach}

%The properties of the dynamics close to equilibrium can be 
%derived using the generic formalism of Section~\ref{subsec:fdt_master}.
In this Section we focus, for simplicity, on the Fokker-Plack 
equation (\ref{fokker-planck}). 
In order to ensure the equilibration of the system at long times 
%$\lim_{t\to\infty} P({\cal C},t) = P_{\sc gb}({\cal C})$,
$P_{\sc gb}$ 
%the Gibbs-Boltzmann probability distribution 
must 
be a stationary solution of the Fokker-Planck equation.
Introducing $P \propto 
\exp(-\beta V)$ in Eq.~(\ref{fokker-planck})  
one realizes that any other ratio between the 
factors in front of the first and second term on the {\sc rhs} 
of the Fokker-Planck equation would not allow for this asymptotic 
solution. 
%Hence, the Einstein relation is a necessary condition 
%to reach an asymptotic stationary state.

We still have to  show that $P_{\sc gb}$ is the actual 
asymptotic solution reached by the dynamic process, 
$\lim_{t\to\infty} P(t) = P_{\sc gb}$.
An easy and elegant proof relies on a mapping between the Fokker-Planck 
and the Schr\"odinger equations~\cite{Parisi-book}. 
Introducing
\begin{equation}
P(q,t) = \psi_0(q) p(q,t) = c \; e^{-\frac{\beta}{2} V(q)}
\, p(q,t)
\label{transformation}
\end{equation}
with $c$ a positive constant, one has 
\begin{equation}
\gamma \frac{\partial p(q,t)}{\partial t} = 
\left[ 
k_B T \frac{\partial^2 }{\partial q^2} -
\left( 
-\frac12 V''(q) + \frac{\beta}{4} (V'(q))^2
\right)
\right]
p(q,t)
=
H_{\sc fp} p(q,t)
\; .
\end{equation}
This is a Schr\"odinger equation in 
imaginary time. Note however that $p(q,t)$ here is 
a probability density and 
plays the role of a wave function
while in true quantum
mechanics it is the modulus squared of the wave function which
has a probability interpretation.
If the term between square brackets 
grows to infinity sufficiently fast when 
$q\to\pm \infty$ the spectrum of the Fokker-Planck Hamiltonian 
$H_{\sc fp}$ is discrete. It is now easy to check that 
$\psi_0(q)$ is the ground state of $H_{\sc fp}$, {\it i.e.}
a positive definite 
eigenvector with eigenvalue $E_0=0$. 
We write
%\begin{equation}
$p(q,t) = \sum_n c_n \psi_n(q) \, \exp(-E_n t)$
%\end{equation}
with $\psi_n(q)$ the eigenvector associated to the eigenvalue $E_n$,
$E_n>0$ when $n>0$. 
When $t\to\infty$ all terms vanish apart from the one with $n=0$,
%\begin{equation}
$\lim_{t\to\infty} p(q,t) = c_0 \psi_0(q) = \psi_0(q)$,
%\end{equation}
where we used 
$c_0=\int dq \psi_0(q) p(q,0) =\int dq P(q,0) =1$. This expression 
implies 
%\begin{equation}
$
\lim_{t\to\infty} P(q,t) = \psi^2_0(q) = c^2 \exp(-\beta V(q))
$
%\; ,
%\end{equation}
and the conservation of probability allows one to compute the normalization
constant, $c^{-2}=\int dq \, \exp(-\beta V(q))$.
Thus $P_{\sc gb}$ is indeed the asymptotic solution
to the Fokker-Planck equation~\cite{Parisi-book}.

Note that this argument assumes that a sufficiently long $t$
($t> t_{\sc eq}$) is reached such that only the $n=0$ term 
survives in the sum. This hypothesis does not hold 
in  the asymptotic analysis
for the relaxing 
models we analyze in the next Sections. In the low-$T$ phase
the equilibration time grows with the size of the system, 
$t_{\sc eq}(N)\to\infty$, while in the analysis 
we only consider times that are finite with respect to $N$. 
Moreover, when non-potential or time-dependent forces are exerted 
on the system, see Eq.~(\ref{lang3}),  the transformation 
(\ref{transformation}) is not sufficient to deal with their effect and 
equilibrium cannot be established.

Just as in usual quantum mechanics one can use an operator notation 
to represent the Fokker-Planck equation. Indeed,
identifying $-i \partial /\partial q$ with the operator 
$\hat p$ the usual commutation relations between 
momentum and coordinate are recovered. The probability distribution $P(q,t)$ 
is then identified with a 
quantum time-dependent ``state'' 
$|P(t) \rangle$. With these new names, the Fokker-Planck 
equation reads
\begin{equation}
\gamma \frac{\partial |P(t)\rangle}{\partial t} = 
\hat H_{\sc fp} |P(t)\rangle
\;\;\;\; \mbox{with} \;\;\;\; 
\hat H_{\sc fp} = \hat p \left( i \hat V'(q) - k_B T \hat p \right)
\; .
\end{equation}
$|P(t) \rangle$ is obtained from the evolution 
of an initial state $|P(0) \rangle$ with the operator 
$\exp(-\hat H_{\sc fp} t)$. 

\subsection{Equilibrium dynamics}
\label{subsec:equilibrium}

All average values (\ref{calG}) can be computed using $P(q,t)$ 
as
%\begin{equation}
$\langle A[q](t) \rangle = 
\int dq \; A[q] \, P(q,t)$.
%\; . 
%\end{equation}
If we pursue the identification with the quantum mechanics formulation
and we associate the bra $\langle - |$ to the ``wave function'' 
identical to $1$ we write the average as
%\begin{equation}
$
\langle A[q](t) \rangle = 
\langle - | \hat A[q] |P(t)\rangle 
$.
%\; . 
%\end{equation}
The normalization of the probability distribution reads 
$\langle -| P(t) \rangle=1$.
Clearly, if the system reached equilibrium at a time $t'$, 
all averages of one-time quantities are time-independent 
henceforth.
 
Any correlation $C_{AB}(t,t')$ for two ``local'' functions of 
the variable $q$ can be expressed as 
\begin{eqnarray}
C_{AB}(t,t') &=& \int dq \int dq' \int dq''
\; A[q] \, T(q',t' \to q,t) \, B[q'] \,
T(q'',0 \to q',t') \, P(q'',0)
\nonumber\\
&=&
\langle - | 
\; \hat A[q] \, \exp(-\hat H_{\sc fp}(t-t')) \, \hat B[q] \,
\exp(-\hat H_{\sc fp}t') \, |P(0)\rangle
\; .
\label{eq-corr-eq}
\end{eqnarray}
In the transition probabilities, $T$, we included the time-dependence 
to clarify their meaning.
In the second line we used the quantum mechanical notation.
If the time $t'$ is longer than the equilibration time, 
the probability density at $t'$ reached the equilibrium one, 
$\int dq'' T(q'',0 \to q',t') \, P(q'',0)=P_{\sc gb}(q')$. Equivalently, 
$|P(t')\rangle=|P_{\sc gb}\rangle$. Two properties follow immediately:

{\it Stationarity:}
Since the transition probability is a function 
of the time-difference only, $T(q,t;q',t')=T(q,q'; t-t')
=\langle q| \exp(-\hat H_{\sc fp}(t-t'))| q'\rangle$ 
the correlation is invariant under translations of time
for $t'>t_{\sc eq}$.

{\it Onsager relations:}
$\langle A[q](t) B[q](t') \rangle = 
\langle A[q](t') B[q](t) \rangle$ for any two observables $A$ and $B$
that depend only on the coordinates. Indeed, 
\begin{eqnarray}
\langle A(t') B(t) \rangle &=& 
\langle -| \hat B e^{- \hat H_{\sc fp}(t-t')} \hat A | P_{\sc gb} \rangle
%\nonumber\\
%&=&
\; = \; 
\langle P_{\sc gb} | {\hat A}^\dag 
\left( e^{- \hat H_{\sc fp}(t-t')} \right)^\dag  {\hat B}^\dag | - \rangle
\nonumber\\
&=&
\langle P_{\sc gb} | \hat A
e^{- \hat H_{\sc fp}^\dag(t-t')}  \hat B | - \rangle
%\nonumber\\
%&=&
\; = \; 
\langle - | e^{-\beta \hat V} \hat A
e^{- {\hat H_{\sc fp}}^\dag(t-t')}  
\hat B e^{\beta \hat V} e^{-\beta \hat V} 
| - \rangle
\nonumber\\
&=&
\langle - | \hat A e^{-\beta \hat V} 
e^{- {\hat H_{\sc fp}}^\dag(t-t')}  e^{\beta \hat V} 
\hat B | P_{\sc gb} \rangle
\; .
\end{eqnarray}
The proof is completed by showing that 
%\begin{eqnarray}
$e^{-\beta \hat V} e^{-\hat H_{\sc fp}^\dag (t-t')} e^{\beta \hat V}  
= 
e^{-\hat H_{\sc fp} (t-t')}$
%\label{relation-FP0} 
%\end{eqnarray}
for all $t-t'$ which is equivalent to 
%\begin{eqnarray}
$
{\hat H_{\sc fp}}^\dag = e^{\beta \hat V}  \hat H_{\sc fp} e^{-\beta \hat V} 
$.
%\; .
%\label{relation-FP}
%\end{eqnarray}
The latter can be checked directly 
using the Fokker-Planck Hamiltonian.
The matrix elements 
%\begin{eqnarray}
$
\langle q | e^{-\hat H_{\sc fp} \tau } |q' \rangle 
= 
T(q', t \to q, t+\tau)
$
and 
%\\
$ 
\langle q | e^{-\hat H_{\sc fp}^\dag \tau } |q' \rangle 
= 
T(q, t \to q', t+\tau)
$
%\end{eqnarray}
yield the transition probabilities and 
the first equation in this paragraph 
%Eq.~(\ref{relation-FP0})
can be recast as 
%\begin{equation}
$
T(q',t\to q,t+\tau) \, e^{-\beta V(q')} = 
T(q,t\to q',t+\tau) \, e^{-\beta V(q)}
$
%\; ,
%\end{equation}
which is {\it detailed balance}. 

Similarly, one proves that the linear response $R_{AB}(t,t')$ 
is also stationary when $P(q',t')=P_{\sc gb}(q')$. 
We represent the instantaneous infinitesimal perturbation $h(t')$ 
as the kick between $t_a-\delta/2$ and $t_a+\delta/2$  
in Fig.~\ref{pert-resp}-left. The Fokker-Planck Hamiltonian
in the presence of the field is
%\begin{equation}
$
{\hat H_{\sc fp}}^h = 
i \hat p \left[ {\hat V}'(q) + h \hat B'[q]+ k_B T i \hat p \right]
$
%\; 
%\end{equation}
while $\hat H_{\sc fp}$ is the Fokker-Planck operator in no field.
The average in a field reads
\begin{eqnarray}
\langle A[q] \rangle_h(t) = 
\langle - | \; \hat A[q] \;  
e^{-\hat H_{\sc fp}\left(t-t'-\frac{\delta}{2} \right)}
\; 
e^{-\hat H^h_{\sc fp}\left[\left(t'+\frac{\delta}{2} \right)
- 
\left(t'-\frac{\delta}{2} \right)
\right]}
%& &
%\nonumber\\
%\times 
e^{-\hat H_{\sc fp}\left(t'-\frac{\delta}{2} \right)}
\; |P(0)\rangle
\; .
\end{eqnarray} 
%In its discrete representation 
and the variation with respect to $h$ yields
\begin{eqnarray}
%R_{AB}(t,t') = 
\frac{\Delta \langle A[q]\rangle(t) }{\Delta h_B(t')}
&=& 
\langle - | \; \hat A[q] \;  
e^{-\hat H_{\sc fp}\left(t-t'-\frac{\delta}{2} \right)}
\left(-\delta \frac{\hat H_{\sc fp}^h - \hat H_{\sc fp}}{\Delta h }\right)
%\nonumber\\
%& &\times
e^{-\hat H_{\sc fp}\left(t'-\frac{\delta}{2} \right)}
\; |P(0)\rangle
\; ,
\end{eqnarray}
with $\Delta h= h\delta $.
The factor between parenthesis equals
$(-i\hat p \hat B'[q] h\delta)/(h\delta)=-i\hat p \hat B'[q]$.
Taking the limit $\delta\to 0$ and evaluating at $h=0$ one has
\begin{equation}
R_{AB}(t,t') = 
\langle - | \; \hat A[q] \;  
\exp\left[-\hat H_{\sc fp}\left(t-t'\right)\right]
\; (-i\hat p \hat B'[q]) \; 
\exp\left(-\hat H_{\sc fp} t' \right)
\; |P(0)\rangle
\; .
\end{equation}
From this expression one recovers several properties of the response:

{\it Causality:} 
The same derivation for $t'> t$ yields 
\begin{equation}
R_{AB}(t,t') = 
\langle - | \; 
\; (-i\hat p \hat B'[q]) \; 
\exp\left[-\hat H_{\sc fp}\left(t-t'\right)\right]
\hat A[q] \;  
\exp\left[-\hat H_{\sc fp} t' \right]
\; |P(0)\rangle
\; .
\end{equation}
Since $\langle - | \hat p =0$, $R(t,t')=0$ for all $t<t'$.

{\it Stationarity:}
When $\exp\left(-\hat H_{\sc fp} t' \right) \; |P(0)\rangle= 
|P_{\sc gb}\rangle$ one has 
\begin{equation}
R_{AB}(t,t') = 
\langle - | \; 
\; \hat A[q] \;  
\exp\left[-\hat H_{\sc fp}\left(t-t'\right)\right]
(-i\hat p \hat B'[q]) \; |P_{\sc gb}\rangle = R_{AB}^{\sc st}(t-t')
\; .
\end{equation}

{\it Response at equal times:}
%\begin{equation}
$
\lim_{t'\to t^-} R_{AB}(t,t') = 
\langle -|\; \hat A'[q] \hat B'[q] \; 
\exp\left[-\hat H_{\sc fp} t' \right] \; |P(0)\rangle
$.
%\; .
%\end{equation}
If $\hat A[q]=\hat B[q]=\hat q$ then 
$\hat A'[q]=\hat B'[q]=1$ and $\lim_{t'\to t^-} R_{AB}(t,t') = 
\langle -|P(t')\rangle=1$ from conservation of probability.

{\it Fluctuation-dissipation theorem:}
We postpone its discussion to
Section~\ref{FDT-FP}.

{\it Onsager relations:}
Using the relation between correlation and responses dictated 
by the {\sc fdt} one finds that the Onsager relations must also 
hold between responses.

\section{The fluctuation -- dissipation theorem ({\sc fdt})}
\label{section:fdt}
\setcounter{equation}{0}
\renewcommand{\theequation}{\thesection.\arabic{equation}}

The fluctuation-dissipation theorem ({\sc fdt}) relates 
the correlations of spontaneous fluctuations 
to the induced fluctuations in equilibrium. It is a model independent 
relation between the linear response and its associated correlation function 
that takes different forms for classical and quantum system. 
The latter reduces to the former when quantum fluctuations become 
irrelevant. In this Section we present several proofs of the {\sc fdt}.
When the equilibration hypothesis is not justified, this relation 
does not necessarily holds. 

\subsection{Static {\sc fdt}}

Many relations between correlations of 
fluctuations and susceptibilities are known in 
statistical mechanics. All these are different statements of the 
{\it static} {\sc fdt}.

Take for instance a perfect gas. The 
fluctuations in the density $\rho=n/\tilde V$
where $n$ is the number of particles 
within a sub volume $\tilde V$ of a system with $N$ particles and 
volume $V$, are defined as: 
$\sigma^2_\rho \equiv 
\langle (\rho-\langle \rho\rangle)^2 \rangle$.  
In the thermodynamic limit $N\to \infty,  V\to\infty$
with $N/V=\overline\rho$ fixed,  
these are related to the isothermal compressibility
$\chi_T=-1/V \partial V/\partial P|_T$ 
via $\sigma^2_\rho = (k_B T) {\overline\rho}^2 \chi_T/{\tilde V}$.
This relation is a form of {\sc fdt}.

For a system in equilibrium with a thermal reservoir at temperature $T$
one has 
\begin{eqnarray}
\chi &\equiv& 
\left. 
\frac{\delta \langle A\rangle_h}{\delta h}
\right|_{h=0} 
=
\beta 
\langle (A-\langle A\rangle )^2 \rangle
\; 
\end{eqnarray} 
for any observable $A$. 
The average $\langle \;\; \rangle_h$
is calculated with the partition function of the system in the 
presence of a small field coupled to $A$ in such a 
way that the Hamiltonian reads $H=H_0-h A$.
For a magnetic system this 
equation relates the magnetic 
susceptibility to the magnetization fluctuations.

\subsection{Dynamic {\sc fdt}}
\label{FDT-FP}

There are several proofs of this theorem. 
Here we 
focus on a Fokker-Planck process. In Section~\ref{subsec:equilibrium}
we computed $R_{AB}$ and $C_{AB}$ in equilibrium. Taking the derivative
of Eq.~(\ref{eq-corr-eq}) with respect to $t'$ one finds 
\begin{eqnarray}
\frac{\partial C_{AB}(t-t')}{\partial t'} 
&=& 
%\langle -| \hat A[q] \exp[-\hat H_{\sc fp}(t-t')] [\hat H_{\sc fp}, 
%\hat B[q]] 
%\exp(-\hat H_{\sc fp}t') |P_{\sc gb}\rangle
%\nonumber\\
%&=& 
\langle -| \hat A[q] \exp[-\hat H_{\sc fp}(t-t')] (-i\hat p k_B T \hat B'[q])
|P_{\sc gb} \rangle 
\nonumber\\
&=& 
k_B T R_{AB}(t-t')
\; 
\label{fdt-classical}
\end{eqnarray}
for $t-t'> 0$.
Note that since in equilibrium the averages of one-time quantities are
constant one can replace $C_{AB}(t-t')$ by the connected correlation
$C^{\sc c}_{AB}(t-t') \equiv C_{AB}(t-t') - \langle A\rangle \langle B 
\rangle$ in 
the left-hand-side ({\sc lhs}) and the {\sc fdt} also reads
%\begin{eqnarray}
$\partial_{t'} C^{\sc c}_{AB}(t-t')
=
k_B T R_{AB}(t-t')
$.
%\; .
%\end{eqnarray}
In integrated form 
\begin{eqnarray}
k_B T 
\chi_{AB}(t-t_w) = C_{AB}(t-t) - C_{AB}(t-t_w) 
%\nonumber\\
%&=&
= C^{\sc c}_{AB}(t-t) - C^{\sc c}_{AB}(t-t_w) 
\; .
\label{int-fdt}
\end{eqnarray}

\subsection{Quantum {\sc fdt}}

Proofs and descriptions of the quantum {\sc fdt} can be found in several 
textbooks~\cite{Kubo,Parisi-book}. Here, 
we express it in the time-domain and in 
a mixed time-Fourier notation
that gives us insight as to how to extend it to the case of glassy 
non-equilibrium dynamics.

If at time $t'$ a quantum system has reached equilibrium 
the density operator $\hat \rho(t')$ is 
just the Boltzmann factor $\exp(-\beta \hat H)/Z$. 
As in Section~\ref{subsec:equilibrium} it is then immediate to show that
time-translation invariance ({\sc tti}) holds 
%\begin{equation}
$C_{AB}(t,t') = C_{AB}(t-t')$. 
%\; .
%\end{equation}
In addition, from the definition of $C_{AB}(t,t')$ in Eq.~(\ref{symm})
one proves the {\sc kms} properties
%\begin{equation}
$C_{AB}(t,t') = C_{BA}(t',t+i\beta \hbar) = C_{BA}(-t-i\beta \hbar, -t')$.
%\; .
%\end{equation}
Assuming, for definiteness, that $t > 0 $ it is easy to 
verify the following equation
\begin{equation}
C_{\{A,B\}}(\tau) + \frac{i\hbar}{2} R_{AB}(\tau) =
C_{\{A,B\}}(\tau^*) - \frac{i\hbar}{2} R_{AB}(\tau^*)
\; ,
\label{FDTtemporal}
\end{equation}
where $\tau = t + i\beta \hbar/2$. This is a way to 
express {\sc fdt} through an analytic continuation to complex times.
In terms of the Fourier transformed $\tilde C_{AB}(\omega)$ 
defined
in Eq.~(\ref{fourier}) the {\sc kms} properties read
$
\tilde C_{AB}(\omega) = \exp(\beta \hbar \omega) \tilde C_{BA}(-\omega) 
$
and lead to 
the following relation between Fourier transforms of symmetrized and 
anti - symmetrized correlations:
%\begin{eqnarray}
$\tilde C_{[A,B]}(\omega) =
\tanh\left( \frac{\beta \hbar \omega}{2} \right) \, 
\tilde C_{\{A,B \}}(\omega)
$.
%\end{eqnarray}
Using now the Kubo relation (\ref{FDTasym}) 
one obtains the quantum {\sc fdt}
\begin{eqnarray}
R_{AB}(t-t') &=& \frac{i}{\hbar} 
\int_{-\infty}^\infty  \frac{d\omega}{\pi} \;  \exp(-i \omega (t-t'))  \; 
\tanh\left( \frac{\beta \hbar \omega}{2} \right) \, 
\tilde C_{\{A,B \}}(\omega)
\; ,
\label{FDTFourier}
\end{eqnarray}
$t\geq t'$. With the representation 
$
\int_0^\infty dt \, \exp(i \omega t)
=
\lim_{\delta\to 0^+} i/(\omega+i\delta) =
\pi \delta(\omega) + i P/\omega
$,
\begin{eqnarray}
\tilde R_{AB}(\omega) &=& - \frac{1}{\hbar} \,
\lim_{\delta\to 0^+} \int_{-\infty}^\infty \frac{d\omega'}{\pi } \; 
\frac{1}{\omega-\omega'+i\delta} \; 
\tanh\left( \frac{\beta \hbar \omega'}{2} \right)  \, 
\tilde C_{\{A,B \}}(\omega')
\end{eqnarray}
from which we obtain the real and imaginary relations
between $\mbox{Im} \tilde R_{AB}(\omega)$,  
$\mbox{Re} \tilde R_{AB}(\omega)$ and 
$\tilde C_{\{A,B \}}(\omega)$.
If 
$\beta\hbar\omega/2  \ll 1$,
$\tanh(\beta\hbar\omega/2 ) \sim \beta\hbar\omega/2 $ and 
Eq.~(\ref{FDTFourier}) becomes the classical {\sc fdt},
Eq.~(\ref{fdt-classical}).

\subsection{Examples}
\label{examples}

\subsubsection{Harmonic oscillator and diffusion}

The simplest example in which one sees the modifications 
of the {\sc fdt} at work is 
a one-dimensional harmonic oscillator with no inertia 
coupled to a white bath. One finds
\begin{eqnarray}
k_B T \frac{R(t,t')}{\partial_{t'} C(t,t')}
&=&
\left[ 1 + \exp\left(-\frac{2kt'}{\gamma}\right) \right]^{-1}
\to 
\left\{
\begin{array}{rcl}
1 \;\;&\mbox{if}& \;\; k>0
\label{osc}
\\
\frac{1}{2} \;\;&\mbox{if}& \;\; k=0
\label{difussion}
\\
0 \;\;&\mbox{if}& \;\; k<0
\label{nosc}
\end{array}
\right.
\end{eqnarray}
for times $t\geq t'$ and $t'\gg \gamma/k$ with $k$ the harmonic constant
of the oscillator. Thus, when there is a confining 
potential and $P_{\sc gb}$ can be defined the {\sc fdt} holds. 
Instead, when the potential is flat ($k=0$) or unbounded from 
below ($k<0$) no normalizable
$P_{\sc gb}$ can be defined and the {\sc fdt} is modified.

If one keeps inertia, the calculations are slightly more involved 
but one can carry them through to show that momenta and coordinates 
behave very differently~\cite{Noelle}. 
Since the probability distribution 
of the momenta very rapidly reaches a Maxwellian
for all values of $k$ these variables 
equilibrate and {\sc fdt} holds for them. This is the reason why the 
kinetic energy of a particle system serves to callibrate the 
external temperature. The coordinates, instead, behave as in 
(\ref{osc}) depending on the value of $k$. 

\subsubsection{A driven system}

Take now a symmetric two-dimensional harmonic oscillator 
$V(x,y)=k/2(x^2+y^2)$ and apply the non-potential force 
$\vec f(x,y) =\alpha (y,-x)$ on it. This force makes a particle 
turn within the potential well and one can check, by direct calculation,
that the {\sc fdt} does not hold.

\subsubsection{No Einstein relation}

If the bath is such that 
the Einstein relation between friction and noise correlation 
does not hold, the {\sc fdt} for the system variables does not 
hold either. Again, this can be easily checked using a
 harmonic oscillator.

\begin{figure}[h]
\centerline{
\psfig{file=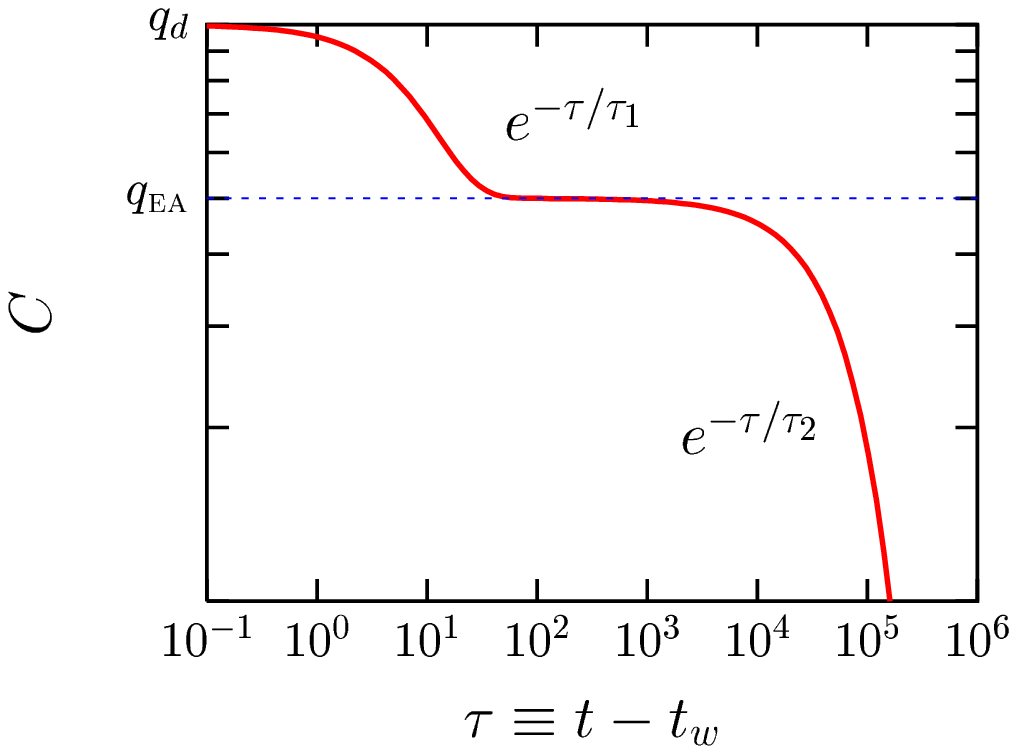,width=8cm}
\hspace{-1cm}
\psfig{file=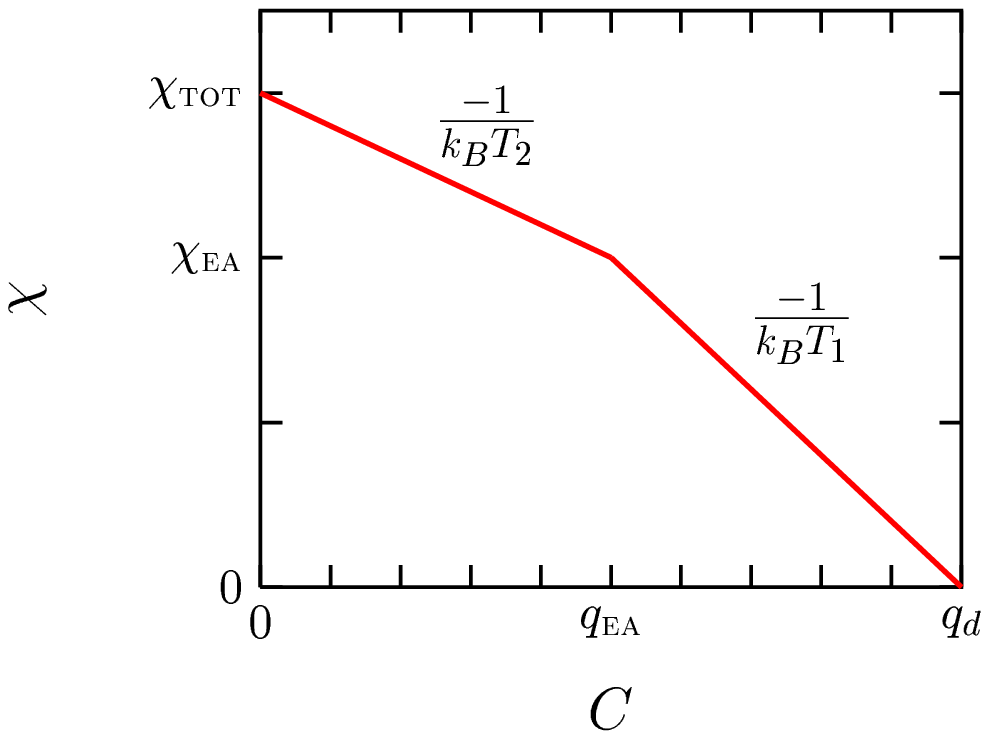,width=8cm}
}
\caption{Left: Decay of the correlation for a harmonic oscillator coupled 
to a mixed bath. Right: paramteric plot of the integrated linear response
against the correlation.}
\label{two-baths}
\end{figure}

\subsubsection{A complex bath}
\label{complexbath}

Let us couple a harmonic oscillator to a complex bath made of two 
parts: a white bath ($\tau_{\xi^{(1)}}\to 0$) 
with friction coefficient $\gamma_1$ 
at temperature $T_1$ and a coloured bath with 
friction kernel $\gamma(t-t') = \gamma_2 \exp[-(t-t')/\tau_{\xi^{(2)}}]$
kept at temperature $T_2$~\cite{Cuku4}. The complex 
bath induces two time-scales (see Section~\ref{time-scales} for its precise 
definition) in the dynamics of the oscillator: the correlation 
is stationary but it decays in two steps, from $q_d$ at equal times 
to an ``Edwards-Anderson parameter'' $q_{\sc ea}$ for time-differences,
$\tau$, that are shorter than a characteristic time 
$\tau_0$ and from $q_{\sc ea}$ to zero for
longer $\tau$'s. The parameters $q_d$, $q_{\sc ea}$ and $\tau_0$ 
are functions of $k$, the friction coefficients $\gamma_1$ and 
$\gamma_2$ and the two temperatures $T_1$ and $T_2$. In 
Fig.~\ref{two-baths}
we display the correlation decay and the parametric plot of the integrated
linear response, $\chi$, against the correlation, $C$, constructed
using $t-t'$ as a parameter that runs from $t-t'=0$ ($C=q_d,\chi=0$)
to $t-t'\to\infty$ ($C\to 0,\chi=\overline\chi$). (See 
Section~\ref{definitions}.)
The {\sc fdt} predicts a linear relation between $\chi$ and $C$ with 
a slope $-1/(k_BT)$ for systems equilibrated with a reservoir at 
temperature $T$. In this problem though ``there is no $T$'' since the 
system is coupled to a bath with two temperatures and two time-scales.
Surprisingly, one finds that the rapid decay 
is controlled by the temperature of the fast bath, $T_1$, while the 
subsequent, slower, decay is controlled by the temperature of the 
slow bath, $T_2$. 
%In Fig.~\ref{two-baths}-right we show the $\chi(C)$ plot.

A similar phenomenon, now self-induced, appears in glassy models.
These interacting systems are coupled to an external white bath 
at temperature $T_1$. Their dynamics is such that several time-scales,
each with its own ``temperature'' is generated. 
We shall show how 
this arises in solvable models in Section~\ref{section:solution}
and we shall prove that the 
{\sc fd} relation can indeed be used to define an ``effective temperature''
in Section~\ref{temp_intro}.

\section{Dynamic generating functionals}
\label{generating-functionals}
\setcounter{equation}{0}
\renewcommand{\theequation}{\thesection.\arabic{equation}}

In this Section we present the functional methods used to 
analyze the dynamics of classical~\cite{MSR} and 
quantum~\cite{Schwinger-Keldysh} models coupled to 
environments. We discuss the relation between the two approaches. 
The generating functionals, with their effective actions, are the 
adequate starting point to apply perturbation theory (when 
it is accepted), self-consistent approximations such as 
the mode-coupling approach, or even more sophisticated techniques
as the functional renormalization group. 

\subsection{Classical models}

For a classical system coupled to a classical environment, we 
use as a starting point a stochastic equation 
with an additive noise
$\xi(t) = {\sc Eq}(q)$.
For instance, if we deal with a massive 
particle governed by the Langevin equation, Eq.~(\ref{lang1})
%\begin{equation}
$
{\sc Eq}(q) = M \ddot q + \int_0^t dt' \, \gamma(t-t') \, \dot q(t') +
V'(q)
$.
%\label{Langevin-rhs}
%\end{equation}

Any averaged observable, 
{\it e.g.} $\langle A[q]\rangle (t)$ in  
Eq.~(\ref{calG}), 
can be computed from the variation of a dynamic 
generating functional ${\cal Z}$ with respect to a time-dependent
source $\eta$:
\begin{equation}
\langle A[q] \rangle(t) = 
\left. \frac{\delta {\cal Z[\eta]}}{\delta \eta(t)} \right|_{\eta(t) =0}
=
\left. 
\frac{\delta }{\delta \eta(t)} 
\int {\cal D} \xi \, P[\xi] \, 
e^{\int dt' \; \eta(t') \, A[q_\xi^{sol}](t')}
\right|_{\eta(t) =0}
\; .
\end{equation}
Since the probability distribution $P[\xi]$ is normalized to one,
${\cal Z}[\eta=0]=1$, and the average is automatically normalized.
The idea is to derive a useful expression for the 
generating functional ${\cal Z}$ by introducing the functional 
identity  
\begin{equation}
1 = \int {\cal D} q \; \delta(q-q_\xi^{sol}) 
= \int {\cal D} q \; \delta(\xi-{\sc Eq}[q]) \, 
\left| \det \left( \frac{\delta {\sc Eq}[q](t)}{\delta q(t')} 
\right)\right| 
\; .
\end{equation}
In the first integral an integration 
and delta function is applied on each time slice and the compact 
notation actually represents
$\int \prod_a [dq(a) \delta(q(a)-q_\xi^{sol}(a))]$. 
The second identity follows from a 
change of variables in the functional integral.
If the stochastic equation
has only one solution, one can eliminate the absolute value in the determinant,
and
\begin{equation}
{\cal Z}[\eta] = \int {\cal D}\xi \, {\cal D}q \, P[\xi]
\, \delta(\xi-{\sc Eq}[q]) \, 
\det \left( \frac{\delta {\sc Eq}[q](t)}
{\delta q(t')} \right)
\, 
\exp\left(\int dt' \; \eta(t') A[q](t') \right)
\; .
\label{generating1}
\end{equation}
This expression can be recast in a more convenient form 
by using the exponential representation of the delta function
\begin{equation}
\delta(\xi-{\sc Eq}[q]) \propto \int {\cal D}i\hat q \, 
\exp \left[ -\int_0^t dt' \, i\hat q(t') 
\left(\xi(t') - {\sc Eq}[q](t') \right) \right]
\label{identity1}
\end{equation}
(the constant of proportionality is numeric, a power of $2\pi$, and 
irrelevant for the calculation of averages, thus we omit it), 
and the determinant through the introduction of a pair 
of fermionic variables~\cite{Zinn}
\begin{equation}
\det \left( \frac{\delta {\sc Eq}[q](t)}{\delta q(t')} \right)
= \int {\cal D}\overline\psi {\cal D}\psi \; 
\exp\left[ - \int_0^t dt' \int_0^t dt'' \; \overline\psi(t')
\, \frac{\delta {\sc Eq}[q](t')}{\delta q(t'')} \, 
\psi(t'') 
\right]
\; .
\label{identity2}
\end{equation}
Having used these identities, the generating functional becomes a
functional integral over $\xi$,$q$,$i\hat q$, $\overline\psi$ and $\psi$. 
Since $\xi$ appears
in quadratic terms of the effective action only, the functional 
integral over $\xi$ can be simply calculated. We obtain
\begin{eqnarray}
&&
{\cal Z}[\eta,\hat\eta] = 
\int {\cal D}q {\cal D}i\hat q 
{\cal D} \overline \psi {\cal D} \psi \; \exp(-S_{\sc eff}) 
\nonumber\\
&&
S_{\sc eff} = 
\int_0^t dt' \int_0^t dt''
\left(
-i\hat q(t') k_B T \theta(t'-t'') \gamma(t'-t'') i \hat q(t'') 
+
\overline\psi(t') 
\frac{\delta {\sc Eq}[q](t')}{\delta q(t'')} \psi(t'') 
\right)
\nonumber\\
& &
\;\;\;\;\;\;\;\;\;\;\;\;\;
-
\int_0^t dt' \; 
\left( i\hat q(t') \, {\sc Eq}[q](t') 
- \eta(t') q(t') -  \hat \eta(t') i \hat q(t') 
\right)
\label{Seff}
\end{eqnarray}
where we have introduced a new 
source $\hat\eta(t)$ coupled to the auxiliary variable $i \hat q(t)$.

In Appendix~\ref{app:relations}
we prove the following very useful relations 
\begin{eqnarray}
R(t,t') &=& \langle q(t) i\hat q(t') \rangle 
\label{Rqihatq}
\\
\frac{1}{2} k_B T \int_0^t dt'' \; \left(\gamma(t',t'')+\gamma(t'',t')\right)
\, R(t,t'') &=&  
\langle q(t) \xi (t') \rangle  
\; .
\label{Rqxi}
\end{eqnarray}
Usually, the kernel $\gamma(t',t'')$ is symmetric 
$\gamma(t',t'')=\gamma(t'',t')$ and the latter relation simplifies. 
For a white noise $\gamma(t',t'')=2\gamma \delta(t'-t'')$ and 
$2 \gamma k_B T R(t,t') =  \langle q(t) \xi (t') \rangle$.

\subsection{Supersymmetry ({\sc susy})}
\label{subsubsec:supersymmetry}

In the white noise limit ${\cal Z}$ in Eq.~(\ref{Seff}) 
can be written in a much more compact form 
if one introduces the super-field formulation of stochastic processes
explained in \cite{Zinn,Gozzi,jorge3}, see Appendix~\ref{app:Grassmann}. 
One first enlarges (space)-time to include two Grassmann coordinates
$\theta$ and $\overline\theta$, {\it i.e.}
$t\to a=(t,\theta,\overline\theta)$.
The dynamic variable $q(t)$ and the auxiliary variable 
$i \hat q(t)$ together with the fermionic ones $\psi(t)$ and
$\overline\psi(t)$ are encoded  in a super-field, 
\begin{equation}
\Phi(a) = 
q(t) +  \overline\theta \psi(t) + \overline\psi(t) \theta
+ i \hat q(t) \theta\overline\theta 
\; .
\end{equation}
With these definitions, 
\begin{equation}
{\cal Z}[\eta] = \int d\Phi \exp\left(-\frac12 \int da \; \Phi(a) 
(-D_{a}^{(2)}) \Phi(a)-\int da \; V[\Phi(a)] + \int da \Phi(a) \eta(a) 
\right)
\label{generating-susy}
\end{equation} 
with $a=(t,\theta,\overline\theta)$,  
$da=dt d\theta d\overline\theta$, 
and the dynamic operator $D_a^{(2)}$ defined as 
\begin{equation}
-D_a^{(2)}= 
2 \gamma  k_B T \frac{\partial^2}{\partial \theta\partial\overline\theta}
+ 2 \gamma \theta \frac{\partial^2}{\partial \theta\partial t} 
- \gamma \frac{\partial}{\partial t} 
- 
M \theta \frac{\partial^3}{\partial \partial \theta \partial t^2}
\; .
\end{equation}
If the model is spherically constrained, 
$-D_a^{(2)} \to -D_a^{(2)} -\mu(a)$ 
with $\mu(a)$ a super
Lagrange multiplier introduced to enforce the constraint. 
The delta function $\delta(a-b)$ is defined in Appendix~\ref{app:Grassmann}
and it satisfies $\int db \delta(a-b) f(b) = f(a)$. 
The super-symmetric  notation allows one to encode in the 
single super correlator
%\begin{equation}
$Q(a,b) \equiv \langle \Phi(a) \Phi(b)\rangle $
%\end{equation}
all correlators and responses. 
The generalization to a system with $N$ degrees of freedom is 
immediate.

\vspace{0.25cm}
\noindent{\it Symmetries}
\vspace{0.25cm}

The properties of the equilibrium
dynamics, {\it i.e.} the invariance under time-translations, 
the fluctuation-dissipation theorem, etc., 
%can be seen as 
are  consequences
of the symmetries of the super-symmetric action. 
The non-equilibrium dynamic solution 
spontaneously breaks these symmetries. 

For a stochastic process with a white noise
the {\sc susy} group is generated by three 
operators \cite{Gozzi,Zinn,jorge3}:
\begin{eqnarray}
\bar D' = \gamma k_B T \partial_\theta + \bar \theta
\gamma \partial_t 
\;\;\;\;\;\;\;\;
D' = \partial_{\bar \theta}
\;\;\;\;\;\;\;\; 
D'\bar D' + \bar D' D
=  \gamma \partial_t 
\end{eqnarray}
with $D'^2\,=\bar D'^2 = 0$. 
We can construct an extension
of this group that acts on two-point (in general $n$-point) functions, as
\begin{eqnarray}
D'(a,b) \equiv D'(a) + D'(b) 
\;\;\;\;&& \;\;\;\; 
\bar D'(a,b) \equiv \bar D'(a) + \bar D'(b) 
\nonumber\\
\left[ D'(a,b), \bar D'(a,b) \right]_+ 
&=&
\partial_{t} + \partial_{t'} 
\; .
\end{eqnarray}
The meaning of the three generators can be understood when they are 
made to act on $Q(a,b)$.
Firstly, causality plus probability conservation imply (irrespective of equilibration):
%\begin{equation}
$D'(a,b) Q(a,b)=0$,
%\; .
%\end{equation}
This equation serves to select the non-vanishing
terms in the super correlator. It imposes
$\langle i\hat q(t) i\hat q(t')\rangle=0$ and the fact that all 
components involving only one $\psi$ and $\overline\psi$ vanish. 
Then
\begin{equation}
Q(a,b) = C(t,t') - (\overline\theta'-\overline\theta) 
\left( \theta' R(t,t') - \theta R(t',t) \right) 
\; 
\end{equation}
(the bifermionic correlator $\langle \psi(t) \overline\psi(t')\rangle$
equals the response $R(t,t')$).
The {\sc susy} action [the exponent
in Eq.~(\ref{generating-susy})] vanishes if evaluated 
on such correlators (when $\eta=\hat\eta=0$). 
The other two generators imply
\begin{eqnarray}
\left(\partial_{t}  + \partial_{t'} \right)Q(a,b) 
=
0 
\;\rightarrow \; {\sc tti} 
\; ,
\;\;\;\;\;\;\;\; 
\bar D'(a,b) Q(a,b)
=
0 \; \rightarrow \; {\sc  fdt}
\; .
\end{eqnarray}

When is a system unable to reach equilibrium? In terms of 
symmetries this question can be addressed as follows.
In driven system the dynamical actions break {\sc susy} explicitly.
They are externally kept far from equilibrium 
by the forcing.
If {\sc susy} is not explicitly broken two possibilities
arise. Either the system evolves from its initial condition
during an out of equilibrium transient in which neither 
stationarity nor {\sc fdt} hold until the equilibration 
time $t_{\sc eq}$ is reached and equilibrium establishes.
In this language,
{\sc susy}  is unbroken by the boundary conditions.
On the contrary, 
if the system never achieves equilibrium
the equilibration time
diverges and cannot be reached in the calculation.
The effect of the
initial conditions  is then to break 
{\sc susy}~\cite{jorge3,andalo} and consequently
violate the equilibrium properties
even for long times.
{\sc susy} is then {\it spontaneously broken}.
The initial conditions play for {\sc susy} 
the same role  played
in ordinary symmetry-breaking by space boundary conditions: 
if the symmetry is spontaneously broken their
 effect extends away from them, in this case for all times.

\subsection{Connection with the replica formalism}

The effective 
action in Eq.~(\ref{generating-susy}) is a kinetic minus a potential
term $V[\Phi]$. When applying the replica trick to compute the 
free-energy a replicated effective potential $V[\phi^a]$ appears.
A connection between the two formalism, that is based on the 
similarity between the zero-dimensional replica space and the 
{\sc susy} one, has been exploited. Roughly speaking, many properties
of the replica overlap $Q^{ab}\equiv N^{-1} \sum_{i=1}^N \langle s_i^a s_i^b 
\rangle$  finds a counteraprt in the dynamic {\sc susy} correlator $Q(a,b)$.
For instance, a summation over a replica index, $\sum_{a=1}^n$ when $n\to 0$, 
translates into an integration over the supercoordinate $\int da$.
For the moment, though, the connection is 
empirical and a formalization of the
relation between the two approaches would be welcome. 

\subsection{Quantum models}
\label{quantum-dyn}

The Schwinger-Keldysh formalism~\cite{Schwinger-Keldysh} 
allows one to analyze the 
real-time dynamics of a quantum system.
The starting point is the time dependent 
density operator
\begin{equation}
\hat \rho_{\sc tot}(t)
=
e^{-\frac{i}{\hbar} \hat H_{\sc tot} t} \; 
\hat \rho_{\sc tot}(0) \;  
e^{\frac{i}{\hbar} \hat H_{\sc tot} t}
\; .
\label{density-op}
\end{equation}
Introducing identities, an element of the time-dependent 
density matrix reads
\begin{eqnarray}
& & \rho(q'',{x_a''}; q', {x'}_a; t) 
=
%\nonumber\\
%& & 
\int_{-\infty}^\infty dQ dQ' dX_a dX'_a \; 
\langle q'',x''_a | \; 
e^{-\frac{i}{\hbar} \hat H_{\sc tot} t} \; | Q, X_a \rangle
\;
\nonumber\\
& &
\;\;\;\;\;\;\;\;\;\;\;\;\;\;\;\;\;\;\;
\times
\langle Q, X_a | \; \hat \rho_{\sc tot}(0) \; | Q', X'_a \rangle
\langle Q',X'_a | \; e^{\frac{i}{\hbar} \hat H_{\sc tot} t}\; 
| q', x_a' \rangle
\; ,
\end{eqnarray}
where $q$ is the coordinate of the particle and $x_a$ are the 
coordinates of the oscillators.
The first  factor is the coordinate representation of 
the evolution operator and it 
can be represented as a functional integral. The third factor
can also be represented in functional form. They read
\begin{eqnarray}
\langle q'',x_a'' | \, e^{-\frac{i}{\hbar} \hat H_{\sc tot} t} \,
| Q ,X_a \rangle
&=&
\int_{q^+(0)=Q}^{q^+(t)=q''} {\cal D} q^+(t) 
\int_{x^+_a(0)=X_a}^{x^+_a(t)={x''_a}} 
{\cal D}{x^+_a}(t)  
\; e^{\frac{i}{\hbar} S^+_{\sc tot}}
\label{pos}\\
\langle Q',X'_a | 
\, e^{\frac{i}{\hbar} \hat H_{\sc tot} t} \, 
| q', {x'_a} \rangle
&=&
\int_{q^-(0)=Q'}^{q^-(t)=q'} {\cal D} q^-(t) 
\int_{x^-_a(0)=X'_a}^{x^-_a(t)=x'_a}
{\cal D}{x^-_a}(t)  
\; e^{-\frac{i}{\hbar} S^-_{\sc tot}}
\; .
\label{neg}
\end{eqnarray}
Interestingly enough, the evolution 
operator in Eq.~(\ref{neg})
gives rise to a path integral going backwards in time,
from $q^-(t)=q', x^-_a(t)=x'_a$ to $q^-(0)=Q', x^-_a(0)=X'_a$. 
The full time-integration can then be interpreted as being closed, going 
forwards from $t_0=0$ to $t$ and then 
backwards from $t$ to $t_0=0$. This motivates the name 
``closed time path formalism''.
A doubling of degrees of freedom $(q^+,q^-)$ 
appeared and it is 
intimately linked to the introduction of Lagrange multipliers 
in the functional 
representation of the stochastic dynamics in the classical limit.

The action $S_{\sc tot}$ has the usual four contributions,
from the system, the reservoir, the 
interaction and the counter-term. The action of the system reads
\begin{equation}
S^+_{\sc syst}[q^+(t),\eta^+(t)] = \int_0^t dt' \, \left[ \, \frac{M}{2} 
\left(\frac{q^+(t')}{dt'}\right)^2 - V(q^+(t')) + 
\eta^+(t') q^+(t') \, \right]
\end{equation}
where we have introduced a time-dependent source $\eta^+(t)$. Similarly,
we introduce a source $\eta^-(t)$ in the path integral going backwards in time.

Since we are interested in the dynamics of the system under
the effect of the reservoir we compute the reduced density matrix
\begin{equation}
\rho_{\sc red}(q'',q';t) = \int_{-\infty}^\infty  dx_a \; 
\langle q'' , x_a | \, \hat \rho_{\sc tot}(t) \, | q' , x_a \rangle 
\; .
\end{equation}

The initial density operator $\hat \rho_{\sc tot}(0)$ has the information 
about the initial state of the whole system. If one assumes that the 
system and the bath are set in contact at $t=0$, $\hat \rho_{\sc tot}(0)$
factorizes
%\begin{equation}
$
\hat \rho_{\sc tot}(0) = \hat \rho_{\sc syst}(0)  \hat \rho_{\sc env}(0)
$.
%\end{equation}
(Other initial preparations, where the factorization does not hold,
 can also be considered and may be more realistic in certain cases.)
If the environment is initially in equilibrium at an inverse 
temperature $\beta$, 
%\begin{equation}
$
\hat \rho_{\sc env}(0) = Z^{-1}_{\sc env} \exp(-\beta \hat H_{\sc env})
$.
%\end{equation}
For a bath made of harmonic oscillators 
the dependence on the bath variables is
quadratic and they can be traced  away 
to yield:
\begin{eqnarray*}
\rho_{\sc red}(q''; q'; t) =
\int_{-\infty}^\infty dQ \int_{-\infty}^\infty
dQ' \int_{q^+(0)=Q}^{q^+(t)=q''} {\cal D} q^+ 
\int_{q^-(0)=Q'}^{q^-(t)=q'} {\cal D} q^-
%\nonumber\\
%& & 
%\;\;\;\;\;\;\;\;\; \;\;\;\;\;\;\;\;\; 
e^{\frac{i}{\hbar} S_{\sc eff}} 
\langle Q | \,\hat \rho_{\sc syst}(0) \, | Q' \rangle
\label{red_quant}
\end{eqnarray*}
with
%\begin{equation}
$S_{\sc eff} = S_{\sc syst}^+ - S_{\sc syst}^- + S_{\sc th}$.
%\; .
%\end{equation}
The last term has been 
generated by the interaction with the environment and it 
reads
\begin{eqnarray}
 S_{\sc th} &=& 
- \int_0^t dt' \int_0^t dt'' \;
\left( q^+(t') - 
q^-(t') \right) \, \eta(t'-t'') \,  
\left( q^+(t'') + q^-(t'') \right) 
\nonumber\\
& & 
+i\int_0^t dt' \int_0^{t'} dt'' \; \left( q^+(t') - 
q^-(t') \right) \, \nu(t'-t'') \,  
\left( q^+(t'') - q^-(t'') \right) 
\; .
\label{Sthermal}
\end{eqnarray}
The noise  and dissipative kernels $\nu$ and $\eta$ are given by 
\begin{eqnarray}
\nu(t) &=&  \int_0^\infty d\omega I(\omega) \, 
\coth\left( \frac12 \beta \hbar \omega \right) \; 
\cos(\omega (t))
\; ,
\label{nu}
\\
\eta(t) &=& \theta(t) \frac{d\gamma(t)}{dt} =  
-\theta(t) \, 
\int_0^\infty d\omega \; I(\omega) \, \sin(\omega (t)) 
\; .
\label{eta}
\end{eqnarray}
In these equations, $I(\omega)$ is the spectral 
density of the bath, already defined in Eq.~(\ref{spectral}).
$\gamma$ is defined in the first identity and as we shall see
below it plays the same role as the friction kernel in classical
problems with colored noise.

Next, we have to choose an initial density matrix for the system.
One natural choice, having in mind the quenching experiments
usually performed in classical systems, is the diagonal 
matrix
%\begin{equation}
$
\langle Q | \; \hat \rho_{\sc syst}(0) \; | Q'\rangle = 
\delta (Q-Q')
$
%\label{initial-syst}
%\end{equation}
that corresponds to a random ``high-temperature'' situation and 
that allows one to simplify considerably $\rho_{\sc red}$:
\begin{eqnarray}
\rho_{\sc red}(q''; q'; t) =
\int_{-\infty}^\infty dQ \int_{q^+(0)=Q}^{q^+(t)=q''} 
{\cal D} q^+(t) \int_{q^-(0)=Q}^{q^-(t)=q'} {\cal D} q^-(t) 
e^{\frac{i}{\hbar} S_{\sc eff}}
\; .
\label{red_quant2}
\end{eqnarray}

Another representation of $\rho_{\sc red}$ 
is obtained by renaming the variables
\begin{eqnarray}
q \equiv \frac{1}{2} (q^+ + q^-)
\;\;\;\;\;\;\;\;\;\;
i\hat q \equiv \frac{1}{\hbar} (q^+ - q^-)
\; .
\end{eqnarray}
and rewriting the effective action as a function of $q$ and 
$i\hat q$. The new form is  
useful to establish contact with the generating functional for 
classical systems. The thermal part simply becomes
\begin{eqnarray*}
 S_{\sc th} = 
- \int_0^t dt' \int_0^t dt'' \;
\hbar i \hat q(t')
\, \eta(t'-t'') \,  2 q(t'')
+i\int_0^t dt' \int_0^{t'} dt'' \; 
\hbar i \hat q(t')
\, \nu(t'-t'') \,  \hbar i \hat q(t'')
\label{Sthermal2}
\end{eqnarray*}
as for a Langevin process in a colored noise.
When $\hbar\to 0$, the full effective action approaches the classical one 
as can be verified by expanding in powers of $\hbar$, 
and keeping only the leading terms.
If, moreover, the limit $\Lambda\to\infty$ is taken, the kernels $\gamma$ and 
$\nu$ become proportional to $\delta$ functions and one recovers a 
white noise. Keeping $\hbar>0$ the rotated version allows one to treat 
classical and quantum problems in parallel.

If one generalizes the above system to be one described by a field
$\phi_i$ with $i=1,N$ components, 
the symmetrized correlation function $C_{ij}(t,t')\equiv\frac{1}{2} \langle  \hat \phi_i(t) \hat \phi_j(t')
 + \hat \phi_j(t') \hat \phi_i(t) \rangle$ 
(where the hats represent operators) is given by
\begin{eqnarray}
C_{ij}(t,t')
 &=&
\left.
\frac{\hbar^2}{2} 
 \left[
\frac{\delta^2}{\delta \eta_i^+(t) \delta \eta_j^-(t')}
+
\frac{\delta^2}{\delta \eta_j^+(t') \delta \eta_i^-(t) }
\right]
\rho_{\sc red}
\right|_{\eta=0}
\; 
\end{eqnarray}
[$C_{ij}(t,t')=C_{ji}(t',t)$].
Using the Kubo formula (\ref{FDTasym}) 
the linear response function
can be expressed in terms of the 
averaged commutator, $R_{ij}(t,t')
= i/\hbar \theta(t-t')
\langle [\hat \phi_i(t),\hat \phi_j(t')] 
\rangle $ and  
\begin{eqnarray}
R_{ij}(t,t')
&=& 
\left.
\frac{\hbar}{i} 
 \left[
\frac{\delta^2}{\delta \eta_i^+(t) \delta \eta_j^+(t')}
+
\frac{\delta^2}{\delta \eta_j^+(t) \delta \eta_i^-(t')}
\right]
\rho_{\sc red}
\right|_{\eta=0}
\; .
\end{eqnarray}
(The hats represent operators.)
It is also useful to write the correlation and response 
in terms of the fields $\phi_i^+$, $\phi_i^-$ and 
their rotated counterparts
\begin{eqnarray}
2 C_{ij}(t,t') 
=
\langle \phi_i^+(t) \phi_j^-(t')  + \phi_j^+(t') \phi_i^-(t)  \rangle
= 
\langle \phi_i(t) \phi_j(t') + \phi_j(t') \phi_i(t) \rangle
\; ,
\;\;\;\;\;
&&
\label{defcorr}
\\
R_{ij}(t,t')
=
\frac{i}{\hbar} \, 
\langle \phi_i^+(t) \, \left( \phi_j^+(t') - \phi_j^-(t') 
\right) \rangle
=
\langle \phi_i(t) \, i\hat \phi_j(t') \rangle
\label{defresp}
\; .
\;\;\;\;\;\;\;\;\;\;\;\;\;\;\;\;\;\;\;\;\;\;\;\;
&&
\end{eqnarray}

\subsection{Average over disorder}

In general one is interested in the evolution of a model
in which the configuration
of disorder is typical. One could either attempt to solve the 
dynamics for one such disorder realization or one can 
assume that the behavior of a typical system is 
described by the averaged behavior over all systems, each
weighted with its probability. Since the former procedure is 
more difficult than the latter one usually studies the 
dynamics averaged over disorder and computes:
\begin{equation}
[\langle A(t)\rangle ]
=
\frac{\int dJ P(J) \int {\cal D} \phi {\cal D} i\hat \phi \;
A[\phi,i\hat \phi] \; e^{-S_{\sc eff}[\phi,i\hat\phi]}}
{\int dJ P(J) \int {\cal D} \phi \int {\cal D} i\hat \phi  
\; e^{-S_{\sc eff}[\phi,i\hat\phi]}}
\; .
\end{equation}
$J$ represents here the random exchanges in Eq.~(\ref{3DEA-hamil}). Similarly,
one can perform an average over a random potential in a problem as 
the one defined in Eq.~(\ref{manifold-hamil}).

One of the advantages of using a dynamic formalism is that 
when the initial conditions are uncorrelated with disorder
there is no need to use the replica trick to average over
disorder~\cite{cirano}. Indeed, the classical generating functional is 
constructed from a path integral that is identical to $1$ 
(and hence independent of disorder) in the absence of sources.
The same holds for the quantum Schwinger-Keldysh generating
functional, ${\mbox{Tr} \hat \rho_{\sc red}(0)}=1$, 
since we have chosen a diagonal density matrix
as the initial condition for the system. 
Thus, 
\begin{equation}
[\langle A(t)\rangle ]
=
\int dJ P(J) \int {\cal D} \phi {\cal D} i\hat \phi \;
A[\phi,i\hat \phi] \; e^{-S_{\sc eff}[\phi,i\hat\phi]}
\end{equation}
and these averages can be simply computed from 
$[{\cal Z}_J]$~\cite{comment-self}. 

If the initial condition {\it is} correlated with the random 
exchanges or the random potential, the situation 
is different. One such example is the study of the {\it equilibrium}
dynamics of a disordered model, {\it i.e.} the study of 
the evolution of initial conditions taken from $P_{\sc gb}$. 
In this case, the use of replicas to
average $\ln {\cal Z}$ is unavoidable and one is 
forced to treat replicated dynamic correlators. For classical models
this has been discussed in \cite{Young}. For quantum problems
the difficulty of the calculation increases since one needs
to work in a mixed real and imaginary time formalism~\cite{unpub}.
The initial density operator is a Boltzmann factor that is represented
with the Matsubara formalism while the real-time dynamics is written
with the Schwinger-Keldysh approach. Mixed correlators and responses
intervene in the dynamic equations.

\section{Dynamic equations}
\label{dynamicequations}
\setcounter{equation}{0}
\renewcommand{\theequation}{\thesection.\arabic{equation}}

In this Section we present three derivations of the dynamic 
equations for the macroscopic order parameters that use 
the classical or quantum dynamic generating functionals as 
starting points. Each method is better adapted for different
kinds of models.

\subsection{A useful derivation for fully-connected models}

\subsubsection{Classical systems}

Even if the use of the {\sc susy} notation is not necessary 
to derive the dynamic equations~\cite{Sozi2}, it is very useful in several 
aspects. Firstly, it allows to establish contact 
with the replicated version of the static partition function 
and the further study of this quantity; secondly, it is very useful as 
a bookkeeping tool; thirdly, it allows us to develop  more sophisticated 
techniques amenable to derive the dynamic equations of models without
fully connected interactions.
For all these reasons, we preferred to 
introduce the {\sc susy} formalism in 
Section~\ref{subsubsec:supersymmetry}
and use it here.

Since for classical models the use of white noises is rather 
generally justified we shall stick to this case. Moreover, we 
shall drop the inertial contribution to further simplify the 
presentation. We analyze here models with $N$ variables
$\vec \phi=(\phi_1,\dots,\phi_N)$ of the kind discussed in 
Section~\ref{interestingproblems}. 
In {\sc susy} notation 
the dynamical generating functional after setting the sources to zero 
reads 
\begin{equation} 
{\cal Z} = \int {\cal D} \Phi 
\; 
\exp 
\left[
- \int \;  da \; \left(
\frac{1}{2}   \sum_i^N   \Phi_i(a) \; ( -D_a^{(2)}-  \mu_s(a)) \; \Phi_i(a) +
V[\vec \Phi]  \right)
\right]
\; 
\end{equation}
with $\vec \Phi$ and $\mu_a$ two super-fields, the latter 
imposing the spherical
constraint ($\mu_s(a) = \mu(t) + \mbox{fermionic} + \hat \mu(t) \theta\overline\theta$, $\hat\mu(t)$ is a Lagrange multiplier that fixes the measure of 
integration and $\mu(t)$ enters the Langevin equation). 
Soft spins with their corresponding potential energy can be
studied in a similar way though their treatment is slightly more 
complicated. 
The potential energy of a rather generic fully connected disordered model 
can be expressed as a series expansion of the form
\begin{equation}
V[\Phi]  
= 
g \; \sum_{r \geq 0}^\infty \;  F_r 
\sum_{i_1 < \dots < i_{r+1} } J_{i_1  \dots  i_{r+1}} 
\; \Phi_{i_1} \dots \Phi_{i_{r+1}}
\; ,
\label{gen-pspin}
\end{equation}
For each $r$ the sum is taken over all possible groups of $r+1$ spins.
The fully-connected character of the model implies that there is 
no notion of distance or geometry. 
$J_{i_1  \dots  i_{r+1}}$ are random interactions taken from a 
Gaussian distribution with zero mean and variance 
%\begin{equation}
$[J_{i_1  \dots  i_{r+1}}^2 ] = (r+1)!/(2 N^r)$,
%\; .
%\end{equation}
just as in the model in Eq.~(\ref{pspsin-hamil})
Thus (\ref{gen-pspin}) is a Gaussian random 
potential with 
\begin{equation}
[V(\vec\Phi(a)) V(\vec\Phi(b))] = N g^2 \sum_{r\geq 0}^\infty F_r^2 
\left( \frac{\vec\Phi(a) \cdot \vec\Phi(b)}{N}\right)^{r+1} 
= N \, {\cal V}^\bullet \left( \frac{\vec\Phi(a) \cdot \vec\Phi(b)}{N}\right)
\; .
\label{pot-corr}
\end{equation} 
The scalar product in the second member is defined as 
$\vec\Phi(a) \cdot \vec\Phi(b)=\sum_i \Phi_i(a) \Phi_i(b)$.
The bullet means that the powers are taken locally in 
the super-coordinates $a$ and $b$ and they do 
not involve an operational product, see Appendix~\ref{app:Grassmann}.
The term $r=0$ corresponds to a random field linearly coupled to 
the spin, the term $r=1$ is quadratic in the fields while for 
$r\geq 2$ we obtain higher order interactions. If
$F_r=F_p\neq 0, p\geq 2$ and all other $F_r=0$ 
one recovers a spherical $p$ spin model. If two parameters are non-zero
one obtains a model with two $p$ spin terms.
The model of a 
particle in an infinite dimensional spherical 
random environment correlated as in (\ref{manifold-hamil}) also falls in this 
category if one can expand the correlator in a power series.

The disordered averaged generating functional reads
\begin{eqnarray}
[{\cal Z}] = \int {\cal D}\Phi 
\,
e^{- \int  da \; 
  \frac{1}{2} \sum_i \Phi_i(a) \; (-D_a^{(2)}- \mu_s(a)) \; \Phi_i(a)
+
 \frac{N}{2}  \int da db \;  
{\cal V}^{\bullet} \left( \frac{\vec\Phi(a) \cdot \vec\Phi(b) }{N} \right) 
} 
\; .
\end{eqnarray}
Introducing the order parameter
$
 Q(a,b)=N^{-1} \sum_{i=1}^N \Phi_i(a) \Phi_i(b)
$
through
\begin{equation} 
1 \propto \int {\cal D}Q {\cal D}i\tilde Q \, 
e^{- \frac{1}{2} \int da \, db 
\left(N  i\tilde Q(a,b)Q(a,b)  - i\tilde Q(a,b) 
\sum_{i=1}^N \Phi_i(a) \Phi_i(b) \right)}
\label{Q111}
\end{equation} 
yields
\begin{eqnarray}
& &
[{\cal Z}] = \int {\cal D}\Phi  {\cal D}Q {\cal D}i\tilde Q
\; \exp \left[ - \frac{1}{2} \int \;  da db\; \; 
\left(
N i\tilde Q(a,b) Q(a,b)  -
N {\cal V}^{\bullet}(Q(a,b))
\right.
\right. 
\nonumber \\
 & & 
\;\;\;\;\;\;\;\;\;\;\;\;\;\;\;\;\;\;
\left. \left.
 -   \sum_{i=1}^N 
\Phi_i(a) \;
 (  -D_a^{(2)}- \mu_s(a)) \;  \delta(a-b)  - 
i\tilde Q(a,b) )   \;                
 \Phi_i(b) 
\right) 
\right]
\; .
\label{gen1}
\end{eqnarray}
(Again we omit irrelevant normalization constants.)
Note that all terms in the exponent are order $N$ {\it if} the 
integrals yield finite contributions. We call the models for which this
is true ``mean-field'' since the saddle-point evaluation of the 
integral when $N\to\infty$ is exact without including fluctuations.
There is however a caveat in this reasoning that we discuss in 
Section~\ref{sompo}.
 
The saddle-point values for the 
Landau fields $Q$ are simply related to correlations 
of the original spins. Indeed, evaluating the generating function 
in Eq.~(\ref{gen1}) with a saddle-point approximation
\begin{equation}
0 = \left. \frac{\delta S}{\delta i \tilde Q(a,b)} \right|_{Q_{\sc sp}}
\;\; \Rightarrow \;\;
N Q_{\sc sp}(a,b) = 
\sum_{i=1}^N \langle \Phi_i(a) \Phi_i(b) \rangle_{\tilde {\cal Z}[Q]}
\; ,
\label{saddle1}
\end{equation}
where the average on the {\sc rhs} is taken with the
generating functional 
\begin{eqnarray}
\tilde {\cal Z}[Q] &\equiv& \int {\cal D}\Phi {\cal D}Q \; 
e^{\int da db \, \frac12 \left[
\sum_{i=1}^N 
\Phi_i(a)
(-D_a^{(2)}- \mu_s(a))  \delta(a-b)                 
 \Phi_i(b) + 
N {\cal V}^\bullet(Q(a,b)) \right]}
\; . 
\end{eqnarray}
Opening up the {\sc susy} notation
Eq.~(\ref{saddle1}) implies, as expected, 
\begin{eqnarray}
& &N C_{\sc sp}(t_1,t_2) = 
\sum_{i=1}^N \langle q_i(t_1) q_i(t_2) \rangle_{\tilde {\cal Z}[Q]}
\; ,
\;\;\;\;\;\;
N \hat Q_{\sc sp}(t_1,t_2) = 
\sum_{i=1}^N \langle i\hat q_i(t_1) i\hat q_i(t_2) \rangle_{\tilde {\cal Z}[Q]}
\; ,
\nonumber\\
& &N R_{\sc sp}(t_1,t_2) = 
\sum_{i=1}^N \langle q_i(t_1) i\hat q_i(t_2) \rangle_{\tilde {\cal Z}[Q]}
\; , 
\;\;\;\;\;
N R^\dag_{\sc sp}(t_1,t_2) = 
\sum_{i=1}^N \langle i\hat q_i(t_1) q_i(t_2) \rangle_{\tilde {\cal Z}[Q]}
\; .
\nonumber
\end{eqnarray}

Going back to Eq.~(\ref{gen1}) we can now shift $i\tilde Q$,
$
{\overline Q}\equiv (-D_a^{(2)}- \mu(t)) \; \delta(a-b) - i\tilde Q(a,b)
\label{shift}
$,
%\end{equation}
and integrate over $\Phi_i$
 \begin{eqnarray*}
[{\cal Z}] = 
\int  {\cal D}Q {\cal D}\overline Q \;
e^{
-  \frac{N}{2}  \int  da db\; 
\left[ 
  Q(a,b) \overline Q(a,b) 
+ (-D_a^{(2)}-\mu(t))\delta(a-b) \; Q(a,b) 
-
{\cal V}^{\bullet}(Q(a,b))
\right]}
e^{-\frac{N}{2}
{\mbox{Tr}} \, {\mbox{Ln}}\,  
\overline Q}            
\; . 
\end{eqnarray*}
Using a saddle-point evaluation, we eliminate $ \overline Q$,
$
[{\cal Z}] = \int {\cal D}Q \; \exp[- N S_{\sc eff}(Q)] 
$,
\begin{eqnarray}
2S_{\sc eff}(Q) =   
\int da db \left[  
  [-D_a^{(2)}- \mu_s(a)] \delta(a-b)\,  Q(a,b) - {\cal V}^{\bullet}(Q(a,b))
\right] -  
{\mbox {Tr}} {\mbox{Ln}} \, Q 
\; .
\label{action-pspin}
\end{eqnarray}
The saddle-point equation over $Q$, $\delta S_{\sc eff}/\delta Q=0$, yields
the dynamic equation 
\begin{equation}
( D_a^{(2)}+\mu(t)) {\bf \delta}(a-b)+Q^{-1}(a,b) + 
{{\cal V}^{\bullet}}'(Q(a,b)) = 0
\; ,
\label{MCeq}
\end{equation}
that takes a more convenient form after multiplying 
operationally by $Q$:
\begin{equation}
( D_a^{(2)}+ \mu(t)) Q(a,b)+\delta(a-b) + 
\int da' \; \Sigma(a,a') Q(a',b) = 0
\; ,
\label{MCeq2}
\end{equation}
with the self-energy defined as
\begin{equation}
\Sigma(a,b) \equiv {{\cal V}^{\bullet}}'(Q(a,b)) = 
g^2 \sum_{r\geq 0}^\infty F_r^2 (r+1) Q(a,b)^{\bullet \, r}
\; .
\label{self-energy-pspins}
\end{equation}
We have recasted the saddle-point dynamic equation in the form of a 
Schwinger-Dyson equation. 
The dynamic field is here a {\sc susy} correlator
that encodes the usual correlation function, the advance and 
retarded linear responses and 
the fourth correlator (that vanishes for causal problems):
\begin{eqnarray*}
&&
{\overline G}_o^{-1}(t) 
R(t,t') 
= \delta(t-t') + 
2\gamma \hat Q(t,t') 
%\nonumber\\
%& & 
%\;\;\;\;\;\;
+
\int dt'' \, [\Sigma(t,t'') R(t'',t')
+ D(t,t'') \hat Q(t''t,t')]
\; ,
\label{schwingerR}
\\
&&
%(M\partial^2_t +\gamma \partial_t + \mu(t)) 
{\overline G}_o^{-1}(t) 
C(t,t') 
= 2\gamma k_B T R(t',t) +
\int dt'' \, \Sigma(t,t'') C(t'',t') +
%\nonumber\\
%& &  
\int dt'' \, 
D(t,t'') R(t',t'')
\; ,
\label{schwingerC}
\\
&&
%(M\partial^2_t + \gamma \partial_t + \mu(t) )
{{\overline G}_o^{-1}}^\dag(t) 
R^\dag(t,t') 
= \delta(t-t') + \int_0^\infty dt'' \, \Sigma^\dag(t'',t) R(t',t'')
%\nonumber\\
%&&
+ \int dt'' \, \hat\Sigma(t,t'') C(t'',t')
\nonumber\\
&& \;\;\;\;\;\;\;\;\;\;\;\;\;\;\;\;\;
+ 2\hat\mu(t) C(t,t')
\; ,
\label{schwingerRdag}
\\
&&
%(M\partial^2_t + \gamma \partial_t + \mu(t)) 
{{\overline G}_o^{-1}}^\dag(t) 
\hat Q(t,t') 
= \int dt'' \, \Sigma^\dag(t,t'') \hat Q(t'',t') 
+
%\nonumber\\
%&&
\int dt'' \, \hat \Sigma(t,t'') R(t'',t') 
+ 2\hat\mu(t) R(t,t')
\; ,
\label{schwingerQ}
\end{eqnarray*}
with 
${\overline G}_o^{-1}(t) \equiv 
M\partial^2_t + \gamma \partial_t + \mu(t)$,
${{\overline G}_o^{-1}}^\dag(t) 
\equiv M\partial^2_t - \gamma \partial_t + \mu(t)$,
$\Sigma^\dag(t,t') = \Sigma(t',t)$
and 
\begin{eqnarray}
\Sigma(t,t'') &=& g^2 \sum_{r\geq 0} F_r^2 \, (r+1) r\, 
C^{r-1}(t,t'') R(t,t'')
\label{kernelSigma}
\\ 
D(t,t'') &=& 
g^2 \sum_{r\geq 0} F_r^2 \, (r+1) \, 
C^{r}(t,t'')
\\
\hat \Sigma(t,t'') &=& g^2 \sum_{r\geq 0} F_r^2 \, (r+1) r\, 
C^{r-1}(t,t'') \hat Q(t,t'')
\label{kernelSigmahat}
\; . 
\label{kernelD}
\end{eqnarray}
We set to zero all fermionic correlators.
We call the above integro-differential equations the Schwinger-Dyson 
equations for $R$, $C$ $R^\dag$ and $\hat Q$, respectively.

Causality can be used to simplify the four Schwinger-Dyson equations
%(\ref{schwingerR}) - (\ref{schwingerQ}) 
considerably. 
For $t' > t$ one has $R(t,t')=0$ while for  
$t > t'$ one has  $R(t',t)=0$. Rewriting the equations for 
$R$ and $R^\dag$ 
%Eqs.~(\ref{schwingerR})and (\ref{schwingerRdag}) 
with these two choices of times one easily sees that 
$\hat Q(t,t')=0$ for all $t$ and $t'$ (note that $\hat Q$ is 
symmetric in $t$ and $t'$) and $\mu_0(t)=0$
for all $t$. Thus, the equation for $\hat Q$
%Eq.~(\ref{schwingerQ})
 vanishes identically when 
causality holds. In the following we search for causal solutions 
and we work with their simplified version.
We loose in this way the possibility of finding solutions that
break causality which are related to instantons~\cite{Zinn}. 
We shall come 
back to this point later. If we focus on the case
$t > t'$ the dynamic equations simplify to
\begin{eqnarray}
\overline G_o^{-1}(t) R(t,t') &=&
%M \partial_{t^2} R(t,t') + \gamma \partial_t R(t,t') = - \mu(t) R(t,t') 
\int_{t'}^t dt'' \; 
\Sigma(t,t'') R(t'',t')
\; ,
\label{eqR2} 
\\
\overline G_o^{-1}(t) C(t,t') &=&
%M \partial_{t^2} C(t,t') + \gamma \partial_t C(t,t') =  - \mu(t) C(t,t') 
 \int_0^{t'} dt'' \; D(t,t'') R(t',t'') 
%\nonumber\\
%&& \;\;\;\;\;\;\;\;\;\;\;\;\;\;\;\;\;\;\;\;\;\;\;\;\;\;\;\;
+\int_0^t dt'' \;  \Sigma(t,t'') C(t',t'') 
\; .
\label{eqC2} 
\end{eqnarray}
In their integrated form they read
\begin{eqnarray}
R(t,t') &=& G_o(t,t') + \int_{t'}^t dt'' \int_{t'}^{t''} dt''' \;
G_o(t,t'') \, \Sigma(t'',t''') \, R(t''',t')
\; ,
\label{intschwingerR}
\\
C(t,t') &=& \int_{0}^t dt'' \int_{0}^{t'} dt''' \;
R(t,t'') \, D(t'',t''') \, R(t',t''')
\label{intschwingerC}
\; ,
\end{eqnarray}
with the propagator given by
%\begin{equation}
$G_o^{-1}(t,t') \equiv \delta(t-t') \overline G_o^{-1}(t)
%\left( M\partial^2_{t} + \gamma \partial_t +
%\mu(t) \right)
$.
%\; .
%\label{freepropagator}
%\end{equation}

The equation for $\mu(t)$ can be derived
from the Schwinger-Dyson equation 
by imposing the spherical constraint through the evaluation 
at $t=t'$.
Multiplying operationally by $G_o^{-1}$ one obtains
\begin{eqnarray}
\mu(t)
&=&
\int_0^t dt'' \left[
\Sigma(t,t'') C(t,t'') + D(t,t'') R(t,t'')
\right]
\nonumber\\
& &
+ M \int_0^t dt'' \int_0^t  dt''' \,
 (\partial_t R(t,t'') ) \,D(t'',t''')\, (\partial_t R(t,t''') )
\\
\nonumber & & + \left. M^2 \left[ \partial_t R(t,s) \partial^2_{s t} C(s,t) -
\partial^2_{s t} R(t,s) \partial_{t'} C(s,t')
\right]\right|_{
s \to 0\,
t \to t'}
\; .
\label{mut1}
\end{eqnarray}
The last two terms are a consequence of having a kinetic term with 
second derivatives. 
It can be easily identified with minus the second-derivative of the 
correlation at equal times by taking the limit $t'\to t^-$ in 
Eq.~(\ref{eqC2}).
Thus
\begin{equation}
\mu(t) 
=
\int_0^t dt'' \left[
\Sigma(t,t'') C(t,t'') + D(t,t'') R(t,t'') 
\right] 
-
M
\left.
\frac{\partial^2 }{\partial t^2} C(t,t') 
\right|_{t'\to t^-}
\; .
\label{mueq}
\end{equation}

One way of deriving  the equation for $\mu(t)$ 
for a Langevin process 
with {\it white noise and no inertia} goes as follows. 
Considering $t>t'$ in the complete Schwinger-Dyson equation for $C$
and taking $t'\to t^-$, and considering 
$t<t'$ in the same equation and taking $t'\to t^+$, one finds
\begin{equation}
\lim_{t'\to t^-} \partial_t C(t,t') =  
\lim_{t'\to t^+} \partial_t C(t,t') - 2 k_B T 
\label{discC}
\end{equation}
where we used $R(t,t'\to t^-) =1/\gamma$.
The derivative of $C$ has a cusp at $t=t'$. The symmetry of the 
correlation function about $t=t'$ implies 
$C(t'+\delta,t')=C(t'-\delta,t')$ and an 
expansion up to first order 
in $\delta$ implies $\lim_{t'\to t^-} \partial_t C(t,t') =  
-\lim_{t'\to t^+} \partial_t C(t',t)$. From Eq.~(\ref{discC}) one has 
$\lim_{t'\to t^-} \partial_t C(t,t') =  - k_B T$. 
Now, one rewrites the complete equation for $C$ exchanging
$t$ and $t'$ and adds this equation to the same equation 
in the limit $t'\to t^-$:
%\begin{eqnarray}
%&&
$
\gamma \lim_{t'\to t^-} \left( \partial_t C(t,t') + 
\partial_{t'} C(t,t') \right) =
-2 \mu(t) 
%\nonumber\\
%&&
%\;\;\;\;\;\;\;\;\;\;\;
%\;\;\;\;\;\;\;\;\;\;\;
+ 
\lim_{t'\to t^-} [
\mbox{rhs eq. for} \; C $$ +
\mbox{rhs eq for} \; C (t' \leftrightarrow t) ]
$.
%\; .
%\label{intermediate-eq}
%\end{eqnarray}
From the discussion above
the {\sc lhs} vanishes and  
%Since $\lim_{t'\to t^-} R(t,t') = 1/\gamma$, 
the {\sc rhs} implies 
\begin{equation}
\mu(t) = k_B T + \int_0^\infty dt'' \left[ \Sigma(t,t'') C(t,t'') 
+ D(t,t'') R(t,t'') \right]
\; .
\label{zt}
\end{equation}

For the spherical $p$ spin model $\mu(t)$ is simply related to 
the energy density ${\cal E}(t)$. Indeed, take the Langevin 
equation evaluated at time $t$,
multiply it by $s_i(t')$, sum over all sites, average over the noise 
and take the limit $t'\to t$. Repeat this
procedure with the Langevin equation evaluated at $t'$ and multiplying 
by $s_i(t)$.  Adding the resulting equations and using  
$ N^{-1}\sum_{i=1}^N \langle  s_i(t) \xi_i(t') \rangle = 
2 \gamma k_B T R(t,t')$
(see Appendix~\ref{app:relations}) we have
%\begin{eqnarray*}
$
\mu(t) = 
-\lim_{t'\to t^-} 
\left\langle \sum_i 
\frac{\delta H_J(\vec s(t))}{\delta s_i(t) } s_i(t') 
\right\rangle
+ k_B T 
$
%\end{eqnarray*}
that for the spherical $p$ spin model becomes
\begin{eqnarray}
\mu(t) = 
- p {\cal E}(t) + k_B T 
\; .
\label{energy-mu}
\end{eqnarray}

%Interestingly enough, the expansion 
%of $R$ and $\partial_t C$ around equal times, $t\to t'^+$, yields
%\begin{equation}
%R(t,t') = -\frac1{k_B T} \partial_t C(t,t') \;\;\;\; \mbox{when} \;\; t'\to t^-
%\end{equation}
%and, close to equal times, {\sc fdt} holds for all temperatures as 
%predicted by the proof in Section~\ref{subsec:bound}.

Thanks to the mean-field character of the model the action is proportional 
to $N$ and the saddle-point evaluation is exact when $N\to\infty$. 
For the fully connected models considered in this Section 
the self-energy is given by a rather simple function of the  
interactions. In Section~\ref{subsec:dilute} we present
a more powerful method that allows us to derive a similar equation
for dilute (as opposed to fully connected) disordered models.
For finite
dimensional problems none of these procedures are exact.
An effective action in terms of {\it local} order parameters 
$Q_i(a,b)$ can be written but the evaluation of the generating functional by 
saddle-point has to include fluctuations~\cite{Sozi3,Claudio0}.

\subsubsection{Quantum models}
\label{subsubsec:quantum-dyneqs}

The similarity between the effective action for classical and quantum 
models can be exploited to derive the dynamic equations of a quantum
system in a very similar manner to what we have just done for classical
models~\cite{Culo}. Even if the {\sc susy} notation is not useful for 
quantum problems we can still use a compact notation. 
We first encode the variables $q,i\hat q$ in a vector. 
The quadratic terms in the action can be condensed into one term 
by introducing the operator
\begin{eqnarray}
{\cal O}p(t,t') 
&=&
\left(
\begin{array}{rl}
\mbox{Op}^{++}(t,t') \;& \; \mbox{Op}^{+-}(t,t') 
\nonumber\\
\mbox{Op}^{-+}(t,t') \; & \; \mbox{Op}^{--}(t,t') 
\end{array}
\right)
=
\{ \mbox{Op}^{\alpha\beta} (t,t')\}
\; ,
\nonumber\\
\;
\nonumber\\
\mbox{Op}^{++}(t,t') &=&
(M \partial^2_t + \mu^+(t)) \, \delta(t-t') - 2 i \nu(t-t') 
\nonumber\\
\mbox{Op}^{+-}(t,t') &=&
2 \eta(t-t') + 2 i \nu(t-t') 
\nonumber\\
\mbox{Op}^{-+}(t,t') &=&
- 2 \eta(t-t') + 2 i \nu(t-t') 
\nonumber\\
\mbox{Op}^{--}(t,t') &=&
-(M \partial^2_t + z^-(t)) \, \delta(t-t') - 2 i \nu(t-t') 
\; 
\end{eqnarray}
in such a way that
\begin{equation}
S_{\sc eff}[{\vec \phi}^+, {\vec \phi}^-]
= 
-\frac12 
\int dt  \int dt' \;
{\vec \phi}^\alpha(t) \,
\mbox{Op}^{\alpha\beta}(t,t') {\vec \phi}^\beta(t') 
- \int dt V[{\vec \phi}^+] + \int dt V[ {\vec \phi}^-]
\end{equation}
where Greek indices label $+,-$ and 
the sum convention is assumed.

Introducing the order parameter $Q^{\alpha\beta}(t,t')= N^{-1} \sum_i
\phi_i^\alpha(t) \phi_i^\beta(t')$ via the identity 
\begin{eqnarray}
1 
%&=& 
%\int \prod^{\alpha\beta} DQ^{\alpha\beta} \;
%     \delta\left( \frac{1}{N}{\vec \phi}^\alpha(t) {\vec \phi}^\beta(t') - 
%Q^{\alpha\beta}(t,t') \right)
%\nonumber\\
\propto 
\int \prod_{\alpha\beta} {\cal D}Q^{\alpha\beta} \; 
{\cal D}i \hat Q^{\alpha\beta} 
e^{- \frac{1}{2\hbar} \, i \hat Q^{\alpha\beta} \, 
\left( {\vec \phi}^\alpha(t) {\vec \phi}^\beta(t')  - 
N Q^{\alpha\beta}(t,t') \right)}
\; ,
\end{eqnarray}
the full action can be rewritten as
\begin{eqnarray}
S_{\sc eff}[{\vec \phi}^+,{\vec \phi}^-]
&=& 
- \frac{1}{2}\int dt  \int dt' \; 
{\vec \phi^\alpha}(t) \,
\left( \, \mbox{Op}^{\alpha\beta}(t,t') 
+  i\hat Q^{\alpha\beta}(t,t') \right)
{\vec \phi}^\beta(t')  
\nonumber\\
& & 
+ \frac{N}{2} \int dt \int dt' \hat Q^{\alpha\beta}(t,t') \, 
Q^{\alpha\beta}(t,t')
+ \frac{N}{2}  \int dt \left( \mu^+(t) - \mu^-(t) \right) 
\nonumber\\
& & 
%+ \frac{i {\tilde J}^2 N}{4 \hbar} \int dt \int dt' 
%\left[ \,
%\left(Q^{++}(t,t') \right)^p +  \left( Q^{--}(t,t') \right)^p
%\right.
%\nonumber\\
%& & 
%\;\;\;\;\;\;\;\;\;\;\;\;\;\;\;\;\;\;\;\;
%-
%\left.
%\left( Q^{+-}(t,t') \right)^p -  \left( Q^{-+}(t,t') \right)^p
%\, \right]
+ \int dt V[\vec \phi^+] - \int dt V[\vec \phi^-] 
\; .
\end{eqnarray}
The stationary-point values of $Q^{\alpha\beta}(t,t')$ 
are related to 
the ``physical'' correlations and responses defined in Eqs.~(\ref{defcorr})  
and (\ref{defresp}) as follows
\begin{eqnarray}
N Q^{++}(t,t') = \overline{\langle {\vec \phi}^{+}(t)  {\vec \phi}^{+}(t') 
\rangle}
=
N \left( C(t,t') - \frac{i\hbar}{2} \; (R(t,t') + R(t',t)) \right)
\; ,
\label{Q++}
\\
N Q^{+-}(t,t') = \overline{ \langle {\vec \phi}^{+}(t)  {\vec 
\phi}^{-}(t') \rangle}
=
N \left( C(t,t') + \frac{i\hbar}{2} \; (R(t,t') - R(t',t)) \right)
\; ,
\label{Q+-}
\\
N Q^{-+}(t,t') = \overline{\langle {\vec \phi}^{-}(t)  {\vec \phi}^{+}(t') 
\rangle}
=
N \left( C(t,t') - \frac{i\hbar}{2} \; (R(t,t') - R(t',t)) \right)
\; ,
\label{Q-+}
\\
N Q^{--}(t,t') = \overline{\langle {\vec \phi}^{-}(t)  {\vec \phi}^{-}(t') 
\rangle}
=
N \left( C(t,t') + \frac{i\hbar}{2} \; (R(t,t') + R(t',t)) \right)
\; ,
\label{Q--}
\end{eqnarray}
with 
%\begin{eqnarray}
$N C(t,t') = \sum_i C_{ii}(t,t')$ and 
$N R(t,t') = \sum_i R_{ii}(t,t')$. 
%&\equiv& \frac{1}{2} 
%\sum_{i=1}^N 
%[\langle \phi_i^+(t) \phi_i^-(t') + \phi_i^-(t) \phi_i^+(t') \rangle]
%\; ,
%\nonumber\\
%N R(t,t') &\equiv& \frac{i}{\hbar} 
%\sum_{i=1}^N [\langle \phi_i^+(t) 
%\left( \phi_i^+(t') -\phi_i^-(t') \right) \rangle]
%\; .
%\end{eqnarray} 
It is easy to check that these functions  satisfy the identity
%\begin{equation}
$
Q^{++}(t,t') +Q^{--}(t,t') -Q^{+-}(t,t')-Q^{-+}(t,t') =0
$.
%\label{identityQo}
%\; .
%\end{equation}
At the classical level this identity reduces to the 
condition $\langle i \hat {\vec s}(t)  i \hat {\vec s}(t')  \rangle = 0$ 
for all pairs of times $t,t'$. In what follows we do not break this 
identity and hence do not allow for solutions that break causality.

The functional  integration over $\phi_i^+(t)$ and $\phi_i^-(t)$ is now 
quadratic and can be performed.
Symmetrizing the operator Op with respect to the greek indices
and times the integral over the fields $\phi^+$ and $\phi^-$ 
amounts to replacing the quadratic term in $i/\hbar S_{\sc eff}$ by
\begin{eqnarray}
-\frac{N}{2}
\int dt  \int dt'
\, \mbox{Tr} \log 
\left( \, \frac{i}{\hbar} \mbox{Op}_{\sc symm}^{\alpha\beta}(t,t') + 
\frac{i}{\hbar} i \hat Q^{\alpha\beta}(t,t') \right)
\; .
\end{eqnarray}

At this stage, all terms in the action depend upon the 
``macroscopic'' quantities 
$i \hat Q^{\alpha\beta}, Q^{\alpha\beta}$ 
and $\mu^\alpha$ and are proportional to 
$N$. 
Since it is easier to write the equations in matrix notation, we encode 
$i \hat Q^{\alpha\beta}$ and 
$Q^{\alpha\beta}$ in two matrices
\begin{eqnarray}
i \hat {\cal Q} = \left( 
\begin{array}{c}
i \hat Q^{++} \;\; i \hat Q^{+-}
\nonumber\\
i \hat Q^{-+} \;\; i \hat Q^{--}
\end{array}
\right)
& \;\;\;\;\;\;\;\;\;\;\;\;\;\;\;\;\;\;\;\;\;\;
{\cal Q} = \left( 
\begin{array}{c}
Q^{++} \;\; Q^{+-}
\nonumber\\
Q^{-+} \;\; Q^{--}
\end{array}
\right)
\; 
\end{eqnarray}
and we define
\begin{eqnarray}
F[{\cal Q}](t,t') 
&\equiv&
g^2 \sum_{r\geq 0} F_r^2 \, (r+1)
\left(
\begin{array}{rl}
(Q^{++}(t,t'))^{r} 
\; &  \;
-(Q^{+-}(t,t'))^{r}
\\
-(Q^{-+}(t,t'))^{r}
\; &  \;
(Q^{--}(t,t'))^{r}
\end{array}
\right)
\label{saddleF}
\; ,
\end{eqnarray}
the saddle-point with respect to $i \hat Q^{\alpha\beta}(t,t')$ 
yields
%\begin{equation}
$i \hat {\cal Q}(t,t') 
= \frac{\hbar}{i} {{\cal Q}^{-1}}(t,t') - {\cal O}p(t,t')
$.
%\; .
%\label{saddlelambda}
%\end{equation}
The matrix and time-operator inverse of  ${\cal Q}$ is denoted 
${\cal Q}^{-1}$. 
The saddle-point equation with respect to $Q^{\alpha\beta}(t,t')$ yields
%\begin{equation}
$
i\hat {\cal Q}(t,t') = 
-\frac{i \, 
%{\tilde J}^2
}{2\hbar} \, F[{\cal Q}](t,t')
$.
%\; .
%\label{saddleq}
%\end{equation}
%Equations~(\ref{saddlelambda}) and (\ref{saddleq}) 
These saddle-point equations
imply, in a compact
matrix and time-operator  notation, 
\begin{equation}
\frac{i}{\hbar} \mbox{Op}_{\sc symm} \otimes {\cal Q}  
= I  - \frac{
%{\tilde J}^2
1}{2\hbar^2} \, F[{\cal Q}] \otimes {\cal Q}
\; ,
\label{saddleQ}
\end{equation}
where $I$ is the identity: 
$I^{\alpha\beta}(t,t')= \delta^{\alpha\beta} \delta(t-t')$
and 
we denote with a cross the standard operational product
in matrix and time (see Appendix~\ref{app:Grassmann}).
%\begin{eqnarray}
%{\cal A} \otimes {\cal B} (t,t')
%&=&
%\left(
%\begin{array}{c}
%\int dt'' \, A^{+\gamma}(t,t'') B^{\gamma +}(t'',t') 
%\;\;\;\;\;\;
%\int dt'' \, A^{+\gamma}(t,t'') B^{\gamma -}(t'',t') 
%\nonumber\\
%\int dt'' \, A^{-\gamma}(t,t'') B^{\gamma +}(t'',t') 
%\;\;\;\;\;\;
%\int dt'' \, A^{-\gamma}(t,t'') B^{\gamma -}(t'',t') 
%\end{array}
%\right)
%\; 
%\end{eqnarray}
%where a sum over $\gamma$ is assumed.
The saddle-point with respect to $\mu^\alpha$ yields
\begin{eqnarray}
i/\hbar &=& 
({\cal O}p + i \hat {\cal Q})^{-1}_{++}(t,t) = 
i/\hbar Q^{++}(t,t)
\; ,
\label{z+}
%\\
%\frac{i}{\hbar} &=& 
%({\cal O}p + i\hat {\cal Q})^{-1}_{--}(t,t)= \frac{i}{\hbar} Q^{--}(t,t)
%\label{z-}
\end{eqnarray}
and similarly for $Q^{--}$. These equations lead 
to the spherical constraint.

The dynamic equations for the auto-correlation and response 
follow from the set of equations (\ref{saddleF})-(\ref{saddleQ})
and the definitions of the dynamic order parameters given in 
Eqs.~(\ref{Q++})-(\ref{Q--}). 
More precisely, the equation of motion for the response function follows from 
the subtraction of the $++$ and $+-$ components of Eq.~(\ref{saddleQ}):
\begin{eqnarray}
& & 
\left( M \partial^2_t + \mu^+(t) \right) R(t,t') + 
4 \int_{t'}^t dt'' \; \eta(t-t'') \, R(t'',t') 
\; = \;\delta(t-t') 
\;\;\;\;\;\;\;\;\;\;\;\;\;\;\;\;\;\;\;\;\;\;
\nonumber\\
& & 
- \frac{g^2}{2 i \hbar} \,
\sum_{r\geq 0} F_r^2 \, (r+1)  
\int_0^\infty dt'' 
\left[
(Q^{++}(t,t''))^{r} - (Q^{+-}(t,t'')^{r}) 
\right] \, 
R(t'',t')
\; ,
%\nonumber\\
\label{eqR1}
\end{eqnarray} 
and the equation of motion for the correlation
follows from the addition of the  $+-$ and 
$-+$ components of Eq.~(\ref{saddleQ}):
\begin{eqnarray}
& & 
\left[ M \partial^2_t + \frac12 \, \left(\mu^+(t) + \mu^-(t) \right) \right]
C(t,t') 
+ 
\frac{i}{2} \left(z^+(t) - z^-(t) \right) \hbar (R(t',t)-R(t,t') )
\;\;\;\;\;\;\;\;\;\;\;
\nonumber\\ 
& & 
-2\hbar \int_0^\infty dt'' \nu(t-t'') R(t',t'') +
4 \int_{0}^t dt'' \; \eta(t-t'') \, C(t'',t') 
\nonumber\\
& & 
=
-\frac{g^2}{2\hbar} \, 
\sum_{r\geq 0} F_r^2 \, (r+1) \mbox{Im}
\int_o^\infty dt'' \;
 \left[
(Q^{++}(t,t''))^{r} Q^{+-}(t'',t') 
\right.
\nonumber\\
&&
\left.
- (Q^{+-}(t,t''))^{r}Q^{--}(t'',t')
 \right]
%\nonumber\\
\label{eqC1}
\; .
\end{eqnarray}
Written in this way, 
Eq. (\ref{eqC1}) is complex. Its imaginary part yields 
%\begin{equation}
$\mu(t) \equiv \mu^+(t)=\mu^-(t)$
%\; .
%\end{equation}
Moreover, since the response is causal, products of advanced $R(t,t'')$ and 
retarded $R(t'',t')$ responses vanish identically for all $t,t''$: 
%\begin{equation}
$R(t,t'')R(t'',t) =0$,
%\;\;\;\;\;\;\; 
$\forall \; t,t''$
%\end{equation}
and one can show that for any integer $k >0$ and any constants
$c_1$, $c_2$, 
%\begin{eqnarray}
$\left[ C(t,t') + c_1 R(t,t') + c_2 R(t',t) \right] ^k =
\left[ C(t,t') + c_1 R(t,t')  \right]^k 
$$+
\left[ C(t,t') + c_2 R(t',t) \right]^k
$
\linebreak
$
%\nonumber\\
- C^k(t,t')
$.
%\; .
%\end{eqnarray}
Using this property one has
%\begin{equation}
$
(Q^{++}(t,t''))^{r} - (Q^{+-}(t,t''))^{r} =
2i \mbox{Im} [ C(t,t'')-\frac{i \hbar}{2} R(t,t'')]^{r}
$
%\end{equation}
and 
%\begin{eqnarray}
$
\mbox{Im}[(Q^{++}(t,t''))^{r} Q^{+-}(t'',t') - 
(Q^{+-}(t,t''))^{r}Q^{--}(t'',t')]
= 
$$
%\nonumber\\
%& & 
2 C(t'',t') 
\linebreak
\times 
\mbox{Im} [ C(t,t'')-\frac{i \hbar}{2} R(t,t'') ]^{r}
-
$$ 
%\nonumber\\
%& & 
\hbar R(t',t'') 
\mbox{Re}[ C(t,t'')-\frac{i\hbar}{2}( R(t,t'')+ R(t'',t))]^{r}
$.
%\; .
%\end{eqnarray}
We can identify the self-energy $\tilde \Sigma$ and the vertex $\tilde D$
as
\begin{eqnarray}
&& 
\Sigma(t,t')
+ 4 \eta(t-t') \equiv \tilde{\Sigma}(t,t')
\;\;\;\;\;\;
D(t,t')- 2 \hbar \nu(t-t') \equiv
\tilde{D}(t,t')
\nonumber\\
%\end{equation}
%with 
%\begin{eqnarray}
&&
%\Sigma(t,t')
%+ 4 \eta(t-t') \equiv 
\tilde{\Sigma}(t,t') =
-\frac{g^2}{\hbar} 
\sum_{r\geq 0} F_r^2 (r+1)
\mbox{Im} [ C(t,t')-(i \hbar)/2 R(t,t')]^{r}
\; ,
\label{sigmatilde}
\\
&&
%D(t,t')- 2 \hbar \nu(t-t') \equiv
\tilde{D}(t,t') =
\frac{g^2}{2}
\sum_{r\geq 0} F_r^2 (r+1)
\mbox{Re}[C(t,t')-(i\hbar)/2( R(t,t')+ R(t',t)) ]^{r}
\; .
\label{Dtilde}
\end{eqnarray}
%For $t\geq t'$, they can be encoded in a single complex equation:
%\begin{equation}
%\tilde D(t,t') + \frac{i \hbar}{2} \, \tilde \Sigma(t,t') 
%=
%\frac{p 
%{\tilde J}^2
%}2 \; \left( C(t,t') +  \frac{i \hbar}{2} \,R(t,t') 
%\right)^{r}
%\; .
%\end{equation}
Note that the total self-energy $\Sigma$ and vertex $D$ are real
and have two contributions of different origin: one arises from the interaction of the 
system and the bath ($\eta$ and $\nu$) and one is caused by the 
non-linearities stemming from the average over disorder
($\tilde \Sigma$ and $\tilde D$). 
%(that we 
%called $\tilde \Sigma$ and $\tilde D$ in Eqs.~(\ref{sigmatilde}) and 
%(\ref{Dtilde})). 

The dynamic equations can then be written in the  compact form 
(\ref{schwingerR}) - (\ref{schwingerQ}). 
It is important to realize that the self-energy $\tilde \Sigma(t,t')$
is proportional to the response function $R(t,t')$, which
in turns implies
%\begin{equation}
$\tilde \Sigma(t,t')= \Sigma(t,t')=0$
for 
$t<t'$.
%\; .
%\end{equation}
This means that the upper limit of integration in Eqs.~(\ref{schwingerR}) and
(\ref{schwingerC}) is $t$, which renders the equations explicitly
causal.
There are no more independent equations for $R$ and $C$. 
The other two equations that can be obtained 
from Eq.~(\ref{saddleQ}) are the 
equation for $R(t',t)$, that is equivalent to 
Eq.~(\ref{schwingerR}), and one equation that identically 
cancels by virtue of the identity between two-point correlators. 

Real and imaginary parts of 
Eqs.~(\ref{z+}) and the one for $Q^{--}$ combined with  
the saddle-point equation for $\lambda$ 
%Eq.~(\ref{saddlelambda}) 
imply the equal-times conditions 
%\begin{eqnarray}
$C(t,t) = 1, \; 
%\; , \;\;\;\;\;
R(t,t) =0$.
%\; .
%\label{diagonal}
%\end{eqnarray}
In addition, from Eq.~(\ref{eqR1}) one obtains that 
the first derivative of the 
response function is discontinuous at equal times:
\begin{eqnarray}
\lim_{t'\to t^-} \partial_t R(t,t') = \frac{1}{M}  
\; ,
\;\;\;\;\;\;
\lim_{t'\to t^+} \partial_t R(t,t') = 0 
\; ,
\label{next-to-diagonal-R}
\end{eqnarray}
while from Eq. (\ref{eqC1}) one obtains that the 
first derivative of the correlation is continuous:
\begin{equation}
\lim_{t'\to t^-} \partial_t C(t,t') = \lim_{t'\to t^+} \partial_t C(t,t') = 0 
\; .
\label{next-to-diagonal-C}
\end{equation}

In conclusion, Eqs.~(\ref{schwingerR}), (\ref{schwingerC}) and (\ref{mueq}) 
are the complete set of equations that determines the dynamics of the 
system.

\subsection{Beyond fully-connected models}
\label{subsec:dilute}

\subsubsection{Classical models}

A very useful formalism to study the {\it statics} of classical
dilute disordered models with the replica trick has been 
introduced by Monasson~\cite{Mo} generalising the previous work of 
Mottishaw and de Dominicis~\cite{ModeDo}. The parallel between the 
static calculation using 
replicas and the dynamic formalism, once expressed in terms 
of superfields, exists also at the level of this approach. 
The presentation in this Section 
follows 
very closely the one in \cite{Secu} where 
the dynamic formalism apt to 
analyze {\it dilute} disordered models was introduced.

The dilute spin-glass, or Viana-Bray model in its Ising version, 
is a spin model defined on a random graph with average connectivity
$\tilde c$ and pair random exchanges 
taken from the probability distribution 
%\begin{equation}
$P(J_{ij}) = (1- \tilde c/N) \, \delta(J_{ij})
+ \tilde c/N [ \frac12 \delta(J_{ij} - J) + 
\frac12 \delta(J_{ij} + J) ] $.
%\label{diluteP}
%\end{equation}
For the sake of simplicity we consider the spherical version of 
this model.
Let us define $c(\Phi)$ as the fraction of sites with super-field
$\Phi_i$ identical to a chosen value $\Phi$
\begin{equation}
c(\Phi) \equiv \frac1N \sum_{i=1}^N \prod_a \delta(\Phi(a) -\Phi_i(a))
\; ,
\label{cPhi-def}
\end{equation}
for all values of the super-coordinate $a$. 
Note that $\int {\cal D}\Phi \, c(\Phi)
%=
%\frac{1}{N} \sum_i \prod_a \int d\Phi(a) \delta(\Phi(a)-\Phi_i(a))
=1$.
Similarly to what we have done when introducing $Q(a,b)$ in Eq.~(\ref{Q111})
we enforce  the
 definition of $c(\Phi)$ in the generating functional by 
introducing an identity in its path integral representation: 
%\begin{eqnarray}
$
1 \propto 
\int {\cal D} c {\cal D} i\hat c \;  
\exp[\int {\cal D}\Phi \, i \hat c(\Phi) 
( N c(\Phi) - \sum_{i=1}^N \prod_a 
\delta(\Phi - \Phi_i) )]
=
\int {\cal D}c \, {\cal D}i\hat c \; 
\exp[\int {\cal D}\Phi N i \hat c(\Phi) c(\Phi) -
\sum_{i=1}^N i \hat c(\Phi_i)]
$.
%\; .
%\end{eqnarray}
We obtain
\begin{eqnarray*}
&&{\cal Z} = \int {\cal D}\Phi_i \, {\cal D}c \,{\cal D}i\hat c \;
\exp\left( \int {\cal D}\Phi \; N i\hat c(\Phi) c(\Phi) -
\sum_{i=1}^N i\hat c(\Phi_i) 
\right.
\nonumber\\
& & 
\;\;\;\;\;\;
\left. 
+\frac{N}2 \int {\cal D}\Phi \;
c(\Phi) \int da \, \Phi(a) (-D_a^{(2)}-\mu_s(a)) \Phi(a) 
- \int da \; V[\Phi(a)]
\right) 
\; .
\end{eqnarray*}
$V[\Phi(a)]=\sum_{ij} J_{ij} \Phi_i(a) \Phi_j(b)$.
Once written in this form, 
the average over disorder can be simply performed 
and the disordered averaged generating functional becomes
$
[{\cal Z}] = 
\int {\cal D}c {\cal D}i \hat c {\cal D}\Phi_i \; \exp\left(-N G\right) 
$
with 
\begin{eqnarray*}
G &=& 
- \int d\Phi \; i \hat c(\Phi) c(\Phi)
+\frac{1}{N} \sum_{i=1}^N i \hat c(\Phi_i) 
+\frac12 \int d\Phi \; c(\Phi) 
\int da \, \Phi(a) (D_a^{(2)}+\mu_s(a)) \Phi(a) 
\nonumber\\
&&
+H_{\sc eff}
\end{eqnarray*} 
and 
%\begin{equation}
$\exp\left(-N H_{\sc eff}\right) = 
\left[ \exp\left(\int da  \; V[\Phi(a)]\right) \right]$
%\; .
%\end{equation}
With the notation $N H_{\sc eff}$ we suggest that 
$H_{\sc eff}$  is of order one. We shall discuss this very important 
issue below.

The second term in $G$ is now the only term where
the $\Phi_i$'s appear and all of them contribute in exactly the 
same form. One can then replace the sum over $i$ by a factor 
$N$ times $\hat c(\Phi')$ and exponentiate the functional integral 
over the representative super-field $\Phi'$,
%\begin{equation}
$
\int {\cal D}\Phi_i \exp\left(-\sum_i i \hat c(\Phi_i) \right) = 
\exp\left(N \ln \int {\cal D}\Phi' \exp\left(-i\hat c(\Phi')\right) \right) 
$.
%\; .
%\end{equation}
The saddle-point equation on $i \hat c$ reads
%\begin{equation}
$i \hat c(\Phi) = -\ln c(\Phi) - 
\ln \left( \int {\cal D}\Phi' e^{-i \hat c(\Phi')} \right) $.
%\; . 
%\end{equation}
Replacing this expression for $i \hat c(\Phi)$ and using the 
fact that $\int {\cal D}\Phi c(\Phi) =1$, 
the generating functional averaged over disorder becomes
%\begin{eqnarray}
$[{\cal Z}] = 
\int {\cal D}c \; \exp[-NS_{\sc eff}(c)]
$
%\nonumber\\
%%\end{equation}
with
\begin{eqnarray*}
S_{\sc eff}(c) = \int {\cal D}\Phi \; c(\Phi) \ln c(\Phi) + 
\frac12 \int {\cal D}\Phi \; 
c(\Phi) \int da \; \Phi(a) (D_a^{(2)} +\mu_s(a)) \Phi(a)
+ H_{\sc eff}
\; .
\end{eqnarray*}
The first term is an entropic contribution, the second term is a kinetic 
energy and the third one is the potential energy.

The difficulty now arises as to how to compute the effective
Hamiltonian $H_{\sc eff}$. For fully connected models  as the ones 
discussed in the previous Section
$H_{\sc eff}$ can be calculated exactly
and one recovers the dynamic 
equations that were already known (see below). 
For dilute disordered models, or models defined on a random graph, 
$H_{\sc eff}$ is still a quantity of order one. In this sense we 
still call them mean-field models.  
This property allows us to pursue the calculation 
with a saddle-point approximation.
$H_{\sc eff}$ is determined by a 
set of iterative approximations. 
For finite dimensional models the situation worsens still  since 
a saddle-point approximation cannot be used without taking into 
account the fluctuations around it.

\vspace{0.5cm}
\noindent{\it Gaussian approximation}
\vspace{0.5cm}

In this Section we assume that we deal with any model such that the 
effective action $S_{\sc eff}$ 
is indeed of order $N$. In a first step we resort 
to a Gaussian approximation, in which one proposes:
\begin{eqnarray}
c(\Phi) &=& 
({\det Q})^{-1/2}
\; e^{-\frac12 \int da db \; \Phi(a) Q^{-1}(a,b) \Phi(b)}
\label{Gaussian-ansatz} 
%\nonumber\\
%&=&
%\sqrt{\det Q^{-1}} \; 
%\exp\left( -\frac12 \, \int da db \; \Phi(a) \, Q^{-1}(a,b) \, \Phi(b) \right)
\; .
\label{gauss-ansatz}
\end{eqnarray}
The denominator
ensures the normalization of  $c(\Phi)$ and 
$Q(a,b)$ is correctly given by the averaged correlator:
%\begin{equation}
$Q(a,b) 
= {\cal N}
\int {\cal D}\Phi \frac1{N} \sum_i \Phi_i(a) \Phi_i(b) 
\;$$ e^{ -\frac12 \int da' db'
\; \Phi(a') Q^{-1}(a',b') \Phi(b') }$
with ${\cal N} = (\det Q)^{-1/2}$. 
%\; .
%\end{equation}
After rather simple manipulations 
the $c$-dependent effective action can be expressed in terms of $Q(a,b)$,
%\begin{eqnarray*}
%2S_{\sc eff}(c) \to 
$
2S_{\sc eff}(Q) = 
\mbox{Tr} \ln Q +
\int da db \; \delta(a-b) (D_a^{(2)}+\mu_s(a)) Q(a,b) 
- 2H_{\sc eff}(Q)
$.
%\; .
%\end{eqnarray*}  
Its variation with respect to $Q$ yields:
\begin{equation}  
\frac{\delta S_{\sc eff}(Q)}{\delta Q} = 0 =
\frac12 Q^{-1}(a,b) + \frac12 (D_a^{(2)}+\mu_s(a)) Q(a,b) 
- 
\frac{\delta H_{\sc eff}(Q)}{\delta Q(a,b)}
\; . 
\end{equation}
Multiplying this equation operationally 
by $Q(b,a')$ (see Appendix~\ref{app:Grassmann}) 
the dynamic equation takes the more
familiar form~(\ref{MCeq2}) 
with 
%\begin{equation} 
$\Sigma(a,a') = -2 \delta H_{\sc eff}(Q)/\delta Q(a,a')
$.
%\; .
%\label{self-energy-dilute}
%\end{equation}

\vspace{0.5cm}
\noindent{\it One example: the infinite range $p$ spin model}
\label{pspin-generic}
\vspace{0.5cm}

The $p$ spin spherical model is a particular case of the 
model in (\ref{gen-pspin}) with 
\begin{eqnarray}
4 H_{\sc eff} 
&=& \frac{1}{N} \sum_{i_1\dots i_p}
\frac{1}{N^{p-1}} 
\left( \int da \Phi_{i_1}(a) \dots \Phi_{i_p}(a)\right)^2 
=
 \int da db \left( \frac1{N} \sum_{i=1}^N \Phi_i(a) \Phi_i(b) 
\right)^p
\nonumber\\
&=&
\int {\cal D}\Phi'_1 \dots {\cal D}\Phi'_p c(\Phi'_1) \dots c(\Phi'_p) 
\int da db 
\; \Phi'_1(a) \Phi'_1(b) \dots \Phi'_p(a) \Phi'_p(b) 
\nonumber\\
&=&
\int da db \; 
\left( \int {\cal D}\Phi' \, c(\Phi') \, \Phi'(a) \Phi'(b) \right)^p 
\end{eqnarray}
Replacing the Gaussian {\it Ansatz} (\ref{Gaussian-ansatz}), 
$H_{\sc eff}$ becomes a simple
function of $Q(a,b)$:
\begin{equation}
H_{\sc eff}(Q) = \frac14 \int da db \;Q^{\bullet p}(a,b)
\; .
\end{equation}
The full expression for the effective action $S_{\sc eff}(Q)$ 
is identical to Eq.~(\ref{action-pspin}). A way to prove 
that the Gaussian {\it Ansatz} (\ref{Gaussian-ansatz})
is exact for this model is to 
check that the {\it exact} equation for $c(\Phi)$ coincides with the 
one obtained from the Gaussian {\it Ansatz} and the saddle-point 
evaluation.

\vspace{0.5cm}
\noindent{\it A second example: the dilute spherical spin-glass}
\label{dilute-spherical}
\vspace{0.5cm}

In this case the effective Hamiltonian $H_{\sc  eff}(c)$
reads
\begin{equation}
-2 H_{\sc eff}(c) = - \tilde c + \tilde c \int {\cal D}\Phi {\cal D}\Phi'
\; c(\Phi) c(\Phi') \, \cosh\left( J \int da \, \Phi(a) \Phi'(a) \right) 
\; .
\end{equation}
With the Gaussian {\it Ansatz}, this expression becomes
\begin{equation}
-2 H_{\sc eff}(Q) = 
- \tilde c + \tilde c/\sqrt{\det(1-J^2 Q^2)}
\; .
\end{equation}
The self-energy
can only be expressed as a series expansion or a functional:
\begin{equation}
\Sigma(a,b) = 
\tilde c J^2 [Q (1-J^2 Q)^{-1}](a,b) = 
\tilde c J^2 \sum_{k=0}^\infty J^{2k} \, Q^{(2k+1)}(a,b) 
\; .
\end{equation}
Using the first expression for $\Sigma$ 
one derives the dynamic equation
\begin{equation}
(\tilde c-1) J^2 Q^2(a,b) + \delta(a-b) = \left(-D_a^{(2)} - \mu_s(a) \right) 
(Q-J^2 Q^3)(a,b) 
\; . 
\end{equation}
Otherwise, if we use the second expression we obtain a dynamic
equation involving a series.
Note that this equation is much more complicated than the 
one for the fully connected $p$ spin spherical  
model. The derivatives act on functionals of the correlator
$Q$ in this case.  
Moreover, the Gaussian approximation is not exact for dilute models.
An iterative method can be implemented
to go beyond this approximation. It was introduced in~\cite{Secu} 
but we shall not describe it here.

\subsubsection{Quantum models}

Once we have presented the method for the classical dynamics, its
extension to a quantum system is simple. The important point to remark
about the previous derivation is that the {\sc susy} notation 
has been used, mostly, as a bookkeeping device. In the quantum case
a {\sc susy} formalism is not useful. Instead,  
it is convenient to 
use the formalism apt to take the classical limit, encoding the 
variables $(q, i\hat q)$ in a two-component (column) {\it vector} $\vec \Phi$, 
and defining a (line) vector $\vec v$:  
\begin{eqnarray}
\vec \Phi_i &\equiv&
\left( 
\begin{array}{c}
q_i 
\nonumber\\
i \hat q_i
\end{array}
\right)
\;\;\;\;\;\;\;\;\;\;
\vec v \equiv \left( 1 \;\; -\frac{i\hbar}{2} \right) 
\; ,
\end{eqnarray}  
the effective action reads
\begin{eqnarray}
S_{\sc eff} = 
S_{\sc kin}(\vec v \cdot {\vec \Phi}_i) - 
S_{\sc kin}({\vec v}^* \cdot {\vec \Phi}_i) +
S_{\sc pot}(\vec v \cdot \vec \Phi_i) - 
S_{\sc pot}({\vec v}^* \cdot \vec \Phi_i) + 
S_{\sc th}(\vec \Phi_i)
\end{eqnarray}
with the thermal part of the action written as 
\begin{eqnarray}
S_{\sc th}(\vec \Phi)
=
\int dt \int dt' \; {\vec \Phi}_i^t(t) {\cal A}(t-t') {\vec \Phi}_i(t')
\; . 
\end{eqnarray}
Calling $a=1,2$ the vector indices, the notation becomes identical
to the {\sc susy} one, with $a$ here playing the r\^ole of  
the coordinate in super-space $a=(t,\theta,\overline\theta)$ 
in the classical problem.
Just as in the previous Section, one introduces 
the identity (\ref{cPhi-def}), where $a$ is now interpreted as 
a vector index, to rewrite the generating functional. The analysis 
of the kinetic and thermal part of the effective action follows the 
same lines as the one presented for the classical problem. 

\subsection{Field equations}

Once we have written the dynamic 
action in terms of $\phi_i$ and $i\hat \phi_i$ 
the ``field equations''  follow from exact properties of the 
functional integration~\cite{Zinn}. Indeed, 
\begin{eqnarray*}
0 &=& \int {\cal D}\phi {\cal D} i\hat \phi \;
\frac{\delta }{\delta i\hat \phi_i(t)} 
e^{- S_{\sc eff}[\phi_i, i \hat \phi_i] + 
\int_C dt' \left( \eta_i(t') \phi_i(t') +
\hat\eta_i(t') i\hat \phi_i(t') 
\right) }
\nonumber\\
&=& \int {\cal D}\phi {\cal D} i\hat \phi \;
\left[ - \frac{\delta S_{\sc eff}(\phi_i,i\hat\phi_i)}
{\delta i\hat \phi_i(t)}  
+ \hat \eta_i(t)
\right]
e^{ - S_{\sc eff}(\phi_i, i\hat \phi_i) +
\int_C dt' \left( \eta_i(t') \phi_i(t') +\hat\eta_i(t') i\hat \phi_i(t') 
\right)}
\; .
\end{eqnarray*}
The subinded $C$ in the integrals stands for ``time contour''
and it can describe the usual integration from the initial time
to infinity for classical models or the close time path for 
quantum ones.
Taking now the variation with respect to the source $i\hat\eta_j(t')$ and 
evaluating at $\eta=i\hat \eta=0$ for all times and components we find
\begin{eqnarray}
0 &=& 
%\left\langle
%\delta(t-t') \delta_{ij}+ 
%\left(  - \frac{\delta S_{\sc eff}(\phi_i,i\hat\phi_i)}
%{\delta i\hat \phi_i(t)}  
%+ i\hat \eta_i(t) \right) i\hat \phi_j(t')
%\right\rangle_{\eta=i\hat\eta=0}
%\nonumber\\
%&=& 
\delta(t-t') \delta_{ij}- 
\left\langle i\hat \phi_j(t') 
\frac{\delta S_{\sc eff}(\phi_i,i\hat\phi_i)}{\delta i\hat \phi_i(t)}
\right\rangle 
\end{eqnarray}
where the brackets denote an average with the measure weighted by 
the dynamic action $S_{\sc eff}$.    
If, instead one takes the variation with respect to $\eta_j(t')$ and later
evaluates at $\eta=i\hat\eta=0$ one obtains:
\begin{equation}
\left\langle i\hat \phi_i(t) 
\frac{\delta S}{\delta \phi_j(t')} \right\rangle
= 0
\; .
\end{equation}

A way to derive dynamic equations for the two-point correlators amounts to
use Wick's theorem and rewrite these averages as a sum over all possible
factorizations in products of two point-functions. This is of course 
exact if the action is quadratic but it is only a Gaussian approximation
for more general models. This kind of derivation has been mainly
used in the study of the dynamics of manifolds in random 
potentials~\cite{Cule}. 

\subsection{The thermodynamic limit and time-scales}
\label{sompo}

It is very important to stress that the dynamic equations derived 
with the saddle-point approximation hold only when  $N\to\infty$ {\it before}
any long-time limit is taken. They describe the dynamics
in finite time-scales with respect to $N$ and they cannot 
capture the crossover from the non-equilibrium relaxation to the 
equilibrium dynamics reached in time scales that diverge with 
$N$ [remember that $t_{\sc eq}(N)$]. 

Previous attempts to study the dynamics of disordered glassy 
systems assumed that these same equations hold for the equilibrium 
dynamics when $N$ is finite and time-scales diverge with $N$~\cite{Sozi}. This
assumption is wrong as shown by several inconsistencies found in 
the solution at low temperatures: (i) 
the asymptotic values of one time-quantities
do not necessarily coincide with the values calculated with the 
equilibrium  distribution. (ii) the solution exhibited 
violates the fluctuation - dissipation theorem. These two 
results are not compatible with equilibrium. 

In order to study the equilibrium dynamics of these models
one should (i) start from random initial conditions but 
reach times that grow with $N$ or (ii) impose equilibrium 
initial conditions. The second route has been implemented
 -- though without solving the full dynamic problem --
by Houghton, Jain and Young~\cite{Young}. They showed 
that in this case one is forced to introduce the replica 
trick to average over disorder. 

The dynamic equations here derived 
are correct when $N\to\infty$ at the outset. 
Since times are always finite with respect to $N$, when  $t_{\sc eq}(N)$ 
diverges with $N$ the dynamics is not forced to reach equilibrium and 
there is no contradiction if the  
solution violates the equilibrium theorems. 

\subsection{Single spin equation}

In the limit $N\to\infty$ one can also 
write the full action $S_{\sc eff}$
in terms of a {\it single} variable. This is at the expense
of modifying the thermal kernel and the interaction term
in a self-consistent way, through the introduction of terms
arising from the non-linear interactions (the vertex
and self-energy, respectively). For a classical model with 
white external noise the single variable equation
reads
\begin{equation}
%\overline G_o^{-1}(t) \phi_i(t)
M \ddot \phi_i(t) + \gamma \, \dot\phi_i(t) + \mu(t) \,  \phi_i(t)  
=
\int_{0}^t dt'' \, \Sigma(t,t'') \; 
\phi_i (t'') + \rho_i(t) +\xi_i(t)
\; .
 \label{9}
\end{equation}
Its generalisation is straightforward.
There are two noise sources in this equation: $\xi_i(t)$ is the 
original white noise while $\rho_i(t)$ is an effective (Gaussian) 
noise with zero mean and 
correlations self-consistently given by 
$\langle \rho_i(t) \rho_j(t') \rangle = \delta_{ij} D(t,t')$. 
The vertex $D(t,t')$ plays the r\^ole 
of the colored noise correlation in a usual Langevin 
equation.
The self-energy $\Sigma(t,t')$ appears here in the place of an 
`integrated friction'. A solution of the problem can 
be attempted numerically using this equation and the self-consistent
definitions of $\Sigma$ and $D$.

This procedure is not particular useful for the analysis of 
``polynomial'' models since
the transformation into a $Q$ dependent effective action can be done 
exactly. It does however 
become useful for dealing with models whose single-spin 
effective action has higher order interaction  terms. An example 
is the quantum {\sc sk} model. This procedure is similar
to the one used in dynamic mean-field theory~\cite{Georges}.

Interestingly enough, as shown in Section~\ref{examples},
a rather flat  harmonic oscillator coupled to 
a bath made of a white and a coloured part at different temperatures
acquires two time-scales controlled by the two temperatures involved.
We see that a similar structure might appear for the glassy system if
the self-energy and vertex self-consistently arrange to 
act  on each degree of freedom as the friction and noise-noise correlator 
of a complex bath. We shall see that this is indeed what happens
to  mean-field models. We believe that a similar mechanism arises in 
finite dimensional glassy models as well~\cite{Cuku4}.

\section{Diagrammatic techniques}
\label{diagrammatic-techniques}
\setcounter{equation}{0}
\renewcommand{\theequation}{\thesection.\arabic{equation}}

In this Section we first  describe the perturbative solution to 
the Langevin process and how it is used to construct series expansions
for the correlations and responses. Self-consistent approximations,
such as the {\it mode coupling} or the {\it self-consistent 
screening}, correspond to a selection of a subset of diagrams from the 
full series. The connection with disordered models is demonstrated.
The presentation is very close to the one 
in~\cite{Mode}. An extension to quantum problems is possible 
using the generating functional formalism.

\subsection{Perturbative solution}

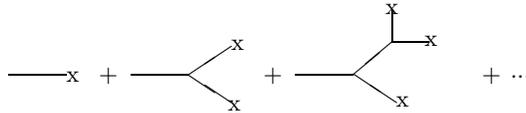
\begin{figure}[ht]
\centerline{
\setlength{\unitlength}{1184sp}%
\begingroup\makeatletter\ifx\SetFigFont\undefined%
\gdef\SetFigFont#1#2#3#4#5{%
  \reset@font\fontsize{#1}{#2pt}%
  \fontfamily{#3}\fontseries{#4}\fontshape{#5}%
  \selectfont}%
\fi\endgroup%
\begin{picture}(10512,2235)(1189,-2311)
\thinlines
\put(1201,-1561){\line( 1, 0){1200}}
\put(3751,-1561){\line( 1, 0){1200}}
\put(7201,-1561){\line( 1, 0){1200}}
\put(4951,-1561){\line( 3, 2){900}}
\put(4951,-1561){\line( 3,-2){900}}
\put(8401,-1561){\line( 6, 5){818.852}}
\put(9226,-886){\line( 0, 1){675}}
\put(9226,-886){\line( 1, 0){750}}
\put(8413,-1544){\line( 3,-2){900}}
\put(2401,-1711){\makebox(0,0)[lb]{\smash{\SetFigFont{9}{10.8}{\rmdefault}{\mddefault}{\updefault}x}}}
\put(3076,-1711){\makebox(0,0)[lb]{\smash{\SetFigFont{9}{10.8}{\rmdefault}{\mddefault}{\updefault}+}}}
\put(5851,-1036){\makebox(0,0)[lb]{\smash{\SetFigFont{9}{10.8}{\rmdefault}{\mddefault}{\updefault}x}}}
\put(5776,-2311){\makebox(0,0)[lb]{\smash{\SetFigFont{9}{10.8}{\rmdefault}{\mddefault}{\updefault}x}}}
\put(9076,-286){\makebox(0,0)[lb]{\smash{\SetFigFont{9}{10.8}{\rmdefault}{\mddefault}{\updefault}x}}}
\put(9901,-961){\makebox(0,0)[lb]{\smash{\SetFigFont{9}{10.8}{\rmdefault}{\mddefault}{\updefault}x}}}
\put(9301,-2236){\makebox(0,0)[lb]{\smash{\SetFigFont{9}{10.8}{\rmdefault}{\mddefault}{\updefault}x}}}
\put(11101,-1711){\makebox(0,0)[lb]{\smash{\SetFigFont{9}{10.8}{\rmdefault}{\mddefault}{\updefault}+}}}
\put(11701,-1636){\makebox(0,0)[lb]{\smash{\SetFigFont{9}{10.8}{\rmdefault}{\mddefault}{\updefault}...}}}
\put(6526,-1711){\makebox(0,0)[lb]{\smash{\SetFigFont{9}{10.8}{\rmdefault}{\mddefault}{\updefault}+}}}
\end{picture}
}
\caption{Terms $O(g^0)$, $O(g^1)$ and $O(g^2)$  
in the perturbative solution to the Langevin equation.}
\label{pert-noise}
\end{figure}

Let us focus on a single scalar degree of freedom, $q$, with potential 
energy
\begin{equation}
V(q) = \frac{\mu(t)}{2} \,  q^2 + \frac{g}{3!}  \, q^3 
\; ,
\label{1}
\end{equation}
and dynamics given by the Langevin Eqs.~(\ref{lang1}) and (\ref{lang2})
in the white noise limit. We take the initial condition $q(t=0)=0$. 
$\mu(t)$ is a time-dependent function that we fix at the end of 
the derivation by requiring $C(t,t)=1$.   
In vector models it is the Lagrange
multiplier that self-consistently imposes a spherical constraint.
Note that this potential is not bounded from below.
Setting 
$G_o(t,t') = [\mu(t) + \gamma \partial_t + M \partial_{t^2}]^{-1}$, 
a perturbative expansion for $q(t)$ in powers of the noise
is easily written as
\begin{equation}
q(t) = 
(G_o \otimes \xi)(t)
-
\frac{g}{2}  \; \left( 
G_o \otimes [ G_o \otimes \xi \bullet G_o \otimes \xi ] \right)(t) + ... 
\label{3}
\end{equation}
where $\otimes$ means a time convolution, 
$(G_o \otimes f)(t) = \int_0^t dt'
G_o(t,t') f(t')$, and $\bullet$ is a simple product at 
equal times. This notation is equivalent to the one used 
in the {\sc susy} formalism, see Appendix~\ref{app:Grassmann}. 
Causality implies
$G_o(t,t')\propto \theta(t-t')$.
If inertia 
can be neglected $G_o(t,t')=
 \exp \left( -\int_{t'}^t d\tau \; \mu(\tau) \right) \theta(t-t')$. 
If one keeps the second-time derivative $G_o(t,t')$ takes a more 
complicated form. 
Equation~(\ref{3}) can be graphically represented 
as in Fig.~\ref{pert-noise}. 
Crosses indicate noise and oriented lines indicate the 
bare propagator $G_o$. 
Each vertex carries a factor $g/2$.
Note that the unknown $q$ is evaluated at time $t$ while the 
noises are evaluated at all previous times.

The expansion for $q$ leads to two 
expansions for the correlation and response. In simple 
words, the former corresponds to sandwiching, {\it i.e.}
averaging over the noise, the usual product of two series 
as the one in Fig.~\ref{pert-noise} evaluated at different 
times $t$ and $t'$. Due to the average over the Gaussian 
noise noise factors have to be taken by pairs.
Let us illustrate this with a few 
examples. 

\begin{figure}[ht]
\centerline{
\input{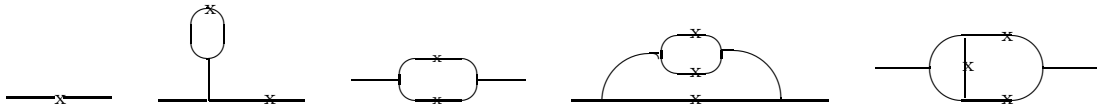}
\hspace{0.2cm}
\setlength{\unitlength}{987sp}%
\begingroup\makeatletter\ifx\SetFigFont\undefined%
\gdef\SetFigFont#1#2#3#4#5{%
  \reset@font\fontsize{#1}{#2pt}%
  \fontfamily{#3}\fontseries{#4}\fontshape{#5}%
  \selectfont}%
\fi\endgroup%
\begin{picture}(3699,2460)(3739,-1636)
\thinlines
\put(4988,127){\oval(824,1276)}
\put(3751,-1561){\line( 1, 0){1200}}
\put(5026,-1561){\line( 0, 1){1050}}
\put(6226,-1561){\line( 1, 0){1200}}
\put(5026,-1561){\line( 1, 0){1200}}
\put(6226,-1636){\makebox(0,0)[lb]{\smash{\SetFigFont{8}{9.6}{\rmdefault}{\mddefault}{\updefault} x}}}
\put(4726,614){\makebox(0,0)[lb]{\smash{\SetFigFont{8}{9.6}{\rmdefault}{\mddefault}{\updefault} x}}}
\end{picture}
\hspace{0.2cm}
\setlength{\unitlength}{987sp}%
\begingroup\makeatletter\ifx\SetFigFont\undefined%
\gdef\SetFigFont#1#2#3#4#5{%
  \reset@font\fontsize{#1}{#2pt}%
  \fontfamily{#3}\fontseries{#4}\fontshape{#5}%
  \selectfont}%
\fi\endgroup%
\begin{picture}(4374,1230)(2389,-3436)
\thinlines
\put(4576,-2836){\oval(1950,1050)}
\put(2401,-2836){\line( 1, 0){1200}}
\put(5551,-2836){\line( 1, 0){1200}}
\put(4426,-2386){\makebox(0,0)[lb]{\smash{\SetFigFont{6}{7.2}{\rmdefault}{\mddefault}{\updefault}x}}}
\put(4426,-3436){\makebox(0,0)[lb]{\smash{\SetFigFont{6}{7.2}{\rmdefault}{\mddefault}{\updefault}x}}}
\end{picture}
\hspace{0.2cm}
\setlength{\unitlength}{987sp}%
\begingroup\makeatletter\ifx\SetFigFont\undefined%
\gdef\SetFigFont#1#2#3#4#5{%
  \reset@font\fontsize{#1}{#2pt}%
  \fontfamily{#3}\fontseries{#4}\fontshape{#5}%
  \selectfont}%
\fi\endgroup%
\begin{picture}(6474,1860)(1489,-4486)
\thinlines
\put(3670,-3211){\oval( 12,  0)[tr]}
\put(3670,-4411){\oval(2838,2400)[tl]}
\put(5251,-4411){\oval(3000,2550)[tr]}
\put(4501,-3248){\oval(1500,976)}
\put(1501,-4411){\line( 1, 0){6450}}
\put(4276,-2836){\makebox(0,0)[lb]{\smash{\SetFigFont{8}{9.6}{\rmdefault}{\mddefault}{\updefault} x}}}
\put(4276,-3811){\makebox(0,0)[lb]{\smash{\SetFigFont{8}{9.6}{\rmdefault}{\mddefault}{\updefault} x}}}
\put(4276,-4486){\makebox(0,0)[lb]{\smash{\SetFigFont{8}{9.6}{\rmdefault}{\mddefault}{\updefault} x}}}
\end{picture}
\hspace{0.2cm}
\setlength{\unitlength}{987sp}%
\begingroup\makeatletter\ifx\SetFigFont\undefined%
\gdef\SetFigFont#1#2#3#4#5{%
  \reset@font\fontsize{#1}{#2pt}%
  \fontfamily{#3}\fontseries{#4}\fontshape{#5}%
  \selectfont}%
\fi\endgroup%
\begin{picture}(5574,1785)(1339,-3511)
\thinlines
\put(4126,-2611){\oval(2850,1650)}
\put(1351,-2611){\line( 1, 0){1350}}
\put(5551,-2611){\line( 1, 0){1350}}
\put(3601,-1861){\line( 0,-1){1500}}
\put(3526,-2686){\makebox(0,0)[lb]{\smash{\SetFigFont{8}{9.6}{\rmdefault}{\mddefault}{\updefault}x}}}
\put(4501,-1936){\makebox(0,0)[lb]{\smash{\SetFigFont{8}{9.6}{\rmdefault}{\mddefault}{\updefault}x}}}
\put(4501,-3511){\makebox(0,0)[lb]{\smash{\SetFigFont{8}{9.6}{\rmdefault}{\mddefault}{\updefault}x}}}
\end{picture}
}
\caption{From left to right: $O(g^0)$, two $O(g^2)$ and two 
$O(g^4)$ terms in the series for $C$. 
The next to last diagram dresses the propagator and the last term  
dresses the vertex. The former is kept in the 
{\sc mca} while the latter is neglected.}
\label{first-order}
\end{figure}

The first term in the expansion is the result of 
averaging two $O(g^0)$ terms 
(first term in Fig.~\ref{pert-noise}):
\begin{eqnarray}
C_o(t,t') = \langle (G_o \otimes \xi)(t) \bullet 
(G_o \otimes \xi)(t') \rangle
=
%\langle
%\int_0^t dt'' \, G_o(t,t'') \xi(t'') 
%\int_0^{t'} dt''' \, G_o(t',t''') \xi(t''')
%\rangle
%\nonumber\\
%&=&
% \int_0^t dt'' \, G_o(t,t'') 
%\int_0^{t'} dt''' \, G_o(t',t''') \, 2 \gamma k_B T \delta(t''-t''')
%\nonumber\\
2 \gamma k_B T  
\int_0^{t'} dt'' \, G_o(t,t'') G_o(t',t'') 
\; ,
&& 
\end{eqnarray}
$t\geq t'$. We depict this term and its contributions to 
more complicated diagrams with a single crossed line, 
see the first graph in Fig.~\ref{first-order}. 
The term $O(g)$, as well as all terms which are odd powers of $g$, vanishes. 
There are two contributions to the 
term $O(g^2)$. One is the result of multiplying a term $O(g^2)$ 
with a term $O(g^0)$ and it is a tadpole, see the second graph 
in Fig.~\ref{first-order}; we assume this term and all its
corrections are included in the contributions from the time-dependent mass
and we henceforth ignore them.
The other 
comes from multiplying two $O(g)$ terms,
see the  third graph in Fig.~\ref{first-order}.
%\begin{eqnarray*}
%\frac{g}{2} (2\gamma k_BT)^2\int_0^t dt'' G_o(t,t'') 
%\int_0^{\min(t',t'')} dt''' G_o(t',t''') G_o(t'',t''')
%\int_0^{t'''} dt^{iv} G^2_o(t''',t^{iv}) 
%\end{eqnarray*}
%and its symmetric contribution where $t$ and $t'$ are exchanged.

Higher order terms are of two types: they either dress the 
{\it propagators}
or they dress the {\it vertices}, see the last two diagrams in
Fig.~\ref{first-order}. These two terms are order
$O(g^4)$. The first one follows from 
averaging two $O(g^2)$ contributions while the 
second one is the result of averaging an $O(g^3)$ and an $O(g)$ term.
The full series yields the {\it exact perturbative expansion}
for $C$.

The series expansion for the response follows
from the relation (\ref{Rqxi}) in the white noise limit.
In graphical terms we obtain it by multiplying the 
series in Eq~(\ref{3}) and Fig.~\ref{pert-noise} evaluated
at time $t$ by a noise evaluated  at time $t'$ and taking the 
average. 

\subsection{The mode coupling approximation ({\sc mca})}
\label{mode-coupling}

The diagrammatic expansions for $C$ and $R$ 
%representation of the correlation and response
%are shown in Fig.~\ref{pert-corr-resp}.
%They  
can be represented analytically by introducing 
the kernels $\Sigma(t,t')$ and $D(t,t')$ through the 
Schwinger-Dyson equations (\ref{intschwingerR}) and (\ref{intschwingerC}) 
in their integral form. Each of them is a compact notation 
for a series of diagrams. These equations are exact perturbatively.
However, for a generic model one cannot compute the kernels
$\Sigma$ and $D$ exactly. 

The mode coupling approach
amounts to approximating the kernels
 $\Sigma(t,t')$ and $D(t,t')$ in the following way.
One takes their 
values at $O(g^2)$ and substitutes in them the bare propagator 
$G_o$ and the bare correlation $C_0$ by their dressed values, {\it i.e.}
by $R$ and $C$ themselves. 
For the model defined in Eq.~(\ref{1}) this 
yields
\begin{eqnarray}
\Sigma(t,t')
&=& 
g^2 \; C(t,t') \, R(t,t') 
\; ,
\label{Sigmap3}
\\
D(t,t')&=&2\gamma k_B T \; \delta(t-t')+ { g^2 \over 2} \; C^2(t,t') 
\label{11}
\label{Dp3}
\; .
\end{eqnarray}
This approximation 
 neglects  ``vertex renormalization'' in the sense that 
all diagrams 
correcting the values of the lines are taken into account
while all diagrams correcting the vertices are neglected. 
For instance, 
one 
keeps the fourth diagram in
Fig.~\ref{first-order} that represents a line correction, 
while leaving aside the fifth diagram drawn in the same figure 
that represents a vertex correction. 

The same procedure can be implemented using the {\sc susy} representation 
of the dynamics. Each line represents the superfield and the super-correlator
follows from the sandwich of two series for the super-field
evaluated at different super-coordinates $a$ and $b$.

The Schwinger-Dyson equations
can be recast, after multiplying by 
 $G_o^{-1}$, into the form (\ref{eqR2}) and (\ref{eqC2})
for a random potential 
(\ref{gen-pspin}) with only one term $r=p=3$. 
Applying the {\sc mca} to the trivial (and ill-defined) model (\ref{1}) we
derived the dynamic equations for the $p=3$ spin spherical model!
On the one hand, this result is worrying since 
it shows that the {\sc mca} can be rather uncontrolled and 
it can generate glassy behavior by itself. On the other hand, 
since the same equations hold in the {\sc mca} of a model
of interacting particles with realistic interactions, this calculation
 allows one to understand why the dynamic equations 
of the {\sc mct} for super-cooled liquids coincide with 
the ones of disordered spin models above $T_d$. 
In the next Subsection we show how the diagrams neglected in the {\sc mca} 
vanish in a disordered model with a large number of components. 
(See also \cite{Zacarelli} for other recent discussions of 
the meaning and range of validity of the 
{\sc mca} and {\sc mct}.)

\subsection{{\sc mca} and disordered models}
\label{mode-coupling-disordered}

The first to notice that the {\sc mca}
for a ``quadratic'' dynamic equation 
corresponds to the exact dynamic equation of a disordered 
problem with a large number of components
was Kraichnan~\cite{Kraichnan} in the context of the 
Navier-Stokes equation. More recently, Franz and Hertz showed that
the ``schematic {\sc mct} equations of the $F_p$ group'' 
for super-cooled liquids 
are identical to those arising from a spin model with pseudo-random
interactions between groups of three spins~\cite{Frhe}. (The schematic
{\sc mct} focus on a chosen wavevector.)

Indeed, for the example chosen in this Section, one easily demonstrates that 
the diagrams retained by the {\sc mca} are precisely those which 
survive if one modifies the initial model (\ref{1}) and 
considers instead the following disordered problem~\cite{Mode}. First, 
let us upgrade $q$ to a vector with $N$ components or ``colors'' $\phi_i$, 
where $i=1,2,...,N$. Second, let us modify the potential energy (\ref{1}) 
into 
\begin{equation}
V(\vec\phi) = 
g \sum_{i< j<k} J_{ijk} \; 
\phi_i \phi_j \phi_k 
\label{derives} 
\end{equation}
with couplings $J_{ijk}$ that are independent
quenched Gaussian random variables of zero mean and variance  
$[J_{ijk}^2]_J =  1/N^{p-1}=1/N^2$.
($p$ is the number of spins in each term in $V$.)
In the large $N$ limit,
the noise and disorder averaged correlation and response 
of this modified model  obey 
Eqs.~(\ref{eqR2}) and (\ref{eqC2}) with 
$\Sigma$ and $D$ given by Eqs.~(\ref{Sigmap3}) and (\ref{Dp3}), respectively.
The fact that these equations are recovered can
be seen either directly on the perturbation theory, 
or using the functional
methods given in Section~\ref{generating-functionals}.
Since we want to stress that the diagrams neglected in the 
{\sc mca} vanish exactly for this model we use here the first 
approach.

The bare propagator is diagonal in the color indices,
${G_o}_{ij}=G_o \delta_{ij}$. The vertex is now proportional 
to the random exchanges $J_{ijk}$. The perturbative solution
to the Langevin equation reads
\begin{equation}
\phi_i(t) = (G_o \otimes \xi_i)(t) -
J_{ijk} \, G_o \otimes \left( G_o \otimes \xi_j \bullet 
G_o \otimes \xi_k \right)(t)
+ \dots 
\;.
\end{equation}
One is interested in computing the self-correlation averaged
over the noise and disorder, $N^{-1}
\sum_{i=1}^N [ \langle \phi_i(t) \phi_i(t')]$. 
The latter average eliminates all terms with an odd number of 
couplings. Similarly, since $J_{ijk} \neq 0$ only if all indices
$i,j,k$ are different, tadpole contributions as the one in 
the second graph in 
Fig.~\ref{first-order} vanish (the noise-noise correlation 
enforces that two indices in the random exchange must coincide).
Finally, one can check that due to the scaling with $N$ of the 
variance of the disordered interactions, 
vertex corrections as the one in the last graph in 
Fig.~\ref{first-order} are sub-leading and 
vanish when $N\to\infty$. Instead, all 
line corrections remain finite in the thermodynamic limit. 
We can check this statement in the
two examples shown in Fig.~\ref{first-order} extended to include 
color indices. 
The vertex correction has four random exchanges that 
due to the averaging over the noise are forced to match as, {\it e.g.}
$J_{ijk} J_{jlm} J_{mni} J_{kln}$ leaving $6$ free-indices.
Averaging over disorder one identifies the indices of two 
pairs of $J's$, {\it e.g.} $i=l$ and $k=m$, this 
yields a factor $(1/N^2)^2$ and, at most,  
it leaves $4$ color indices over which we have to sum from $1$ to $N$
($i,j,k,n$). We have then an overall
factor $1/N^4 \times N^4= 1$ and this term vanishes when one normalises the 
correlation by $N$.
Instead, in the line correction, after averaging over the noise,
we are left with $6$ free indices, {\it e.g.} 
$J_{ikj} J_{klm} J_{lmn} J_{inj}$, the average over the noise only imposes
$k=n$ in its most convenient contribution, and the overall factor is 
$1/N^4 \times N^5=N$. This term contributes to the normalisaed 
global correlation.

Interestingly enough, the equivalence between 
 the {\sc mca} and a disordered system 
extends to an arbitrary {\it non-linear} coupling 
$F(q)$. Expanding $F$ in a power series 
%\begin{equation}
$F(q)= \sum_{r=2}^\infty {F_r \over r!} q^r$
%\end{equation}
the {\sc mca} leads to 
\begin{eqnarray}
\Sigma(t,t') &=& g^2 \; \sum_{r=2}^\infty {F_r^2 \over (r-1)!} \;  
C^{r-1}(t,t') \; R(t,t')
\label{sigmagen}
\; ,
\\
D(t,t') &=& 2 \gamma k_BT \delta(t-t') + 
g^2 \; \sum_{r=2}^\infty {F_r^2 \over r!} \; C^r(t,t')
\; .
\label{mcgen}
\end{eqnarray}
[Note that for $r$ odd, there appears an additional ``tadpole'' contribution
in Eq.~(\ref{sigmagen}), which we have assumed again that 
it has been re-absorbed into the mass term $\mu(t)$.] 
The dynamic equations can 
also be obtained as the exact solution 
of the Langevin dynamics of $N$ continuous spins
$\phi_i$ interacting through the potential
\begin{equation}
V_J[{\vec \phi}]
=
g \sum_{r \geq 2}^\infty F_r 
\sum_{i_1 < \dots < i_{r+1}} J_{i_1 \dots i_{r+1}} \; 
\phi_{i_1} \dots \phi_{i_{r+1}}
\end{equation}
where $J_{\alpha_1,..\alpha_{r+1}}$ are quenched 
%symmetrical and otherwise 
independent Gaussian variables with zero mean and 
$
[(J_{\alpha_1,..\alpha_{r+1}})^2] \propto N^{-r}  
\label{13}
$.
Therefore the {\sc mc} equations for a single dynamic 
variable in contact with a heat reservoir and 
under an arbitrary nonlinear potential $F(q)$ 
describe exactly a fully-connected 
spin-glass problem with arbitrary
multi-spin interactions or
a particle evolving in an $N$ dimensional space 
in a quenched random potential $V[{\vec \phi}]$
with a Gaussian 
distribution with zero mean and  
variance~(\ref{pot-corr})~\cite{Mefr,Cule}. 
Let us note that in order to be well defined, the model given 
by $V$ must be supplemented by a constraint 
preventing the field $\phi_i$ from exploding in an unstable direction 
set by the coupling tensor $J_{i_1 \dots i_{r+1}}$. This 
problem is cured by imposing the spherical constraint
%\begin{equation}
$\sum_{i=1}^N \phi_i^2(t) = N C(t,t) 
\equiv N 
%\; .
%\label{sph}
%\end{equation}
$.

The extension of the mapping to a space dependent $\phi({\vec x},t)$ (or to a 
multicomponent field) is straightforward. Several 
interesting physical examples 
involve an equation of the type:
\begin{eqnarray*}
{\partial \hat \phi({\vec k},t) \over \partial t} = -(\nu k^2 + \mu) 
\hat \phi({\vec k},t)
- \sum_{r=2}^\infty \sum_{{\vec k}_1,..{\vec k}_r} {F_r \over r!}
 {\cal L}_r({\vec k}|{\vec k}_1,.....{\vec k}_r)
\hat\phi({\vec k}_1,t)....\hat\phi({\vec k}_r,t) + \xi({\vec k},t) 
\nonumber\\
\end{eqnarray*}
where $\hat \phi({\vec k},t)$ is the Fourier transform of 
$\phi({\vec x},t)$, and 
$\xi({\vec k},t)$ a Gaussian noise such that 
$\langle \xi({\vec k},t) \xi({\vec k'},t')\rangle = 2\gamma k_B 
T \delta({\vec k+\vec k' }) \delta(t-t')$. The 
Kardar-Parisi-Zhang ({\sc kpz}) equation \cite{HH}
corresponds to $r=2$, $ {\cal L}_2({\vec k}|{\vec k}_1,{\vec k}_2)=[{\vec k}_1\cdot
 {\vec k}_2] \
\delta({\vec k}_1 + {\vec k}_2 + {\vec k})$, 
while domain coarsening in the $\phi^4$ theory
corresponds to $r=3$, 
${\cal L}_3({\vec k}|{\vec k}_1,{\vec k}_2,{\vec k}_3)= 
\delta({\vec k}_1 +
{\vec k}_2 + {\vec k}_3 + {\vec k})$, with a
negative $\mu$ \cite{Bray}. 
The Navier-Stokes equation is similar to the {\sc kpz} case
with, however, an extra tensorial structure due to the 
vector character of the velocity field.
The correlation and response functions now become ${\vec k}$ dependent,
%\begin{eqnarray}
$\delta^d({\vec k}+{\vec k'}) C({\vec k},t,t') 
= \langle \tilde \phi({\vec k},t) \tilde \phi({\vec k'},t') \rangle 
$ and 
%\\
%\label{15-a}
$
\delta^d({\vec k}+{\vec k'}) R({\vec k},t,t') = 
\langle {\partial \tilde \phi({\vec k},t)/\partial \xi({\vec k'},t')} \rangle 
$.
%\label{15-b}
%\end{eqnarray}
The generalized {\sc mc} equations 
then read (assuming that the structure factors 
%${\cal L}_r({\vec k}|{\vec k}_1,.....{\vec k}_r)$ 
are invariant 
under the permutation
of ${\vec k}_1,...,{\vec k}_r$):
\begin{eqnarray}
\Sigma({\vec k},t,t') &=& g^2 \; \sum_{r=2}^\infty {F_r^2 \over (r-1)!} 
\sum_{{\vec k}_1,..{\vec k}_r} {\cal L}_r({\vec k}|{\vec k}_1,.....{\vec k}_r)
{\cal L}_r({\vec k}_r|{\vec k}_1,.....{\vec k})
\nonumber
\\
& & C({\vec k}_1,t,t') 
...
C({\vec k}_{r-1},t,t') R({\vec k}_r,t,t')
\label{16-a}
\\
D({\vec k},t,t')&=& 2\gamma k_B T \; 
\delta(t-t') + g^2 \; \sum_{r=2}^\infty {F_r^2 \over r!} 
\sum_{{\vec k}_1,..{\vec k}_r} \left({\cal L}_r({\vec k}|{\vec k}_1,.....{\vec k}_r)\right)^2 
\nonumber
\\
& & C({\vec k}_1,t,t')...C({\vec k}_r,t,t')
\label{16-b}
\end{eqnarray}
where $\Sigma({\vec k},t,t')$ and $D({\vec k},t,t')$ are defined 
in analogy with
Eqs.~(\ref{sigmagen}) and (\ref{mcgen}). 

\subsection{{\sc mca} for super-cooled liquids and 
glasses}

In the last 20 years the {\sc mca} has been much used in the study of 
super-cooled liquids.
Starting from the realistic 
interactions between the constituents of a liquid, 
G\"otze {\it et al}~\cite{Gotze}
used the {\sc mca} together with an assumption of 
equilibrium to derive a dynamic equation for the density-density correlator.
This analysis lead to the {\it schematic mode coupling theory} 
({\sc mct})~\cite{Le}  of  super-cooled
liquids and generalisations~\cite{Be}
(with no reference to wave-vector dependence) and
to more sophisticated versions that include a dependence on 
space~\cite{Gotze}. The difference
between these models lies on the form of the kernels $\Sigma$ and $D$.
Kirkpatrick, Thirumalai and Wolynes~\cite{Kithwo} 
realized in the late 80s that 
the schematic mode coupling equation~\cite{Le} is identical to the dynamic
equation for the spin-spin correlator in the 
disordered Potts or $p$ spin model, building a bridge between the study 
of structural and spin glasses. Why these models and not {\sc sk}?
This will become clear when we present their dynamic and 
static behavior.

In this Section we explained why the dynamic equation 
of a disordered model and the one stemming from a {\sc mca} of 
a model with more realistic interactions co\"{\i}ncide:
the terms neglected in the latter vanish exactly in the former.
The example studied here serves also to signal the danger in 
using a {\sc mca}. One could conclude that a trivial model has a highly 
non-trivial dynamics, this being generated by the approximation itself.

In the derivation of the dynamic equations presented in this 
Section no assumption of equilibrium was used. Therefore, these 
equations hold also in the low temperature 
phase where equilibrium is lost. It is then natural to propose
that the dynamics of the $p$ spin spherical model below $T_d$ 
schematically describes the dynamics 
of glasses just as its dynamics above $T_d$ yields the schematic 
{\sc mct} of super-cooled liquids~\cite{review}. To go
beyond the schematic theory while still keeping a single mode 
description (as in ~\cite{Be}) 
one simply has to  consider $p_1 + p_2$ spherical disordered models.
Moreover, the dynamics of a manifold in a random potential is described by 
dynamic equations with a $\vec k$ dependence that goes beyond the single 
mode {\sc mct}. 

Recently, Latz showed how the generic 
dynamic equations derived in this Section 
can also be obtained starting from the microscopic fluid system
and using the {\sc mc} approximation though with no equilibration 
assumption~\cite{Latz}. 
Alternative derivations of
mode-coupling equations are discussed in~\cite{Zacarelli}

Kawasaki and Kim~\cite{KK} derived the same schematic {\sc mc}
equation using a non-disordered
quadratic Hamiltonian for densities and velocities
complemented with random non-linear dynamic equations. 
Interestingly enough, in this case the {\sc mct} arises 
from a model with {\it trivial} statics and complex dynamics.
Tunning the ratio between the number of density variables and velocity 
variables they even managed to include the so-called ``hopping term'' 
that softens the dynamic transition~\cite{Gotze,Vo}. 
Is worth noting that in Kawasaki and Kim's model
this term {\it is not} due to 
thermally activated processes but to the effect of the velocity-like
variables through the complex dynamics.

\section{Glassy dynamics: Generic results}
\label{sec:lowT}
\setcounter{equation}{0}
\renewcommand{\theequation}{\thesection.\arabic{equation}}

Before presenting the explicit solution to the mean-field 
models we state some generic features 
of the low-$T$ dynamics that we believe hold in general.
  
\begin{figure}[ht]
\centerline{
\psfig{file=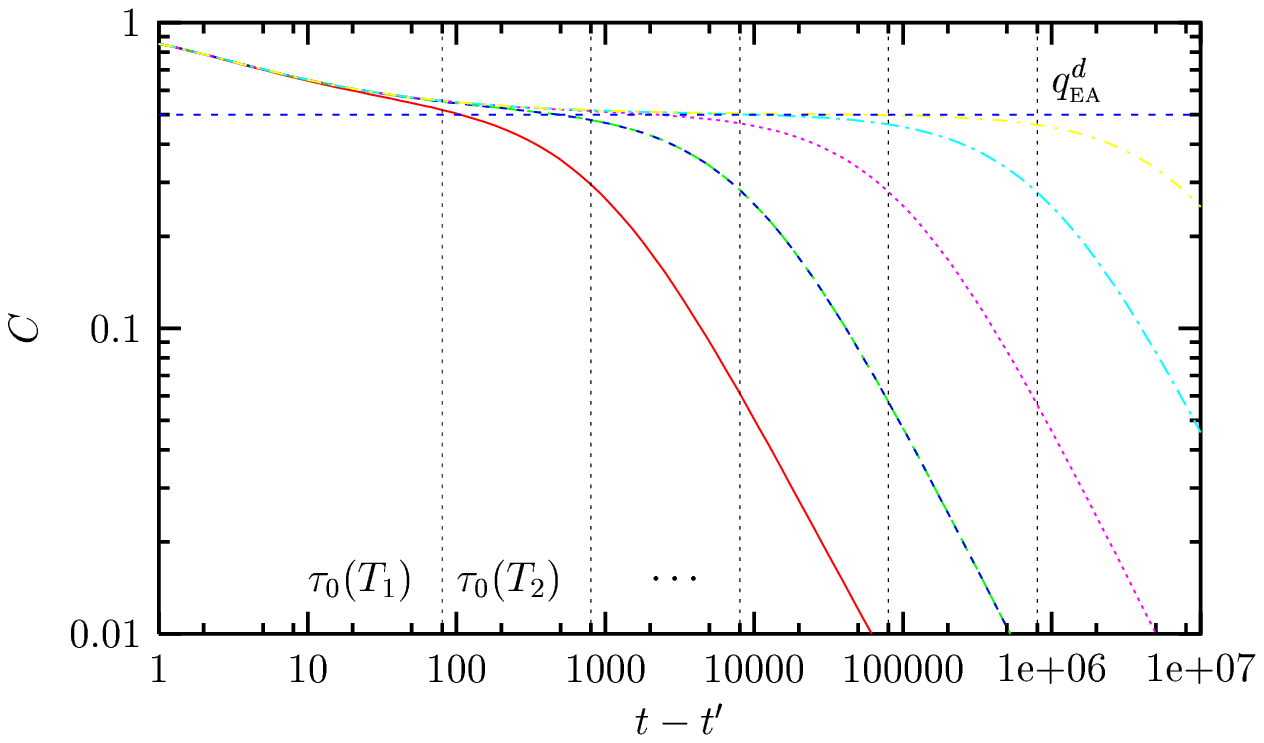,width=8cm}
\psfig{file=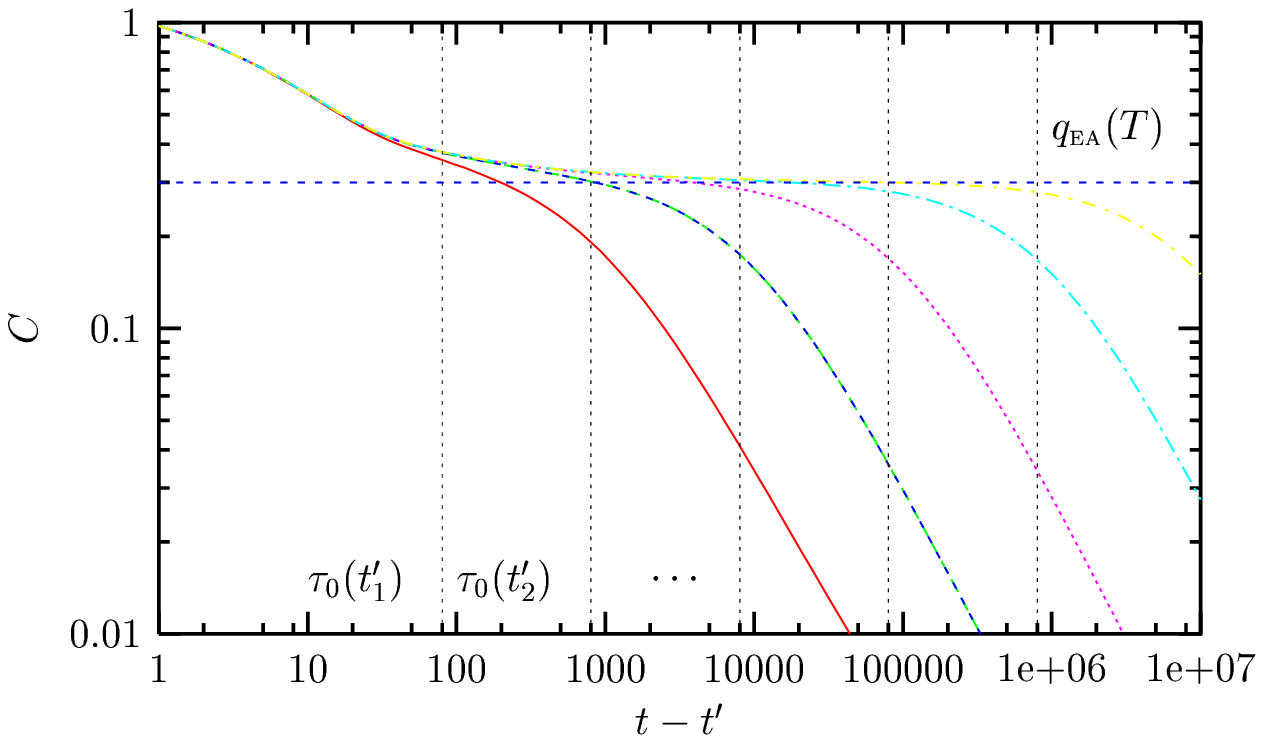,width=8cm}
}
\vspace{-0.5cm}
\caption{Left: Sketch of the decay of the stationary 
correlations in the high $T$
phase close to $T_d$, $T_1 > T_2 > \dots$. Right: Sketch of the decay of
the aging 
correlations in the low $T$ phase, at fixed $T< T_d$, $t'_1< t'_2< \dots$}
\label{sketch-scales-highT}
\end{figure}

\subsection{The weak-ergodicity breaking scenario}
\label{subsec:web}

Figure~\ref{sketch-scales-highT}-right shows a sketch of the 
decay of the correlation as obtained from the numerical
solution to the dynamic equations for the mean-field 
models (see Section~\ref{numerical-sol}). It
develops a separation of time scales in 
the long $t'$ limit. It first approaches a plateau at
$q_{\sc ea}$ in a stationary manner and it then decays below 
this value with an explicit waiting-time dependent form.
For each waiting-time there is a sufficiently long 
$t$ such that the correlation decays to zero. 
These properties are included in the {\it weak-ergodicity breaking} ({\sc web})
scenario that states that, for $t\geq t'$,  $C$
decays in such a way that 
\begin{eqnarray}
 \lim_{t'\to\infty} C(t,t') 
&=&
q_{\sc ea} + C_{\sc st}(t-t')
\\
\lim_{t-t'\to \infty} C_{\sc st}(t-t') 
= 0 
\;\;\;\; &\Rightarrow & 
\;\;\;\; 
\lim_{t-t'\to \infty} \lim_{t'\to\infty} C(t,t') 
= q_{\sc ea}
\label{limitqq}
\\
\lim_{t \to\infty} C(t,t') &=& 0
\;\;\;\;\; \mbox{at fixed} \; t'
\label{limit0}
\; .
\end{eqnarray}
Equation~(\ref{limitqq}) 
defines the Edwards-Anderson order parameter,
$q_{\sc ea}$.
For finite $t'$  there is a crossover between two time-scales controlled
by a waiting-time dependent characteristic time $\tau_0(t')$
that is a growing function of $t'$ 
whose precise form depends on the model.
For large $t\geq t'$ such that $t-t'$ is small 
with respect to $\tau_0(t')$,
the correlation function first decays from 1 to $q_{\sc ea}$ 
in a {\sc tti} manner.  At longer $t-t'$ it 
goes further below $q_{\sc ea}$ to eventually reach $0$ 
in a manner that depends both
upon $t$ and $t'$ (the aging effect). This behavior suggests
the presence of at least two time-sectors in which the dynamics is
stationary and non-stationary, respectively. We shall see that 
the number of time-scales, or more precisely correlation scales, 
depends on the model.

 We write $C$  as the 
sum of a stationary
and an aging contribution:
\begin{eqnarray}
C(t,t') &=& C_{\sc st}(t-t') +C_{\sc ag}(t,t')
\; .
\end{eqnarray} 
The matching conditions at equal times 
between $C_{\sc st}$ and $C_{\sc ag}$ are
%\begin{eqnarray}
$C(t,t)=1$ implying
$C_{\sc st}(0) +C_{\sc ag}(t,t)=1$ 
%\end{eqnarray}
with
%\begin{eqnarray}
$C_{\sc st}(0)=1-q_{\sc ea}$ 
%\;\;\;\;\;\;\;\;\;\;\;\;\;\;
and 
$C_{\sc ag}(t,t) = q_{\sc ea}$.
%\end{eqnarray}
Together with 
Eq. (\ref{limitqq})  
they ensure that in the two-time sector in which
$C_{\sc st}$ decays from $1-q_{\sc ea}$ to $0$, 
$C_{\sc ag}$ is just a constant $q_{\sc ea}$. 
Instead, in the two-time sector in which $C_{\sc ag}$ decays from
$q_{\sc ea}$ to $0$, $C_{\sc st}$ vanishes identically.

The name {\sc web}~\cite{trap-model,Cuku1} reflects  
the fact that for short time-differences the 
system
behaves as if it were trapped in some region of phase 
space of ``size'' $q_{\sc ea}$ --
suggesting ergodicity breaking. However, it is always
able to escape this region in a time-scale $\tau_0(t')$ 
that depends upon its age $t'$. Hence, trapping is gradual and 
ergodicity breaking is {\it weak}. $q_{\sc ea}$ 
depends on temperature when $T< T_d$.

We have already described, phenomenologically, such a separation 
of time-scales in the decay of correlation functions when we discussed 
the domain growth problem and glassy dynamics 
in Section~\ref{interestingproblems}. 
The first term describes in 
this case the fast fluctuations within domains while the second term 
characterises the domain growth itself. A second example where 
such a separation of time-scales occurs are the trap models in phase space.  
The first term corresponds to the dynamics within the traps while the second
describes the wandering of the system from trap to trap. In glasses,
the first term corresponds to the rapid rattling of each particle within 
its cage while the second one describes the destruction of the cages
and hence the structural relaxation.

In driven models rendered stationary by a weak perturbation we also find a 
separation of time-scales with $\tau_0$ increasing with weaker 
strengths of the perturbation. We can also propose that $C$ and $R$ 
separate in two terms, both being stationary but evolving in 
different time-scales.

In classical purely relaxational models governed by a 
Langevin equation with no inertia the correlation functions are monotonic with
respect to both times $t$ and $t'$, as it is easily checked 
numerically. Inertia introduces oscillations
and the decay can be non-monotonic. The magnitude of the oscillations 
depends upon the relative
value of the mass $M$ with respect to the other parameters in the problem.
However, for a wide choice of parameters
the oscillations appear only in the stationary regime, the aging dynamics
having a monotonic decay towards zero. This is relevant since it
allows one to use the general properties of monotonic correlation
functions proven in \cite{Cuku2} and discussed in Section~\ref{time-scales}
to find the two-time scaling of $C_{\sc ag}(t,t')$.

\subsection{The weak long-term memory scenario}
\label{subsec:wltm}

Regarding the response function, we propose a similar separation in two
terms:
\begin{equation}
R(t,t') = R_{\sc st}(t-t') + R_{\sc ag}(t,t')
\label{Rtwo}
\end{equation}
with
%\begin{eqnarray}
$R_{\sc st}(t-t') \equiv \lim_{t'\to\infty} R(t,t')$.
%\; .
%\end{eqnarray}
The matching conditions close to equal times are different for 
a model with or without inertia. In the former case, 
%\begin{equation}
$R(t,t) = 0$,  
%\;\;\;\;\;\;\;\;\;\;\;\;\;\;\;
$
R(t,t^-) = 1/M$
%\; , 
%\end{equation}
while in the latter, using the Ito convention,
%\begin{equation}
$R(t,t) = 0 
\; ,
%\;\;\;\;\;\;\;\;\;\;\;\;\;\;\;
R(t,t^-) = 1/\gamma
$.
%\; . 
%\end{equation}
In both cases the equal-times condition implies
%\begin{eqnarray}
$R_{\sc st}(0)=0$,  
%\;\;\;\;\;\;\;\;\;\;\;\;\;\;\; 
$
R_{\sc ag}(t,t) = 0
$
%\; ,
%\end{eqnarray}
while the next-to-main diagonal values yield
%\begin{eqnarray}
$
R_{\sc st}(\delta)=1/M, \;  
%\;\;\;\;\;\;\;\;\;\;\;\;\;\;\;
R_{\sc ag}(t,t-\delta) = 0
$
%\; ,
%\nonumber\\
and 
$
R_{\sc st}(\delta)=1/\gamma
\; , 
%\;\;\;\;\;\;\;\;\;\;\;\;\;\;\;  
R_{\sc ag}(t,t-\delta) = 0
$,
%\; ,
%\end{eqnarray}
respectively.

The response tends to zero when times get far apart,
and its integral over a {\it finite} time-interval as well:
\begin{equation}
\lim_{t\to \infty} R(t,t') =0 
\; ,
\label{shortR}
%\end{equation}
\;\;\;\;\;\;\;\;\;\;\;\;\;
%\begin{equation}
 \lim_{t\to\infty} \int_0^{t'} dt'' \, R(t,t'') = 0
\;\;\;\;\;\;\;\;\;\;\;\;\;\forall \; \mbox{fixed} \; t'
\; .
\label{finitetimes}
\end{equation}
These properties imply
\begin{eqnarray}
\lim_{t-t'\to\infty} \lim_{t'\to\infty} R(t,t') = 0 
&\Rightarrow& 
\lim_{t-t'\to\infty} R_{\sc st}(t-t')=0 
\; ,
\;
\lim_{t\to\infty} R_{\sc ag}(t,t') = 0
\; .
\label{limit0R}
\end{eqnarray}
However, the contribution of the response to the dynamic equations
and to other measurable quantities is not trivial.  Examining the 
integral of the response function over a growing time interval one
finds that even if the response vanishes, it yields a contribution 
to the integration. Figure~\ref{sketch-fdt-time-freq-domain}-left 
shows the integrated linear response (\ref{int-linear-response}).
%\begin{equation}
%\chi(t,t') = \int_{t'}^t dt'' \; R(t,t'')
%\end{equation}
Using (\ref{Rtwo})
\begin{equation}
\chi(t,t') = \int_{t'}^t dt'' \; 
\left[ R_{\sc st}(t-t'') + R_{\sc ag}(t,t'') \right]
=\chi_{\sc st}(t-t') + \chi_{\sc ag}(t,t')  
\; .
\label{separationR}
\end{equation}
If, for long enough $t'$, the contribution of the 
second term in (\ref{separationR}) were negligible, $\chi(t,t')$ should be a
stationary quantity. Instead, for all $t'$s studied
and for $t$ long enough 
one clearly sees a waiting-time
dependence that can only come
from the integration of the second term. This is a
{\it weak long-term memory} ({\sc wltm}), the system has an ``averaged'' 
memory of its past.

When a system is in equilibrium, the response is simply 
related to the correlation via {\sc fdt}. We then assume
(and test on the dynamic equations) that
the dynamics in the stationary 
regime satisfies {\sc fdt}:
\begin{eqnarray}
R_{\sc st}(\tau) &=& \frac{1}{k_B T} \frac{dC_{\sc st}(\tau)}{d\tau}
\;\;\;\;\;\;\tau \geq 0 
\; ,
\nonumber\\
R_{\sc st}(\omega) &=& - \frac{2}{\hbar} \lim_{\epsilon\to  0^+} 
\int \frac{d\omega'}{2\pi} \; \frac{1}{\omega-\omega'+i \epsilon} \,
\tanh\left( \frac{\beta\hbar\omega'}{2}\right) \, C_{\sc st}(\omega')
\end{eqnarray}
in a classical and quantum problem, respectively.
One can formally prove that {\sc fdt} has to 
hold for any generic
relaxing model fro short time-differences~\cite{Cudeku}, 
see Section~\ref{subsec:bound}.
For longer time-differences, when $C_{\sc ag}$ and $R_{\sc ag}$ 
vary in time while 
$C_{\sc st}$ and $R_{\sc st}$ have decayed to zero, one cannot 
assume the validity of 
{\sc fdt} and, as we shall see, the equations have a solution that explicitly
modifies {\sc fdt}.

\subsection{Slow time-reparametrization invariant dynamics}
\label{slowdynamics}

We have already mentioned that the correlations decay monotonically
(only below $q_{\sc ea}$ if $M\neq 0$). 
The final insight coming from the numerical solution to the 
full equations is that
the dynamics becomes slower and slower for fixed waiting-time and as 
$t-t'$ 
increases. In the stationary regime
$\partial_{t^2} [C(t,t'),R(t,t')]$ and
$\partial_{t^2} [C(t,t'),R(t,t')]$
are not negligible with respect to the terms in 
the {\sc rhs}
of Eqs.~(\ref{eqR2}) and (\ref{eqC2}).  On the contrary,
in the second decay below $q_{\sc ea}$, $C$ and $R$ 
decay in a much 
slower manner such that, 
$\partial_t C(t,t') \ll -\mu(t) C(t,t')$
and 
$\partial_{t^2} C(t,t') \ll -\mu(t) C(t,t')$ (similarly for $R$), 
and the time-derivatives can be neglected.

We choose the following strategy to solve the equations in the long $t'$ limit
where a sharp separation of time-scales can be safely assumed. 
First, we take advantage of the fact that one-time quantities approach 
a limit, as one can verify numerically, and write the asymptotic form 
of Eq.~(\ref{zt}) for $\mu_\infty$. The integrals on the {\sc rhs} are
approximated using the separation of $C$ and $R$ in two terms that vary in
different time-scales that we assume are well-separated. We detail
this calculation below. As regards to the equations for $C$ and $R$, we proceed
in two steps. 
On the one hand, we choose $t-t'$ short in such a way that $C>q_{\sc ea}$ and we write 
the dynamic equations for $C_{\sc st}$ and $R_{\sc st}$.
On the 
other hand, we take $t$ and $t'$ widely separated so as $C<q_{\sc ea}$ and we 
write the dynamic equations for $C_{\sc ag}$ and $R_{\sc ag}$. 
In this way we double the number of 
unknown functions and equations but we simplify the problem enough 
as to make it solvable.

Once the time-derivatives are neglected 
and the integrals are approximated as we explain in 
Section~\ref{subsec:aging-regime}
the aging equations become invariant under reparametrizations of time
$t \to h(t)$ that transform the two-point functions 
as
\begin{equation}
C_{\sc ag}(t,t') \to 
C_{\sc ag}(h(t),h(t')) \; , \;\;\;\;
R_{\sc ag}(t,t') \to [d_{t'} h(t')] \; R_{\sc ag}(h(t),h(t')) \; .
\label{rpg}
\end{equation}
This is not an exact invariance of the dynamic equations. It is
only generated when dropping the time-derivatives. This invariance was first 
noticed by Sompolinsky~\cite{Sozi} in his study of the equilibrium
dynamics (see also~\cite{Levetal} and it later appeared in the nonequilibrium
dynamics~\cite{Cuku1,Cuku2,Mefr,Cuku4,quantum-others}.
We shall 
see that this approximation forbids us to solve completely the 
dynamic equations, in particular, to fix the time scaling
 (select $h(t)$).

\subsection{Correlation scales}
\label{time-scales}

Take three ordered times
%\begin{equation}
$t_3 \geq t_2 \geq t_1$.
%\label{ordered-times}
%\; .
%\end{equation}
%igure~\ref{sketch-traj} is a sketch of a trajectory in phase space, the
%four points on the line represent the instantaneous configurations
%of the system. 
The correlations  are
%\begin{equation}
$C(t_i,t_j) = \frac1{N} \sum_k \langle s_k(t_i) s_k(t_j)  \rangle 
\equiv
%\cos(\hat{\vec s(t_2), \vec s(t_1)}) = 
\cos \theta_{ji}
\; .$
%\end{equation}
The monotonicity of the decay of the correlations with respect to the 
longer time (keeping the shorter time fixed) and the shorter time (keeping the 
longer time fixed) allows us to derive general properties 
that strongly constrain the possible scaling forms. Indeed, one 
can relate any three
correlation functions via {\it triangle relations}~\cite{Cuku2} 
constructed as follows.
Using the fact that the decay is monotonic, one can 
invert the relation between 
correlation and times to write, for example,
%\begin{equation}
$t_2 = g(C(t_2,t_1), t_1) $
%\end{equation}
with $g: [0,1]\times[0,\infty] \to [0,\infty]$.
This allows us to rewrite $C(t_3,t_1)$ as 
\begin{equation}
C(t_3,t_1) = 
C(g(C(t_3,t_2),t_2), t_1) = C(g(C(t_3, t_2), g(C(t_2,t_1),t_1), t_1)
\; .
\end{equation}
We now define a real function $f(x,y)$, $f: [0,1]\times[0,1] 
\to [0,1]$, by taking the limit $t_1\to\infty$ while keeping 
the intermediate correlations fixed
\begin{eqnarray}
\lim_{
\begin{array}{c}
t_1 \to\infty
\nonumber\\
C(t_2,t_1) \; \mbox{and} \; C(t_3,t_2) \; \mbox{fixed}
\end{array}
}
C(t_3,t_1) 
&=& f(C(t_3,t_2),C(t_2,t_1))
\; .
\end{eqnarray}
The fact that the limit exists is a reasonable working assumption.
This function
completely characterizes the  correlations and their 
scales in the asymptotic
limit. 
(Note that we defined $f$ using the correlation between the 
longest time and the intermediate as the first argument.)

%\begin{figure}[h]
%\centerline{
%%\psfig{file=.ps,width=4cm}
%}
%\vspace{0.25cm}
%\caption{Sketch of a trajectory in phase space. Four 
%instantaneous configuration are indicated by the points.}
%\label{sketch-traj}
%\end{figure}

\subsubsection{Properties}
\label{subsec:properties}

The definition of the function $f$, as well as the properties 
shown in this Subsection, are model independent.
The form taken by $f$ for each model is determined by the 
dynamic equations.

\vspace{0.25cm}
\noindent{\it Time reparametrization invariance}
The function $f$ is invariant under reparametrizations 
of time that satisfy (\ref{rpg}).

\vspace{0.25cm}
\noindent{\it Associativity}
%\vspace{0.25cm}
Take now four times $t_4 \geq t_3 \geq t_2 \geq t_1$. The correlation 
between $t_4$ and $t_1$ can be written in two ways
\begin{eqnarray}
C(t_4,t_1) &=& f(C(t_4,t_2), C(t_2,t_1)) = 
f(f(C(t_4,t_3), C(t_3,t_2)), C(t_2,t_1))
\; ,
\nonumber\\
C(t_4,t_1) &=& 
f(C(t_4,t_3), C(t_3,t_1)) = f(C(t_4,t_3), f(C(t_3,t_2), C(t_2,t_1)))
\; .
\nonumber
\end{eqnarray} 
Thus $f$ satisfies
$f(f(x,y),z) = f(x,f(y,z))$, {\it i.e.} 
it is an associative function.

\vspace{0.25cm}
\noindent{\it Identity.}
%\vspace{0.25cm}
If one takes $t_1=t_2$
%, $C(t_2,t_1)=1$. Then
\begin{equation}
C(t_3,t_1) = f(C(t_3,t_2),C(t_2,t_1)) = 
f(C(t_3,t_1),C(t_1,t_1))= f(C(t_3,t_1),1) 
\; ,
\end{equation}
for all $C(t_3,t_1) \in [0,1]$. 
Equivalently, if one takes $t_2=t_3$
%, $C(t_3,t_2)=1$. Then
\begin{equation}
C(t_3,t_1) = f(C(t_3,t_2), C(t_2,t_1)) = f(C(t_3,t_3), C(t_3,t_1)) =
f(1,C(t_3,t_1))
\; ,
\end{equation}
for all $C(t_3,t_1) \in [0,1]$. The correlation at equal times
acts as the identity since $x=f(x,1)$ and $y=f(1,y)$ for 
all $x,y \in [0,1]$. 

\vspace{0.25cm}
\noindent{\it Zero}.
%\vspace{0.25cm}
Taking $t_3$ and $t_2$ and much larger than $t_1$ 
in such a way that 
$C(t_2,t_1)\sim 0$ and $C(t_3,t_1)\sim 0$
while $C(t_3,t_2)>0$, 
\begin{equation}
0 \sim
C(t_3,t_1) = f(C(t_3,t_2), C(t_2,t_1)) \sim f(C(t_3,t_2),0)
\; . 
\end{equation} 
Equivalently, taking $t_3 \gg t_2$ and $t_1$, 
then $C(t_3,t_2)\sim 0$  and 
$C(t_3,t_1)\sim 0$ while  
$C(t_2,t_1)>0$ and one has 
\begin{equation}
0 \sim 
C(t_3,t_1) = f(C(t_3,t_2), C(t_2,t_1)) \sim f(0,C(t_2,t_1))
\; . 
\end{equation}
The minimum correlation acts as a zero of $f(x,y)$ since $0=f(x,0)$ 
and $0=f(0,y)$ for all $x,y \in [0,1]$.  (This property can 
be easily generalised if the correlation approaches a non-zero 
limit.)

\vspace{0.25cm}
\noindent{\it Bound.}
%\vspace{0.25cm}
Given that we assume that the system drifts away in 
phase space, $C(t_2,t_1)$
decays as a function of $t_2$ for $t_1$ fixed, and 
$C(t_2,t_1)$
increases as a function of $t_1$ for $t_2$ fixed. This 
property implies
\begin{eqnarray}
y = f(1,y) \geq f(x,y) \;\;\;\forall y, \; x<1
\; , 
\;\;\;\;\;\;\;
%\nonumber\\
x= f(x,1) \geq f(x,y) \;\;\;\forall x, \; y<1
\; .
\end{eqnarray}
Therefore
%\begin{equation}
$f(x,y) \leq \min(x,y)$.
%\end{equation}

\vspace{0.25cm}
\noindent{\it Forms for $f$}
%\vspace{0.25cm}
In \cite{Cuku2} we proved that
\begin{eqnarray}
f(x,y) &=& \jmath^{-1} \left(\jmath(x) \jmath(y) \right)
\;\;\;\;\; \mbox{Isomorphic to the product}
\label{isoproduct}
\\
f(x,y) &=& \min(x,y) 
\;\;\;\;\; \;\;\;\;\; \;\;
\mbox{Ultrametricity}
\label{ultrametricity}
\end{eqnarray}
are the only possible forms that satisfy the properties of 
$f$ shown above.
Note that for $\jmath$ equal to the identity the first type of function
becomes simply $f(x,y)=x y$, hence the name. 
It is also possible to prove that the first kind of function 
(\ref{isoproduct})
is only compatible with the time scaling~\cite{Hazewinkel,Cuku2}
\begin{equation}
C(t_2,t_1) = \jmath^{-1} \left( \frac{h(t_2)}{h(t_1)}\right)
\label{onescale}
\end{equation}
with $h(t)$ a monotonically growing function.
The actual correlation can have a piecewise form.
Here, instead of reproducing the proofs given in \cite{Cuku2}
we explain these statements reviewing
the scaling forms found for some physical systems
and in the analytic solution to mean-field
models.

\vspace{0.25cm}
\noindent\noindent{\it Examples: domain growth}
\vspace{0.25cm}

The correlation decays in two steps, 
see the right panel in Fig.~\ref{sketch-scales-highT}
and for $C>q_{\sc ea}=m_{\sc eq}^2$ the decay is stationary:
%In order to simplify the discussion
%let us assume that this decay is simply exponential
%\begin{equation}
%C_{\sc st}(t_2,t_1) = q_{\sc ea} +(1-q_{\sc ea}) \, \exp(-\alpha (t_2-t_1))
%\; .
\begin{equation}
C_{21} \equiv C(t_2,t_1) = q_{\sc ea} + C_{\sc st}(t_2-t_1) 
\; ,
\label{form1}
\end{equation}
and it can be put in  the form (\ref{onescale}) using $h(t)=\exp(\ln t))$
and $\jmath^{-1}(x)=q_{\sc ea}+C_{\sc st}(x)$.  
Any three correlation satisfying 
(\ref{form1}) also verify 
$t_3-t_1 = C_{\sc st}^{-1}(C_{31}-q_{\sc ea})
= t_3-t_2 + t_2 - t_1
= 
C_{\sc st}^{-1}(C_{32}-q_{\sc ea}) + C_{\sc st}^{-1}(C_{21}-q_{\sc ea})  
$ that implies
\begin{eqnarray}
C_{31} = 
C_{\sc st}[
C_{\sc st}^{-1}(C_{32}-q_{\sc ea}) + C_{\sc st}^{-1}(C_{21}-q_{\sc ea})  
]+q_{\sc ea}
\; .
\end{eqnarray}
This equation is equivalent to (\ref{isoproduct}).
This means that any three correlations above 
$q_{\sc ea}$ can be related with an $f$ that is isomorphic
to the product, see (\ref{isoproduct}),  with 
$\jmath_{\sc st}^{-1}(x) = C_{\sc st}(\ln x) +q_{\sc ea}$ and  
$\jmath_{\sc st}(x) = \exp(C^{-1}_{\sc st}(x-q_{\sc ea})$.

When the times are such that the domain walls move, the self-correlation
decays below $q_{\sc ea}$ in an aging manner, with 
\begin{equation}
C_{21} \equiv C(t_2,t_1) = C_{\sc ag}(t_2,t_1) = 
\jmath_{\sc ag}^{-1} \left( \frac{{\cal R}(t_2)}{{\cal R}(t_1)} \right)
\; ,
\label{form2}
\end{equation}
$\jmath_{\sc ag}^{-1}(1)=q_{\sc ea}$ and $\jmath_{\sc ag}^{-1}(0)=0$.
It is obvious that any three correlations below 
$q_{\sc ea}$ also satisfy (\ref{isoproduct}) 

Take now $t_3=t_2+\tau_{32}$ with $\tau_{32} < \tau_0(t_2)$ 
and $C_{32}>q_{\sc ea}$, 
and $t_3$ and $t_2$ sufficiently larger than $t_1$
($t_3=t_1+\tau_{31}$ with $\tau_{31} > \tau_0(t_1)$
and $t_2=t_1+\tau_{21}$ with $\tau_{21} > \tau_0(t_1)$)
such that $C_{31}<q_{\sc ea}$ and $C_{32}<q_{\sc ea}$. One has
\begin{eqnarray*}
C_{31} &=& \jmath_{\sc ag}^{-1} 
\left( \frac{{\cal R}(t_3)}{{\cal R}(t_1)}\right)
=
\jmath_{\sc ag}^{-1} \left( 
\frac{{\cal R}(t_3)}{{\cal R}(t_2)}
(\jmath_{\sc ag} \otimes \jmath_{\sc ag}^{-1}) 
\left( \frac{{\cal R}(t_2)}{{\cal R}(t_1)} \right) 
\right)
\nonumber\\
&=& 
%=
\jmath_{\sc ag}^{-1} \left( \frac{{\cal R}(t_3)}{{\cal R}(t_2)} 
\jmath_{\sc ag}(C_{\sc 21}) 
\right) = C_{21}
\; .
\end{eqnarray*}
The last idendity is a consequence of 
${\cal R}(t_3)/{\cal R}(t_2)\sim 1$ since for a sufficiently
small $\tau_{32}$, 
${\cal R}'(t_2) \tau_{32}/{\cal R}(t_2) \ll 1$. 

Thus, when the times are such that two correlations, say with values 
 $a$ and $b$, are both 
greater than $q_{\sc ea}$ one explores the dynamics in the 
stationary regime and $f(a,b)$ is isomorphic to the product. When 
they are both smaller that $q_{\sc ea}$ one explores the dynamics
in the aging coarsening regime and again $f(a,b)$ is isomorphic to the 
product though with a different function $\jmath$. Finally,
if $a> q_{\sc ea}$ and $b < q_{\sc ea}$, $f(a,b)=\min(a,b)$ and 
one finds dynamic ultrametricity.

The structure discussed in the context of the domain growth problem 
is indeed generic. Some special values of the correlation
act as ``fixed points'' of $f(a,a)$, $f(a,a)=a$. A
``correlation scale'' spans the values of correlations 
comprised between two subsequent fixed points.
Within a correlation scale $f$ is isomorphic to the product.
Any two correlations falling into different correlation 
scales are related by an ultrametric $f$.
In the domain growth example 
$1$, $q_{\sc ea}$ and $0$ are fixed points that are 
simple to visualize physically. In more abstract models as
the {\sc sk} spin-glass the form of $f$ is 
more involved, with a stationary scale between $1$ and $q_{\sc ea}$ 
and a dense set fixed points, hence correlation scales, that fill 
the interval  $[0,q_{\sc ea}]$.

\vspace{0.25cm}
\noindent\noindent{\it Scaling functions}
\vspace{0.25cm}

\begin{figure}[ht]
\centerline{
\psfig{file=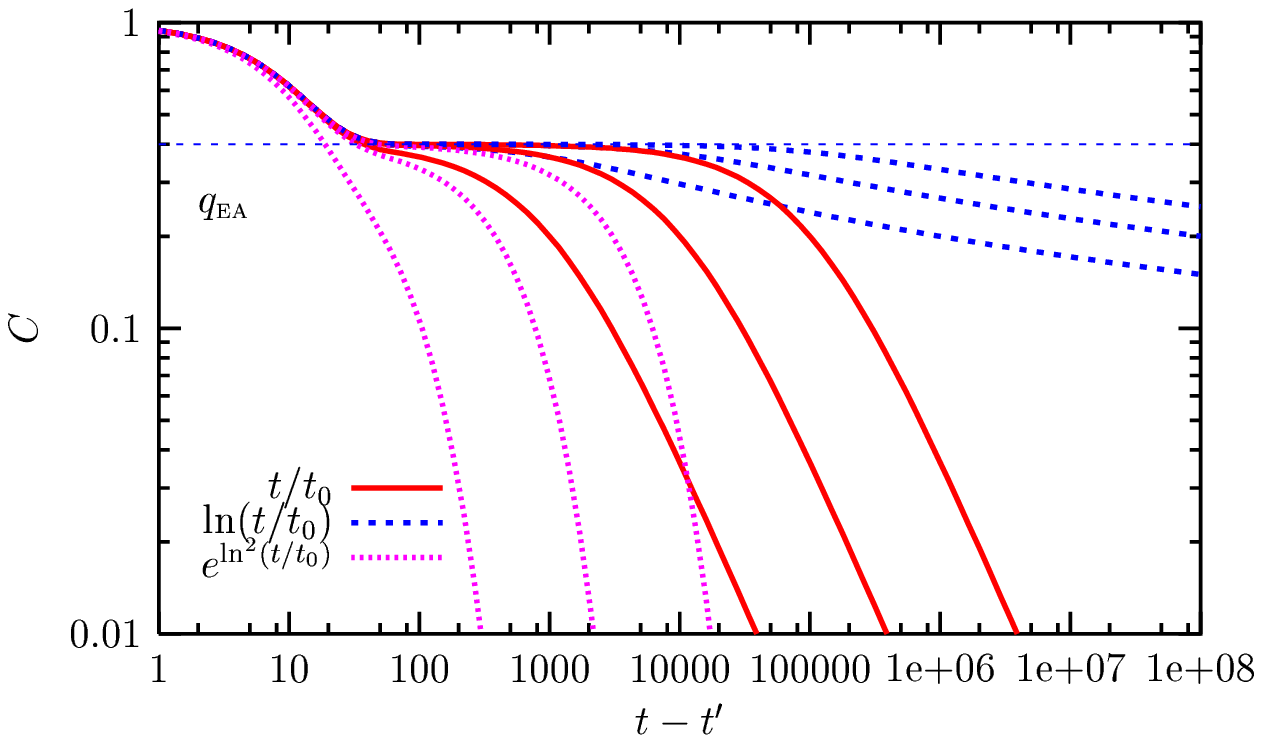,width=8cm}
\hspace{-1cm}
\psfig{file=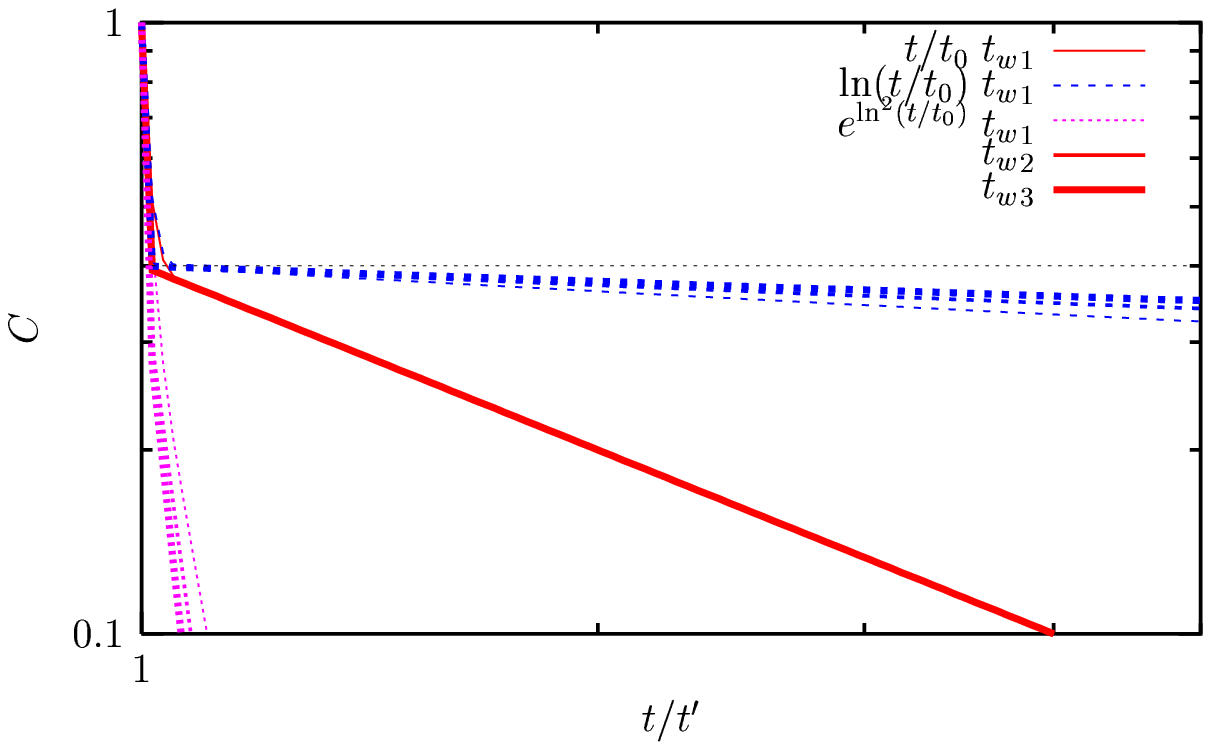,width=8cm}
}
\caption{Comparison between three $h(t)$s, power law, enhanced 
power law and logarithm. Plot of 
$C(t,t') = (1-q_{\sc ea}) \exp(-(t-t'))+ q_{\sc ea} h(t)/h(t')$
against the time-difference $t-t'$ (on the left) and against
the ratio of times $t/t'$ (on the right) for three 
waiting times. Note the drift of the curves 
in the right panel. For the logarithmic law (sub-aging) the curves 
drift towards the left for increasing waiting-time. Instead, 
for the enhanced power law (super-aging) 
the curves drift to the right 
for increasing waiting-time. For the power law (simple aging)
the scaling is perfect. In real systems the decay of
the stationary part towards $q_{\sc ea}$ is much slower than 
exponential (typically power law with a small exponent) and 
the separation of time-scales is not so neat.  
}
\label{cfr-scaling}
\end{figure}

Most solvable models, numerical data and experimental results
can be described with only two correlation scales, a stationary and a slow
one. Several scaling functions $h(t)$ for the slow decay
have been proposed in the 
literature. In the following we summarize and discuss the main ones.
In Fig.~\ref{cfr-scaling} we compare the
decay of the correlation from $q_{\sc ea}$ 
for three of the four laws discussed below. 

\vspace{0.2cm}
\noindent{Power law:} $h(t) = a t^\alpha$.
%\vspace{0.2cm}
This is the simplest scaling also called {\it simple aging}.
Ferromagnetic domain growth realizes this form with $\alpha=1/2$
for non conserved dynamics and $\alpha=1/3$ for conserved dynamics~\cite{Bray}.
Several solvable model have simple aging, an example being the 
classical spherical $p=2$ model~\cite{p=2-statics,p=2}.
In~\cite{Cuku1} it was conjectured that a power law also characterized the
aging dynamics of the fully connected $p$ spin-model with $p\geq 3$. 
This was later
confirmed with the algorithm of Kim and Latz~\cite{KimLatz} that allows
one to reach much longer times. Aging below $T_c$ 
in the simplest trap model also scales with this law~\cite{trap-model}. 
The molecular dynamic simulations of Lennard-Jones mixtures 
show this type of scaling too. Note that for all $\alpha$, $C$
scales as a function of $t_2/t_1$.

\vspace{0.2cm}
\noindent{Enhanced power law:} $h(t) = \exp(\ln^\alpha(t/t_0))$  
%\vspace{0.2cm}
This law yields the most accurate description of spin-glass experimental data.
The exponent $\alpha$ typically takes a possibly $T$-dependent 
value about $2$~\cite{experiments}.

\vspace{0.2cm}
\noindent{Stretched exponential}: $h(t) = \exp[(t/t_0)^\alpha]$  
%\vspace{0.2cm}
This law has been proposed to describe the slowing down of the full 
correlation above the 
critical temperature. As far as we know, no aging model that 
satisfies a scaling (\ref{onescale}) with a 
stretched exponential has been found yet.

\vspace{0.2cm}
\noindent{Logarithm}: $h(t)=\ln^\alpha(t/t_0)$
%\vspace{0.2cm}
In the Fisher and Huse droplet model for spin-glasses, activated dynamics
is assumed and the domains are found to grow as 
${\cal R}(t) \sim \ln(t/t_0)$.   
This leads to $C(t_2,t_1) \sim g(\ln(t_2/t_0)/\ln(t_1/t_0))$.
However, this law does not fit the aging experimental 
data~\cite{experiments}.

\begin{figure}[ht]
\centerline{
\psfig{file=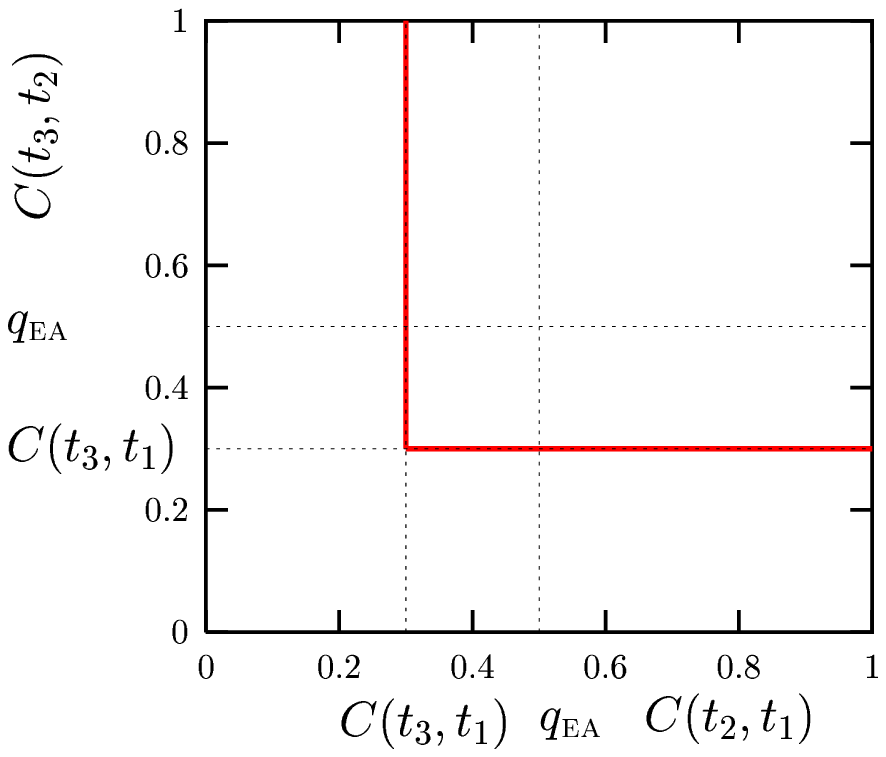,width=8cm}
\psfig{file=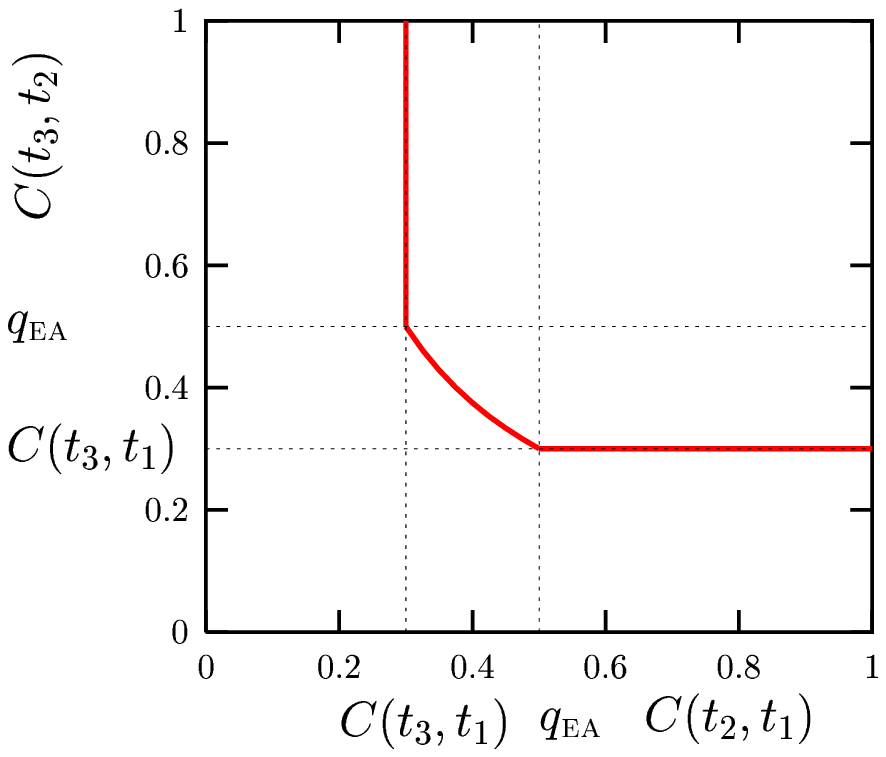,width=8cm}
}
\caption{Sketch of a check of ultrametricity using the parametric plot 
$C(t_3,t_2)$ against $C(t_2,t_1)$ for $C(t_3,t_1)=0.3< q_{\sc ea}$ 
fixed.
On the left an ultrametric model, on the right a model with 
two correlation scales.
}
\label{ultram}
\end{figure}

\vspace{0.2cm}
\noindent{Dynamic ultrametricity:}
%\vspace{0.2cm}
Even though it 
seems mysterious at first sight there is
a simple graphical construction that allows one to test 
it. Take two times $t_3 > t_1$ such that $C(t_3,t_1)$ equals some 
prescribed value, say $C(t_3,t_1)=0.3=C_{31}$. 
Plot now $C(t_3,t_2)$
against $C(t_2,t_1)$ using $t_2$, $t_1 \leq t_2 \leq t_3$, as  
a parameter. Depending on the value of $C_{31}$ with respect 
to $q_{\sc ea}$ we find two possible plots. If $C(t_3,t_1)> q_{\sc ea}$,
for long enough $t_1$,  
the function $f$ becomes isomorphic to the product. Plotting then 
$C(t_3,t_2)$ for longer and longer $t_1$, the construction approaches a 
limit in which $C(t_3,t_2)
=
\jmath^{-1} (\jmath(C_{31})/\jmath(C(t_2,t_1)))$.
If, instead, $C_{31}< q_{\sc ea}$, in the long $t_1$ limit the 
construction approaches a different curve. We sketch in Fig.~\ref{ultram}
two possible outcomes of this construction. 
On the right, we represent a model with two 
correlation scales, ultrametricity holds between them and within 
each of them $f$ is isomorphic to the product. 
On the left  instead
we represent a model such that dynamic ultrametricity holds 
for all correlations below $q_{\sc ea}$. The construction approaches, 
in the long $t_1$ limit, the broken curve depicted in the sketch.

The {\sc sk}  spin-glass~\cite{Cuku2} and the dynamics of 
manifolds in an infinite dimensional embedding space in the 
presence of a random potential with long-range 
correlations~\cite{Mefr,Cukule}
have ultrametric decays everywhere within the aging regime. This 
scaling is also found 
in the trap model at the critical temperature~\cite{Bertin}. 
Dynamic ultrametricity in finite dimensional
systems has been search numerically. There is some 
evidence for it in the $4d$EA model. In $3d$ instead the numerical data 
does not support this scaling~\cite{Picco,ultrametricity-num}. 
Whether this is due to the short times
involved or if the scaling asymptotic 
is different in $3d$ is still an open question.

\subsubsection{Definition of a characteristic time}

Expanding the argument in  (\ref{onescale}) for $t_2=t_1+\tau$ with 
$\tau \ll t_1$ one finds, to leading order,
%\begin{equation}
%\frac{h(t_1)}{h(t_2)} = 1 - \tau \, \frac{h'(t_1)}{h(t_1)} + 
%O\left(\tau^2 \left( \frac{{h'}^2(t_1)}{h^2(t_1)} + 
%\frac{h''(t_1)}{h(t_1)}\right) \right)
%\; .
%\end{equation}
%To leading order
\begin{equation}
\frac{h(t_1)}{h(t_2)} = 1 - \frac{\tau}{t_c(t_1)} \, 
\;\;\;\;\;\;\;\;\;
t_c(t_1) \equiv \left( \frac{h'(t_1)}{h(t_1)} \right)^{-1}
\; ,
\label{expansion}
\end{equation} 
with $O\left(\tau^2 \left({h'}^2(t_1)/h^2(t_1) + 
h''(t_1)/h(t_1)\right) \right)$ corrections.
The characteristic time $t_c(t_1)$ is given by 
\begin{eqnarray}
t_c(t_1) &=&
\left\{
\begin{array}{cc} 
t_1/\alpha \;\;\;\;\;\; & \mbox{Power law}
\nonumber\\
t_1/[\alpha \ln^{\alpha-1}(t_1/t_0)]
\;\;\;\;\;\; & \mbox{Enhanced power law}
\nonumber\\
t_1 \left(t_0/t_1\right)^\alpha
\;\;\;\;\;\; & \mbox{Stretched exponential}
\nonumber\\
t_1 \, \ln(t_1/t_0)
\;\;\;\;\;\; & \mbox{Logarithm}
\end{array}
\right.
\end{eqnarray}
Note that $t_c(t_1)$ is defined close to the limit of equal times
and (\ref{expansion}) does not make sense for 
large $\tau$. Rather often 
in the literature, the scaling variable $
x=\tau/t^a_1 
$
has been used even for large values of $\tau$. 
This scaling is incompatible with the general properties of the 
triangular relations recalled in 
Section~\ref{subsec:properties} if the exponent $a$ is 
larger than $1$~\cite{jorge2}. See the right panel in 
Fig.~\ref{cfr-scaling}
to see the different trends of these scalings when plotted
as functions of $t/t'$.

For the power law $t_c(t_1)$ scales just as $t_1$. In the cases of the
stretched exponential and the enhanced power law, $t_c(t_1)$ has a slower 
growth than the linear dependence iff $\alpha>0$ in the first case and 
$\alpha>1$ in the second. This behavior has been called {\it sub-aging}.
For the logarithm $t_c(t_2)$ grows
faster than linearly.
This function belongs to a different class that we called 
{\it super aging}~\cite{experiments}.

\subsection{Modifications of {\sc fdt}}

One of the most important outcomes of the analytic solution to 
the mean-field glassy models~\cite{Cuku1,Cuku2} is the need to modify the 
fluctuation--dissipation relations between linear responses, 
$R(t,t_w)$, and their partner correlations between spontaneous 
fluctuations, $C(t,t_w)$, when $T<T_d$.
In this Subsection we discuss different ways of presenting 
the modification of {\sc fdt} expected in rather 
generic systems with slow dynamics. 

\subsubsection{Time domain}

\begin{figure}
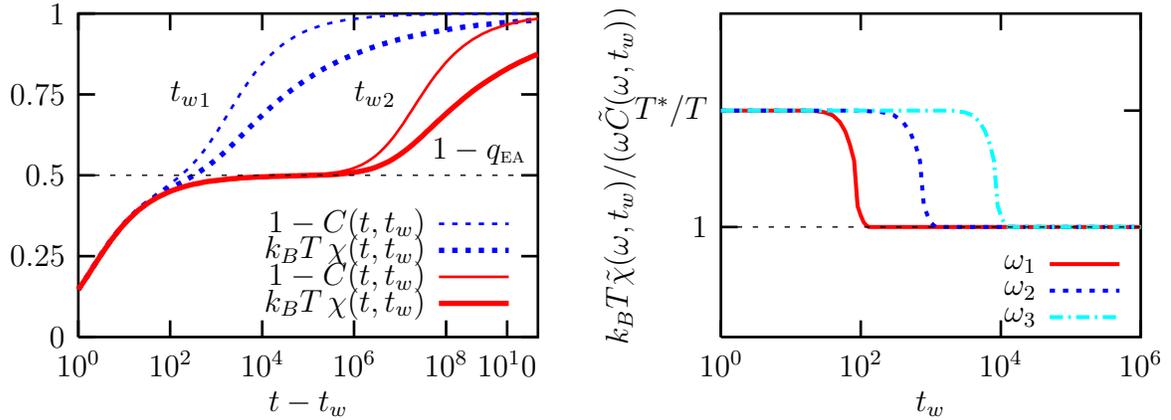

\centerline{
\input{sketch-fdt-time-domain.pslatex}
\input{sketch-fdt-freq-domain.pslatex}
}
\caption{Left: sketch of the modification of {\sc fdt} in the time-domain.
Right: sketch of the modification of {\sc fdt} in the frequency domain 
for a glassy system, $\omega_1>\omega_2>\omega_3$.}
\label{sketch-fdt-time-freq-domain}
\end{figure}

The {\sc fdt} is a linear relation between $\chi(t,t_w)$ and 
$C(t,t_w)$ for any pair of times $(t,t_w)$, see Eq.~(\ref{int-fdt}).   
In early simulations of the $3d${\sc ea} 
model~\cite{suecos} as well as in the analytic solution to 
fully-connected disordered models a modification of this relation 
below $T_d$ appeared~\cite{Cuku1}. Plotting $k_B T \chi(t,t_w)$ and $1-C(t,t_w)$ 
for $t_w$ fixed as a function of $\ln (t-t_w)$ one typically obtains 
the pair of curves schematically shown on the left panel of 
Fig.~\ref{sketch-fdt-time-freq-domain}. 
The two functions go together until $t-t_w$ reaches a characteristic 
time $\tau_0(t_w)$ and they then depart demonstrating that {\sc fdt} does not 
hold beyond this time-scale. The characteristic time $\tau_0(t_w)$ 
is of the order of the time needed to reach the plateau in the 
correlation function
(this holds for mean-field models but it is not certain
in finite dimensional systems).
Summarizing
\begin{eqnarray}
t-t_w < \tau_0(t_w) \;\;\; && \mbox{ {\sc fdt} holds in the fast scale}
\; ,
\\
t-t_w > \tau_0(t_w) \;\;\; && \mbox{ {\sc fdt} is modified in the slow scale}
\; ,
\end{eqnarray}
with $\tau_0(t_w)$ an increasing function of $t_w$ that depends
on the system considered (see Fig.~\ref{sketch-scales-highT}).

\subsubsection{Frequency domain}
\label{frequency-domain}

As explained in Section~\ref{average-thermal} taking a 
Fourier transform with respect
to the time-difference while keeping $t_w$ fixed allows one to 
work in a mixed frequency-time domain. Since many experimental 
set-ups are prepared to apply ac-fields it is particularly important
to predict the aspect {\sc fdt} modification have when using 
these parameters. The condition $t-t_w < \tau_0(t_w)$ to explore the 
fast relaxation 
roughly translates into $\omega^{-1} < \tau_0(t_w)$, {\it i.e.} for a fixed 
waiting-time high frequencies
are required. The longer the waiting time the lower the frequency one has 
to use to see this scale since $\tau_0(t_w)$ increases with $t_w$. 
Instead, when $t-t_w > \tau_0(t_w)$ one has 
$\omega^{-1} > \tau_0(t_w)$, and very low frequencies are 
needed to explore the 
slow scale.  These conditions imply 
\begin{eqnarray}
\omega \tau_0(t_w) > 1 \;\;\; && \mbox{{\sc fdt} holds in the fast scale}
\; ,
\nonumber\\
\omega \tau_0(t_w) < 1 \;\;\;  && \mbox{{\sc fdt} does not hold in the slow scale}
\; .
\end{eqnarray}
Reversing the argument above, 
if one weakly perturbs the sample with an ac-field of a fixed 
frequency $\omega_1$ at a chosen time $t_w$, one can 
follow the deviation from {\sc fdt} using  $t_w$ as the control
parameter. This procedure yields the solid line on the right
panel of  Fig.~\ref{sketch-fdt-time-freq-domain}. Choosing now
a lower frequency $\omega_2 (<\omega_1)$ the crossover from the slow
to the fast regime occurs at a larger value of $t_w$. One obtains then 
the dotted curve on the right panel of Fig.~\ref{sketch-fdt-time-freq-domain}. 
So on and so forth, the smaller the frequency of the applied ac-field the 
longer the slow regime lasts and the longer one sees deviations from 
{\sc fdt}. (Note that the probe does not modify the dynamics.)
In the Figure we chose to sketch the behavior of a system
with only two-time scales, in which the {\sc fdt} ratio takes two 
constant values separated at single breaking point in which the 
correlation reaches the plateau value $q_{\sc ea}$. 
This procedure is commonly employed experimentally, see 
Section~\ref{subsec:structural}
where we discuss the measurements of Grigera and Israeloff
for glycerol~\cite{Gris}. 

\subsubsection{Time-reparametrization invariant formulation}
\label{definitions}

A more interesting way of displaying the modification of the {\sc fdt}
has been suggested by the analytic solution to the mean-field models 
discussed in Section~\ref{pspin-solution}. One of its advantages 
is that it allows one
to classify the systems into sort of ``universality classes'' 
according to the form the {\sc fdt} modification takes. 

The analytic solution is such 
that, in the asymptotic limit in which the waiting-time $t_w$ 
diverges after $N\to\infty$, the integrated linear 
response approaches the limit
\begin{eqnarray}
\lim_{
\begin{array}{c}
t_w\to\infty\\
C(t,t_w)=C
\end{array}
} 
\chi(t,t_w) = \chi(C)
\; 
\label{FDT}
\end{eqnarray}
when $t_w$ and $t$ diverge while keeping the correlation between 
them fixed to $C$~\cite{Cuku2}. 
Deriving this relation with respect to the waiting time
$t_w$, one finds that the opposite of the inverse of the slope
of the curve $\chi(C)$ is a parameter that replaces temperature in 
the differential form of the {\sc fdt}.
Thus, using Eq.~(\ref{FDT}) one defines
\begin{equation}
k_B T_{\sc eff}(C) \equiv -(\chi'(C))^{-1}
\; ,
\label{teff_def}
\end{equation}
that can be a function of the correlation. Under certain
circumstances one can show that this quantity has the properties of a 
temperature~\cite{Cukupe} 
in the sense to be described in Section~\ref{temp_intro}.

One of the advantages of this formulation is that,
just as in the construction of triangle relations,
times have been ``divided away'' and the relation (\ref{FDT}) 
is invariant under the reparametrizations
of time defined in Eq.~(\ref{rpg}). 

\begin{figure}[ht]
\centerline{
\psfig{file=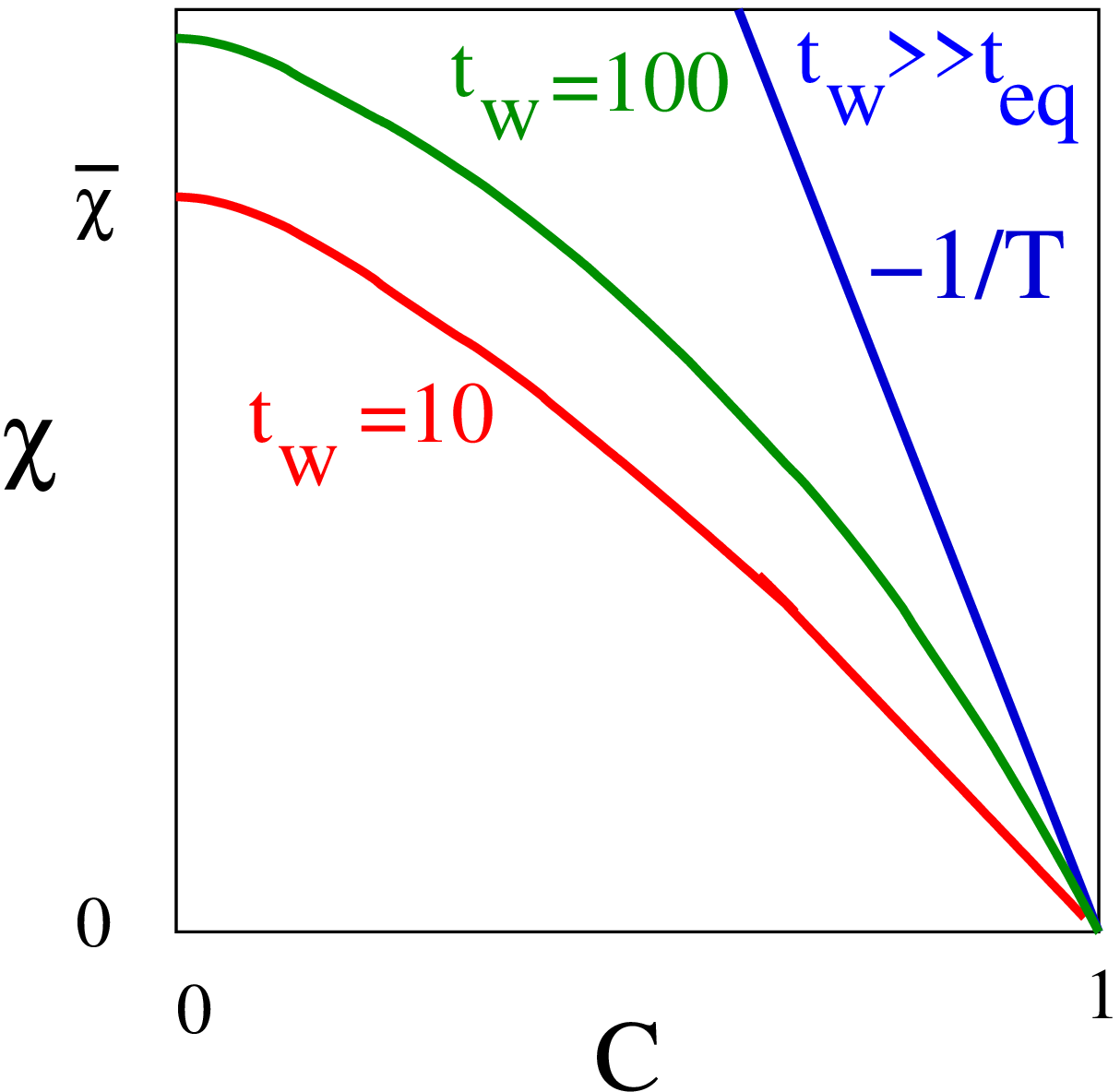,width=5cm}
\hspace{1cm}
\psfig{file=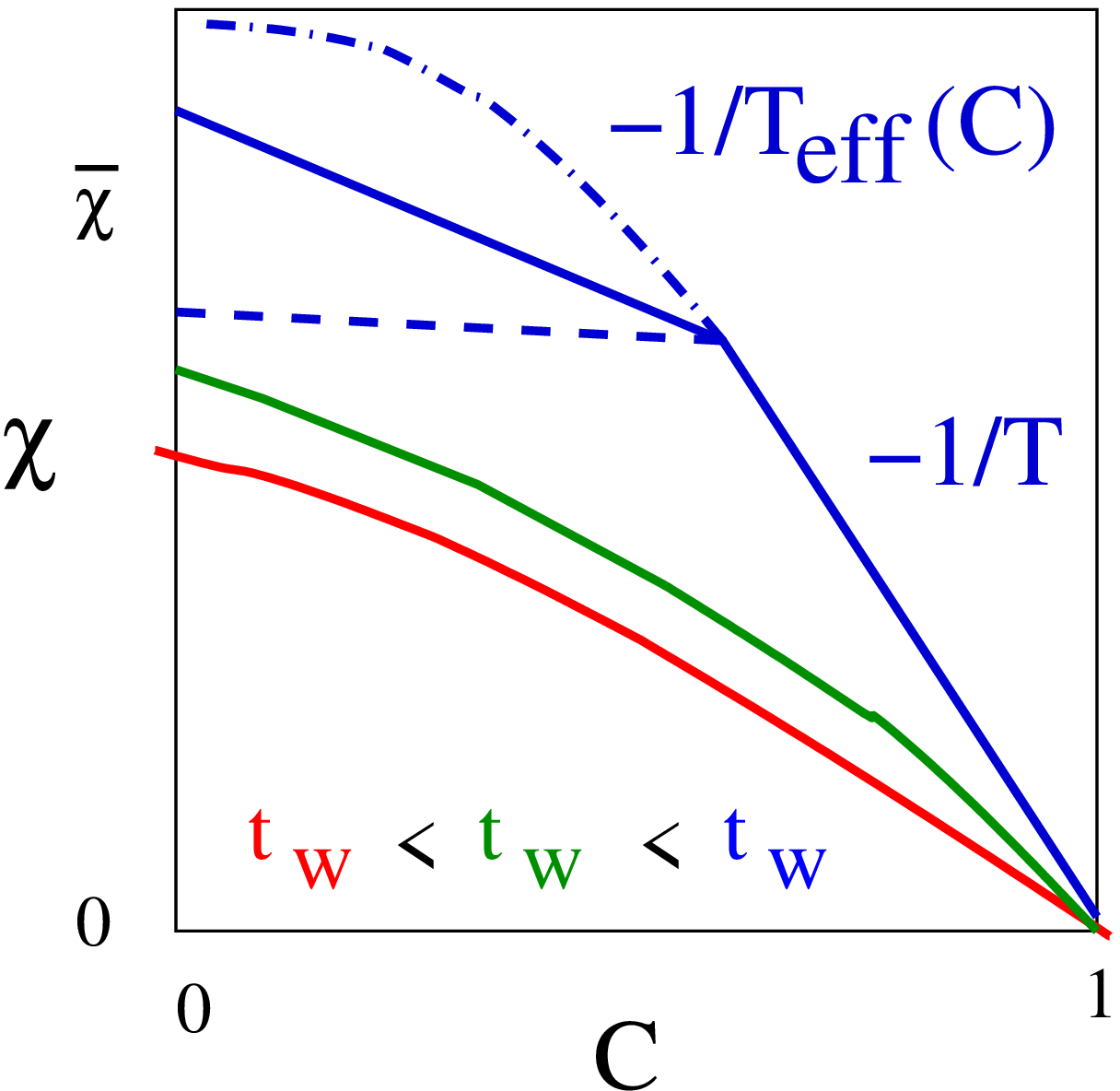,width=5cm}
}
\caption{
The asymptotic behavior of the integrated linear response against the 
correlation in a parametric plot, for fixed waiting time and using $t$ 
as a parameter. Left: behavior in equilibrium. 
Right: behavior in a slowly relaxing system out of equilibrium.
See text for an explanation.}
\label{FDT-high-low}
\end{figure}

Equation~(\ref{FDT}) is easy to understand graphically. 
Let us take a waiting time
$t_w$, say equal to $10$ time units after the preparation of the 
system (by this we mean that the temperature of the environment
has been set to $T$ at the initial time) and trace 
$\chi(t,t_w)$ against $C(t,t_w)$ using $t$ as a parameter 
($t$ varies between $t_w$ and infinity). If we choose to work with a 
correlation that is normalized to one at equal times, the parametric
curve starts at the point $(C(t_w,t_w)=1,\chi(t_w,t_w)=0$) and it 
arrives at the point $(C(t\to\infty,t_w)\to \overline C,\chi(t\to\infty,t_w)=
\overline \chi$). Without loss of generality we can assume that the 
correlation decays to zero, $\overline C=0$. 
This first curve is traced in red in 
Figs.~\ref{FDT-high-low}. Now, let us choose a longer waiting time, say 
$t_w=100$ time units, and reproduce this construction.
One finds the green curves in Figs.~\ref{FDT-high-low}.
Equation~(\ref{FDT}) states that if one repeats this construction 
for a sufficiently long waiting time, the parametric curve
approaches a limit $\chi(C)$, represented by the blue curves.

%** MODIFY THESE FIGS ***

When the system equilibrates with its environment, the construction 
approaches a straight line with slope
$-1/(k_B T)$  as predicted by the {\sc fdt}. This is the result shown in 
the left panel of Fig.~\ref{FDT-high-low}. 
Instead, for non-equilibrium systems evolving slowly 
the asymptotic limit is different, it is given by a {\it curve} 
$\chi(C)$. For solvable fully-connected models one distinguishes 
three families, as drawn in the right panel of Fig.~\ref{FDT-high-low}. 
They correspond to some systems 
undergoing domain growth~\cite{p=2} ({\it e.g.} the $O(N)$ model in $d=3$ when 
$N\to\infty$),
 systems behaving like structural glasses~\cite{Cuku1,Cule,Cukule}
({\it e.g.} the $p$-spin model) and 
spin-glasses~\cite{Cuku2,Mefr,Cule,Cukule} 
({\it e.g.} the {\sc sk} model). 
Several numerical studies in more realistic models of the 
three cases tend to confirm this classification~\cite{numerics}. 
However, two provisos are 
in order. First, one has to be 
very cautious about the numerical results given the very short time
scales and rather small system sizes 
accessible in simulations~\cite{Babe}. Second, as shown in 
Section~\ref{fdt-domain-growth}, 
at least one system that undergoes domain growth, the ferromagnetic 
chain, has a non-trivial $\chi(C)$ like the one found for the {\sc sk} 
model. 

We have already found these asymptotic $\chi(C)$ curves 
when we discussed the dynamics of a (flat) harmonic oscillator in 
contact with a complex bath made of subsystems with different 
characteristic times and temperatures (Section~\ref{complexbath}). Here we 
claim that the same structure arises in a glassy model coupled to
a white-bath. Different values of the effective temperature are
self-generated in the system.

This plot is invariant under reparametrisations of time
$t\to h(t)$ acting on the two-point functions
as in~Eqs.(\ref{rpg}). A different choice of the 
functions $h$ only changes the speed with which the
$\chi(C)$ curve is traced but not its form.

\subsubsection{{\sc fdt} part}
\label{subsec:bound}

The formalism used in Section~\ref{section:fdt} to derive the {\sc fdt} can 
be used to obtain a bound on {\sc fdt} violations~\cite{Cudeku}.
Indeed, one bounds the difference between response and variation
of the correlation with 
the Cauchy-Schwartz inequality leading to
\begin{equation}
\left|k_B TR(\tau+t_w,t_w) - \left. 
\partial_{s} C(\tau+t_w,s)\right|_{s=t_w} \right| \leq
c \sqrt{-d_{t_w} {\cal H}(t_w)}
\label{bound}
\end{equation}
where $c$ is a constant and 
${\cal H}(t_w)\equiv \int dq P(q,t_w) (E(q) - k_B T\ln P(q,t_w))$ 
is a positive definite function that monotonically decreases towards 
the free-energy when the system eventually equilibrates~\cite{Kubo}.
One finds a similar bound for 
Kramers processes and a generalization that includes the power input 
when time-dependent or non-potential forces are applied. 
For systems such that $d_{t_w} {\cal H}(t_w)\to 0 $ sufficiently fast 
when $t_w\to\infty$  the bound implies that 
the {\sc lhs} vanishes in this limit. This can be achieved in 
two ways: either each term is finite and the difference between them
vanishes or each term tends to zero independently. The former 
possibility is what happens in the fast regime where {\sc fdt} holds.
The latter holds in the slow regime where both the response and the 
variation of the correlation are very small but the relation between 
them does not follow {\sc fdt}. One derives 
a more useful bound by integrating 
(\ref{bound}) over time:
\begin{equation}
|k_B T\chi(\tau+t_w,t_w) - C(\tau+t_w,t_w) + C(t_w,t_w)| \leq
c \int_{t_w}^{\tau+t_w} dt' \; \sqrt{-d_{t'}{\cal H}(t')} 
\; .
\label{bound2}
\end{equation}
The terms in the {\sc lhs} are now always finite while the value 
of the {\sc rhs} depends on the relation 
between the time-difference $\tau$ and the waiting-time $t_w$.
For sufficiently short $\tau$ such that the {\sc rhs} vanishes {\sc fdt}
has to be satisfied in its integrated form. This result 
explains the existence of a common straight-line
with slope $-1/(k_BT)$ in the nonequilibrium curves 
in Fig.~\ref{FDT-high-low}. For sufficiently 
long $\tau$ such that the {\sc rhs} takes a finite 
value {\sc fdt} can be violated. In this second scale 
a departure from the straight line of slope $-1/(k_B T)$ can occur
and it is indeed what happens in systems with slow non-equilibrium 
dynamics, see the right panel in Fig.~\ref{FDT-high-low}.
One sees how a separation of time-scales in the dynamics influences 
the {\sc fdt} violations.

In driven systems the bound
depends on the 
power input and only vanishes in the limit of vanishing applied forces.
The {\sc fdt} is not even enforced in the fast scale and deviations
start as soon as $C$ decays from $1$. However, as we shall see below, 
the modification of {\sc fdt} follow a very similar pattern to the one 
shown in Fig.~\ref{FDT-high-low} with the strength of the applied force 
playing a similar role to the one of the waiting-time here.

\subsubsection{Diffusion}

In these Lectures we focus on models with a bounded 
self-correlation for an observable with zero average 
that is normalised at equal times. If the averaged observable
does not vanish but the equal-time correlation reaches a 
time-independent limit one can still use the simple 
self-correlation in the generalisations of {\sc fdt}. However, 
in more general diffusive model with 
an unbounded time-dependent equal-time correlator
it is more 
natural to compare the behaviour of the ``displacement'' 
$\Delta(t,t') \equiv C(t,t)+C(t',t') -2 C(t,t')$ 
(that vanishes by definition at equal times) to the 
linear response. In normal diffusion these are linked 
by $R(t,t') = 1/(2k_B T) \Delta(t,t')$. In glassy models 
like the massless manifold in a random potential and 
others this relation is modified~\cite{Cukule,yoshino,Noelle,Lale}.

\section{Solution to mean-field models}
\label{section:solution}
\setcounter{equation}{0}
\renewcommand{\theequation}{\thesection.\arabic{equation}}

In this Section we turn our attention to the solution 
to the Schwinger-Dyson equations derived in previous Sections. 
We start by describing the simplest numerical 
algorithm that solves these equations and we next briefly discuss the 
asymptotic analytic solution at high temperatures. Next
we describe in quite detail the solution at low $T$.

\subsection{Numerical solution}
\label{numerical-sol}

One can attempt a numerical solution to the 
set of causal integro-differential equations
(\ref{eqR2}), (\ref{eqC2}) together with the equation for
the Lagrange multiplier $\mu(t)$. One of the  
questions we would like to explore  is
whether they encode
a non-equilibrium evolution as the one described in the Introduction and 
Section~\ref{interestingproblems}. 

The correlation $C(t,t')$ and response
$R(t,t')$ are two-time quantities, that is, they depend on 
$t$ (which physically corresponds to the time of observation) and
$t'$ (which corresponds to the age of the system). 
In the simplest algorithm one discretises the two-times plane with a 
{\it uniform grid}, $t'=j \delta$ and $t=i \delta$. The correlation and 
response on the diagonal and the next-to-main diagonal of the two-times
plane $(i,j)$ are known, see Eqs.~(\ref{next-to-diagonal-R}) 
and (\ref{next-to-diagonal-C}), for all times. 

The time-derivatives $\partial^2_{t}C(t,t')$ and 
$\partial^2_{t}R(t,t')$ in their discretized form 
are used to update the two-point functions.
Due to causality, to advance one time step, the integrals only need values of 
$C$ and $R$ that are already known. 
This algorithm is simple and efficient but it is severely limited by the
computer storage capacity. 
Since one has to store $C$ and $R$ 
for all previous time steps, the memory used grows as
$i^2$ and this number
becomes rather quickly prohibitive.
In standard {\sc pc}s one can use 
$i_{\sc max} \sim 10^4$, get an acceptable
precision for $\delta \leq 0.1$ and reach 
total times of the order of $10^3$. 

In the quantum case
the presence of  non local kernels $\eta$ and $\nu$,  that
appear convoluted with $C$ and $R$, renders the numerical solution harder.
The larger the cut-off $\Lambda$, 
the smaller the iteration  step $\delta$ we need
to compute these integrals with a good precision. The maximum
total time is of the order of $10^2$ in this case.

A different starting point for a numerical solution is 
the single variable equation (\ref{9}). This route was followed
by Eissfeller and Opper for spin-glasses~\cite{Eiop} and it is usually
used in the so-called dynamic mean-field theory~\cite{Georges}.
Again, this method is limited by the storage capacity.

The knowledge of the qualitative features of the solution helps one 
devising a more performant algorithm with a {\it variable two-time grid}.
As we shall see from the analytic solution, 
$C$ and $R$ vary rapidly when times are near the diagonal $j=i$ and 
decay very slowly far from it. Kim and Latz have exploited 
this property and wrote such an algorithm for the spherical $p$ spin
model reaching total times of the order of $10^8$~\cite{KimLatz}.

Finally, one can think of an iterative search where one 
starts from a trial form of $C$ and $R$ and uses the
dynamic equations to extract the new form. One can expect to 
obtain the solution by repeating this
procedure until the iteration converges to 
a fixed point. This method would allow one to look for solutions of the 
full set of Schwinger - Dyson equations that 
break causality.

The numerical solution for the causal problem, 
found with the simple uniform grid, has 
been of great help in 
deriving the asymptotic analytic solution.
In the 
following we describe how this solution builds up.

\subsection{Solution at high temperatures}
\label{subsec:highT}

At high temperature the system equilibrates with its 
environment since
\begin{equation}
t_{\sc eq}(N\to\infty,T) = \mbox{finite}
\; .
\end{equation}
 The mere existence of an asymptotic limit implies that 
one-time quantities as, {\it e.g.}, the energy  
density, ${\cal E}(t)$, or the Lagrange multiplier, $\mu(t)$,
have to approach an asymptotic limit, $\lim_{t\to\infty}{\cal E}(t)
= {\cal E}_\infty$ and $\lim_{t\to\infty}\mu(t)
=\mu_\infty$. In equilibrium ${\cal E}_\infty={\cal E}_{\sc eq}$ 
and similarly 
for all one-time quantities.
Two time-quantities, as $C$ and $R$, depend on times but 
only through time differences as 
explained in Section~\ref{sec:prob}.

To solve the high $T$ dynamics one first assumes that
after a transient equilibrium is reached and a solution 
of the form 
$\mu(t) \to \mu_\infty$, 
\begin{eqnarray}
C(t,t') \to C_{\sc st}(t-t') \; , \;\; \;\;
R(t,t') \to R_{\sc st}(t-t')  
%\nonumber\\ 
%\;\;\;\; \mbox{with} \;\;\;\; 
%R_{\sc st}(t-t') = \frac{1}{k_B T} \, \theta(t-t') \, 
%\partial_{t'} C_{\sc st}(t-t') 
%\; ,
\label{ansatz-highT}
\end{eqnarray}
with $R_{\sc st}$ and $C_{|sc st}$ related by {\sc fdt}, 
for long waiting-times $t'$ and all time-differences $t-t'$, exists.
These properties also apply to $D$ and $\Sigma$ that behave as a
correlation and a response, respectively.
This {\it Ansatz} should solve Eqs.~(\ref{eqR2}) and (\ref{eqC2}) 
when $T>T_d$, with $T_d$ the dynamic critical temperature. 
In order to prove it
we take $t'$ long and we assume that we can separate the integrals in
Eqs.~(\ref{eqC2}) and (\ref{eqR2}) in  a {\it preasymptotic} 
and an {\it asymptotic} 
contribution,
\begin{equation}
\int_0^\infty dt'' \; \cdots \approx   
\int_0^{t_{\sc eq}} dt'' \; \cdots +  \int_{t_{\sc eq}}^{\infty} dt'' \; 
\cdots
\; .
\end{equation}
Next, we assume that the two-point functions
decay as fast as to ensure that all preasymptotic 
contributions vanish, {\it e.g.}
%\begin{equation}
$\int_0^{t_{\sc eq}} dt'' \; A(t,t'') B(t',t'') \sim 0  $
%\end{equation}
when $t'$ and $t\geq t'$ are in the asymptotic regime. 
Using the {\it Ansatz} (\ref{ansatz-highT}) and this assumption 
the integrals in the {\sc rhs} of Eq.~(\ref{eqC2}), for a 
classical problem,  read
\begin{eqnarray*}
&& 
\int_{t_{\sc eq}}^{t'} dt'' \; 
D_{\sc st}(t-t'')  \,
\frac{1}{k_B T} \frac{\partial C_{\sc st}(t'-t'')}{\partial t''} 
+
\int_{t_{\sc eq}}^t dt'' \;  
\frac{1}{k_B T} \frac{\partial D_{\sc st}(t-t'')}{\partial t''} \, 
C_{\sc st}(|t''-t'|) 
%&&
\nonumber\\
&&=
\frac1{k_B T}
\int_{t_{\sc eq}}^{t'} dt'' \; 
\frac{\partial}{\partial t''}
\left[ D_{\sc st}(t-t'') C_{\sc st}(t'-t'') \right]
+
%\nonumber\\
%&&
%\;\;\;\;\;\;\;\;\;\;\;\;\;\;\;\;\;\;
\frac{1}{k_B T}
\int_{t'}^t dt'' \;  
\frac{\partial D_{\sc st}(t-t'')}{\partial t''} \, 
C_{\sc st}(t''-t')  
\end{eqnarray*}
The first integral in the {\sc rhs} is a total derivative and it
can be readily evaluated, it yields
%\begin{equation}
$D_{\sc st}(t-t') C_{\sc st}(0) - 
D_{\sc st}(t-t_{\sc eq}) C_{\sc st}(t'-t_{\sc eq})
\approx  
D_{\sc st}(\tau)  $
%\end{equation} 
where we assumed that $t$ and $t'$ are well in the asymptotic regime 
in such a way that $C_{\sc st}(t'-t_{\sc eq})\sim 0$, and we 
defined $\tau\equiv t-t'$.  Integrating by parts the last integral 
in the {\sc rhs} one finally obtains the high $T$ equation 
for the correlation
\begin{eqnarray}
%M d_{\tau^2} C_{\sc st}(\tau) +
%\gamma d_\tau C_{\sc st}(\tau) &=&  
G_o^{-1}(\tau) C_{\sc st}(\tau) = 
\frac{1}{k_BT} D_{\sc st}(0)
C_{\sc st}(\tau) 
-\frac{1}{k_BT} \int_0^\tau d\tau' \; D_{\sc st}(\tau-\tau') 
d_{\tau'}C_{\sc st}(\tau')
\; 
\label{eq-MCT}
\end{eqnarray}
with $G_o^{-1}(\tau) = M d_{\tau^2}+\gamma d_\tau+\mu_\infty$.
One can check that Eq.~(\ref{eqR2}) coincides
with Eq.~(\ref{eq-MCT}) under the same assumptions. To prove this 
statement one has to integrate Eq.~(\ref{eqR2}) with respect to 
$t'$ from $t_{\sc eq}$ to $t'$ taking care of the fact that $t'$ 
appears in the lower limit of the integral. 

Equation~(\ref{eq-MCT}) for the spherical $p$ spin model 
coincides with the schematic {\sc mc} equation~\cite{Gotze,walter}.
This equation has a decaying solution above a sharp 
critical temperature that we call 
$T_{\sc mct}=T_d$ where the assumptions of 
{\sc tti} and {\sc fdt} are justified. After a 
short transient 
(eliminated by the limit $t'\gg t_{\sc eq}$) the system equilibrates 
with its environment even if the thermodynamic 
limit has already been  taken.
At very high $T$  the decay to zero is very fast and typical 
of, say, a high-$T$
liquid. Closer to  $T_d$, however, a very interesting structure appears. The 
solution takes the form sketched in the left panel in 
Fig.~\ref{sketch-scales-highT}.
In a logarithmic scale one sees a two step relaxation develop with a first 
relatively 
quick decay towards a plateau at a value that we call 
$q_{\sc ea}$ and next a slower relaxation
towards zero. The length of the plateau increases when temperature approaches 
$T_d$ from above and it diverges at $T_d$. At $T_d$
the height of the plateau, $q^d_{\sc ea}$, follows from the 
asymptotic analysis of Eq.~(\ref{eq-MCT}). If one loosely considers 
$q^d_{\sc ea}$ 
to be an order parameter, the high temperature
analysis yields $q^d_{\sc ea} > 0$ [see Eq.~(\ref{qead})] 
and the transition is {\it discontinuous}.
It is important to stress that, as we shall see below, this does not mean that 
the model has a first order thermodynamic transition. All susceptibilities 
are continuous when going across $T_d$ even though $q^d_{\sc ea}>0$.  
In the mode-coupling literature these transitions are called type B. 

The details of the asymptotic analysis of the schematic
{\sc mc} equation and 
its relation with the behavior of real systems has been discussed at
length in the literature (see, {\it e.g.} \cite{walter}). 
We shall not develop it here. With the purpose of future comparison
with the low-$T$ solution we just recall that the approach and 
departure from the plateau ({\it beta} relaxation) occurs with two 
power laws:
\begin{eqnarray}
C_{\sc st}(\tau) \sim q_{\sc ea}^d + c_a \tau^{-a} + \dots 
\;\;\;\;\;\;\;\;\;\;
C_{\sc st}(\tau) \sim q_{\sc ea}^d - c_b \tau^{b} + \dots 
\label{highTexp0}
\end{eqnarray}
given by 
\begin{equation}
\frac{1}{k_B T_d} 
\frac{\Gamma^2(1+b)}{\Gamma(1+2b)} = 
\frac{1}{k_B T_d} 
\frac{\Gamma^2(1-a)}{\Gamma(1-2a)} = 
\frac{1}{2} 
\frac{{\cal V}'''(q_{\sc ea}^d)}{({\cal V}''(q^d_{\sc ea}))^{3/2}}
\; .
\label{highTexp}
\end{equation}

A similar analysis can be done for a quantum model.

\subsection{Solution at low-$T$}
\label{pspin-solution}

Three families of mean-field models have been found so far.
In this Section we present the solution to the spherical 
mean-field descriptions of ferromagnetic domain growth 
and structural glasses in some detail.  
We use a generic notation that 
allows us to treat the classical and quantum problem 
simultaneously.
The presentation follows~\cite{Culo}. By the end of this Subsection we 
discuss the generalisation of these results to models of ``spin-glass''
type, models with spatial dependence and the effect of 
different microscopic dynamics.

The numerical solution to the dynamic equations at low $T$
shows no evidence for an arrest 
in the waiting-time dependence of the decay of $C$ and $R$. 
In this regime of temperatures,
\begin{equation}
t_{\sc eq}(N,T<T_d) \to\infty 
\end{equation}
and the equations do not admit the choice of a  
$t' > t_{\sc eq}$. In order to consider the crossover 
towards the equilibration regime one should revisit the 
derivation of the dynamic equations allowing for $N$ 
finite. This program has not been pursued in the literature
and it remains one of the most interesting open problems in the 
field.

\subsubsection{The Lagrange multiplier}

We approximate the integral in Eq.~(\ref{mut1})
by separating its support in three intervals
\begin{equation}
t'': 0\to \delta_0\;, \;\;\; 
t'':\delta_0 \to \Delta_t\; , \;\;\;
t'':\Delta_t \to t
\label{interval1}
\; .
\end{equation}
The first time-interval contains only finite times $t''$. Hence, 
all correlations and responses 
of the form $C(t,t'')$ and $R(t,t'')$ vanish due to 
Eqs.~(\ref{limit0}) and (\ref{limit0R}).
In the last time-interval 
$t''$ is close to $t$ in the sense that correlations 
of the kind $C(t,t'')$ are of the form $C_{\sc st}(t-t'')+q_{\sc ea}$
and similarly for the responses. Finally, in the intermediate time-interval the
$C$ and $R$ vary in the aging regime. Of course, we are sloppy
in that we do not precise what are the values of $\delta_0$ and $\Delta_t$.
The definitions of correlation scales given in 
Section~\ref{time-scales} correct this imprecision
exchanging the time limts by limits in the correlation.
Within these assumptions the asymptotic value of $\mu(t)$ is given by 
\begin{eqnarray}
\mu_\infty&=&
A_\infty 
+ 
q_{\sc ea} \int_0^\infty d\tau' \;  \Sigma_{\sc st}(\tau') + 
\tilde D_{q_{\sc ea}}
\int_0^\infty d\tau' R_{\sc st}(\tau')
\nonumber\\
& & 
+ \int_0^\infty d\tau' \left[ \; \Sigma_{\sc st}(\tau') C_{\sc st}(\tau') + 
D_{\sc st}(\tau') R_{\sc st}(\tau') \; 
\right]
+ \mbox{Last}
\label{eqmuinfty}
\end{eqnarray}
(see Appendix~\ref{integrals}). $\Sigma$ and $D$ are made of two terms, 
one contribution from the bath and one contribution from the interactions
(see Eqs.~(\ref{sigmatilde}) and (\ref{Dtilde})).
We called Last a term that equals 
$- M \left. \partial^2_\tau C_{\sc st}(\tau)\right|_{\tau\to 0}$
in a model with inertia (classical or quantum) and simply $k_B T$
in classical models without inertia.
$A_\infty$ is the aging contribution:
\begin{eqnarray}
A_\infty 
&=& 
\lim_{t\to\infty} \int_0^t dt'' 
\left[ \; \Sigma_{\sc ag}(t,t'') C_{\sc ag}(t,t'') + 
D_{\sc ag}(t,t'') R_{\sc ag}(t,t'') \; 
\right]
\; . 
\end{eqnarray}
The bath does not contribute to the integrals in $A_\infty$  
when the kernels $\eta$ and $\nu$ 
decay sufficiently fast to zero as to yield vanishing integrals.
This is trivially true for a white noise. It can be a working 
assumption for colored noises based on a weak limit of the 
strength of the coupling to the noise (see, however, \cite{quantum_bath}). 
More precisely, we are neglecting terms of the form
%\begin{equation}
$\lim_{t\to\infty} 
\int_0^t dt'' \, A(t-t'') B(t,t'')
$
%\end{equation}
where $A$ is either $\nu$ or $\eta$ and $B$ is either 
$C_{\sc ag}$ or $R_{\sc ag}$.  
In this case
\begin{eqnarray}
A_\infty 
&=& 
\lim_{t\to\infty} \int_0^t dt'' 
\left[ \; \tilde \Sigma_{\sc ag}(t,t'') C_{\sc ag}(t,t'') + 
\tilde D_{\sc ag}(t,t'') R_{\sc ag}(t,t'') \; 
\right]
\; . 
\end{eqnarray}
The second and third terms in Eq.~(\ref{eqmuinfty}) 
come from the constant (non-zero) limit of
the first decay of the correlation
$q_{\sc ea}\equiv \lim_{t-t'\to\infty} \lim_{t'\to\infty} C(t,t') $ and
the vertex
%\begin{equation}
$
\tilde D_{q_{\sc ea}} 
\equiv \lim_{t-t'\to\infty}\lim_{t'\to\infty}\tilde D(t,t')
%\label{limitvertexFDT} 
%\; .
%\end{equation}
$.
For the classical spherical $p$ spin model 
%\begin{equation}
$
\tilde D_{q_{\sc ea}}=\frac{
%{\tilde J}^2\, 
p }{2} \, q_{\sc ea}^{p-1}
$
%\end{equation}
and this equation also holds for its quantum extension 
if we use $\lim_{\tau\to\infty} R_{\sc st}(\tau) \ll  q_{\sc ea}$, 
a property of the {\sc wltm} scenario. The integral over the stationary
parts can be simply performed using {\sc fdt} for classical problems
but they cannot in quantum problems.

\subsubsection{The stationary regime}
\label{stationaryregime}

If $(t,t')$ are such that $C(t,t') > q_{\sc ea}$, 
$C(t,t')=q_{\sc ea}+C_{\sc st}(t-t')$ and 
$R(t-t')=R_{\sc st}(t-t')$. 
The Schwinger-Dyson equation for $R$ in this time sector reads
\begin{equation}
\left( M \partial^2_\tau + \mu_\infty \right) R_{\sc st}(\tau) 
=
\delta(\tau) + \int_0^\tau d\tau' \, \Sigma_{\sc st}(\tau-\tau') 
R_{\sc st}(\tau') 
\end{equation}
and it keeps the same form as in the high-temperature phase, apart 
from the fact that the constant $\mu_\infty$ has contributions 
from the aging regime.
The Schwinger-Dyson equation for $C$ reads
\begin{eqnarray}
&& \left( M \partial^2_\tau + \mu_\infty \right) (q_{\sc ea} + C_{\sc st}(\tau)) 
=
A_\infty 
+ q_{\sc ea} \int_0^\infty d\tau' \, \Sigma_{\sc st}(\tau') 
+ \tilde D_{q_{\sc ea}}
 \int_0^\infty d\tau' \,  R_{\sc st}(\tau')
\nonumber\\
& & 
\;\;\;\;\;\;\;\;\;\;\;\;\;\;
+ 
\int_{-\infty}^\infty d\tau' \left[ \Sigma_{\sc st}(\tau+\tau') 
C_{\sc st}(\tau') + D_{\sc st}(\tau+\tau') R_{\sc st}(\tau') \right]
\; .
\label{Cst}
\end{eqnarray}
One can now Fourier-transform both equations
\begin{eqnarray}
R_{\sc st}(\omega) 
&=&  
\frac{1}{-M \omega^2 + \mu_\infty -  \Sigma_{\sc st}(\omega)}
\label{Rstat}
\; ,
\nonumber\\
\left(- M \omega^2 +\mu_\infty \right) C_{\sc st}(\omega) 
+ \mu_\infty q_{\sc ea} \delta(\omega) 
&=&
\left(
A_\infty +  q_{\sc ea} \Sigma_{\sc st}(\omega) + 
 \tilde D_{q_{\sc ea}}
R_{\sc st}(\omega)
\right)
\delta(\omega) 
\nonumber\\
& & 
+ \Sigma_{\sc st}(\omega) C_{\sc st}(\omega) + 
D_{\sc st}(\omega) R_{\sc st}(-\omega) 
\; .
\nonumber
\end{eqnarray}
The formal solution to the equation for $C_{\sc st}$ is
\begin{eqnarray*}
C_{\sc st}(\omega)
=
\left(-\mu_\infty q_{\sc ea} + A_\infty +  q_{\sc ea} 
\Sigma_{\sc st}(\omega)
+  \tilde D_{q_{\sc ea}} R_{\sc st}(\omega) \right)
\delta(\omega) R_{\sc st}(\omega)
+
D_{\sc st}(\omega) |R_{\sc st}(\omega)|^2 
\; .
\end{eqnarray*}
The first term on the {\sc rhs} 
 has an imaginary and a real part.
The imaginary part vanishes identically since, due to {\sc fdt}, 
both ${\mbox{Im}} R_{\sc st}(\omega)$ and  
${\mbox{Im}} \Sigma_{\sc st}(\omega)$ are 
proportional to $\tanh\left(\beta\hbar\omega/2\right)$
which is zero at $\omega=0$ for classical and quantum problems.
Concerning the real part of this first term, 
as we have assumed that $C_{\sc st}(\tau)$ goes to zero for 
$\tau\to\infty$, we need to impose 
the self-consistent condition
\begin{equation}
-\mu_\infty q_{\sc ea} + A_\infty +  q_{\sc ea} \Sigma_{\sc st}(\omega=0)+
 \tilde D_{q_{\sc ea}} R_{\sc st}(\omega=0)=0
\; .
\label{cond2}
\end{equation}
This is the condition that fixes $q_{\sc ea}$. 
We shall find it again in the next section as the matching condition 
between the stationary and aging regimes. 
The final equation for $C_{\sc st}(\omega)$ is
\begin{equation}
C_{\sc st}(\omega) = D_{\sc st}(\omega) |R_{\sc st}(\omega)|^2
\label{Cstat}
\; .
\end{equation}
One can check that these calculations are consistent with the results from 
$\mu_\infty$.
Actually, the integrals in equation for $\mu(t)$
involving the stationary parts can be 
evaluated
with the help of the equations for 
$R_{\sc st}$ and $C_{\sc st}$, Eqs.~(\ref{Rstat}) and (\ref{Cstat}), 
and yield once again Eq.~(\ref{cond2}).  

Similarly to the high-temperature case one can now show that {\sc fdt} for 
$\tilde \Sigma_{\sc st}$ and $\tilde D_{\sc st}$ implies {\sc fdt} for 
$R_{\sc st}$ and $C_{\sc st}$. 
The remainder of the proof, {\it i.e.} 
to show that  {\sc fdt} between $R_{\sc st}$ 
and $C_{\sc st}$ 
implies {\sc fdt} between $\tilde \Sigma_{\sc st}$ and $\tilde D_{\sc st}$ 
depends only upon the 
form of $\tilde \Sigma_{\sc st}$ and $\tilde D_{\sc st}$ as functions of 
$R_{\sc st}$ and $C_{\sc st}$ and is not modified from 
the one discussed in Section~\ref{subsec:highT}.
   
\subsubsection{The aging regime}     
\label{subsec:aging-regime}

If we now choose the times $t,t'$ to be well-separated so as to have 
$C(t,t') = C_{\sc ag}(t,t')\leq q_{\sc ea}$ and $R(t,t') = R_{\sc ag}(t,t')$,
the {\sc web} and {\sc wltm}
hypotheses allow us to throw the second time
derivatives on the {\sc lhs}. We assume that their contribution is 
much weaker than the one of each of the integral terms on the {\sc rhs}. 
This is an assumption 
that we have to verify at the end of the calculation, once the solution for 
$C_{\sc ag}$ and $R_{\sc ag}$ is known. It corresponds to the 
over-damped limit.
%\begin{eqnarray}
%M \partial^2_t C_{\sc ag} 
%\ll {\mbox{Terms in the}} \;\; {\sc rhs}_{\sc c}(t,t')
%\; ,
%\nonumber\\
%M \partial^2_t R_{\sc ag} \ll {\mbox{Terms in the}}\;\; {\sc rhs}_{\sc r}(t,t')
%\; .
%\end{eqnarray}

Using the approximation
described in Appendix~\ref{integrals},
the equation for $R$ in the aging regime becomes
\begin{eqnarray}
&&
\mu_\infty R_{\sc ag}(t,t') 
=
\tilde \Sigma_{\sc ag}(t,t')  \int_0^{\infty} d\tau' R_{\sc st}(\tau') 
+
R_{\sc ag}(t,t') \int_0^{\infty} d\tau' \, \Sigma_{\sc st}(\tau')
\nonumber\\ 
& & 
\;\;\;\;\;\;\;\;\;\;\;\;\;\;\;\;\;
+
\int_{t'}^t dt'' \, \tilde \Sigma_{\sc ag}(t,t'') R_{\sc ag}(t'',t')
\; 
\end{eqnarray} 
and we call it the $R_{\sc ag}$-eq.
Similarly, the equation for $C$ becomes
\begin{eqnarray}
&& 
\mu_\infty C_{\sc ag}(t,t')  
=
C_{\sc ag}(t,t') \int_0^{\infty} d\tau' \Sigma_{\sc st}(\tau') 
+ 
\tilde D_{\sc ag}(t,t') \int_0^{\infty} d\tau' R_{\sc st}(\tau') 
\nonumber\\
& & 
\;\;\;\;\;\;\;\;\;\;
+
\int_0^t dt'' \; \tilde \Sigma_{\sc ag}(t,t'') C_{\sc ag}(t'',t') +
\int_0^{t'} dt'' \; \tilde D_{\sc ag}(t,t'') R_{\sc ag}(t',t'') 
\; 
\end{eqnarray}
and we call it the $C_{\sc ag}$-eq. 
In all integrals over the slow regime we neglected the contributions of the 
noise kernels $\eta$ and $\nu$ and we approximated 
$\Sigma_{\sc ag}(t,t') \sim \tilde \Sigma_{\sc ag}(t,t')$ and
$D_{\sc ag}(t,t') \sim \tilde D_{\sc ag}(t,t')$
(again, see \cite{quantum_bath} for a discussion on the effect of a strong 
bath). 

\subsubsection{The Edwards-Anderson parameter}
\label{subsec:qea}

The Edwards-Anderson parameter, $q_{\sc ea}$, is determined 
self-consistently from 
the matching of $\lim_{t\to\infty} C_{\sc ag}(t,t)= \lim_{t-t'\to\infty} 
\lim_{t'\to\infty} C(t,t')=q_{\sc ea}$. 
Taking the limit $t'\to t^-$ in the $R_{\sc ag}$-eq and $C_{\sc ag}$-eq one 
obtains
\begin{eqnarray} 
\mu_\infty R_{\sc ag}(t,t) 
=
\tilde  \Sigma_{\sc ag}(t,t) \int_0^\infty d\tau' \; R_{\sc st}(\tau')
+
R_{\sc ag}(t,t) \int_0^\infty d\tau' \; \Sigma_{\sc st}(\tau')
\; ,
&&
\label{hola}
\\
\mu_\infty q_{\sc ea} 
=
A_\infty +
q_{\sc ea} \int_0^\infty d\tau' \; \Sigma_{\sc st}(\tau') +
\tilde D_{\sc ag}(t,t) \int_0^\infty d\tau' \; R_{\sc st}(\tau')
\; .
&&
\label{Cq}
\end{eqnarray}

The first equation admits the solution $R_{\sc ag}(t,t)=0$ since 
$\tilde \Sigma_{\sc ag}(t,t)$ is proportional to $R_{\sc ag}(t,t)$
-- see Eq.~(\ref{sigmatilde}). 
This corresponds to the high-temperature solution 
where there is no aging regime. 
Here we concentrate on the other possibility.
%, that is to say when
%\begin{eqnarray}
%\mu_\infty 
%&=& 
%\frac{\tilde \Sigma_{\sc ag}(t,t)}{R_{\sc ag}(t,t)} 
%\int_0^\infty d\tau' \; R_{\sc st}(\tau') + 
%\int_0^\infty d\tau' \; \Sigma_{\sc st}(\tau')
%\; .
%\end{eqnarray}
%In actual fact, we can further approximate this equation 
%by using one of the assumptions already used all over this section that 
The response becomes smaller and smaller as time passes  
-- though its integral over 
an infinite interval gives a finite contribution. If 
we neglect all terms
that are proportional to $R_{\sc ag}(t,t)$ with respect to terms that are 
proportional 
to $q_{\sc ea}$, only the first term in the power expansions of
 $\tilde \Sigma$ and $\tilde D$ 
survive and
\begin{equation}
\left(\tilde \Sigma/R \right)_{q_{\sc ea}} \equiv 
\lim_{t\to\infty} 
\frac{\tilde \Sigma_{\sc ag}(t,t)}{R_{\sc ag}(t,t)}
\;\;\;\;\;\;\;\;\;\;\;\;\;\;\;\;\;\;\;
\tilde D_{q_{\sc ea}} \equiv \lim_{t\to\infty} \tilde D_{\sc ag}(t,t) 
\end{equation}
that for the $p$ spin model become
\begin{equation}
\left(\tilde \Sigma/R \right)_{q_{\sc ea}}
=
\frac{
%{\tilde J}^2 \, 
p(p-1)}{2} \; q_{\sc ea}^{p-2} 
\;\;\;\;\;\;\;\;\;\;\;\;\;\;\;\;\;\;\;
\tilde D_{q_{\sc ea}} 
= 
\frac{
%{\tilde J}^2 \, 
p}{2} \; q_{\sc ea}^{p-1}
\; ,
\label{Dqstat}
\end{equation}
in accord with the large $\tau$ limit of the stationary  values 
(see Section~\ref{stationaryregime}).
Equations (\ref{hola}) and (\ref{Cq}) become
\begin{eqnarray}
\mu_\infty 
&=& 
\left(\tilde \Sigma/R \right)_{q_{\sc ea}} 
\int_0^\infty d\tau' \; R_{\sc st}(\tau') + 
\int_0^\infty d\tau' \; \Sigma_{\sc st}(\tau')
\; ,
\label{secondq}
\\
\mu_\infty q_{\sc ea} 
&=&
A_\infty +
q_{\sc ea} \int_0^\infty d\tau' \; \Sigma_{\sc st}(\tau') +
\tilde D_{q_{\sc ea}}  \, \int_0^\infty d\tau' \; R_{\sc st}(\tau')
\; .
\label{firstq}
\end{eqnarray} 
The second equation 
is the same as the one arising from the end of the 
stationary regime, Eq.~(\ref{cond2}).

From Eqs.~(\ref{Rstat}) and 
(\ref{Cstat}) one derives 
\begin{equation}
\int_0^\infty d\tau \; R_{\sc st}(\tau) 
= 
R_{\sc st}(\omega=0) = \frac{1}{\mu_\infty - \Sigma_{\sc st}(\omega=0)}
\label{secondqq}
\; ,
\end{equation}
and 
\begin{eqnarray}
1 &=& \left( \tilde \Sigma/R\right)_{q_{\sc ea}} \, R_{\sc st}^2(\omega=0)
\; .
\label{eq-qea}
\end{eqnarray}
We remind that the factor $R_{\sc st}^2(\omega=0)$ can be written
in terms of the stationary correlation function using {\sc fdt}; therefore
this is a closed equation for the correlation that determines $q_{\sc ea}$.
In the case of the $p$-spin model it reads
\begin{eqnarray}
1 &=& \frac{
%{\tilde J}^2 
p(p-1)}{2} q_{\sc ea}^{p-2} 
\left(\frac{1}{\hbar} \; 
P \int_{-\infty}^\infty 
\frac{d\omega'}{\omega'} \, 
\tanh\left(\frac{\beta\hbar\omega'}{2}\right) \, 
C_{\sc st}(\omega') \right)^2
\; .
\end{eqnarray}
In the classical case, the integral can be readily computed
and the final equation for $q_{\sc ea}$ is
\begin{equation}
\frac{ 
%{\tilde J}^2 \, 
p(p-1)}{2} \; q_{\sc ea}^{p-2} (1-q_{\sc ea})^2 =
(k_B T)^2
\; ,
\label{qea-class}
\end{equation}
that coincides with the result for the purely relaxational 
dynamics~\cite{Cuku1}. 
For $p\geq 3$ fixed, $q_{\sc ea}$ is a function of temperature. 
 Equation~(\ref{qea-class}) can be solved 
graphically. The {\sc lhs} has a bell shape. It vanishes   
at $q_{\sc ea}=0,1$ and it reaches a maximum at 
$q_{\sc ea}^{\sc max}=(p-2)/p$. The equation has two  solutions for 
all temperatures $(k_B T)^2 < (k_B T^{\sc max})^2 = 
p(p-1)/2 \, [(p-2)/p]^{p-2}
\, (2/p)^2$, these merge ar $T^{\max}$ and disappear for higher $T$'s.
The physical solution corresponds to the branch on the right of the 
maximum, the one that continues the solution $q_{\sc ea}=1$ at $T=0$.
The minimum value of $q_{\sc ea}$ is reached at the dynamic critical
temperature $T_d(<T^{\sc max})$, where 
$q^d_{\sc ea}\equiv q_{\sc ea}(T_d) > q^{\sc max}_{\sc ea}$.

\subsubsection{Fluctuation - dissipation relation}

In order to advance further we have to relate the 
response to the correlation.
If we {\it assume} that 
\begin{equation}
R_{\sc ag}(t,t') = \frac{1}{k_B T^*} 
\frac{\partial C_{\sc ag}(t,t')}{\partial t'} 
\; ,
\label{FDT6}
\end{equation}
with $T^*$ the value of an effective temperature (see Section~\ref{temp_intro})
that  
is determined by Eqs.~(\ref{firstq}) and (\ref{secondqq})  
%\begin{equation}
$0 = A_\infty - \frac{q_{\sc ea}}{R_{\sc st}(\omega=0)} + 
\tilde D_{q_{\sc ea}} \, R_{\sc st}(\omega=0)$.
%\; .
%\end{equation}
Using Eq.~(\ref{FDT6}) and the equivalent relation between
$\tilde \Sigma_{\sc ag}$ and $\tilde D_{\sc ag}$, we obtain
%\begin{eqnarray}
$A_\infty
=
(k_B T^*)^{-1} \, \lim_{t\to\infty} 
\left( \tilde D_{\sc ag}(t,t) C_{\sc ag}(t,t) \right)
=
(k_B T^*)^{-1} \, q_{\sc ea} \tilde D_{q_{\sc ea}} 
$
%\; 
%\end{eqnarray}
and
\begin{equation}
\frac{1}{k_B T^*} \, 
=
\frac{(p-2)}{q_{\sc ea}} R_{\sc st}(\omega=0)
\,
= \sqrt{ \frac{2 (p-2)^2}{p(p-1) } } \; q_{\sc ea}^{-p/2}
\; .
\label{Xfinal}
\end{equation}
In the classical limit $T/T^*=(p-2)(1-q_{\sc ea})/q_{\sc ea}$ and the result 
in~\cite{Cuku1} is recovered.
Note that both in the classical and quantum case, 
$T^*\to \infty$ if $p=2$.
Since the case $p=2$ is formally connected to ferromagnetic domain 
growth in $d=3$ (in the mean-field approximation) there is
no memory neither in the classical nor in the quantum 
domain growth. 

The {\it Ansatz} in Eq.~(\ref{FDT6}) solves classical {\it and} quantum
aging equations. The modification of the {\sc fdt} in this regime
became thus classical even when quantum fluctuations exist. This is 
an interesting sort of decoherent effect that will become clearer
when we shall discuss the interpretation of this results in 
terms of effective temperatures.

Using Eq.~(\ref{FDT6}) for all values of $C$ below $q_{\sc ea}$ we
assumed that there is only one aging correlation scale in the problem.
Interestingly enough, one do a more general analysis using the 
formalism described in Section~\ref{time-scales} and find that the dynamic
equations force the solution to have only one aging correlation 
scale~\cite{andalo}.
 
\subsubsection{Discontinuous classical transition}

The classical dynamic critical point $(T_d,\hbar=0)$ 
can arise either when $q_{\sc ea} \to 0$ or when $T^* \to T$.
For the $p$ spin model, using Eqs.~(\ref{qea-class}) and (\ref{Xfinal}) 
the latter holds and~\cite{Cuku1}
\begin{equation}
(k_B T_d)^2 = \frac{p\, (p-2)^{p-2}}{2 \, (p-1)^{p-1}} 
\;\;\;\;\;\;\;\; q_{\sc ea}^d = \frac{p-2}{p-1}
\; .
\label{qead}
\end{equation}
The transition is {\it discontinuous} since the order 
parameter $q_{\sc ea}$ jumps at $T_d$. However, it is still of {\it second
order} thermodynamically since there are no thermodynamic 
discontinuities, all susceptibilities being continuous across
$T_d$. For instance, 
\begin{equation}
\lim_{t\gg t_w} \chi(t,t_w) = \frac{1}{k_B T} (1-q_{\sc ea}) + 
\frac{1}{k_B T^*} q_{\sc ea} \to \frac1{k_BT} \;\; \mbox{when} \;\; 
T\to T^* \;\; \mbox{at} \;\; T_d 
\; . 
\end{equation}

The dynamic transition occurs at a value $T_d$ that is higher than 
the static transition temperature $T_s$. 
The latter is fixed as the temperature 
where replica symmetry breaking occurs
(using the standard prescription to fix the parameters in the 
Parisi {\it Ansatz} to compute the free-energy density)~\cite{Crso}. 
This feature is an explicit realisation of the 
discussion on $T_g$ and $T_0$ in Section~\ref{interestingproblems}
They are sharp in this model.

\subsubsection{The classical threshold level}

The asymptotic energy density reads
%\begin{equation}
$
{\cal E}_{\sc \infty} 
= -\frac{1}{p} 
\int_0^\infty dt'' [\Sigma(t,t'') C(t,t'') $$
+ D(t,t'') R(t,t'') ]
$
%\end{equation}
where we used Eq.~(\ref{energy-mu}). 
Replacing the solution found above we obtain
\begin{equation}
{\cal E}_{\sc \infty} = -\frac1{2} 
\left[ 
\frac{1}{k_B T} \left(1-q_{\sc ea}^p \right)
+ 
\frac{1}{k_B T^*} q_{\sc ea}^p 
\right]\equiv {\cal E}_{\sc th}
\; .
\end{equation}
If one compares this expression with the equilibrium energy density, 
found studying the partition function~\cite{Crso}, one discovers
that~\cite{Cuku1}
\begin{equation}
{\cal E}_\infty = {\cal E}_{\sc th} > {\cal E}_{\sc eq}
\label{inequality}
\; .
\end{equation}
Thus, the non-equilibrium dynamics does not approach the equilibrium 
level asymptotically but it reaches a {\it threshold} level that 
is extensively higher than equilibrium (note that the inequality
(\ref{inequality}) holds for the  energy density).
The name {\it threshold} is motivated by a similarity with {\it percolation}
(in phase space) that we shall discuss in 
Section~\ref{section:tap}~\cite{Cuku1}.

\subsubsection{Two $p$ models}

In Section~\ref{subsec:qea} we took the limit $t'\to t^-$, or 
equivalently, $C_{\sc ag} \to q_{\sc ea}^-$ in the
equations for the slow part of the response and the correlation
and this lead us to Eqs.~(\ref{eq-qea}) and (\ref{Xfinal}) for $q_{\sc ea}$
and $T^*$. Let us now take subsequent variations
of this equation with respect to the 
correlation and evaluate them in the same limit. 
It is easy to see that if we neglect the 
contributions 
from the integral between $t'$ and $t$, assuming that 
the integrands are analytic in this limit, we get new equations
linking $T^*$ and $q_{\sc ea}$ that, for generic models,
are not compatible. Indeed, as we shall 
see below, the pure spherical $p$  spin model is the only one 
for which the solution is given by
an analytic function $\jmath^{-1}(x)$ when $x\to 1^-$.

The way out of this contradiction is to propose that the 
correlation approaches the plateau at $q_{\sc ea}$ with a
power law decay and that it departs from it with another non-trivial
power law~\cite{Cule,Cukule}:
\begin{eqnarray}
C_{\sc st}(t-t') &=& (1-q_{\sc ea}) + c^{(1)}_a (t-t')^{-a} + 
c^{(2)}_a (t-t')^{-2a} + \dots
\dots 
\label{exponenta}
\\
C_{\sc ag}(t,t') &=& q_{\sc ea} - 
c^{(1)}_b  \left( 1- \frac{h(t')}{h(t)}\right)^b -
c^{(2)}_b  \left( 1- \frac{h(t')}{h(t)}\right)^{2b}
+ \dots 
\label{exponentb}
\end{eqnarray} 
with $c^{(i)}_a$ and $c^{(i)}_b$ constants. 
If the exponent $b$ is smaller than one, the integrals
generated by taking derivatives with respect to $C_{\sc ag}$ 
do not 
vanish when $t'\to t^-$. The expansion of the stationary 
and aging equations around $q_{\sc ea}$ fix
the exponents $a$ and $b$. One finds~\cite{Cule}
\begin{equation}
\frac{1}{k_B T^*}\frac{(\Gamma(1+b))^2}{\Gamma(1+2b)}
=
\frac{1}{k_B T}\frac{(\Gamma(1-a))^2}{\Gamma(1-2a)}
=
\frac{1}{2} \frac{{\cal V}'''(q_{\sc ea})}{({\cal V}''(q_{\sc ea}))^{3/2}}
\;
\label{eq-exponents}
\end{equation}
that are to be confronted to Eqs.~(\ref{highTexp0}) and (\ref{highTexp}) 
for the high $T$ behavior. We recall that ${\cal V}(C)$ is the 
correlation of the random potential. Importantly enough, the exponents $a$ and $b$ 
are now $T$-dependent and they are related via an equation in which 
$T^*$ enters.

\vspace{0.2cm}
\noindent {\it Classical spherical} $p$ {\it spin model} 
\vspace{0.2cm}

Since ${\cal V}(C)=C^p/2$ using Eqs.~(\ref{eq-qea}) and (\ref{Xfinal})
to fix $T^*$ and $q_{\sc ea}$ 
one finds $(\Gamma(1+b))^2/\Gamma(1+2b)=1/2$
and $b=1$ for all $T<T_d$. The exponent $a$ interpolates 
between $a= 1/2$ at $T\to 0$ and $a= 1$ at $T\to T_d$ since
$(\Gamma(1-a))^2/\Gamma(1-2a)=T/(2T^*)$. 
 
\vspace{0.2cm}
\noindent {\it Classical mixed} $p_1 + p_2$ {\it spherical spin model}
\vspace{0.2cm}

For adequate choices of the coefficients 
in ${\cal V}(C)=a_1/2 C^{p_1} + a_2/2 C^{p_2}$ (see below)
one finds $T$-dependent exponents $a(T)$ and $b(T)$. 

\vspace{0.2cm}
\noindent {\it Ultrametric limit} 
\vspace{0.2cm}

It is interesting to notice that $(\Gamma(1+b))^2/\Gamma(1+2b)$
is bounded by one. Thus, Eq.~(\ref{eq-exponents}) constrains the
random potentials for which a solution with only two
correlation scales exists. For a particle in a power-law correlated random
potential
one sees the transition towards an ultrametric-like 
solution arrives when the potential goes from short-range to 
long-range correlated~\cite{Cule}. To our knowledge this has not been 
found in a static calculation.
An interpretation of the exponents $a$ and $b$, and this consequence,
in terms of the
properties of the {\sc tap} free-energy landscape is
not known either. 

\subsubsection{{\sc sk} model and similar}
\label{sk-solution}

A different family of models, to which the {\sc sk} model 
belongs, are solved by an ultrametric {\it Ansatz},
$C_{31}=f(C_{32},C_{21})$,
for all correlations below $q_{\sc ea}$. 
The $\chi(C)$ 
plot yields a non-trivial curve (instead of a straight line) 
for $C\in [0,q_{\sc ea}]$. The transition is continuous
$q_{\sc ea}^d=0$. These models are called type A in the 
{\sc mct} literature.

Indeed, for a generic disordered model with random potential 
correlated as in Eq.~(\ref{pot-corr}), one finds that the solution 
is ultrametric if and only if~\cite{Cule}
\begin{equation}
\frac{{\cal V}'''(C)}{{\cal V}'''(q_{\sc ea})} 
\left( \frac{{\cal V}''(q_{\sc ea})}{{\cal V}''(C)}\right)^{3/2}
< 1
\; .
\end{equation}
This bound constrains, for instance, the values of the 
coefficients in a polynomial random potential
for which the dynamic solution is ultrametric. The {\sc fdt} is 
modified with a $C$ dependent factor given by
$T/T_{\sc eff}(C)=q_{\sc ea}{\cal V}'''(C) 
\sqrt{{\cal V}''(q_{\sc ea})}/(4({\cal V}''(C))^{3/2})$.

\subsubsection{Mode dependence}

The models we solved so far have no spatial dependence. 
The manifold problem (\ref{manifold-hamil}) has an internal
structure that leads to a mode-dependence. This model has 
been solved for generic potential correlations~\cite{Cukule}. 
We summarize the outcome without presenting
its detailed derivation.
All modes are slaved to one in the sense that one has to solve 
for the dynamics of one of them and the mode-dependence follows 
from an algebraic equation.
The value of the effective temperature
does not depend on the mode. The mathematical 
reason for this is the slaved structure of the equations. 
The physical reason is that
all interacting observables evolving in the same time-scale have 
to partially equilibrate and acquire the same effective temperature
(see Section~\ref{temp_intro}). 
The height of the plateau, $q_{\sc ea}$, is a $k$ dependent quantity. 
The approach to it and departure from it also depends on $k$ 
but only via the prefactors; the exponents $a$ and $b$, see
Eqs.~(\ref{exponenta}) and (\ref{exponentb}, are
the same for all modes.

Mode-couling equations including a wave-vector dependence have been
derived by Latz using the Mori-Zwanzig formalism; the structure of 
the solution to these equations shares the properties just 
described~\cite{Latz}.

\subsubsection{Quantum fluctuations}

The simplest effect of quantum fluctuations is to introduce oscillations
in the first step of relaxation. These disappear at long enough 
time-differences and they are totally suppressed from the second decay,
that superficially looks classical~\cite{Culo,quantum-others}. 

The Edwards-Anderson parameter $q_{\sc ea}$ depends upon $T$ {\it and} 
$\hbar$. As expected, quantum fluctuations introduce further fluctuations 
in the stationary regime and they decrease the value of $q_{\sc ea}$,
$q_{\sc ea}(T,\hbar\neq 0)< q_{\sc ea}(T,\hbar\to 0)$. 

The modification of {\sc fdt} in the quantum model is of a rather 
simple kind: $R_{\sc ag}$ and $C_{\sc ag}$ are
related as in the classical limit. For the quantum extension of the 
$p$ spin model  there are two  
correlation scales, one with the temperature of the 
environment, $T$, the other with another value of the 
effective temperature, $T^*$, 
that depends on $T$, $\hbar$ and the characteristics of 
the environment. This is a kind of decoherent effect.

As regards to the transition from the glassy to the liquid or
paramagnetic phase, an interesting effect appears. Keeping all 
other parameters fixed, the plane $(T,\Gamma\equiv \hbar^2/(JM))$ is 
separated in these two phases by a line that joins the classical
dynamic critical point ($T_d,\Gamma=0$) and the quantum dynamic 
critical point ($T=0,\Gamma_d)$. 
Close to the classical dynamic critical point the transition is 
discontinuous but of second order thermodynamically until it reaches 
a tricritical point where it changes character to being of first 
order. This behavior is reminiscent of what 
has been reported for
the quantum spin-glass studied in \cite{Aeppli}. 

A still more dramatic effect of quantum mechanics
is related to the very strong role played by the quantum environment 
on the dynamics of a quantum system. Indeed, the location of the 
transition line depends very strongly on the type of quantum 
bath considered and on the strength of the  coupling between system and 
environment~\cite{quantum_bath}.

\subsubsection{Driven dynamics}

The effect of non potential forces can be mimicked with a 
force as the one in (\ref{non-symmetric-force})~\cite{Cukulepe,Ludo}
where the strength of the force, $\alpha$, is analogous to the 
shear stress $\sigma$.
For strengths that are not too strong, the dynamics presents a 
separation of time scales with a fast approach to the plateau and 
a slow escape from it that is now, however, also stationary.
Indeed, after a characteristic time $t_{\sc sh}$ the full 
dynamics becomes stationary though the system is still far from 
equilibrium. One defines a structural relaxation time, $\tau_\alpha$, 
as the time needed to reach, say, a correlation equal to a half. 
One relates the structural relaxation to the viscosity via 
$\eta\equiv \int dt \, C(t)$. 
The scaling of $\eta$ with the shear rate $\dot\gamma\equiv \sigma/\eta$
has been successfully confronted to the behavior 
in rheological experiments in super-cooled liquids and 
glasses~\cite{Ludo}. In terms of the general scalings discussed in 
Section~\ref{time-scales}, the correlations are characterised by 
two different functions $\jmath$, one for the fast decay towards
the plateau and another for the slow decay from the plateau, while 
the functions $h(t)$ are simple exponentials.

Interestingly enough, from the study of 
{\sc fdt} violations above (though close to)
and below $T_d$, when the forcing is weak, one extracts a 
still well-defined slope of the 
$\chi(C)$ plot when $C$ evolves in the slow 
scale~\cite{Cukulepe,JLBarrat,Ludo}. 
This means that an effective temperature can also be identified 
for these systems kept explicitly out of equilibrium 
(see also \cite{Zamponi}). 

Oscillatory forces, as the ones used to perturb granular matter,
have a different effect. Aging is not stopped in a finite region
of the phase diagram ($T$-strength of the force-frequency of the 
force)~\cite{Becuig}. An effective temperature can still be defined 
as the slope of the $\chi(C)$ plot drawn using stroboscopic time,
with a point per oscillatory cycle.

\section{Modifications of {\sc fdt} in physical systems} 
\label{Modifs}
\setcounter{equation}{0}
\renewcommand{\theequation}{\thesection.\arabic{equation}}

In this Section we discuss 
the realizations of the modifications of {\sc fdt} in each of 
the physical systems presented  in Section~\ref{interestingproblems}.

The asymptotic curves in Fig.~\ref{FDT-high-low}  have a rather peculiar form.
They are linear with slope $-1/(k_B T)$ when the correlation 
decreases from $1$ to the plateau value $q_{\sc ea}$. After 
this breaking point, when the correlation decays towards zero, 
the curve is non-trivial, taking the three forms described in 
Section~\ref{definitions} for ferromagnetic domain-growth (in $d>d_c$),
structural glasses and spin-glasses. The separation between these 
two parts is  sharp when the dynamics has a sharp separation of time-scales. 
In Section \ref{subsec:bound} we gave a formal explanation for 
the existence of the first scale where {\sc fdt} holds for any 
relaxing system. Here and in the two next Subsections we give more 
intuitive arguments for the validity of {\sc fdt} in the fast 
regime in the context of the physical systems discussed in the Introduction.

\begin{figure}[h]
\centerline{
\psfig{file=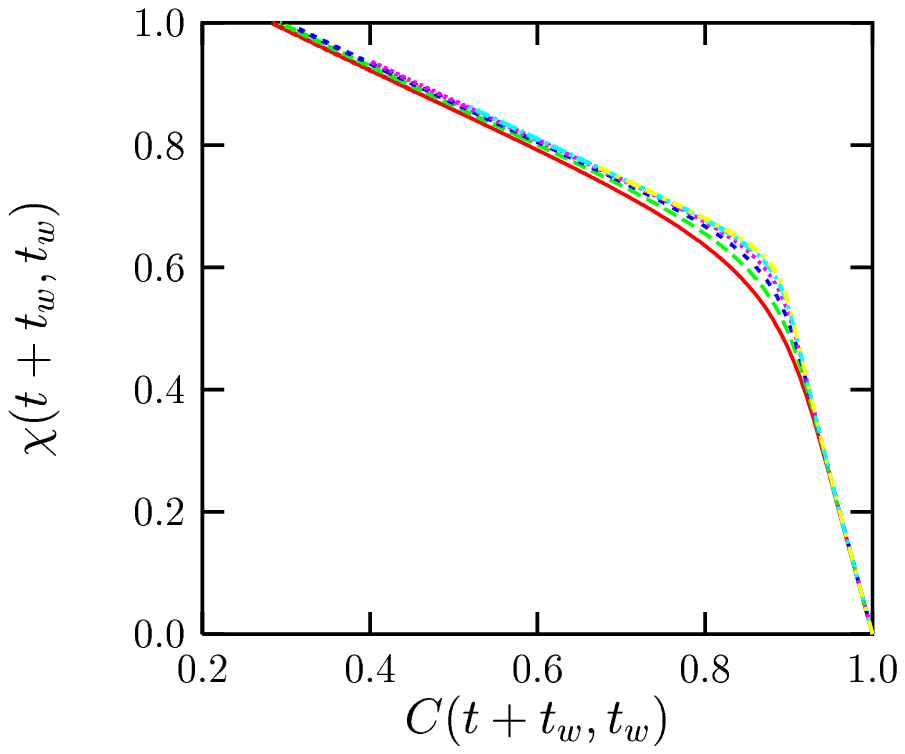,width=6cm}
\psfig{file=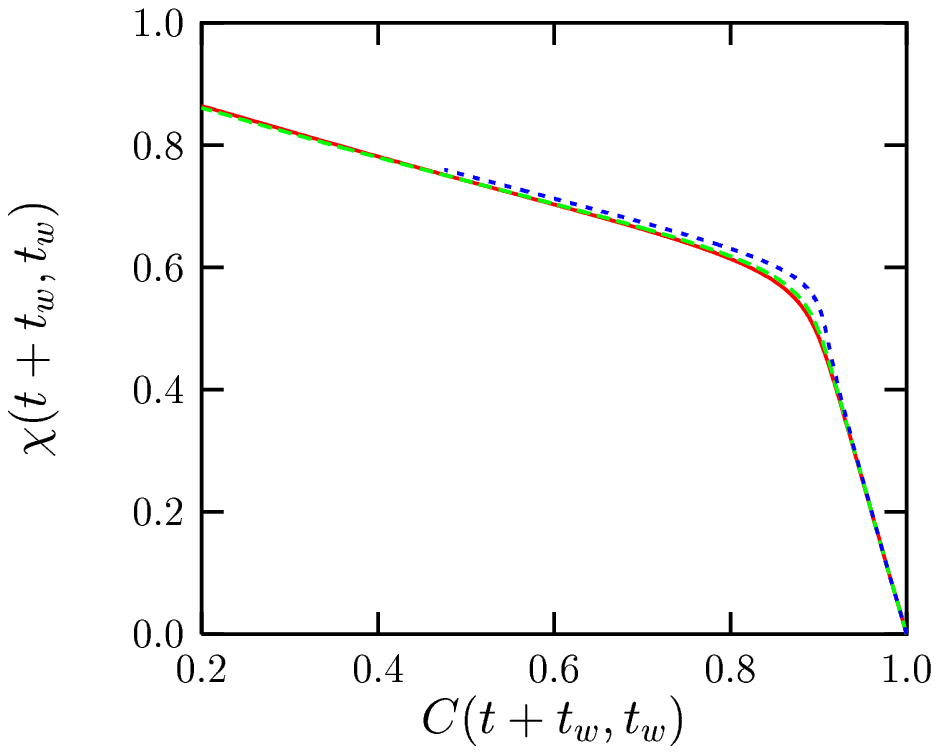,width=6cm}
}
\centerline{
\psfig{file=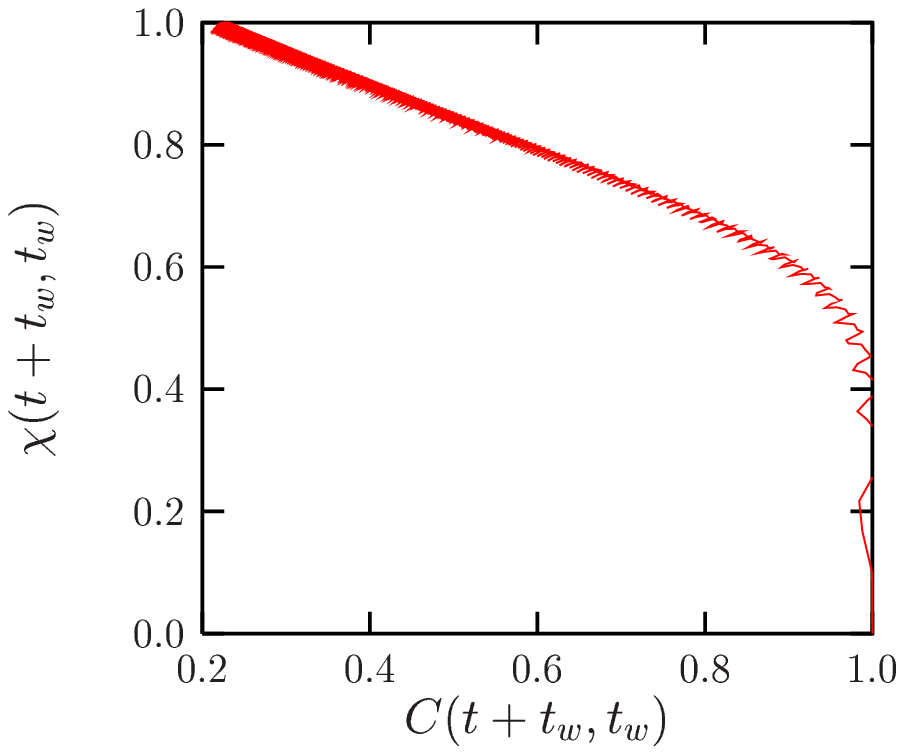,width=6.5cm}
\psfig{file=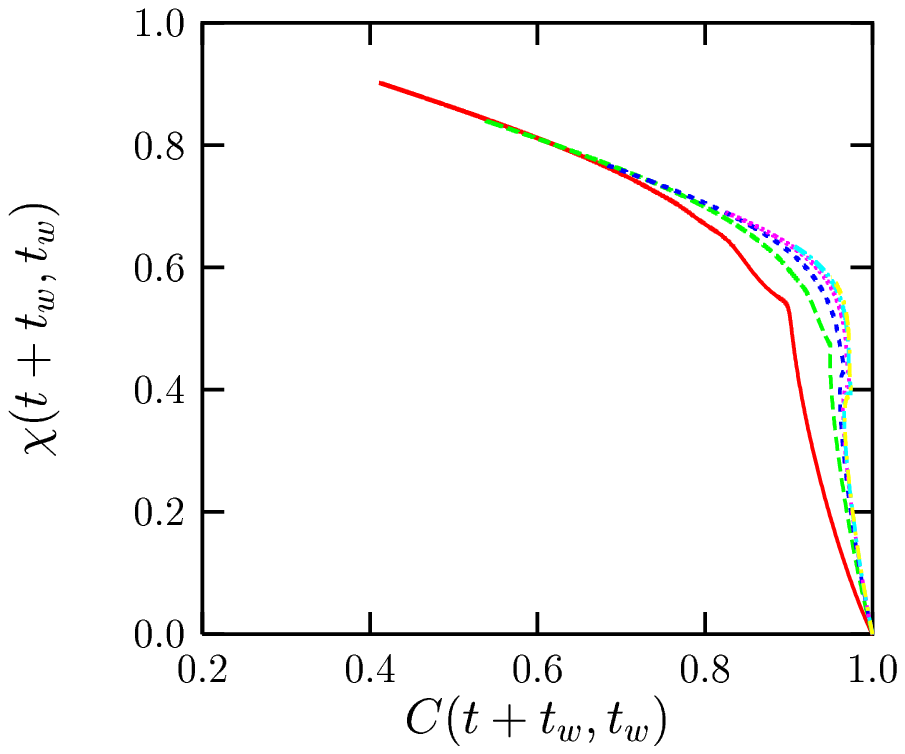,width=5.75cm}
}
\caption{The parametric $\chi(C)$ plot for the $p$-spin model:
relaxational dynamics
(upper left panel)~\cite{Cuku1}, with asymmetric interactions
(upper right panel)~\cite{Cukulepe}, under the effect of an oscillatory field 
(lower left panel)~\cite{Becuig} and with quantum fluctuations 
(lower right panel)~\cite{Culo}. The different curves on each panel 
correspond to different waiting-times.}
\label{chiC-fig}
\end{figure}

For the sake of comparison we show in Fig.~\ref{chiC-fig} the form of the 
$\chi(C)$ plot for a $p$ spin model adapted to mimic a structural 
glass (the original model), a sheared liquid or glass, 
vibrated granular
matter and a quantum glass.

\subsection{Domain growth}
\label{fdt-domain-growth}

The separation of time-scales is easy to visualize in the case of a system 
undergoing domain growth.
If the two times $t$ and $t_w$ are not very different, the domain
walls do not move between $t$ and $t_w$ and the dynamics consists 
of spin flips within the domains due to 
thermal fluctuations. This dynamics is identical to the equilibrium 
dynamics since the system can be thought of as being 
a patchwork of independent equilibrated finite systems of linear 
size ${\cal R}(t_w)$. It is then natural that the {\sc fdt} holds in this 
time scale. On the contrary, if $t$ grows in such a way that $\tau$
becomes of the order of ${\cal R}(t_w)$, the domains grow and the 
non-equilibrium
dynamics takes place. 
 The motion of the domain walls in the presence of an external
 perturbing random field introduced to measure the 
staggered response is due to 
two competing factors: on the one hand, the system tends to diminish 
the curvature of the interfaces due to surface tension, on the other 
hand the random field tends to pin the domain walls in convenient 
places. 
The full response of the walls is approximately~\cite{Corberi} 
\begin{equation}
\chi_I(\tau+t_w,t_w) \approx \rho(\tau+t_w) \chi_s(\tau+t_w,t_w)
\approx {\cal R}^{-1}(\tau+t_w) \chi_s(\tau+t_w,t_w)
\end{equation}
where $\rho(\tau+t_w)$ is the density of interfaces and 
$\chi_s(\tau+t_w,t_w)$
is the integrated response of a single wall. The contribution of a 
single interface depends on dimensionality and, for a ferromagnetic
Ising model with first neighbor interactions and non-conserved 
order parameter it has been estimated to be 
\begin{eqnarray}
\chi_s(\tau+t_w,t_w) &=& t^{-\alpha} \;\;\;\;\;\;\;
\mbox{with} \;\;\; \alpha =  
\left\{
\begin{array}{lll}
(3-d)/4 \;\;&  \;\; d<3
\nonumber\\
%0.33 + 0.066\, \ln(\tau) \;\;& \;\; d=3
%\nonumber\\
0 \;\;& \;\; d>3
\end{array}
\right.
\end{eqnarray}
and $0.33 + 0.066\, \ln \, \tau$ in $d=3$~\cite{Corberi}.
Thus, below the critical dimension $d_c=3$ the response of a 
single interface {\it grows} indefinitely with the time-difference. 
The same trend though with a different value of $d_c$ and with 
slightly different exponents has been obtained for a continuous 
spin model. Still, for all $d>1$, $\rho$ decays sufficiently
fast as to compensate the growth of $\chi_s$, $\chi_I$ vanishes 
and the integrated linear response function gets stuck at the value 
reached at the end of the stationary time scale, 
$\chi_{\sc eq}=(k_BT)^{-1}(1-m_{\sc eq}^2)$. In the slow regime the 
parametric plot is then given by a flat straight line with vanishing
slope, see the dashed line 
 in the right panel of  
Fig.~\ref{FDT-high-low}. Instead, the case $d=1$ behaves in a totally
different way: $\rho(\tau+t_w)$ still decays as $(\tau+t_w)^{-1/2}$ while 
$\chi_s$ grows as $\tau^{1/2}$, thus, in the regime of times 
such that $\tau/t_w$ is finite the interfaces do contribute to the 
integrated response and the $\chi(C)$ curve is non-trivial. 
As explained in \cite{Corberi} these results can be 
easily interpreted as follows. 
When $d>d_c$ the curvature driven mechanism
dominates and the interface response decreases with $\rho$. 
When $d<d_c$ instead this mechanism is weaken while the
field driven motion, and consequently the single interface response,
become more important. In the limit $d=1$ the curvature driven mechanism
disappears, $\chi_s$ compensates exactly the decay in 
$\rho$ and $\chi_I$ is non-trivial~\cite{Corberi}.   

Besides the qualitative arguments just presented, 
the ferromagnetic Ising chain is completely solvable and very 
instructive~\cite{Godreche-Luck,Lipiello,Fielding}.
The transition occurs at $K=J/(k_BT)\to \infty$ and this limit 
can be reached either by letting $T\to 0$ or $J\to \infty$. 
The latter is better adapted to compute the integrated linear response and 
one finds
\begin{equation}
\chi(C) = \frac{\sqrt{2}}{\pi} \; \mbox{arctan}
\left[ 
\sqrt{2} \mbox{cot} \left(\frac{\pi}{2} C \right)
\right]
\; .
\end{equation}
The fast regime is eliminated for this choice of parameters
and the full $\chi(C)$ curve is given by this equation.
This model is a concrete example of a system undergoing 
domain growth that has a non-trivial $\chi(C)$. 
For finite $J$ (or finite $T$) the equilibration time is 
finite and, for $t_w\geq t_{\sc eq}$ the trivial $\chi(C)=1/(k_BT)$ 
must be reached. However, for fixed $t_w\leq t_{\sc eq}$ one still finds
a very rich structure: the master curve $\chi(C)$ 
corresponding to $J\to\infty$ is followed from $C(t_w,t_w)=1$ 
down to $C_{\sc eq}=C(t_{\sc eq}, t_w)$. For longer time-differences,
$\tau+t_w > t_{\sc eq}$ the system equilibrates and $\chi(C)$ departs 
from the master curve and approaches the point $(0,1/(k_BT))$. 
The point of departure $C_{\sc eq}$ depends on $J$ and $t_w$. For fixed
$t_w$, $C_{\sc eq}$ increases with increasing $J$; for fixed $J$, 
$C_{\sc eq}$ increases with increasing $t_w$. 
Corberi {\it et al} also 
argued that even if the form of $C$ and $R$
depend on the microscopic dynamics, the $\chi(C)$ curve 
should be universal. 

Note an unexpected feature of this result:
when $J\to\infty$, 
even if the correlation and response vary in a single time-scale
with a simple aging scaling, the $\chi(C)$ relation is a continuous
function. This property poses some problems for the 
interpretation of the slope of the $\chi(C)$ plot as an effective 
temperature (see Section~\ref{temp_intro}) 
as well as the relation between statics and dynamics
discussed in Section~\ref{connection_with_eq}.
It would be interesting to identify the generic origin of these
``contradictions''.

For some time it has been argued that systems undergoing domain growth 
cannot have a non-trivial $\chi(C)$. This example, even though
in $d=1$, demonstrates that this is not true. It would be extremely 
interesting to construct a coarsening model in higher dimensions
with a non-trivial contribution to the integrated response coming
from the interfaces in such a way that $\chi(C)$ be non-trivial.
%Moreover, it has been demonstrated that in some coarsening systems
%the approach to the 
%asymptotic (piecewise with a flat part) $\chi(C)$ can be so slow 
%that for almost all accessible times the $\chi(C)$ curve can 
%look different~\cite{}

\subsection{Structural glasses}
\label{subsec:structural}

For glassy systems, where no such growth of order has been 
identified yet, the form of the {\sc fdt} modification is 
different from the one found for domain
growth. These systems still show a separation of time-scales
in the sense that the correlation decays in two sharply
separated steps as suggested by numerical studies. The asymptotic parametric
curve has the form of the solid line in Fig.~\ref{FDT-high-low}.
One can argue in terms of cage motion to get an 
intuitive interpretation of why {\sc fdt} holds in the 
fast correlation scale. Indeed, in this correlation-scale
the time-different $t-t_w$ is so short that each particle 
moves within a rather ``solid'' cage formed by its neighbors.  
Loosely speaking, the cages act as a confining potential 
on each particle. The rapid motion is again due to thermal fluctuations
and the dynamics is like the one expected in equilibrium: 
it is then no surprise that {\sc fdt} holds.
For the moment there is no easy interpretation for the form of the 
second part of the parametric curve. Why does it have a non-zero
constant slope or, equivalently, a single 
finite value of the effective temperature? This result was obtained
using fully-connected models and it 
was later checked numerically in a number of more realistic glassy 
models~\cite{numerics}. 

\begin{figure}[h]
\centerline{
\includegraphics[scale=0.55,angle=0]{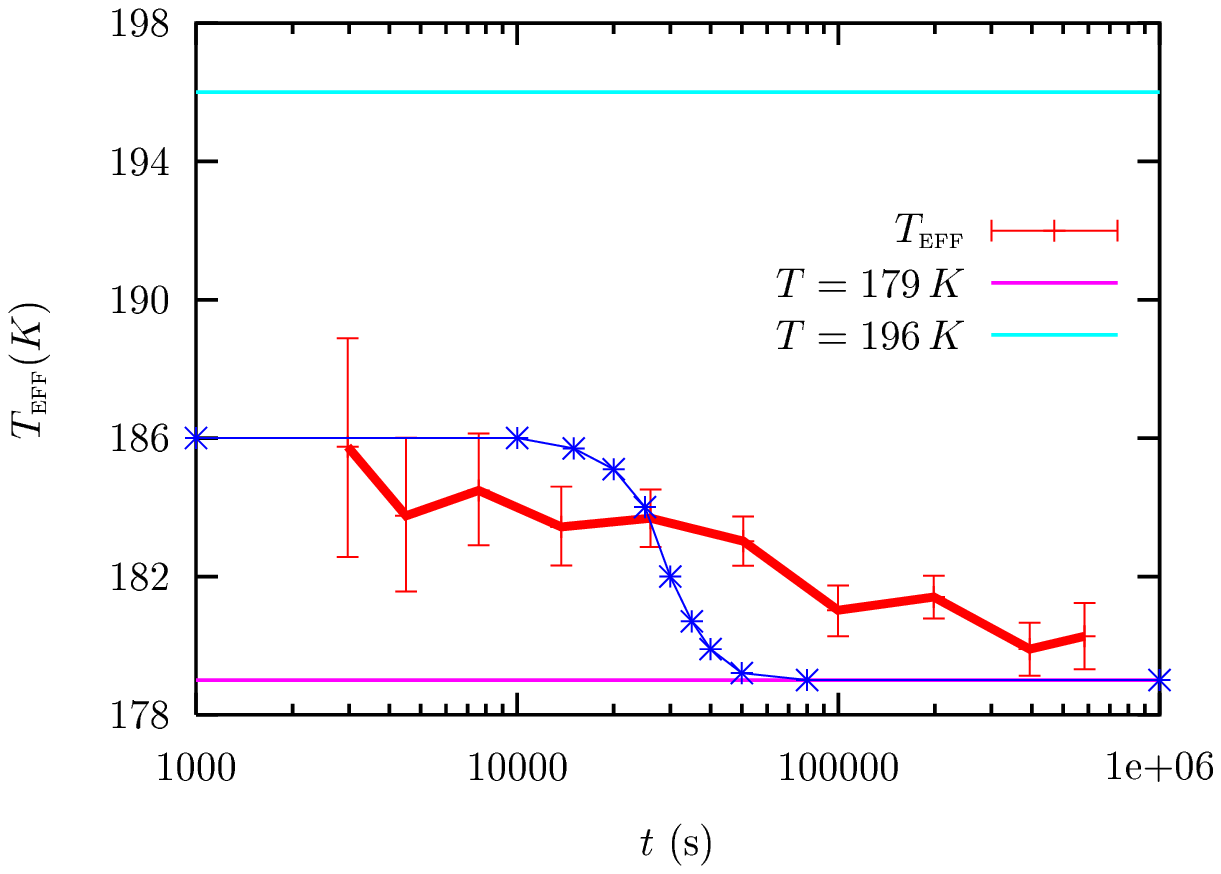}
\hspace{-0.5cm}
\includegraphics[scale=0.40,angle=0]{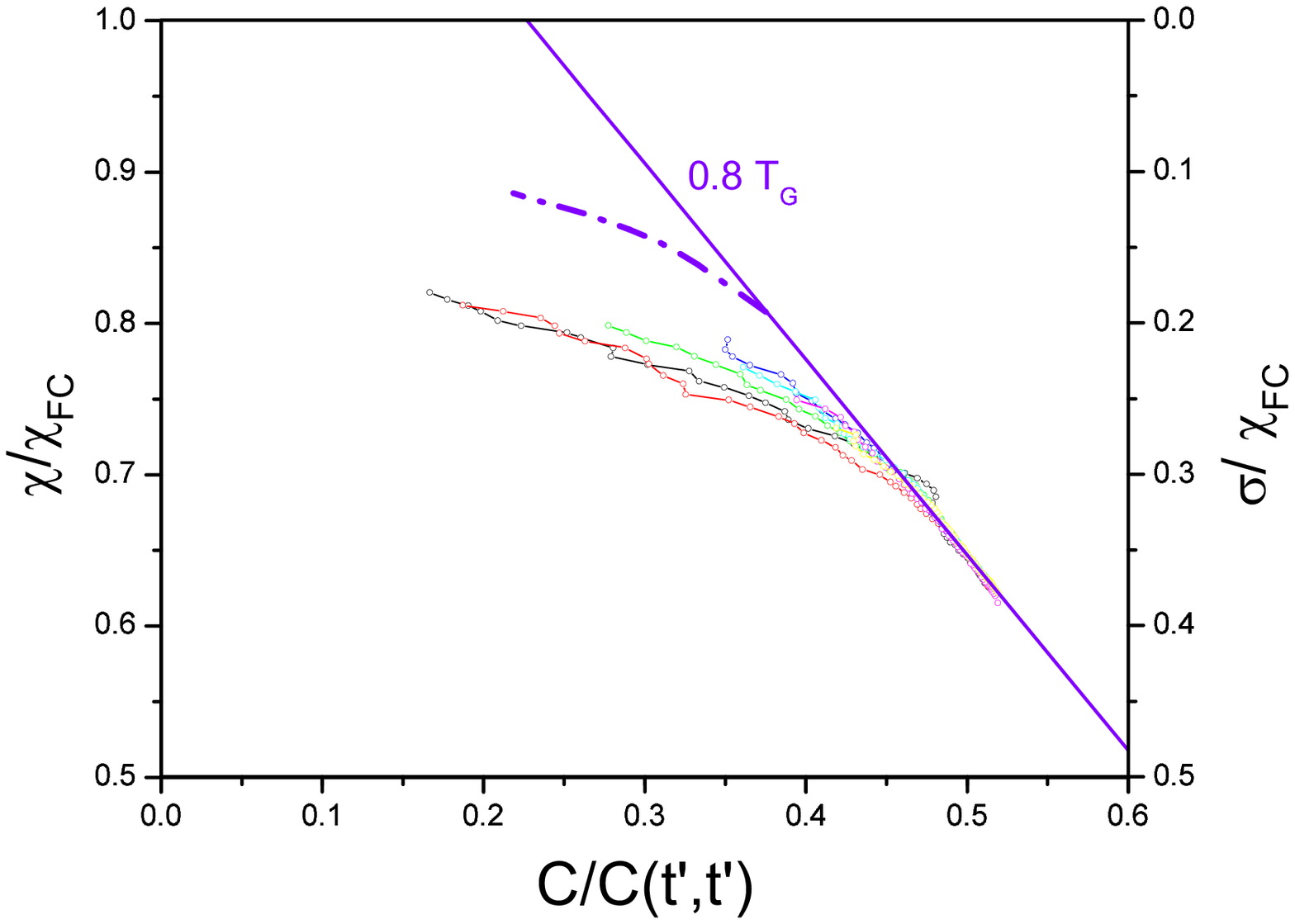}
}
\label{tomas}
\caption{
Left: the waiting-time evolution of the effective temperature of 
glycerol~\cite{Gris}. Right: the parametric $\chi(C)$ plot for 
thiospinel, an insulator spin-glass~\cite{Didier}.
}
\end{figure}

Two sets of experiments using laponite~\cite{Ciliberto} and 
glycerol~\cite{Gris} have investigated the modifications of {\sc fdt} in
glasses. 
The former is explained in~\cite{Ciliberto}.
In the latter Grigera and Israeloff~\cite{Gris} monitored the time-evolution 
of the {\sc fd} ratio
for glycerol at $T=179.8\, K$, 
the glass transition being at $T_g=196\, K$. The measurements
were done at fixed frequency $\omega=7.7Hz$ and the results 
presented in the manner described in Section~\ref{frequency-domain}. 
For a perfect time-scale separation the curve should have a 
step-like form as sketched in the right panel of 
Fig.~\ref{sketch-fdt-time-freq-domain} and also 
included with line-points in Fig.~\ref{tomas}. The experience
shows that the {\sc fd} ratio 
evolves very slowly from $T^*$ to $T$ with a very long transient 
between one and the other. Measurements at lower frequencies should
yield a more sharp separation between the two steps.

\subsection{Spin-glasses}
\label{subsec:fdt-spin-glasses}

The parametric curve for the fully-connected {\sc sk} model for spin-glasses
is given by the curve with a varying slope in the right panel of 
Fig.~\ref{FDT-high-low}. 
This result corresponds to having a succession of temporal scales
each one with an effective temperature, $T_{\sc eff}(C)$.
The question as to whether this behavior 
also applies to the finite dimensional case remains open.
The only results
available for the moment and, most probably, for a long while
are numerical and experimental. 
For the time-scales
explored, the parametric curves obtained have a very mild 
curvature. In order to decide beyond doubt if the asymptotic plot
is curved it is 
necessary to perform a very careful analysis of the times and
sizes explored (see~\cite{Babe}).

Very recently Herisson and Ocio studied the evolution of the 
correlations in the 
magnetization fluctuations (noise) and the thermoremanent magnetization
(integrated response) 
of thiospinel, an insulator spin-glass~\cite{Didier}. 
Their aim was to obtain the 
parametric curve $\chi(C)$. In the right panel of Fig.~\ref{tomas} we show 
their results 
%for $T=xx < T_g=xx$ and $t_w=xx$.
It is important to note that even the 
experimental data is very far away from the expected nonequilibrium 
asymptote that is estimated to be given by the dotted-dashed curve 
in the figure.
(for more details on this experiment see~\cite{Didier}.)

\subsection{Rheology}
\label{subsec:fdt-sheared}

In Section~\ref{interestingproblems} 
we have already explained that a non potential 
external force might stop aging.
When the strength of this force is small, the separation of 
time-scales is still present but the dynamics becomes stationary. 
This  fact makes the dynamics much simpler when observed in the time
domain but it still captures some of the interesting features 
of non-equilibrium. For instance, the correlation 
function of a super-cooled liquid under homogeneous shear
approaches a non-equilibrium stationary state and the 
parametric construction of Fig.~\ref{FDT-high-low} yields the same 
results where the waiting-time dependence is replaced by a 
shearing rate dependence. Thus, the red curve corresponds 
to a rather strong shearing rate, the green curve to a shearing rate 
of intermediate magnitude and the blue curves to the limit of 
vanishing shearing rate depending on the underlying system studied. 
These results were first obtained
by solving the dynamics of a fully-connected model with 
non-symmetric interactions as the one introduced in 
Eqs.~(\ref{non-symmetric-force}) and 
(\ref{lang3})~\cite{Cukulepe,rheology_theor}.
The numerical study of Lennard-Jones
mixtures under uniform shear performed by Berthier and J-L Barrat 
completed the study of this framework~\cite{JLBarrat} 
(see also \cite{Zamponi} and~\cite{Ludo}).

In Section~\ref{temp_intro} we present the interpretation
of the modification of {\sc fdt} in terms of self-generated
effective temperatures. Let us use this language here to explain these
results and motivate further studies in other systems with different
microscopic dynamics. Within the effective temperature interpretation,
we see that $T_{\sc eff}(C)> T$ control the slow 
relation. In slightly more technical terms, the correlation scales in which 
the time derivatives of the correlation are negligible with respect to 
the correlation itself
evolve according to 
a temperature that is
given by the modification of the {\sc fdt} relation.
This fact suggests that the effective temperatures
should appear in 
systems in which the microscopic dynamics is not necessarily thermal
but in which a separation of time-scales rapid-slow is self-generated
as time passes. 

\subsection{Vibrated models and granular matter}
\label{subsec:fdt-vibrated}

In fact, a similar modification to {\sc fdt} has also been 
observed for glassy models  driven by a time-dependent
oscillatory force that mimics the perturbations used to 
move granular matter~\cite{Becuig}. In this case, since the 
perturbation introduces its own characteristic time $t_c\propto 
1/\omega$ it is more convenient to present the data using
stroboscopic time, {\it i.e.} using a single point for each 
cycle. Modifications of {\sc fdt} in models for granular  matter
were studied in~\cite{Barrat-Ed,Makse,Barrat-granular}.

\subsection{Driven vortex systems}

The effective temperature has also been observed in 
the transverse motion of a drive vortex system~\cite{Koltonetal}.
Very interestingly, $T_{\sc eff}$ shares many quantitative properties 
with the ``shaking temperature'' of Koshelev and Vinokur~\cite{Valerii}.

\subsection{Quantum fluctuations}
\label{subsec:fdt-quantum}

Even more spectacular are the results for glassy models in which 
quantum fluctuations are important and keep a separation of fast-slow 
time-scales~\cite{Culo,quantum-others}. 
The fast scale is fully controlled by the quantum dynamics and the 
{\sc fdt} takes the complicated quantum form described in 
Section~\ref{section:fdt}.
In the slow scale though the quantum {\sc fdt} is no longer valid and 
it is replaced by a modified classical form in which the deviations
from the classical {\sc fdt} depend on the strength of quantum 
fluctuations. The dynamics in the 
slow scale superficially looks classical. This result was found in the 
solution to mean-field models (quantum extensions of the $p$ spin
and {\sc sk}, fully connected $SU(N)$). 
It is important to notice that Montecarlo
simulations of quantum problems in real-time are not possible. 

\subsection{Systems of finite size: preasymptotic behavior  }

The previous discussion shows that the correlation scales (time-scales)
play a very important role in the global behavior of the
systems. For relaxational systems we argued that the asymptotic 
parametric curve appears in the limit 
%\begin{equation}
$\lim_{t_w\to\infty} \lim_{N\to\infty}$.
%\end{equation}
The thermodynamic limit that is taken first ensures that the
equilibration time, $t_{\sc eq}$, diverges and cannot be reached by the 
waiting-time. For a systems of finite size, $N<+\infty$, 
the equilibration time is a function of $N$. In consequence,
the limit curve $\chi(C)$ has a lifetime that is bounded by
$t_{\sc eq}$. When the waiting-time becomes of the same order of magnitude
as $t_{\sc eq}$, the curve $\chi(C)$ starts changing to approach the
equilibrium asymptote, the straight line of slope $-1/(k_B T)$.
The way in which the approach to equilibrium    is achieved is not 
known in general. It is a very difficult problem even for fully-connected 
models and there is little hope to solve a problem in sufficient detail
as to be able to determine all  the crossovers. This might be possible 
for the $O({\cal N})$ model of ferromagnetic growth or for the spherical
{\sc sk} disordered model. Crisanti and Ritort~\cite{Crisanti}
analysed the crossover to equilibrium in the $p$ spin model 
with numerical simulations.

\subsection{Critical dynamics}

Godr\`eche and Luck studied the coarsening dynamics of 
ferromagnetic models quenched 
to the critical point $T_c$. Spatial correlations develop in the 
system but only up to a length scale that grows as $(\tau+t_w)^{1/z_c}$
with $z_c$ the dynamic critical exponent. 
The equilibrium magnetization, $m_{\sc eq}$,  and,
hence, $q_{\sc ea}$, vanish. 
Still, the dynamics is highly non-trivial. For finite though long
$t_w$ a stationary regime for $\tau\ll t_w$ and an 
{\it interrupted aging} regime for $t_0 \ll t_w \approx \tau+t_w$ 
can be identified even if they are not clearly separated by a 
plateau in the correlation at a finite $q_{\sc ea}$. 
In the former regime, the correlation and response are stationary and 
satisfy {\sc fdt}. In the latter
\begin{eqnarray}
C(t,t_w) &\approx& t_w^{-2\beta/\nu z_c} F_C\left(\frac{t}{t_w}\right)
\; , 
\;\;\;\;\;\;\;\;
%\nonumber\\
\chi(t,t_w) \approx t_w^{-2\beta/\nu z_c} F_\chi\left(\frac{t}{t_w}\right)
\; ,
\end{eqnarray} 
with $\beta$ and $\nu$ the usual static critical exponents
and $F_C$ and $F_\chi$ two scaling functions. 
As $t_w$ increases the stationary regime arrives up to lower values of $C$
and when $t_w$ grows to infinity the full decay is stationary 
as in equilibrium.
Since the interrupted aging part of $C$ and $R$ 
decay only algebraically with $t_w$ the $\chi(C)$ plot
for finite though long $t_w$ is very rich with a non-trivial
functional form. Similar results were obtained for the critical
dynamics of the XY model~\cite{Ludo-XY}.

\subsection{Connection with equilibrium }
\label{connection_with_eq}

The relation 
\begin{equation}
\left. -T \frac{d^2\chi(C)}{dC^2}\right|_{C=q}
=
\lim_{h\to 0} P_h(q) 
\; ,
\label{eq-dyn}
\end{equation}
between the static and non-equilibrium 
dynamic properties of slowly decaying systems, 
where $\chi(C)$ is the non-equilibrium relation between integrated 
response and correlation function as defined in Eq.~(\ref{FDT}) and 
$P_h(q)$ is the probability distribution of overlaps~\footnote{The overlaps
are given by the correlation between two equilibrated configurations.} in the 
perturbed Gibbs state, 
 became apparent from the analytic solution to some mean-field models.

 This relation holds in the exact 
solution to the {\sc sk} model~\cite{Cuku2} and the problem of a finite 
dimensional manifold embedded in an infinite dimensional space
in the presence of a random potential with long-range 
correlations~\cite{Mefr,Cukule}.
It is not verified in the exact solution to the $p$ spin model
and the manifold in a random potential with short range correlations.
The reason for this discrepancy is that for the latter models
the dynamics in the $t_w\to\infty \; N\to\infty$ limit approaches the 
threshold and not equilibrium. More precisely, all 
generalized susceptibilities, and in particular the energy density, 
approach a limit 
$\lim_{t_w \to\infty N\to\infty} \chi^{\sc gen}(t_w) 
= \chi^{\sc gen}_{\sc th} \neq \chi^{\sc gen}_{\sc eq}$. 
In all cases the dynamics occurs in a region of phase space that
is different from the one sampled in equilibrium and, for 
$p$ spin models and  the like it 
has different statistical properties. For the former models, 
even if still the region of phase space explored by the dynamics is
different from the one corresponding to the equilibrium states,
its statistical properties can be thought to be somehow equivalent,
since $\lim_{t_w \to\infty N\to\infty} \chi^{gen}(t_w) 
= \chi^{gen}_{th} = \chi^{gen}_{eq}$, 
and the relation (\ref{eq-dyn}) holds.

More recently, Franz, M\'ezard, Parisi and Peliti proposed that 
the connection (\ref{eq-dyn}) goes beyond mean-field and applies to 
finite $d$ systems in which 
all dynamic susceptibilities converge to their equilibrium values, 
linking in this way the (easy to measure) 
non-equilibrium dynamics properties of realistic models to the 
(hard to measure) equilibrium properties of the same models.
Since a threshold level as the one found for the $p$ spin
model cannot subsist for ever in finite $d$,
they argued that the validity of  Eq.~(\ref{eq-dyn}) should be 
rather generic. 

Several comments are in order. Corberi {\it et al} showed 
that there is a non-trivial nonequilibrium  dynamic
$\chi(C)$ that does not satisfy Eq.~(\ref{eq-dyn}) in one 
dimensional coarsening systems where the interface response
does not vanish asymptotically (see Section~\ref{fdt-domain-growth}). 
Even if the hypothesis of convergence of the generalized 
susceptibilities is not verified in this model, it provides 
a very simple example where the relation (\ref{eq-dyn}) does not 
hold. It might be possible to extend this 
result to domain growth 
in $d>1$ with an interface geometry such that 
the domain wall response does not vanish 
asymptotically. This problem deserves further study.

Based on numerical simulations, Marinari {\it et al} 
claimed~\cite{Marinari} that the relation (\ref{eq-dyn}) is verified in 
the $3d${\sc ea}. As pointed out by 
Berthier, Holdsworth and Sellitto in the context of the 
{\sc xy} model~\cite{Ludo-XY} and by A. Barrat 
and Berthier~\cite{Babe} in the context of the $2d$ and $3d$ 
{\sc ea} models, 
one has to be extremely careful when extrapolating the 
numerical results obtained for finite waiting-time
out of equilibrium and finite size in equilibrium.
Indeed, these authors showed that one can tune the 
finite waiting-time and the finite size to have 
a relation like (\ref{eq-dyn}) well before the 
asymptotic limits are reached 
and even in 
the trivial phase of the {\sc xy}  and $2d${\sc ea} models.

Finally, it is worth stressing that a non-trivial $\chi(C)$ has been
found in explicitly out of equilibrium situations 
for which equilibrium is trivial as, for instance, in rheological 
measurements of super-cooled liquids or in the long-time 
dynamics of super-cooled liquids before equilibration is reached.
Indeed, non-trivial $\chi(C)$ 
curves have been found in glassy model above the putative $T_s$ 
in Lennard-Jones systems and at finite $T$ in models 
for which $T_s=0$ like the kinetically constrained 
lattice models~\cite{kinetic2,kinetic3}.

In conclusion, 
a better determination of which are the conditions under which (\ref{eq-dyn})
holds is necessary.

\section{Effective temperatures}
\label{temp_intro}
\setcounter{equation}{0}
\renewcommand{\theequation}{\thesection.\arabic{equation}}

Temperature has remained an ill-defined concept until the
development of thermodynamics and statistical mechanics.
Evidently, the fact that a quantity, called temperature, 
must characterize the sensation of coldness or warmth has been known 
since the old times.
However, temperature was usually 
confused with heat.  

The thermodynamics
is an empirical theory based on four 
postulates that has been 
developed to determine some properties of the macroscopic objects 
without knowing the details of their constituents and 
interactions. After $t_{\sc eq}$, an isolated and finite system
reaches an equilibrium macroscopic state that can be characterized by  
a small number of parameters, the state variables. Temperature is one 
of these parameters. The first law states that energy is 
conserved in an isolated system
after having established the equivalence between heat and work.
The zeroth law
states that if two systems are in 
thermal equilibrium with a third, then they are also in equilibrium  
between them. The temperature is determined by an auxiliary measurement.
One sets a thermometer in contact with the system, waits until 
thermal equilibrium is established and then determines the temperature
by calibrating the reading of the thermometer. If one repeats this
procedure with a second system, equilibrated with the first one, 
the first law ensures that the thermometer will itself 
be in equilibrium with the second system and, consequently, its reading
will yield the same temperature. Hence, all systems in thermal 
equilibrium among them are at the same temperature.

The statistical mechanics establishes a bridge between the mechanical 
description of the microscopic constituents of the system and 
its macroscopic behavior. It yields a precise sense to the concept 
of temperature. To illustrate this statement, let us take an isolated 
system in the microcanical ensemble with volume $V$ and internal energy 
$U$. The entropy of the system is defined as 
$S(U)=k_B \ln \Omega(U)$ with $\Omega(E)dE$ the number of 
accessible states with mean energy between   
$E$ and $E+dE$.
The microcanonical definition of temperature is 
given by $1/(k_BT) \equiv \partial S(E)/dE$, evaluated at $E=U$.
The development of statistical mechanics allows one to show 
that this definition is equivalent to the thermodynamic one. 
One can equally find it with the canonical and 
grandcanonical formalisms.   

The previous paragraphs describe the behavior of systems in thermal 
equilibrium. What can one say about the systems that evolve out of 
equibrium? Can one define a temperature for them? Or is it at least possible 
for a subclass of nonequilibrium systems? If this holds true, can one 
use this definition as  a first step towards the development of 
a thermodynamics and a statistical mechanics for systems far from 
equilibrium?

Hohenberg and Shraiman discussed the possibility
of defining a temperature for certain systems out of equilibrium~
\cite{Hohenberg} using the modification of {\sc fdr}.  
In Ref.~\cite{Cukupe} we critically studied the definition of such 
an {\it effective temperature}, $T_{\sc eff}$, we 
insisted on the need to reach a regime with slow dynamics
(related to small entropy production) to be able to define
such  a ``state variable'' 
and we demonstrated the importance of analyzing it in 
separated time-scales.
In this Section we show in which sense the expected 
 properties of a temperature are 
satisfied by $T_{\sc eff}$ and we display some 
numerical tests.

\subsection{Thermodynamical tests}

\subsubsection{How to measure a temperature}
\label{measure-temperature}

In normal conditions,
the temperature of an object is measured by coupling 
it to a thermometer during a sufficiently long time interval 
such that all heat exchanges between thermometer and system
take place and the whole system equilibrates.

\begin{figure}[h]
\centerline{\includegraphics[scale=0.25,angle=0]{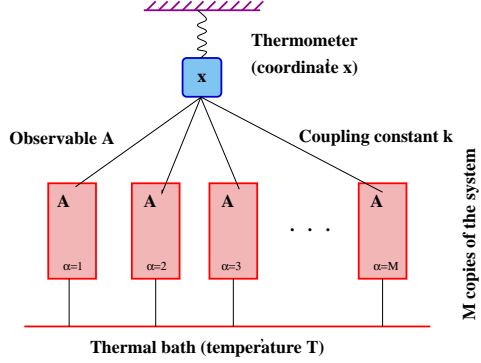}
}
\nobreak
\vskip 0.5cm
\caption{
Coupling between a thermometer and the observable $A$ of the system.}
\label{thermometer}
\end{figure}

\noindent
Let us call $t=0$ the instant when thermometer and system 
are set in contact. For simplicity, 
we choose to describe the thermometer with a single variable
$x$. In order to have a statistical measure of the 
object's temperature~\cite{Jean-Louis,Cuku4} we couple
the thermometer to $M$ independent copies of the system.
Each system is characterized by a variable 
$\vec \phi_\alpha$, $\alpha=1,\dots,M$. The energy of the total system is
\begin{eqnarray*}
E_{\sc tot}
=
m \frac{{\dot x}^2}{2}+ V(x)
+
\sum_{\alpha=1}^M E({\vec \phi}^\alpha)
-
\frac{k}{\sqrt{M}} x
 \sum_{\alpha=1}^M A[{\vec \phi}^{\alpha}]
\; ,
\end{eqnarray*}
with $V(x)$ the potential energy of the isolated thermometer
and  $E({\vec \phi}^\alpha)$ the one of the $\alpha$th isolated system.
Note the similarity between this coupled model and the 
treatment of system+environment done in 
Section~\ref{systemsincontactwithreservoirs} and the 
discussion on the harmonic oscillator coupled to a complex bath
in Section~\ref{complexbath}.
For a given value of 
$M$, the last term acts as an infinitesimal field $k x/\sqrt{M}$ 
that is coupled to the observable 
$A[{\vec \phi}]$ of each copy. The equation of motion of the thermometer 
reads 
\begin{eqnarray}
m {\ddot x}(t) = -\frac{\partial V(x)}{\partial x(t)} - \frac{k}{\sqrt{M}}
\sum_{\alpha=1}^M A[{\vec \phi}^{\alpha}](t)
\; .
\label{eqmotionx}
\end{eqnarray}
Again for simplicity we choose an observable $A$ with vanishing mean
$\langle A[{\vec \phi}] \rangle_{k=0} =0$ 
where  the angular brackets
represent the average over different histories of the system (see
Section~\ref{average-thermal}) or the average over
different systems, {\it i.e.}
$\langle f \rangle = 1/M \sum_{\alpha=1}^M f_\alpha$.
When the averaged observable does not vanish 
we use the difference between the observable and its average
as the fundamental quantity. The index $k=0$ 
indicates that the average is taken in the absence of the thermometer.
We denote $\langle \;\;\; \rangle_k$ the averaged in the presence of
the thermometer.
Equation~(\ref{eqmotionx}) can be rewritten as 
\begin{equation}
m {\ddot x}(t) = 
-\frac{\partial V(x)}{\partial x(t)} + k^2 \int_0^t 
ds \, R(t,s) x(s) + \rho(t)
\label{lange3}
\end{equation}
with $R(t,s)$ the linear response of the observable  
$A(\vec \phi)$ to the change in energy
$E_{\sc tot} \to E_{\sc tot} 
-k /\sqrt{M} \, x \, A(\vec \phi)$ performed at time $s$.
The force $\rho(t)$ is  a sum of $M$ 
independent random variables and due 
to the central limit theorem,  it becomes a Gaussian variable 
with vanishing average and variance
\begin{equation}
\langle \rho(t) \rho(s) \rangle_k = 
k^2 C(t,s)=k^2 \langle A[\vec \phi](t)
A[\vec \phi](s)\rangle_{k=0}
\;
\end{equation}
for large $M$
at first order in $k$. Thus, the evolution of the thermometer is determined 
by a Langevin-like equation [{\it cfr.} Eq.~(\ref{lang1})] with 
a correlated noise $\rho(t)$ and a retarded friction 
generated by the coupling to the systems.
For a generic system out of equilibrium there is no relation between
$R$ and $C$. For the problems we are 
interested in there is one. Next we explore the consequences of the 
modification of {\sc fdt} for the reading of the thermometer.

If the systems are equilibrated with their environments 
{\sc fdt} holds and it ensures that $R$ 
is related to 
$C$ by $R(t,s) = $$1/(k_B T) \partial_{s} C(t,s) \;\theta(t-s)$, 
with $T$ the temperature of the thermal bath. The thermometer is then 
coupled to an equilibrated  colored bath and it will eventually reach
equilibrium with it. The reading of the thermometer is defined from the 
value of its asymptotic internal energy and it has to be callibrated
from the 
characteristics of the thermometer that, of course, must be known
before starting the measurement. If one takes a simple harmonic oscillator
as a thermometer, one proves that the internal energy approaches 
$k_B T$. This is the result expected from equipartition 
since the oscillator has only two degrees of freedom (position and momentum). 

Imagine now that the systems are glassy of the type discussed in 
Section~\ref{section:solution}. If one studies a system 
with two correlation scales such that 
\begin{eqnarray*}
&&
R(t,s) = 
\left\{
\begin{array}{ll}
R_{\sc st}(t-s) & \;\mbox{if} \;t-s \ll s
\\
\frac{1}{t} \; R_{\sc ag}\left(\frac{s}{t}\right) 
& \; \mbox{if} \;\frac{s}{t}=O(1)
\end{array}
\right. 
\;\; 
C(t,s)=\left\{
\begin{array}{ll}
C_{\sc st}(t-s) & \; \mbox{if} \; t-s \ll s
\\
C_{\sc ag}\left(\frac{s}{t}\right) 
& \; \mbox{if} \; \frac{s}{t}=O(1)
\end{array}
\right. 
%\end{eqnarray*}
%avec 
%\begin{eqnarray*}
\nonumber\\
&&
R_{\sc st}(t-s) = \frac{1}{k_B T} \partial_s C_{\sc st}(t-s)
\;, \;\;\;\;\;\;\;\;
R_{\sc ag}\left(\frac{s}{t} \right) 
= \frac{t}{k_B T^*}\partial_s C_{\sc ag}\left(\frac{s}{t} \right) 
\end{eqnarray*}
with $T$ the temperature of the thermal bath and 
$T^*$ a different value read from the modified {\sc fdt} relation.
The reading of the thermometer, or its asymptotic internal energy density,
is found to be~\cite{Cukupe} 
\begin{equation}
E_{\sc therm} = 
\frac{\omega_0 \tilde C(\omega_0,t_w)}{{\tilde \chi}''(\omega_0,t_w)}
\end{equation}
where $\omega_0$ is the characteristic frequency of the thermometer,
$\tilde C(\omega_0,t_w)$ is the Fourier transform of the
 correlation function with respect to the time-difference
and ${\tilde \chi}''(\omega_0,t_w)$ is the out of phase susceptibility
defined in Eq.~(\ref{chi1}).

If the characteristic frequency 
$\omega_0$ is very high, the thermometer evaluates the 
evolution during the first step of the relaxation  
$t-s\ll s$, one finds $E_{\sc therm}=k_B T$  and 
one identifies $T$ with the temperature of the system. Instead, 
if the characteristic frequency is very low, the thermometer 
examines the behavior of the system in the long time scales,
$s/t=O(1)$ in the example, one finds 
$E_{\sc therm}=k_B T^*$  and one identifies $T^*$ as the 
temperature of the system. 

One can easily generalize this discussion to a problem with 
many correlation scales, each with a different value of the 
 effective temperature.
In order to measure them it is sufficient to tune the 
characteristic frequency of the oscillator to the 
desired scale.

\begin{figure}[ht]
\centerline{
\psfig{file=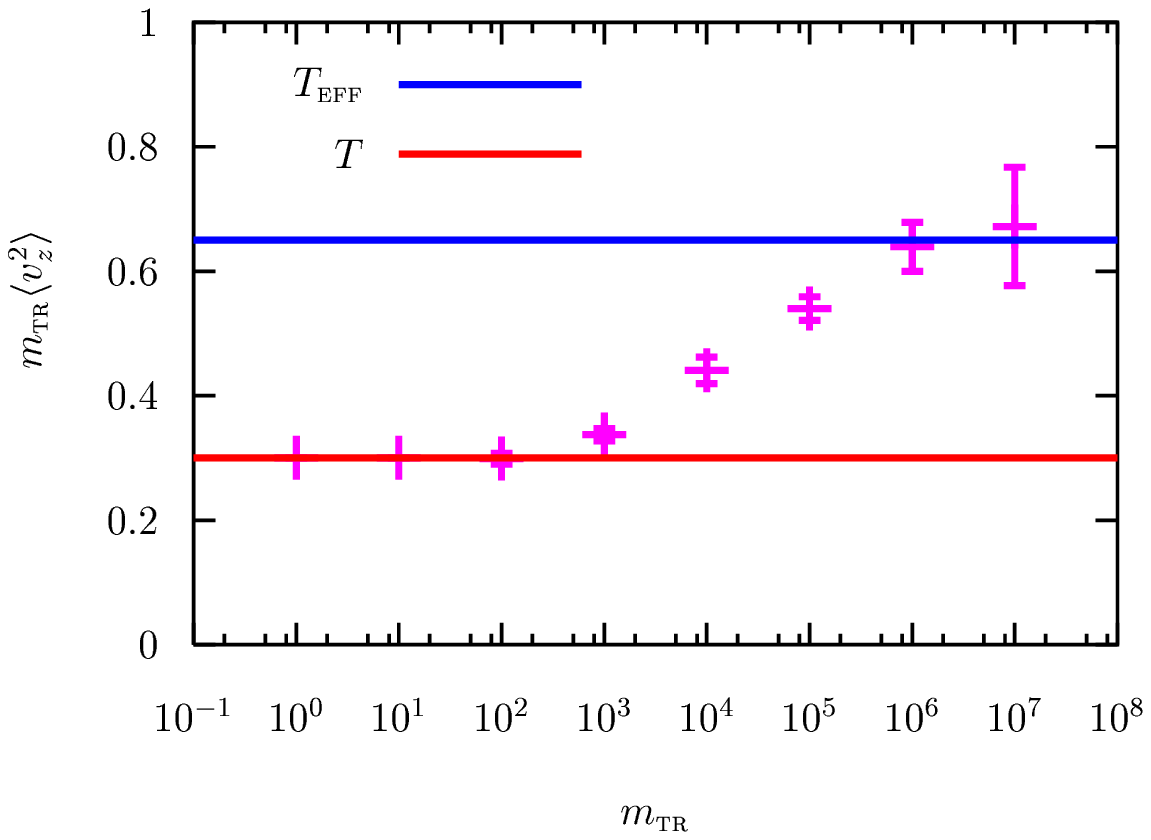,width=7cm}
\psfig{file=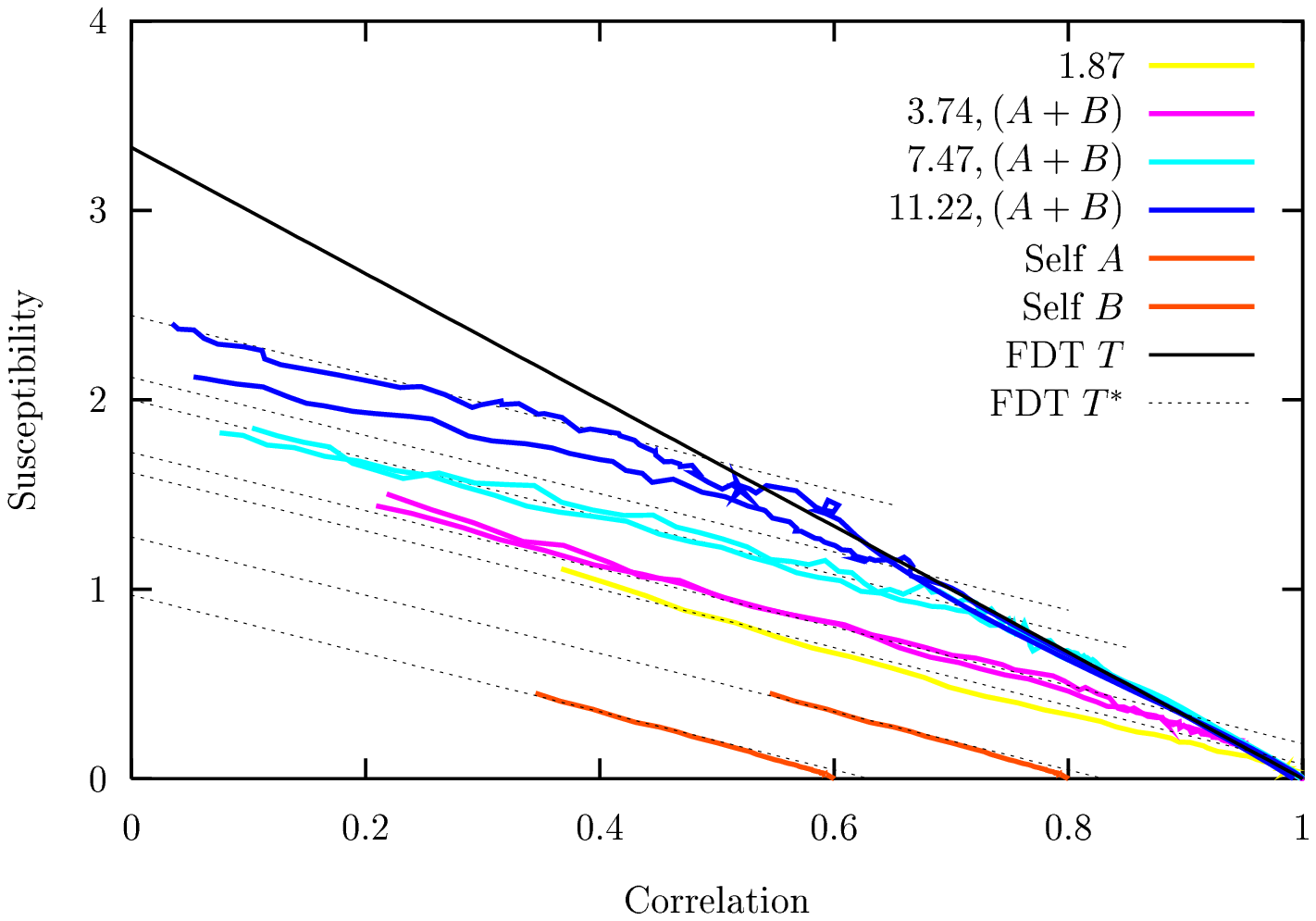,width=6.5cm}
}
\vspace{-0.5cm}
\caption{
A Lennard-Jones liquid
with two types of particles,  $A$ and $B$, in its liquid phase
under homogeneous shear.
Left: the asymptotic value of the transverse component of the 
averaged kinetic energy of the tracer particle normalized by $k_B$, 
$T_{|sc eff} = m_{\sc tr} \langle v^2_z\rangle /k_B$, against its mass.
Right:  Parametric plot 
between the linear response integrated over a time-interval of length
$t-t_w$, $\chi(k;t-t_w)$, and the incoherent scattering function
$C(k;t-t_w)$. The four first curves correspond to 
$k=1.87, 3.74, 7.47, 11.22$. 
The self-displacement is also shown.  
The equilibrium straight line of slope $-1/(k_B T)$ 
is represented by a black line while the 
black dotted parallel straight lines have slopes $-1/(k_BT^*)$.
From Ref.~\cite{JLBarrat}.
}
\label{wave-vectors}
\label{tracer}
\end{figure}

A concrete realization of this measurement 
corresponds to a generalization 
of the Brownian motion experience of Perrin in which one follows the 
evolution of a tracer immersed in the fluid. 
If the latter is equilibrated at temperature $T$,
after a short equilibration period, each component of the 
 averaged kinetic energy of 
the tracer approaches  $k_B T/2$. Instead, if the fluid evolves
out of equilibrium, one can choose the mass of the particle (that plays
the role of the characteristic frequency in the previous discussion)
in order to examine the rapid dynamics, or the slow dynamics. 
The tracer acquires an averaged kinetic energy 
$E_{\sc kin} = k_B T^*/2$ with $T^*$ 
the value of the effective temperature in the time scale explored.  
Berthier and J-L Barrat~\cite{JLBarrat} performed this measurement in a
numerical experiment using a dense liquid modeled 
as in Eq.~(\ref{lennard-jones}) under homogeneous shearing
in contact with a thermal bath at temperature $T$ as a 
fluid. They chose a Lennard-Jones particle as a tracer  with mass
$m_{\sc tr}$. 
%The left panel in 
%Fig.~\ref{tracer} represents a sketch of an instantaneous 
%configuration of the binary liquid and the tracer. 
The left panel 
shows the asymptotic value of the transverse component 
of the averaged kinetic energy (transverse to the shearing rate
direction) as a function of the mass of the tracer. One 
clearly sees how the effective temperature 
$T^*=m_{\sc tr} \langle v^2_z\rangle /k_B$ 
interpolates between the value $T$ at small masses 
and the value $T^*>T$ for large masses. The trend can be easily 
understood. A very light tracer easily reacts to the quick 
bombardment of the particles in the fluid and it feels the temperature 
of the bath via the rapid scale of relaxation. 
A heavy tracer instead can only move via large rearrangements of the fluid  
and these correspond to the slow part of the relaxation.
The transition between the 
two values is very smooth and  it occurs over several orders of magnitude
in the mass. The consistency of the explanation, {\it i.e.} the relation 
between $T^*$ and the {\sc fdt} relation, will become explicit in 
Section~\ref{zerothlaw}.

A complementary discussion on thermometric measurement of 
$T_{\sc eff}$ and, in particular, of the effect of not having 
well-separated time-scales, appeared in~\cite{Exartier-these,raphael}.

\subsubsection{Zeroth law}
\label{zerothlaw}

A temperature, even if it is defined out of equilibrium,
should control the direction of heat flows and the partial equilibration
between observables in interaction. (Two observables interact if and 
only if the crossed linear response $R_{AB}(t,t')$, as defined in 
Eq.~(\ref{linear-response}), does not vanish.)

With a pair of observables $A$ and $B$ one constructs two self and two cross
correlations and responses.
In each correlation scale
the {\sc fdt} relation is modified by a constant factor with which one
constructs the effective temperature.
If the notion of an effective temperature is correct, two 
interacting observables 
that evolve in the same time-scale should acquire the same value of 
$T_{\sc eff}$. 
The equilibration between different observables is well defined only
in the limit of small heat, or energy, exchanges. This limit is achieved 
in a free relaxing system when the waiting time becomes very long 
or in a driven system rendered stationary when the nonconservative
force vanishes. It is only possible for observables 
evolving in the same time-scale (note that even a single observable can 
have several values of the effective temperature when examined on 
different time-scales). We call {\it partial equilibrations} those 
arriving in the same time-scale. In contrast, if the two observables  
$A$ and $B$ do not interact they should evolve in different
time-scales.

The property of partial equilibration has been proven analytically for 
the fully connected models solved 
in Section~\ref{section:solution}~\cite{Cukupe} and, 
later, for all perturbative 
approximation of more realistic models under the assumption of 
there being a slow dynamics with a sharp separation of 
time-scales~\cite{Cuku4,Cuku5}.
It can be seen numerically as follows. Take, {\it e.g.}, two
$p$ spin models with different exchange strengths, $\tilde J_1$ and 
$\tilde J_2$ in contact with the same thermal bath at temperature $T$   
and couple them  linearly with a term $\alpha
\sum_{i=1}^N s_i \sigma_i$ where $s_i$ and $\sigma_i$ are the
dynamic variables of each system and $\alpha$ is the coupling constant.
(Other couplings are, of course, possible.)
When $\alpha=0$ the two systems evolve independently and their
$T_{\sc eff}$'s in the aging scale take different values.
If we now let the coupled system $(\alpha > 0$) 
evolve it will reach an asymptotic limit (roughly when the energy of 
the full system decays very slowly). In this limit   
we trace the {\sc fdt} parametric plots for the two self
and cross $C$ and $R$'s. We observe that if 
$\alpha < \alpha_c$ the cross responses vanish and the effective 
temperatures still take different values in the aging  scales that 
are now forced to evolve with different scalings. Instead, 
if $\alpha>\alpha_c$, the cross responses do not vanish,
the two systems lock and $T_{\sc eff}$ in the aging scale
acquires the same value for both self and cross {\sc fd} relations.
Note that for the manifold, all modes have the same value of the
effective temperature in the same correlation scale. This is 
achieved internally and it means that all modes are in interaction.
%In the left panel of Fig.~\ref{wave-vectors} 
%we show the parametric {\sc fdt} plot for 
%two coupled $p=3$ spin model and $\alpha > \alpha_c$.

Partial equilibrations
have been tested numerically by  Berthier and 
J-L Barrat~\cite{JLBarrat} in the dense homogeneously 
sheared liquid discussed in Section~\ref{measure-temperature}
(partial equilibrations simply relaxation Lennard-Jones mixture were 
studied by Kob and J-L Barrat in \cite{numerics}).
The relevant correlators 
for this system are the wave vector dependent 
incoherent scattering functions  and their associated responses.  
Berthier and Barrat measured these two-time functions 
for the same and different species and 
several values of the wave vector.
%, $k=1.87, 3.74, 7.47, 11.22$. 
Figure~\ref{wave-vectors} shows the parametric {\sc fdt} plots.
All the curves have a broken line form with a first part with 
slope $-1/(k_B T)$ and a second part with a {\it common}
slope  $-1/(k_B T^*)$. 
The value $T^*$ is 
identical to the result found monitoring the kinetic energy of the 
tracer (see Section~\ref{measure-temperature}) demonstrating that
the scenario is consistent. Note that in this case we cannot tune
the coupling between different wave vectors as we did 
when we externally coupled the two $p$ spin model. The fact that 
$T_{\sc eff}$ in the slow scale is the same 
for all wave vectors studied shows that they interact. This is also
seen from the fact that the cross responses do not vanish. 
Coupled oscillator models with parallel or sequential Montecarlo
dynamics do and do not partially equilibrate, respectively.
~\cite{Garriga}. In the latter case the thermal conductivity is 
very small and one is  in situation in which, effectively,  
$\alpha<\alpha_c$.

Intriguingly, the fact that each time-scale has its own 
effective temperature has been shown to fail in two models. These are 
the trap model~\cite{Fielding} and the $1d$ Ising chain at 
$T=0$~\cite{Fielding,Corberi,Lipiello} when special perturbations
are applied. In both cases one can construct observables 
that evolve in the same time-scale but have different values of the 
effective temperatures.
In the first case, the
failure might be adscribed to the fact that the model does not 
have an equilibrium state. 
%Indeed, the condition of having slowly 
%varying one-time quantities is not met by the model.
This is not the case in the second model though. 
The origin of the failure might then be related to having used 
special observables, see the discussion in \cite{Fielding}.
Still another pathology was already signaled for this problem: the global 
correlation and response decay in a single time-scale
but the asymptotic $\chi(C)$ is a curve contradicting the notion of 
a correlation-scale dependent effective temperature.
It is quite
plausible that this problem is also related to the 
failure of the relation between dynamics and statics discussed
in Section~\ref{connection_with_eq}.
A good understanding of the 
conditions under which this property holds is still lacking.
  Experiments in laponite show
some mismatch between the value of $T_{\sc eff}$ found with mechanical 
and electrical measurements. The reason for this is 
unclear~\cite{Ciliberto}.

One can also verify that the heat exchanges go from the higher values of 
$T_{\sc eff}$ to lower ones. However, it is still not clear why one 
can have a partial equilibration to a {\it higher} value of $T_{\sc eff}$ 
than those of the two independent observables, as it 
has been obtained analytically for some solvable fully-connected 
models.

\subsubsection{Auxiliary thermal baths}
\label{auxiliarybaths}

A supplementary verification of the thermodynamic character of the
effective temperature comes from the study of the action of 
complex thermal baths on the dynamics of simple systems, as discussed in 
Section~\ref{auxiliarybaths}, or glassy systems. 

We found that an arbitrarily weak  auxiliary 
bath with $\gamma(0)$ small has an important effect 
provided it is sufficiently slow and its 
temperature is within the range of values of the effective temperatures 
of the slow dynamics. 
The (slow) time dependence of all $C$ and $R$'s
are affected by  a time rescaling $t \rightarrow K(t)$. This is 
such that the time-scale which has an effective temperature 
equal to the temperature of the auxiliary bath, say $T_1$, 
is locked to the time-scale of the bath.
In particular, if we couple an aging system with a bi-valued 
$T_{\sc eff}= (T,T^*)$ to a composite stationary 
bath with two time scales, $\tau_1\to 0$ and $\tau_2 > 0$ and two temperatures
$T$ and $T_2$, respectively, we find
that the system becomes 
stationary if $T_2>T^*$, and is hardly affected 
if $T_2<T^*$. More generally, 
an aging  system with multiple effective temperatures, $T_{\sc eff}(C)$,\
becomes partially stationary for  
all the time-scales with $T_2>T_{\sc eff}(C)$, but still ages 
for time-scales with $T_2<T_{\sc eff}(C)$. We found this behavior in
Montecarlo simulations of the $3d${\sc ea} model~\cite{Cuku5}.

\subsection{Temperature fixing by {\sc susy} breaking}

For any model one can derive a set of integro-differential 
equations that couple all two-point functions. These admit a 
representation in terms of the super-correlator $Q(a,b)$.
In the asymptotic limit in which we
neglect the time-derivatives, these equations 
are invariant under any change of supercoordinates, $a$, that have
unit superjacobian~\cite{Frku}. Fixing the time-reparametrisation 
(the function $h(t)$ in each correlation scale) and the value of 
the effective temperature corresponds to breaking this large 
symmetry group to a subgroup~\cite{Cuku5}.

\subsection{Fictive temperatures}

``Fictive temperatures'' are phenomenological concepts that have been 
used to describe experimental data~\cite{fictive} (see also~\cite{Edwards-T}). 
The basic idea has been to claim that,
when crossing $T_g$, the system remembers its equilibrium configuration
before the quench and it remains effectively at 
a higher temperature $T_f$. Several refined prescriptions to extract 
time and preparation dependences of $T_f$ have been proposed. 
The relation between $T_{\sc eff}$ and $T_f$ has been discussed in 
\cite{Cukupe}. Here we simply stress that even if
the same basic idea can be used to interpret $T_{\sc eff}$, 
the latter has the properties of a  thermodynamic temperature
while this is not obvious for $T_f$. 

\subsection{Nonequilibrium thermodynamics}
\label{thermo}

Once equipped with a {\it bonafide} effective temperature 
the natural next step is to try to construct a thermodynamics
for such systems out of equilibrium. This has been proposed 
by Nieuwenhuizen~\cite{Nieu}  based on two previous results: (1) 
similar constructions done using the fictive temperatures by 
Tool, Moynihan and others~\cite{fictive}. (2) the relation between the 
dynamics 
and the {\sc tap} approach that we shall discuss in Section~\ref{section:tap}.
The idea is to define generalized thermodynamic potentials in which 
$T_{\sc eff}$ 
intervenes as a supplementary parameter (one can also include other state 
variables as an effective 
pressure, etc.). For the fully connected 
models these potentials have a precise meaning based on the {\sc tap}
analysis (Section~\ref{section:tap}). 
Their validity for more realistic models is still an open 
problem.

\subsection{Statistical mechanics}
\label{statmech}

%\vspace{0.25cm}
\noindent{\it Edwards measure}
\vspace{0.25cm}

In the 80s Edwards proposed that the stationary properties of 
dense granular matter under a weak forcing can be determined 
with a {\it flat} distribution of the blocked configurations,
{\it i.e.} those in which no grain is able to move) at the chosen 
density~\cite{Ed}. Note that 
this is possible in granular matter since the external temperature 
is irrelevant. Following this ``microcanonical'' approach the 
logarithm of the number of blocked configurations at constant 
density defines an entropy from which one obtains a compactivity
by simple derivation with respect to the density. Extending this
prescription one can define an Edwards' temperature for soft systems
by considering 
blocked  configurations at fixed energy and computing
 $1/(k_B T_{\sc edw}) =
\partial S_{\sc edw}(E)/\partial E$.

Edwards' temperature can be calculated explicitly for  
the fully connected models at zero external temperature (again
using techniques discussed in Section~\ref{section:tap}). 
Surprisingly, one finds that its actual value coincides
with the value taken by $T_{\sc eff}$ defined
from the modification of {\sc fdt} in a purely dynamic calculation.
This result has boosted the study of solvable models in low dimensionalities
analytically and more realistic systems numerically to check if the 
coincidence goes beyond the fully-connected case. 
Until now, several models where Edwards' distribution yield very good 
results have been exhibited~\cite{Barrat-Ed,others-Ed} while some others
where it fails have also been shown ({\it e.g.} the random field
Ising model~\cite{Barrat-Ed}). 
In the latter cases one finds a well-defined 
$T_{\sc eff}$ 
while Edwards' measure fails to give the correct results.
The result of including additional constraints on the configurations 
counted have also been analysed and naturally improve the results 
derived from Edwards' measure. 
This area of research is very active  and more work is necessary
to determine the limits of validity of Edwards' proposal.

The definition of Edwards' measure is unambiguous since it is a zero
temperature entity based on 
the study of minima of the potential energy. An extension 
to finite $T$ necessitates the analysis of a 
free-energy landscape. This can be done
analytically for fully-connected models 
as explained in Section~\ref{section:tap}. In finite $d$ cases 
for which one is forced to use numerical techniques
it is not, however, obvious how to define, let alone compute, 
the entropy $\Sigma(\beta,f)$.

\vspace{0.25cm}
\noindent{\it Inherent structures}
\vspace{0.25cm}

The {\it inherent structure} approach of Stillinger and 
Weber~\cite{Stillinger}
has been recently revisited 
and extended to study dynamic properties of glass formers and 
glasses~\cite{Sciortino}.
The original approach, even if static, 
is close to the ideas of Edwards. At the static level 
it consists in dividing
the partition function in inherent structures, or pockets of 
configurations around minima of the {\it potential energy}, and transforming
the sum into sums over minima and the configurations associated to them.
In order to connect with the dynamics, one computes the entropy of 
the inherent structures and then derives an inherent structure 
temperature
with a similar prescription to the one of Edwards. The sampling of the 
inherent structures numerically is a non-trivial issue.

A critical analysis of the applicability of this approach at finite 
temperatures has been presented by Biroli and Monasson~\cite{Bimo} 
who stressed 
the need of analyzing the free-energy, instead of the potential energy, 
at finite $T$. They also compared with the {\sc tap} analysis discussed in 
Section~\ref{section:tap}. Having said this, the use of the inherent 
structure approach has been rather successful when confronted to 
$T_{\sc eff}$~\cite{Sciortino}.

\section{Metastable states}
\label{section:tap}

The aim of the {\sc tap} approach is to define and study 
a free-energy density as a function of the relevant order 
parameters~\cite{tap}. Its stationary points are the  
(stable and unstable) metastable states. 
For a fully connected Ising magnet the {\sc tap} free-energy
density simply depends on the averaged global magnetization and, below $T_c$, 
has the form drawn on the right panel of Fig.~\ref{aging}. The vanishing 
global magnetization is not a good order parameter for disordered
spin systems
and one is forced to introduce all local ones. The approach 
can be extended to describe classical and quantum 
systems in and out of equilibrium~\cite{Biroli1,Biroli2}. Under different
names similar approaches appear in other branches of condensed matter 
physics. The {\it density functional theory} of 
the inhomogeneous electron gas in solids or the {\it dynamic mean-field
theory} of strongly correlated electrons are very similar in spirit to 
the {\sc tap} approach~\cite{Georges}.

The clearest way of deriving the {\sc tap} free-energy density 
for disordered models is based on two ingredients: one identifies 
an adequate perturbative expansion, and 
one Legendre transforms the standard free-energy density 
with respect to a well-chosen set of parameters. 
When these two choices are correctly done, 
only a finite number of terms in the expansion do not vanish for 
typical configurations of disorder for infinite
dimensional models. In finite $d$  one can attempt
a large $d$ expansion~\cite{Georges-Yedidia}.

For spin-glasses
at low-$T$ the {\sc tap} free-energy density has an infinite
number of stationary points. The {\sc tap} approach gives us access to  
the complete organization of metastable states of any type of stability 
(minima, saddles of all types, maxima).
In particular,  this has been 
analysed in great detail for the $p$ spin model and  the like 
since we expect it to mimic what occurs in real 
glassy systems though in an exaggerated manner.

In the following we first derive the {\sc tap} free-energy 
density for a generic spin model with Ising or spherical variables.
Next we summarize the consequences of this approach
for the $p$ spin spherical model. 
We also briefly explain how the number of metastable states
is computed and how this depends on the parameters in the model.
We explain how the approach is generalized to deal with dynamic
properties. This full set of results gives us a complementary 
view of the non-equilibrium dynamics and serves as the
basis for an image of the dynamics of finite $d$ glassy 
models.

\subsection{Equilibrium}
\label{replicas}

Before describing the {\sc tap} approach let us summarize the 
equilibrium behavior of the mean-field models as 
derived from the 
analysis of its disordered averaged  free-energy density
$-\beta [f]_J = [\ln Z]_J/N$ with the replica trick.
Again the three 
classes, ferromagnetic-like ($p=2$), glassy-like ($p\geq 3)$~\cite{Crso} 
and spins-glass-like ({\sc sk})~\cite{Mepavi} 
models have different characteristics.

For the spherical $p$ spin model with $p\geq 3$~\cite{Crso} 
the standard Parisi {\it Ansatz}~\cite{Parisi} 
yields a static transition
at $T_s$ from a paramagnetic phase where a {\it replica symmetric Ansatz}
solves the problem to a glassy phase where a {\it one step 
replica symmetry breaking Ansatz} is needed. At low $T$'s, 
the breaking point parameter, $x$, and the Edwards-Anderson 
order parameter, $q_{\sc ea} \equiv N^{-1} \sum_i [\langle s_i\rangle^2]$, 
in the Parisi matrix are
fixed by requiring that the free-energy density be a maximum.
This yields a static transition 
temperature, $T_s$, given by the set of equations
$T_s=y \sqrt{p/(2y)}(1-y)^{p/2-1}$ and $2/p=-2y (1-y+\ln y)/(1-y)^2$.
One can check that $T_s < T_d$. The static transition
is discontinuous in the sense that $q_{\sc ea}$ 
jumps at $T_s$ but it is of second order thermodynamically
since all susceptibilities are continuous across $T_s$. 
The equilibrium energy density at $T=0$ 
is given by the implicit equation
$\sqrt{2y/p} = 
[-{\cal E}(T=0)-\sqrt{-{\cal E}^2(T=0)-{\cal E}^2_{\sc th}}]/(p-1)
$. One checks that ${\cal E}_{\sc eq} < {\cal E}_\infty={\cal E}_{\sc th}$

A modified prescription to
determine the breaking point, $x$, in the one step {\it Ansatz}
yields different results. 
If instead of extremizing with respect to $x$ one requires that
the solution be {\it marginally stable} by setting the 
{\it replicon eigenvalue} of the matrix of quadratic 
fluctuations~\cite{Bray-Moore-replicon}
to zero, leads to a different solution with a dynamic meaning.
(All other eigenvalues are positive.)
Namely, one finds
a transition from the paramagnetic to the glassy phase at 
a higher temperature that coincides with the dynamic 
critical temperature found with a purely dynamic calculation, $T_d$.
Moreover, $q_{\sc ea}$ coincides with the 
dynamic one and the breaking point $x$ yields the ratio $T/T_{\sc eff}$.
Consequently, all  one-time quantities as the asymptotic energy density
and generalized susceptibilities coincide with the ones found 
dynamically ({\it e.g.} ${\cal E}={\cal E}_{\sc th}$). The fact that 
with the condition of marginal stability we access the region of 
phase space reached dynamically is due to thte fact that with both 
we search for the path of flat direction formed by the threshold 
level, as explained below. 

When $p=2$ instead~\cite{p=2-statics}, 
the replica solution below $T_s$ is {\it replica 
symmetric} with a non-vanishing $q_{\sc ea}$ and marginally stable. 
The static transition is continuous, dynamic and static transition temperatures
co\"{\i}ncide, $T_s=T_d$, and $x=T/T_{\sc eff}=0$~\cite{p=2}. 
The gap between the threshold and the equilibrium 
level collapses and the static energy density is the value reached
dynamically, ${\cal E}_{\sc eq}={\cal E}_\infty$. 
Still the nonequilibrium dynamics does not 
stop in finite times with respect to $N$.

For {\sc sk} the equilibrium calculation needs a {\it full replica 
symmetry breaking solution}~\cite{Mepavi}. This can be interpreted as being 
equivalent to having a sequence of correlation scales in the 
nonequilibrium dynamics. The static transition is continuous,
the static and dynamics $q_{\sc ea}$ are identical, the relation 
in (\ref{eq-dyn}) holds  
and all one-time quantities coincide with the equilibrium values. 
The static solution is also marginally stable.

From these three paragraphs one observes that the structure of the 
static and dynamic solution of purely potential mean-field problems 
is indeed very similar. One can propose a ``dictionary''
\begin{eqnarray*}
\mbox{{\sc rs}} &\Leftrightarrow& 
2\;\; \mbox{corr. scales, the aging one with} 
\;\; T_{\sc eff}=T^* \to \infty  
\nonumber\\
\mbox{1 step {\sc rsb}} 
&\Leftrightarrow& 
2\;\; \mbox{corr. scales, the aging one 
with}\;\; T_{\sc eff}=T^* < + \infty
\nonumber\\
\mbox{full {\sc rsb}} &\Leftrightarrow& \infty \;\; \mbox{corr. scales,  
the aging ones with} 
\;\; T_{\sc eff}(C) < +\infty 
\end{eqnarray*}
while remembering that the replica solution that describes the 
statistical properties of the region of phase space reached 
dynamically has to be determined using the condition of marginally
stability -- and not equilibrium. This connection between replicas  
and dynamics has been
extensively exploited to use the replica trick as a 
substitute for dynamics. In fact, partial information about dynamics
as the dynamic critical temperature, the value of $T_{\sc eff}$, etc.
is thus obtained. One has to keep in mind though that many aspects of 
the dynamics go beyond static calculations. Features 
like the existence of an effective temperature have been shown in 
models with trivial 
statics~\cite{kinetic2,kinetic3} or above the dynamic transition when 
non-potential or time-dependent forces are included~\cite{Ludo,Becuig}.

\subsection{Static {\sc tap} approach}
\label{subsec:static-tap}

In the introduction to this Section we announced that two 
choices facilitate the derivation of the {\sc tap} free-energy 
density. In order to use a good perturbative expansion we 
weight a part of the original Hamiltonian with 
a parameter $\alpha$. For the $p$ spin model we simply propose
\begin{equation}
H(\alpha,\vec s) = \alpha \sum_{i_1< \dots < i_p} 
J_{i_1,\dots, i_p} s_{i_1}\dots s_{i_p}
\; .
\end{equation}
In other models we weight the higher order interactions with $\alpha$
without modifying the quadratic terms. The idea is to expand in 
powers of $\alpha$ and set $\alpha=1$ at the end of the calculations
to recover the original model~\cite{Plefka,Georges-Yedidia}.
 
The second important step in the derivation is the choice of the
order parameters to use in the Legendre transform of the 
free-energy density. For spin models these are the averaged 
local magnetizations $m_i$ and a global spin constraint
$l$. The Legendre transform reads
\begin{equation}
-\beta\Gamma(\beta,\alpha;m_i,l;h_i,\lambda) =
\ln \mbox{Tr}_{S_i} e^{-\beta H(\alpha,\vec s)
- \sum_{i=1}^N h_i (s_i-m_i) 
-\frac{\lambda}{2} \sum_{i=1}^N (s_i^2 -l) }
\; .
\label{Legendre1}
\end{equation}
The trace represents a sum over all configurations of the spins,
{\it e.g.} $\mbox{Tr}_{S_i}=\prod_{i=1}^N  \sum_{s_i=\pm 1}$ for Ising 
and $\mbox{Tr}_{S_i}=\int_{-\infty}^\infty \prod_{i=1}^N  ds_i$ for
spherical variables. Requiring stationarity with respect to the 
Langrange multipliers $h_i$ and $\lambda$ one obtains
\begin{eqnarray}
m_i \equiv \langle s_i \rangle
\;\;\;\mbox{and} \;\;\;
l \equiv \frac1{N} \sum_{i=1}^N \langle s_i^2 \rangle
\; ,
\label{lagrange-impose}
\end{eqnarray}
 for all values of $\alpha$. 
The angular brackets indicate the statistical
average over the spins using the full Botzmann weight in 
(\ref{Legendre1}). Since the {\sc rhs}s depend on $h_i$ and 
$\lambda$ these equations can be inverted to yield
%\begin{eqnarray}
$h_i(\beta,\alpha; m_i,l)$
and 
$\lambda(\beta,\alpha;m_i,l)$.
%\; ,
%\end{eqnarray}
that one substitutes 
in (\ref{Legendre1}) to write $\Gamma$ as a function of 
$(\beta, \alpha;m_i,l)$.
%\begin{eqnarray} 
%&&-\beta\Gamma(\beta,\alpha;m_i,l) =
%\ln \mbox{Tr}_{S_i} e^{-\beta H(\alpha,\vec s)
%- \sum_{i=1}^N h_i (s_i-m_i)
%\right.
%\nonumber\\
%& &  
%\left.
%-\frac{1}{2} \lambda  \sum_{i=1}^N (s_i^2 -l) 
%}
%\right)
%\; .
%\label{Legendre2}
%\end{eqnarray}
Henceforth we drop the parameter dependence and 
simply note $\Gamma$.
The introduction of the parameter $\alpha$ allows us to
express $-\beta\Gamma
%(\beta,\alpha;m_i,l)
$ as a power series in 
$\alpha$:
\begin{equation}
-\beta\Gamma =
\sum_{n=0}^\infty 
\left.
\frac{\partial^n(-\beta\Gamma
%(\beta,\alpha;m_i,l)
)}{\partial\alpha^n}
\right|_{\alpha=0}
\, \frac{\alpha^n}{n!}
\;.
\end{equation}
For the $p$ spin model and  the like
this is equivalent to a high $T$
expansion. In other cases it is an expansion around an 
``equivalent'' free theory. Interestingly enough only a 
finite number of terms contribute to the series for fully connected 
models.
For finite dimensional cases a truncated series yields an approximation
around mean-field.

The zero-th order term depends on the nature of the spins considered:
\begin{eqnarray*}
-\beta\Gamma^{\sc ising}
%(\beta;\alpha=0;m_i,l) 
&=&
-\sum_{i=1}^N \left[ \frac{1+m_i}{2} \ln \left( \frac{1+m_i}{2} \right)
+\frac{1-m_i}{2} \ln \left( \frac{1-m_i}{2} \right)
\right]
\; ,
\\
-\beta\Gamma^{\sc sph}
%(\beta;\alpha=0;m_i,l) 
&=&
\frac{N}{2} \ln \left( 1- \frac{1}{N} \sum_{i=1}^N m_i^2\right)
\; .
\end{eqnarray*}
These are the entropies of $N$ independent Ising or 
spherical spins constrained to have local magnetizations $m_i$. 
The first order term is proportional to
\begin{eqnarray}
&&
\left.
\frac{\partial(-\beta\Gamma
%(\beta;\alpha;m_i,l)
)}{\partial \alpha} 
\right|_{\alpha=0}
=
\sum_{i_1 \leq \dots \leq i_p} J_{i_1\dots i_p} 
\langle s_{i_1} \dots s_{i_p} \rangle_{\alpha=0}
-\left. \sum_{i=1}^N \frac{\partial h_i
%(\beta,\alpha;m_i,l)
}
{\partial\alpha}
\right|_{\alpha=0} 
\langle s_i -m_i\rangle_{\alpha=0} 
\nonumber\\
& & 
- \left. \sum_{i=1}^N 
\frac{\partial \lambda
%(\beta,\alpha;m_i)
}{\partial\alpha}
\right|_{\alpha=0} \sum_{i=1}^N \langle s^2_i -l\rangle_{\alpha=0} 
%\nonumber
%\\
%&& 
=
 \sum_{i_1 \leq \dots \leq i_p} J_{i_1\dots i_p} 
m_{i_1} \dots m_{i_p}
\label{first}
\; .
\end{eqnarray}
In the first equality, the last two terms on the {\sc rhs}
vanish due to 
Eqs.~(\ref{lagrange-impose}). The average in the first term
factorizes since it has to be taken with the 
free-action ($\alpha=0$). 
The quadratic term in the expansion is proportional to 
\begin{eqnarray}
\left.
\frac{\partial^2(-\beta\Gamma
%(\beta;\alpha;m_i,l)
)}{\partial \alpha^2} 
\right|_{\alpha=0}
=
\left\langle \left( \sum_{i_1\dots i_p} 
Y_{i_1\dots i_p}\right)^2 \right\rangle^{c}_{\alpha=0}
\end{eqnarray}
with 
$Y_{i_1\dots i_p}= J_{i_1\dots i_p} s_{i_1} \dots s_{i_p} - 
(s_{i_1} - m_{i_1}) m_{i_2} \dots m_{i_p} - \dots 
- m_{i_1} \dots m_{i_{p-1}} (s_{i_p} - m_{i_p})$.  
This term has been computed using the following identities. First, the 
variation of $-\beta\Gamma
%(\beta;\alpha;m_i,l;h_i,\lambda)
$ in 
Eq.~(\ref{Legendre1}) with respect to $m_i$ and $l$ yields
%\begin{equation}
$h_i = 
\frac{\partial (-\beta\Gamma
%(\beta;\alpha;m_i,l;h_i,\lambda)
)}{\partial m_i}$
and
$\lambda = 
\frac{2}{N} 
\frac{\partial (-\beta\Gamma
%(\beta;\alpha;m_i,l;h_i,\lambda)
)}{\partial l}
$.
%\end{equation}
Taking the variation 
with respect to $\alpha$ and 
evaluating at $\alpha=0$ one has
\begin{eqnarray}
\left. 
\frac{\partial h_i}{\partial \alpha}\right|_{\alpha=0} = 
\left. 
\frac{\partial^2 (-\beta\Gamma
%(\beta,\alpha;m_i,l;h_i,\lambda)
)}{\partial m_i\partial\alpha}
\right|_{\alpha=0}
\; ,
%\\
\;\;\;\;\;\;\;\;\;
\left. 
\frac{\partial \lambda}{\partial\alpha} \right|_{\alpha=0} = 
\frac{2}{N}
\left. 
\frac{\partial^2 
(-\beta\Gamma
%(\beta,\alpha;m_i,l;h_i,\lambda)
)}{\partial l\partial\alpha}
\right|_{\alpha=0}
\; .
\end{eqnarray}
Now, the contributions $O(N)$ are proportional to 
$J_{i_1,\dots,i_P}^2$ and these can be estimated by replacing
its value by $p!/(2N^{p-1})$
since $J_{i_1,\dots,i_p}\sim \sqrt{p!/(2N^{p-1})}$. 
As done for the first order term we
factorize the thermal averages evaluated at $\alpha=0$ and 
\begin{eqnarray}
%\frac{\beta^2}{2} 
\left.
\frac{\partial(-\beta\Gamma^{\sc ising}
%(\beta;\alpha=0;m_i,l)
)}{\partial \alpha^2}
\right|_{\alpha=0} 
&=&
%\beta^2
2N \left( 1-q_{\sc ea}^p - p (q_{\sc ea}^{p-1}-q_{\sc ea}^p) \right) 
\\
%\frac{\beta^2}{2} 
\left.
\frac{\partial(-\beta\Gamma^{\sc sph}
%(\beta;\alpha=0;m_i,l)
)}{\partial \alpha^2}
\right|_{\alpha=0}  
&=&
\frac{\beta^2}{2} 
N \left( l^p-q_{\sc ea}^p - p (l q_{\sc ea}^{p-1}-q_{\sc ea}^p) \right) 
\end{eqnarray}
%*** CHECK LAST EQ ***
where we introduced the overlap or (static) Edwards-Anderson parameter, 
\begin{equation}
q_{\sc ea} \equiv \frac1{N} \sum_{i=1}^N m_i^2
\label{overlap}
\; .
\end{equation}

Higher order terms in the series expansion are sub-leading in 
$N$ and do not contribute in the thermodynamic limit. Thus, 
for these mean-field models the {\sc tap} free-energy density is
made of three terms: the zero-th order has an entropic origin, 
the first order is the interaction term
in the mean-field approximation which is exact for fully connected models, 
and  the second order is the reaction or Onsager term.

\subsection{The {\sc tap} equations}
\label{subsec:tap-eq}

The variation of the {\sc tap} free-energy density, 
$-\beta\Gamma
%(\beta,\alpha;m_i,l)
$, with respect
to $m_i$ (and $l$ for the spherical model) yields the 
{\sc tap} equations. For the spherical model one finds~\cite{Kupavi}
\begin{eqnarray} 
\frac{m_i}{(1-q_{\sc ea})} = \beta p \sum_{(i_2 \leq \dots \leq i_p) \neq i}
J_{i,i_2,\dots,i_p} m_{i_2} \dots m_{i_p} - \frac{\beta^2 p (p-1)}{2}  
q_{\sc ea}^{p-2} (1-q_{\sc ea}) m_i  
&&
\label{eq-tap1}
\\
\lambda = 
\frac{1}{1-q_{\sc ea}} + \frac{p\beta^2}{2} \left( 1-q_{\sc ea}^{p-1}\right)
\; .
\;\;\;\;\;\;\;\;\;\;\;\;\;\;\;\;\;\;\;
\;\;\;\;\;\;\;\;\;\;\;\;\;\;\;\;\;\;\;
\;\;\;\;\;\;\;\;\;\;\;\;
&&
\label{eq-tap2}
\end{eqnarray}
The study of these equations is simplified by defining the 
angular variables  $\sigma_i\equiv m_i/\sqrt{q_{\sc ea}}$ that verify 
the spherical constraint $\sum_i \sigma^2_i\equiv 1$. 
Multiplying Eqs.~(\ref{eq-tap1}) 
%in terms of $\vec \sigma$, multiplying 
%them 
by $\sigma_i$ and summing over $i$ we rewrite them 
in terms of the zero-temperature energy density 
${\cal E}\equiv -\frac{1}{N} \sum_{i_1< \dots <  i_p}
J_{i_1,\dots,i_p} \sigma_{i_1} \dots \sigma_{i_p}$:
\begin{eqnarray}
 {\cal E} \sigma_i &=& - \sum_{i_2< \dots <  i_p}
J_{i,i_2\dots,i_p} \sigma_{i_2} \dots \sigma_{i_p}
\; .
\label{eq-tap3}
\end{eqnarray}
The overlap $q_{\sc ea}$ is related to ${\cal E}$ by 
\begin{equation}
\beta p {\cal E} = - \frac{1}{q_{\sc ea}^{(p-2)/2} (1-q_{\sc ea})} \left[ 
1+ \beta^2 \frac{p(p-1)}{2} (1-q_{\sc ea})^2 q_{\sc ea}^{p-2}
\right]
\; .
\label{eq-tap4}
\end{equation} 
Equation~(\ref{eq-tap3}) does not depend on $T$ while 
(\ref{eq-tap4}) does. 
The multiplicity of solutions at a given ${\cal E}$ is entirely
determined by Eq.~(\ref{eq-tap3}), see Section~\ref{subsec:complexity}.
The existence or not of these solutions 
has to be tested with Eq.~(\ref{eq-tap4}). 
The remaining Eq.~(\ref{eq-tap2}) fixes 
$\lambda$.

Conveniently rewritten Eq.~(\ref{eq-tap4}) is quadratic 
and yields
\begin{equation}
q_{\sc ea}^{(p-2)/2} (1-q_{\sc ea}) = 
\frac{k_B T}{p-1} 
\left[ -{\cal E} \pm \sqrt{{\cal E}^2 - {\cal E}^2_{\sc th}} \right]
\;\; \mbox{with} \;\;
{\cal E}_{\sc th} \equiv - \sqrt{\frac{2(p-1)}{p}}
\; .
\label{eq-q}
\end{equation}
This equation admits a real solution only if 
${\cal E} < {\cal E}_{\sc th}$, the {\it threshold} energy density at 
$T=0$. The minus (plus) sign correspond to a minimum (maximum)
of the free-energy. The physical $q_{\sc ea}$ is then associate to the 
minus sign. With this choice, $q_{\sc ea}$ takes its maximum value on 
the theshold and then monotonically decreases until reaching
its minimum value on the equilibrium level, 
$q({\cal E}_{\sc th}) \geq q_{\sc ea}({\cal E}) \geq 
q_{\sc ea}({\cal E}_{\sc eq})$. 

For fixed $\beta$ and ${\cal E}$ 
the {\sc lhs} in Eq.~(\ref{eq-q}) has a bell shape form.
When $T\to 0$,
$q_{\sc ea}\to 1$ and $\beta(1-q_{\sc ea})$ is finite. 
The maximum is located at $q_{\sc ea}^* = (p-2)/p$ and it 
has a height $2/p \; ((p-2)/p)^{(p-2)/2}$. 
Each solution at $T=0$ that corresponds
to a given ${\cal E}$  can be followed to finite $T$
until it disappears at $T^*({\cal E})$ when the {\sc lhs} reaches 
the maximum. 
One can check that the {\sc tap} solutions do not merge nor
bifurcate as a function of $T$. Then at any temperature $T$
we label the {\sc tap} solutions with their associated
zero temperature energy density ${\cal E}$.

Plugging ${\cal E}_{\sc th}$ in Eq.~(\ref{eq-tap4}) we 
find that the overlap $q_{\sc ea}$ is given by the equation
\begin{equation}
\frac{p (p-1)}{2} \, 
q^{p-2}_{\sc th} (1-q_{\sc th})^2 = (k_B T)^2
\; .  
\end{equation}
The threshold energy density ${\cal E}_{\sc th}$ and the
overlap $q_{\sc th}$ coincide with the asymptotic value of the dynamic 
energy density  and the Edwards-Anderson parameter at $T=0$, 
respectively. Thus, the nonequilibrium dynamics approaches asymptotically
the threshold level.

\subsection{Stability of, and barriers between, the {\sc tap} solutions}
\label{subsec:stability-tap}

The spectrum of the free-energy Hessian around a stationary point 
of the {\sc tap} free-energy is a shifted semi-circle.
The lowest eigenvalue $\lambda_{\sc min}$ is greater than zero for 
sub-threshold free-energy densities (or zero-temperature 
energy density). This means that for ${\cal E} < {\cal E}_{\sc th}$
typical stationary states are minima. Instead, near the threshold
$\lambda_{\sc min}$ drops to zero as~\cite{Cuku1} 
\begin{equation}
\lambda_{\sc min}(f,T) \sim \frac{p}{q_{\sc th}} \; (f_{\sc th}-f) 
\end{equation}
and the stationary states on the threshold are typically
marginal in the sense that they have many flat directions.

Even if the connection between dynamics and the static {\sc tap} free-energy
landscape is not obvious {\it a priori}, it has been proposed
and used in several works~\cite{Kithwo,Cuku1}. The formalism in
Section~\ref{subsec:dyn-quantum-tap} and \cite{Bi} 
settles it on a firm ground.
If one imagines that the dynamics can be viewed as the displacement 
of a representative point in the free-energy landscape, not only the
organization minima, saddles and maxima has to be known but also
how do the barriers between these stationary points scale with $N$.
Few results about barriers exist and, in short, they are the 
following. 
The barriers between threshold states vanish when 
$N\to\infty$~\cite{Kula,Cagipa2} and there is no sharp separation between 
them. This has been proven by analyzing the constrained complexity, 
related to the number of threshold states that have an overlap $\tilde q$
with a chosen one ($\tilde q\equiv N^{-1}
\sum_{i=1}^N \sigma_i \tau_i$ where $\vec \sigma$ and $\vec \tau$ are the 
two configurations). This complexity decreases with increasing
values of $\tilde q$ and it vanishes as a power law 
at $\tilde q = q_{\sc th}$.
This means that one can find threshold states
that are as similar as required to the chosen one.
The threshold level is then a series of flat connected 
channels. The non-equilibrium dynamics starting from random (typical)
initial conditions approaches this level asymptotically
and it never stops since the system drifts in a slower and slower manner 
as time evolves following these flat directions~\cite{Cuku1}.
On the other hand, 
the barriers between equilibrium and metastable states have been 
estimated to be $O(N)$~\cite{Kupavi,Cagipa2}.
One can guess that the barriers between sub-threshold {\sc tap} states 
also scale with $N$. For finite $N$ the dynamics should penetrate below
the threshold and proceed by thermal activation. An Arrhenius-like
time will then be needed to descend from one level to the next.
Naturally one should see another kind of separation of time-scales 
develop. Simulations of the $p$ spin model for finite $N$ confirm 
the existence of metastable states below the threshold. This is most
clearly seen following the evolution of the energy density 
of the model weakly perturbed with a non-potential force. One sees
periods of trapping in which the energy-density is fixed to a given 
value and periods in which the system escapes the confining state
and surfs above the threshold until being trapped in a new 
state~\cite{Cukulepe,Becuig,Ludo} (see also~\cite{Crisanti}
a numerical study of the finite $N$ $p$-spin model).

\subsection{Index dependent complexity}
\label{subsec:complexity}

At low $T$'s the number of 
stationary points 
is exponential in $N$. This suggests to 
define the {\it complexity} $\Sigma_J({\cal E})$:
\begin{equation}
\Sigma_J({\cal E}) \equiv \lim_{N\to\infty} N^{-1} 
\ln {\cal N}_J({\cal E}) 
\; ,
\end{equation}
where ${\cal N}_J({\cal E})$ is the number of solutions with energy 
density ${\cal E}$. Actually, one can refine the study 
by grouping the stationary points of the {\sc tap} 
free-energy density into classes according to the number of
unstable directions. Thus, minima are saddles of index $0$, 
saddles with a single unstable direction have index $1$ and so
on and so forth. The complexity 
of each kind of saddle is
${\Sigma_J}_k({\cal E}) \equiv \lim_{N\to\infty} N^{-1} 
\ln {{\cal N}_J}_k({\cal E})$, where $k$ denotes the index of the 
considered saddles. For the 
spherical $p$ spin model,  
their average over disorder,
$\Sigma_k\equiv [{\Sigma_J}_k]$, 
%is given by~\cite{Cagipa}:
%\begin{eqnarray*}
%\Sigma_k({\cal E}) = \frac{(k+2) q_+^2}{2p(p-1)} 
%-\frac{k q_-^2}{2p(p-1)} + \frac{k+2}{2} 
%\ln\left(-p {\cal E}+q_+\right) - \frac{k}{2} 
%\ln\left(-p {\cal E}+q_-\right)  +A({\cal E})
%\label{complexity_k}
%\end{eqnarray*}
%with $A({\cal E}) = 1/2 -1/2 \ln(p/2) -{\cal E}^2$ and 
%$
%q_{\pm} = \frac{p}{2} \left( {\cal E} \pm 
%\sqrt{{\cal E}^2 -{\cal E}_{\sc th}^2}\right)
%$.
%The complexities 
are ordered in such a way that~\cite{Cagipa} 
\begin{eqnarray}
\begin{array}{rcl}
\Sigma_0({\cal E}) > \Sigma_1({\cal E}) > \Sigma_2({\cal E})  > \dots
&
\;\;\; \mbox{for all} \;\;\; & {\cal E} < {\cal E}_{\sc th} \;\;\;
\nonumber\\
\Sigma_0({\cal E}) = \Sigma_1({\cal E}) = \Sigma_2({\cal E})  = \dots
& \;\;\; \mbox{if} \;\;\; &  {\cal E} = {\cal E}_{\sc th} \;\;\;
\end{array}
\end{eqnarray}
%and they merge at ${\cal E}_{\sc th}$. 
Thus, when 
${\cal E} < {\cal E}_{\sc th}$ minima are exponentially
dominant in number with respect to all other saddle points.
Moreover, one proves that the complexities vanish at 
a $k$-dependent value of ${\cal E}$. The complexity of minima
is the last one to disappear at 
${\cal E}_{\sc eq}$.

%\begin{figure}[h]
%\centerline{
%\psfig{file=complexity.ps,width=10cm}
%}
%\caption{The complexities of minima and saddles with index $1,2,3$
%for the $p$ spin spherical model with $p=3$~\cite{Cagipa1}.}
%\label{complexity}
%\end{figure}

\subsection{Weighted sums over {\sc tap} solutions}
\label{subsec:weigthed}

The thermodynamics at different $T$ is determined by the
partition function, ${\cal Z}_J$.  
De Dominicis and Young~\cite{Dedoyo} showed that the  
equilibrium results obtained with the replica trick or 
the cavity method are recovered from the {\sc tap} 
approach when one writes ${\cal Z}_J$
as a weighted sum over the {\sc tap} stationary states.

Indeed, if one divides phase space
into pockets of configurations that surround the stationary points
of the {\sc tap} free-energy one can carry out the statistical sum 
on each sector of phase space:
\begin{equation}
Z_J= \sum_{\vec s} \exp(-\beta H_J(\vec s)) = 
\sum_{\alpha} \sum_{{\vec s}^\alpha}\exp(-\beta H_J(\vec s^\alpha))
\end{equation}
where the index $\alpha$ labels the {\sc tap} solutions and 
$\sum_{{\vec s}^\alpha}$ represents a restricted sum over the 
configurations that belong to the pocket associated to the 
solution $\alpha$. In order to make this separation precise one 
needs to assume that the barriers separating the pockets 
diverge in such a way to avoid ambiguities when associating a
configuration to a {\sc tap} solution. 
The restricted sum is related to the free-energy of the {\sc tap} solution
%\begin{equation}
$\exp(-\beta F_J({\vec m}^\alpha)) = 
\sum_{{\vec s}^\alpha}\exp(-\beta H_J(\vec s^\alpha))$
%\; ,
%\end{equation}
and the partition function becomes
%\begin{equation}
$Z_J= \sum_\alpha \exp[-\beta F_J({\vec m}^\alpha)]$.
%\; .
%\label{partition-tap}
%\end{equation}
Thus, any statistical average can be computed using the 
weight 
%\begin{equation}
$\exp[-\beta F_J({\vec m}^\alpha)]/Z_J$.
%\end{equation}
The sum 
over solutions
can be replaced by an integral over free-energies if one introduces
their degeneracy ${\cal N}_J(\beta,f)$ and their associated complexity, 
$\Sigma_J(\beta,f)=N^{-1} \ln{\cal N}_J(\beta,f) $, 
with $f\equiv F/N$, 
\begin{equation}
Z_J
%= \int dF_J \; {\cal N}(F_J) \exp(-\beta F_J({\vec m}^\alpha))
=
\int df \; \exp\left[ -N (\beta f - \Sigma_J(\beta,f)) \right]
\; .
\label{partition-tap2}
\end{equation}
The complexity is a self-averaging quantity, we then 
write $\Sigma_J(\beta,f)=\Sigma(\beta,f)$.
When $N\to\infty$ the integral is dominated by the solutions 
that minimize the generalized free-energy, $\beta f - \Sigma(\beta,f)$. 
This is achieved either by 
$f  = f_{\sc min}$ with $\Sigma(\beta,f_{\sc min})=0$ 
or by states that do not have the minimum
free-energy if their complexity is finite, in which case~\cite{Remi}
\begin{equation}
\beta = \left. \frac{\partial \Sigma(\beta,f)}{\partial f} \right|_{f^*}
\;\;\; \Rightarrow \;\;\; Z_J \sim \exp[-N (\beta f^* - \Sigma(\beta, f^*))] 
\; .
\label{eq:f-tap}
\end{equation}
Minima of the generalized free-energy density with higher ($f>f^*$)
or lower ($f<f^*$)
free-energy density are metastable.

From the analysis of the partition function one distinguishes 
three temperature regimes~\cite{Kithwo,Remi}:

\vspace{0.2cm}
%\begin{equation}
\noindent{\it Above the dynamic transition} 
$T> T_d= \sqrt{\frac{p (p-2)^{p-1}}{2 (p-1)^{p-1}}}.$
%\end{equation}
\vspace{0.2cm}

The paramagnetic or liquid solutions $m_i=0$, $\forall i$, 
dominate the partition sum, ${\cal Z}\sim \exp(-\beta f^{\sc pm})$. 
States with $m_i\neq 0$
exist in this range of temperatures but they do not 
dominate the sum.

\vspace{0.2cm}
\noindent{\it Between the static and dynamic transitions}
$T_s < T <  T_d.$
\vspace{0.2cm}

The paramagnetic state is fractured into an exponential in $N$
number of  non-trivial {\sc tap} solutions with $m_i\neq 0$. 
Their free-energy density and the partition sum are 
given by Eq.~(\ref{eq:f-tap}). Each of these states has a rather 
high free-energy that is counterbalanced by the entropic contribution.
Besides the states that dominate the partition sum, a very large number of 
other metastable states, with higher and lower free-energy density, 
also exist but are thermodynamically irrelevant.
It turns out that $f^*$ coincides with the extrapolation of $f^{\sc pm}$ from 
$T> T_d$ to this temperature region even if the paramagnetic
solution does not exist. Note that in this temperature regime the 
standard replica calculation fails since it tells us that the 
equilibrium state is the simple paramagnet. Refinements on this 
method are able to extract more precise information about the
non-trivial states contributing to equilibrium~\cite{Cagipa2}. 

\vspace{0.2cm}
\noindent{\it Below the static transition}
$T< T_s$.
\vspace{0.2cm}

An infinite though not exponential in $N$ number of states 
with the minimum free-energy density dominate the sum. 
The complexity $\Sigma(\beta,f)$ vanishes and this is 
associated to an entropy crisis. This is similar to the 
argument used by Kauzmann to justify the existence of a 
dynamic crossover at $T_g>T_s$ since, otherwise, the projection 
of the difference between the liquid and the crystal 
entropy would vanish at $T_K$. In this sense,
this model realizes the Kauzmann paradox at $T_s$.

\subsection{Accessing metastable states with replicas}

The replica trick can be improved to access the
non-trivial metastable states existent between $T_s$ and 
$T_d$~\cite{Kith,Remi,Frpa} using a ``pinning-field'' or a 
``cloning method''. Let us sketch how the latter works.
Consider
$x$ copies or clones of the systems coupled by an attractive,
infinitesimal (but extensive) interaction. 
The free energy for the system of $x$ clones reads:
\begin{equation}\label{reltaprep0}
f_{Jx}=\lim_{N\rightarrow \infty }
\frac{-1}{\beta N}\ln {\cal Z}_{Jx}=
\lim_{N\rightarrow \infty }\frac{-1}{\beta N}
\ln \int df \exp[-N (\beta xf-\Sigma(\beta,f))]
\end{equation}
using the formalism described in the previous Subsection.
On the other hand, the free-energy density of the $x$ clones
can also be computed
with the replica approach:
\begin{equation}\label{reltaprep1}
f_{Jx}=
%\lim_{N\rightarrow \infty }\frac{-1}{\beta N}\ln {\cal Z}_{x}=
\lim_{N\rightarrow \infty }\frac{-1}{\beta N}[\ln {\cal Z}_{Jx}]=\
\lim_{N\rightarrow \infty,n\rightarrow 0 }\frac{-1}{\beta N n}
\ln [{\cal Z}_{Jx}^{n}]
\; 
\end{equation} 
where we used the fact that the free-energy density is self-averaging.
Since the attractive coupling between the $x$ clones is infinitesimal,
the computation of the {\sc rhs} of Eq.~(\ref{reltaprep1}) reduces
to the calculation of $\lim_{n'\rightarrow 0}(x/n')
\ln [{\cal Z_{J}}^{n'}]$,
where the replica symmetry between the $n$ groups of $x$-replicas 
($n'=nx$) is {\it explicitly} broken. When the system is in the 
replica symmetric phase ($T_{s}<T$),
the problem becomes one where one has to study one-step replica 
symmetry breaking solutions non-optimized with respect to $x$:  
\begin{equation}
\label{reltaprep}
-\lim_{N\rightarrow \infty}\frac{1}{\beta N }
\ln \int df \exp[-N (\beta xf-\Sigma(\beta,f))]
={x}\; {\mbox{Extr}}_{q_{\sc ea}}f_{\sc rep}(q_{\sc ea};\beta,x)
\; ,
\end{equation}
where $f_{\sc rep}$ is the free-energy-density computed by using replicas, 
 $q_{\sc ea}$ is the Edwards-Anderson parameter and 
$x$ is the breakpoint or the size of the blocks in the replica matrix. 
(For simplicity we consider that the inter-state overlap 
$q_{0}$ equals zero~\cite{Mepavi}.) 
Since the integral on the 
{\sc lhs} of Eq.~(\ref{reltaprep}) is dominated by a saddle point 
contribution, one finds
 that, for a given $T$, fixing the value 
 of $x$ one selects the states with a given free-energy-density 
$F$.
The 
relationship between $f$ and $x$ reads
\begin{equation}\label{rela:xf}
\beta^*\equiv \beta x=\frac{\partial\Sigma(\beta,f) }{\partial f } 
\; .
\end{equation}  
Note that within this framework one does not 
optimize with respect to $x$.
Instead, $x$ is a free parameter and, by changing the value of $x$, 
one selects different groups of metastable states.
Inversely, choosing a value of the free-energy density 
one determines an ``effective temperature'' $T^*\equiv T/x$. This value
coincides, indeed, with the dynamic result for $T_{\sc eff}$ in the 
aging scale when $f=f_{\sc th}$. Otherwise, it gives a
free-energy level dependent effective temperature. If one pursues the 
empirical relation between this parameter and the dynamic effective
temperature this result means that when the system evolves in sufficiently
long time-scales as to penetrate below the 
threshold it takes different values of $T_{\sc eff}$ depending on 
the deepness  it reached.

M\'ezard and Parisi generalised this approach to search for a Kauzmann 
critical temperature ($T_K=T_s$) and characterise the thermodynamic
properties of the glassy phase below $T_s$ in finite $d$ interacting
particle systems. The mean-field nature of the approach has been 
stressed by Thalmann~\cite{Th} who showed that there is no lower
critical dimension.

\subsection{Dynamics and quantum systems}
\label{subsec:dyn-quantum-tap}

The derivation of the {\sc tap} equations presented 
in Section~\ref{subsec:static-tap} can be generalized to treat other
problems as the real-time dynamics~\cite{Biroli1} of the same models
or their quantum extensions~\cite{Biroli2}. Again, the procedure can 
be made easy if one correctly chooses
the perturbative expansion and the order parameters. 
To study the real-time dynamics of the classical problem 
one is forced to Legendre transform with respect to the 
time-dependent local 
magnetization, the two-time  
correlation and the response~\cite{Biroli1}. 
The presentation is further simplified when 
one uses the {\sc susy} notation introduced in 
Section~\ref{subsubsec:supersymmetry}
that renders the dynamic formalism very close to the static one.
For quantum problems in equilibrium one uses the Matsubara representation 
of the partition function and then Legendre transforms with 
respect to the imaginary-time correlation as well as the local 
magnetizations~\cite{Biroli2}. In order to derive {\sc tap} equations
for a quantum problem in real-time one should use the 
Schwinger-Keldysh
formalism.

The derivation and study of dynamic {\sc tap} equations justifies the 
interpretation of the asymptotic non-equilibrium dynamics in terms 
of the local properties of the {\sc tap} free-energy density landscape.
Biroli~\cite{Biroli1} showed that the dynamics in the {\sc tap} 
free-energy landscape
is in general non-Markovian due to the presence of memory terms
in the dynamic {\sc tap} equations. In the very long-time limit and for 
random initial conditions the contribution of these terms 
vanishes and one proves that the dynamics is a relaxation following 
flat directions in the {\sc tap} free-energy landscape 
(the {\it threshold}) as proposed in \cite{Cuku1,Kula}. 

\section{Conclusions}
\label{conclusions}

We discussed the behaviour of a family of 
disordered models that yield a mean-field 
description of the glass transition and 
dynamics of super-cooled liquids and glasses.
The relevance of these models to describe structural
glasses was signaled and explained by Kirkpatrick,
Thirumalai and Wolynes in the 80s~\cite{Kithwo}. Their nonequilibrium 
dynamics and hence the connection with other systems
far from equilibrium started to develop more recently~\cite{Cuku1}.

In short, their behaviour is the following.
The dynamic transition arises when the partition 
function starts being dominated by an exponentially
large number of metastable states yielding a finite complexity. 
The static transition
instead is due to an entropy crisis, {\it i.e.} it occurs 
when the complexity vanishes and the number of states is no 
longer exponential in $N$, just as in the Adams-Gibbs-di Marzio
scenario~\cite{Adams}. These transitions mimic,
in a mean-field way, the crossover to the glassy phase 
at $T_g$ and the 
putative static transition at $T_o$ (or $T_K$) of fragile 
glasses, see Fig.~\ref{viscosity}-left~\cite{Kithwo}. 

The equilibrium dynamics close
and above $T_d$ coincides with the one obtained with the 
mode-coupling approach~\cite{Kithwo}. 
It describes the relaxation of super-cooled
liquids and it contains its most distinctive feature of having a 
two step decorrelation. The first step is 
ascribed to the motion of particles within the cages made
by their neighbours while the second one is 
the structural relaxation related to the destruction
of the cages.

Below $T_d$ the equilibration time diverges with the size of the system and 
the models do not equilibrate any longer with their environments
(unless one considers times that grow with the size of the 
system)~\cite{Cuku1}. 
This is very similar to 
the situation encountered in real systems below $T_g$. The experimental 
time-window is restricted and one is not able to equilibrate the 
samples any longer below $T_g$. Aging 
effects as shown in Fig.~\ref{agingandrheology} are captured. The correlations 
still decay to zero but they do in a waiting-time dependent manner. Their
 decay also occurs in two steps separated by a temperature-dependent 
plateau at a value related to the size of the cages. One can interpret
their stiffness as increasing with the age of the system given that 
the beginning of the 
structural relaxation is delayed and slowed down for longer waiting-times.
 
The nonequilibrium dynamics below $T_d$ approaches a 
threshold level of flat directions in phase space and it never 
goes below this level in finite times with respect to the 
size of the system~\cite{Cuku1}. 
The aging dynamics corresponds to the slow drift 
of the point representing the system in the slightly tilted set of 
channels that form the threshold. The motion that is transverse to the channels
is related to thermal fluctuations and the first stationary step of the 
relaxation towards $q_{\sc ea}$, that charcaterises then 
the transverse ``size'' of the channel. 
The longitudinal motion along the channels is related to the 
structural relaxation. The tilt is proportional to the magnitude of the 
time-derivatives and these become less and less important as time passes.
In more generalitiy one interprets the long but finite time 
nonequilibrium dynamics following saddles that are
the borders between basins of attraction of more stable states in phase 
space~\cite{Kula}.

For times that scale with the size of the system, $N$, the sharp 
dynamic transition 
is avoided, the system penetrates below the threshold via activation 
and it approaches equilibrium in much longer time-scales. 
Metastable states below the threshold are typically 
minima~\cite{Kupavi,Cagipa} (the fact that they are local minima can 
be checked studying the dynamics 
with initial conditions set to be in one 
of them~\cite{Babume,Frpa,Cukulepe}). 
This structure allows one to describe the cooling rate effects 
described in Fig.~\ref{viscosity}-right. For large but finite 
$N$ and sufficiently slow cooling rate, the system 
penetrates below the threshold via activation when this 
is facilitated by $T$, 
{\it i.e.} when passing near $T_d$. To which level 
it manages to arrive (roughly speaking to which of the curves 
in the figure) depends on how long it stays close to $T_d$. 
The slower the cooling rate the lower level the system 
reaches with the ideal ``equilibrium'' glass corresponding 
to an infinitely slow cooling~\cite{eqglass}.

The region of phase space reached asymptotically in the thermodynamic 
limit is the threshold 
of flat directions. The replica analysis of the partition function 
gives an alternative way of determining its statistical 
porperties. Indeed, by evaluating the partition function on a 
marginally stable saddle-point in replica space
one selects the threshold ``states''.
Dynamic information such as the value of $q_{\sc ea}$ is thus 
obtained with a pseudo-static calculation. Other facts as, 
for instance, the scaling of
the correlation are not accessible in this way. 
 
One of the hallmarks of the glassy nonequilibrium dynamics 
is the modification of the relation between correlations
and responses, namely, the fluctuation-dissipation theorem.
In mean-field models for structural glasses one finds that
the integrated linear response is in linear relation with the 
associated correlation with a proportionality constant that
takes the equilibrium value $1/(k_B T)$ when the correlation
is above the plateau and it takes a different value
$1/(k_B T^*)$ when it goes below the plateau~\cite{Cuku1}.
This behaviour has been found in a number of finite dimensional
glassy models numerically~\cite{numerics,kinetic2,kinetic3,lattice-models}.
%A more complicated dependence $\chi(C)$ is found in mean-field 
%spin-glasses~\cite{Cuku2,Mefr}. 

The behavior just described corresponds to a family of mean-field 
disordered models to which the $p$ spin models with $p\geq 3$ and the Potts
glass belong. Other two families exist and they are related to 
ferromagnetic domain growth and spin-glasses. Two representative models 
are the spherical $p$ spin model with $p=2$ and the {\sc sk} model,
respectively. They are characterised by different scalings of the 
correlations in the aging regime and by different forms of the 
modification of {\sc fdt}. The classification in families according 
to the nonequilibrium behaviour has a static counterpart given
by the structure of replica symmetry breaking in the low-$T$ phase,
see Section~\ref{replicas}.

The modification of {\sc fdt} allows one to define an
observable and correlation-scale dependent  
effective temperature~\cite{Cukupe}. 
Fast observables like the kinetic energy are equilibrated with the 
environment and the effective temperature equals the thermal bath temperature
for them. Other observables though show different values of the effective
temperature depending on the time-scales on which one investigates them.
The effective temperature has a thermodynamical meaning 
even if defined out of equilibrium.
In particular, it can be directly read with a thermometer coupled 
to the desired observable and a zero-th law holds for interacting observables
that evolve in the same time-scale.
As one should have expected the effective temperature 
shares some of the qualitative features of the phenomenological
fictive temperatures~\cite{fictive}. For instance, 
a system that is quenched from high 
temperatures has effective temperatures that take higher values 
than the temperature of the bath, etc. At the mean-field level, when 
$N\to\infty$, it is history independent but one expects it to depend 
on the preparation of the sample for finite size and 
finite dimensional systems. (This is in close relation to the  
discussion above on cooling rate effects.) There is still no 
precise determination
of which are the necessary conditions a nonequilibrium system has to fulfill
to ensure the existence of well-behaved effective temperatures. 
A clear condition are the need to 
reach a dynamic regime in which the dynamics is slow and heat 
exchanges are weak.

Once the effective temperature has been identified one interprets the 
behaviour in the low $T$ phase as follows: the system adjusts 
to a situation in which each observable sees two baths, one is the 
white external one and the one characterising the fast 
motion of the particles, the other is coloured and at 
a different temperature $T^*$. The latter is  
generated by the interactions. In more complex systems -- as 
mean-field spin-glasses --
the asymptotic regime might be multithermalised with several
time-scales each with its own value of the effective temperature.
These results, first derived explicitly
for $p$ spin fully-connected models~\cite{Cukupe} actually hold for any 
resummation of the pertubative approach that keeps an infinite
subset of diagrams (the {\sc mca} being one such 
example). The structure of time-scales and values of the 
effective temperature is related to the breaking of 
supersymmetry down to 
a residual group~\cite{Cuku5}.

\begin{figure}
\centerline{
\psfig{file=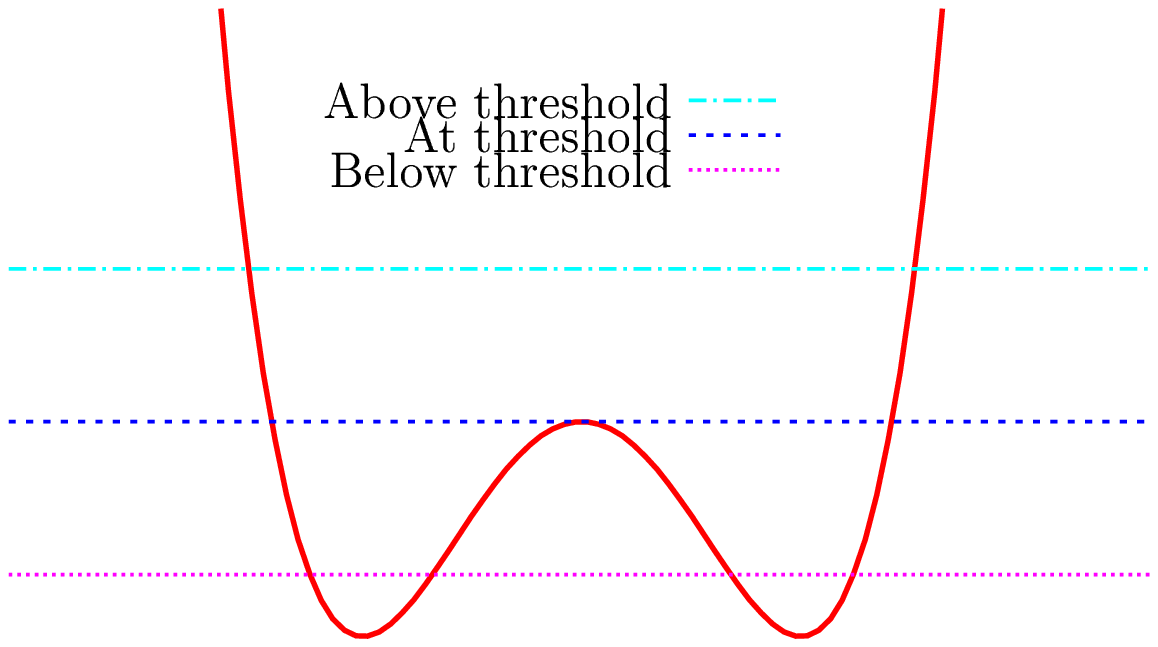,width=6cm}
\hspace{-0.25cm}
\psfig{file=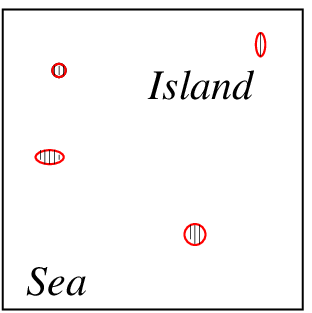,width=2.7cm}
\hspace{0.25cm}
\psfig{file=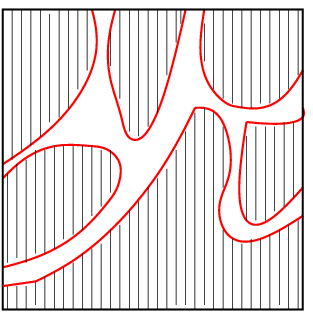,width=2.7cm}
\hspace{0.25cm}
\psfig{file=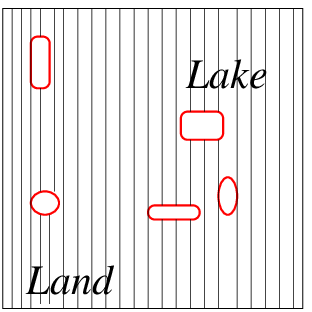,width=2.7cm}
}
\caption{Left: a $1d$ simplified sketch of the free-energy density.
Three top views of the free-energy landspcape:
above, at and below the threshold.}
\label{sketch-free-energy}
\end{figure}

The structure of the free-energy landscape
can be computed exactly for mean-field models in general, 
and for the spherical  $p$ 
spin model in particular~\cite{Kupavi,Cuku1,Cagipa,Cagipa2,Cagipa3}. 
We expect its main features to be reproduced -- at least in a 
smoothen way -- in real glassy systems. The free-energy landscape
at fixed and low $T$ has a structure as the one roughly sketched in 
Fig.~\ref{sketch-free-energy}. A pictorial image of the aging process
can be quite helpful to understand it. Imagine that one fills
phase space 
with water whose level reaches a free-energy density value, say, 
$f$. 
At high levels of the water, {\it i.e.} for high 
free-energy densities, the landscape
has only some few isolated stationary states. Looking at the landscape 
from above one only sees some maxima that 
are represented as islands in the second panel in the figure. 
Lowering the water level the islands grow in size and some of them 
merge: land bridges develop.
Lowering still the water level,  
it eventually reaches a threshold, that corresponds to  $f=f_{\sc th}$,
where land percolates. One is left with a labyrinthic path of 
water as 
drawn schematically in the third panel that represents a top view of the 
landscape. This level is ``marginal'' since the bottom of the 
 water channels is almost completely flat.
Draining water from the system the 
``connectivity'' of paths is reduced until the water level goes 
below the threshold, $f< f_{\sc th}$,  where minima dominate.
In the fourth panel  we represent them as lakes immersed in 
land. Lowering the water level one sees the sizes of the 
lakes diminish and some of them dry. 
These minima exist until the lowest level, $f=f_{\sc eq}$. 
A ``gap'' in free-energy density separates the threshold 
and the  equilibrium levels.

This picture allows us to give a natural interpretation 
of the nonequilibrium
dynamics following a quench. Initially, the system is in a 
configuration typical of high-$T$, thus, its initial ``free-energy
density'' is very high. This corresponds to a high level of
water that fills the landscape. As time passes, water abandons the 
landscape in such a way that the quantity of water progressively 
diminishes lowering its level. The system's configuration can be 
associated to a ship and its evolution to the 
displacements of the ship sailing on the water.
Initially, the water level is very high
and the ship can move very rapidly far away from 
its initial position. It only sees some very few 
isolated islands that it simply avoids along its motion
and the dynamics is very fast.
As time passes the water level goes down. Roughly speaking 
we can associate the speed of drainage with the magnitude of the 
rate of change of the energy-density.
When it 
approaches the threshold the available path becomes a 
series of rivers forming a very intricate network. The 
ship can still follow this network without remaining trapped
in any confining region. Its motion, however, gets slower and 
slower. In finite times with respect to $N$ the water level 
does not go below the threshold. But for longer 
times that scale with $N$ it does. When such long times are 
attained the ship remains trapped in lakes. For still longer times 
the higher lakes dry and, if the ship got trapped in one
of them it must be transported through the land to 
reach other lakes at lower levels. This action represents an
activated process.
Part of this image was introduced by Sibani and Hoffmann 
phenomenologically~\cite{Siho}. The $p$ spin-models and the like 
realise it explicitly. All quantitative features of the 
landscape here described with words have been, or in principle can be,
calculated analytically. The use of this image has been ultimately justified
by the dynamic {\sc tap} approach of Biroli~\cite{Biroli1}.

The value taken by the effective temperature is in direct 
relationship with the structure of the free-energy landscape.
Indeed, again for $p$-spin model and the like, 
it has been shown analytically that the asymptotic value $T^*$ reached 
for long but finite times with respect to $N$ is given by 
$\beta^* =\partial \Sigma(\beta,f)/\partial f|_{f_{\sc th}}$,
with $\Sigma$ the complexity~\cite{Remi}. 
For even longer times such that the 
system penetrates below the threshold one expects the effective 
temperature to take different values related to the complexity 
at lower free-energy density levels. The Edwards-Anderson 
parameter, 
$q_{\sc ea}$, also changes since $q_{\sc ea}(f)$. In the longest time-scale
such that equilibrium is reached and $q_{\sc ea}$ equals the equilibrium 
value also obtained with a replica calculation using the standard
maximization prescription to determine the breaking point parameter $x$.
This result is intimately related to 
Edwards' flat measure for granular matter~\cite{Ed,jorge} and also to 
the more recent use of a flat measure over inherent 
structures~\cite{Stillinger} 
to describe the nonequilibrium dynamics of 
glasses~\cite{Sciortino,Crisanti,Cavagna}. 
Note that the these, being defined using the {\it potential energy-density}
landscape, are valid only at zero temperature (see, 
{\it e.g.}~\cite{Bimo}). However, extensive numerical checks recently 
performed suggest that the approach, even if not obviously correct at finite 
$T$, yields a very good approximation~\cite{Sciortino}.

Within this picture two distinct regimes would appear in the
low-$T$ isothermal dynamics of real systems: a mean-field-like one when the
system approaches a pseudo-threshold of flat directions in phase space
and a slower activated regime  in which  the system jumps
over barriers to relax its excess energy density and very slowly progress
towards equilibrium. How and if the aging properties in the first 
and second regime resemble is a very interesting open problem. 

The existence of a threshold  plays a fundamental 
role in explaining several  features of many experimental 
observations in such diverse systems as driven 
granular matter, the rheological properties of complex 
liquids and glasses, etc. Just to cite two examples, 
trapping and Reynolds dilatancy 
effects in granular matter~\cite{jorge,Becuig} 
as well as the existence of a 
static yield stress and thixotropic behaviour 
in some rheological experiments~\cite{Ludo} 
can be interpreted in terms of
threshold and subthreshold states.
These features support the claim that this free-energy structure
exists in real physical systems.  Moreover, maybe not surprisingly,
this structure also appears in optimisation problems such as 
{\sc xor-sat} and {\sc k-sat} that can be mapped to dilute 
$p$-spin models at zero temperature. In this context
the control parameter is the number of requirements over the 
number variables, $\alpha$, and the static transition,
$\alpha_s$, is related to the sat-unsat transition 
while the dynamic transition, $\alpha_d<\alpha_s$
corresponds to the value where greedy algorithms fail to find 
the existing solutions~\cite{Martin}.

All these arguments can be adapted to include quantum fluctuations
The statics is studied with the 
Matsubara replicated partition function~\cite{Cugrsa}, metastability with an
extension of the {\sc tap} approach~\cite{Biroli2} and the real-time 
dynamics with the Schwinger-Keldysh formalism~\cite{Culo,quantum-others}. 
The picture that 
arises is very similar to the one above 
with some intriguing new ingredients as the 
emergence of truly first order transitions close to the quantum
critical point~\cite{Niri,Cugrsa}, highly non-trivial effects due to the 
quantum environments~\cite{Biroli2}, a waiting-time dependent
quantum-to-classical crossover in the dynamic scaling, etc.

The models we studied in these notes have quenched random interactions.
Real glassy systems of the structural type do not. One may wonder if 
this is an important defficiency of the approach or if similar results 
can be obtained for models with no disorder. A large variety of 
models of mean-field type, or defined on large $d$ spaces, with no 
explicit quenched disorder and having the same phenomenology have 
been introduced in recent years~\cite{models-without-dis,Mepa}.
Finite $d$ models with similar, eventually interrupted, 
dynamic behaviour have 
also been exhibited~\cite{lattice-models,kinetic1,kinetic2,kinetic3}.  
Their existence supports the belief that 
the scenario here summarized goes beyond simple modelling. 
Indeed, it is at the basis of several conjectures for the 
behaviour of other nonequilibrium systems with slow dynamics
that have been later checked numerically. It has also motivated
several experimental investigations in a variety of systems.

\section{Perspectives}

We would like to end these notes by mentioning some of the 
directions 
for future development in this area. 
Within the ``mean-field'' approach there remain at least two important
open technical problems to complete their analytic solution:

\vspace{0.2cm}
\noindent
{\it Matching.} Having approximated the dynamic equations
as explained in the text we are not able to determine
the complete scaling form of correlations and responses. In models
like the $p$ spin spherical we cannot fix the 
scaling function $h(t)$. Going beyond the asymptotic
solution requires to solve the matching of the solution 
at short time-differences with the one at long time-differences.
This is a very tough mathematical problem.

\vspace{0.2cm} 
\noindent
{\it Dynamics at infinite time-scales.} 
In order to penetrate below the threshold in $p$ spin spherical models
and approach 
equilibrium one has to consider times that grow with $N$ and 
include instanton solutions to describe the activated dynamics below $T_d$.
This will smoothen the dynamic transition and convert it into a 
crossover. It will also possibly allow the mode-coupling and similar 
approaches to yield more accurate quantitative predictions. 

\vspace{0.2cm}
Even though the models we discussed are mean-field
we believe that the general picture developed holds beyond this 
limit. Some numerical and experimental tests support 
this belief. Still, one would like to justify this claim 
theoretically. Some of the lines of research that are  
now being followed with this aim are:

\vspace{0.2cm}
\noindent{\it Dynamics of dilute disordered models.}
These are disordered spin models on random 
graphs. Even if still mean-field they include some 
spatial fluctuations that one can study numerically and 
analytically~\cite{Secu,Mori}. Moreover, their relation to 
optimisation problems make them interesting {\it per se}.

\vspace{0.2cm}
\noindent{\it Dynamic heterogeneities.}
Supposedly these are nanoscale regions in supercooled 
liquids and glasses that are very important in determining 
the glass transition and the dynamics. They have been studied 
numerically and experimentally so far and one would like to 
have a theoretical model including and describing them.
A step in this direction was followed in~\cite{Claudio0}
where a sigma-model-like description for the spin-glass problem 
was introduced. This lead to proposing that the distribution 
of the coarse-grained local correlations 
and responses should be constrained to follow the global 
relation $\chi(C)$; this proposal was checked numerically 
in~\cite{Claudio} for the $3d${\sc ea} and studies on other 
disordered models are underway. The study of these distributions in 
real glasses as well as the development of a complete analytical 
description are problems that deserve further study.

\vspace{0.2cm}
\noindent{\it Dynamic functional renormalisation group ({\sc frg}).}
It would be very interesting to extend the {\sc frg} to attack 
nonequilibrium situations with aging dynamics.

\vspace{0.2cm}
\noindent
{\it {\sc fd} relations for finite $d$ systems.}
Many questions about the form of the relations between global 
responses and correlations in real systems can be raised. 
Do all systems undergoing domain growth 
in $d>1$ have two-time scales with the slower one characterised by 
$T^*\to\infty$? 
Can a situation as the one found for the {\sc sk} model be realised,  
{\it i.e.} does any real system have an
effective temperature with more than two values ($T$ and $T^*$)?

\vspace{0.2cm}
\noindent{\it Relations between statics and dynamics.}
A link between the nonequilibrium {\sc fd}
relations and Parisi's $P(q)$ was found in some 
mean-field models (note that it does not hold for models with a 
threshold as the $p$ spin). It was then argued that it should  
hold in finite $d$ systems under certain assumptions~\cite{FMPP}.
The long-lasting debate about the nature of the spin-glass 
phase has now been rephrased in nonequilibrium terms, the 
question now being what is the form of the asymptotic $\chi(C)$ 
plot. Recents experiments address this problem~\cite{Didier}. 
Analytical results for finite $d$ models are yet not available.

\vspace{0.2cm}
\noindent
{\it Thermodynamics and statistical mechanics out of equilibrium.}
From a more generic viewpoint, the development of a thermodynamics
and statistical mechanics for models that evolve slowly out of 
equilibrium is a very important issue. Some progress in both directions
has been made recently. One should try to establish these proposals in a less 
phenomenological way.

\vspace{0.2cm}
All these problems are challenging and very interesting. Many more could be 
added to this short list. We can expect to see progress in this very active 
area of research in the near future. 

%\newpage
\appendix

\section{Generalized Langevin equations}
\label{sec:langevin_general}
\setcounter{equation}{0}
\renewcommand{\theequation}{A.\arabic{equation}}

In this Appendix we derive a generalized Langevin equation starting from 
time-reversal microscopic equations for the motion 
of the system {\it and} the constituents of the 
bath. We use the simplest image of a thermal bath that 
is given by an ensemble of harmonic oscillators 
with masses $m_\alpha$ and frequencies $\omega_\alpha$, 
$\alpha=1,\dots,N$
and, for 
simplicity, we consider a system made of a single particle with mass $M$. 
We consider the one-dimensional case to
simplify the notation.
The generalization to 
more complex systems and/or to more 
complicated baths and higher dimensions 
is straightforward~\cite{Feve}. 
We call $q,p$ and 
$x_\alpha,p_\alpha$, $\alpha=1,\dots,N$ the positions and momenta of 
the particle and oscillators, respectively. 
The Hamiltonian of the total system is of the form (\ref{Htot}), 
\begin{eqnarray*}
H &=& 
\frac{p^2}{2M} + V(q) + \frac12 \sum_{\alpha=1}^N
\frac{c^2_\alpha}{m_\alpha\omega_\alpha^2}  q^2
+
 \sum_{\alpha=1}^N \frac{p_\alpha^2}{2m_\alpha} +   
\frac{1}{2} \sum_{\alpha=1}^N m_\alpha
\omega_\alpha^2 x_\alpha^2 
- \sum_{\alpha=1}^N  c_\alpha q x_\alpha 
\; .
\end{eqnarray*}
All these terms have been discussed in Section~\ref{staticproperties}.
Hamilton's equations for the particle are
\begin{eqnarray}
\dot q(t) = \frac{p(t)}{M}
\; , \;\;\;\; && \;\;\;\;
\dot p(t) =  
- \frac{\partial V(q)}{\partial q(t)}  
- \sum_{\alpha=1}^N \frac{c_\alpha^2}{m_\alpha\omega_\alpha^2} q(t)
+ \sum_{\alpha=1}^N c_\alpha x_\alpha(t)
\; ,
\label{eq-p0}
\end{eqnarray}
while the dynamic equations for each member of the environment read
\begin{eqnarray}
\dot x_\alpha(t) = \frac{p_\alpha(t)}{m_\alpha}
\; , \;\;\;\; & & \;\;\;\;
\dot p_\alpha(t) =  -m_\alpha\omega_\alpha^2 x_\alpha(t) + c_\alpha q(t)  
\; ,
\end{eqnarray}
showing that they are all forced massive harmonic oscillators. 
These equations are readily solved yielding
\begin{equation}
x_\alpha(t) = x_\alpha(0) \cos(\omega_\alpha t)  + \frac{p_\alpha(0)}{m_\alpha\omega_\alpha}
\sin(\omega_\alpha t)
+
\frac{c_\alpha}{m_\alpha\omega_\alpha}
\int_0^t dt' \sin(\omega_\alpha(t-t')) q(t')  
\end{equation} 
with $x_\alpha(0)$ and $p_\alpha(0)$ the initial coordinate
and position at time $t=0$ when the particle has been set in contact
with the bath. 
The replacement of this expression
in the last term on the {\sc rhs} of Eq.~(\ref{eq-p0}),
implies
\begin{eqnarray}
\dot p(t) =
- \frac{\partial V(q)}{\partial q(t)} + \eta(t) - \int_0^t dt' \, 
\gamma(t-t') \dot q(t')
- \sum_{\alpha=1}^N \frac{c^2_\alpha}{m_\alpha\omega_\alpha^2} 
\cos(\omega_\alpha t) q(0) 
\; ,
\label{eq:langevin_generalized}
\end{eqnarray}
with the kernel $\gamma$ given by 
\begin{equation}
\gamma(t-t') = \sum_{\alpha=1}^N \frac{c_\alpha^2}{m_\alpha\omega^2_\alpha} \cos(\omega_\alpha
(t-t'))
\; ,
\end{equation}
and the time-dependent force $\eta$ given by
\begin{equation}
\eta(t) = \sum_{\alpha=1}^N \frac{c_\alpha}{m_\alpha\omega_\alpha} 
p_\alpha(0) \sin(\omega_\alpha t)
+
 \sum_{\alpha=1}^N c_\alpha x_\alpha(0) \cos(\omega_\alpha t)
\; .
\end{equation} 
%The force ${\bf \eta}$
%depends on the initial conditions $\{p_\alpha(0),x_\alpha(0)\}$. 
Usually, the environments are
made of ensembles of {\it equilibrated} entities  
at a chosen temperature $T$. Then, $\{p_\alpha(0),x_\alpha(0)\}$
are initially distributed 
according to 
%the Gibbs-Boltzmann-Maxwell distribution:
\begin{equation}
P(\{p_\alpha(0),x_\alpha(0)\}) = 
\frac{\exp\left(-\beta H_{\sc env}[\{p_\alpha(0),x_\alpha(0)\}]\right)}
{\int \prod_{\alpha=1}^N dp_\alpha(0) dx_\alpha(0) \, 
\exp\left(-\beta H_{\sc env}[\{p_\alpha(0),x_\alpha(0)\}] \right)}
\; .
\end{equation}
It is convenient to assume that equilibrium distribution is shifted 
with respect to the coupling to the particle at the initial time; this 
allows one to eliminate
the last term in Eq.~(\ref{eq:langevin_generalized}). As when including
the counter-term, we 
choose
\begin{equation}
H_{\sc env} = \sum_\alpha \left[ 
\frac{m_\alpha\omega_\alpha^2}{2} 
\left( x_\alpha - \frac{c_\alpha}{m_\alpha\omega_\alpha^2} q(0) \right)^2
\frac{p_\alpha^2}{2m_\alpha} \right]
\; .
\end{equation}
%This implies that the force $\eta$ is itself a Gaussian random variable 
%with 
%\begin{eqnarray*}
%\langle \eta(t) \rangle &=&  
%\sum_{\alpha=1}^N \frac{c^2_\alpha}{m_\alpha\omega_\alpha^2} q(0)
%\cos(\omega_\alpha t)
%\, , 
%\\
%\langle (\eta(t)-\langle \eta(t) \rangle)  
%        (\eta(t') -\langle \eta(t')\rangle)
%\rangle &=&  
%k_B T \; \sum_{\alpha=1}^N \frac{c_\alpha^2}{m_\alpha \omega_\alpha^2}
%\cos(\omega_\alpha(t-t')) 
%\; .
%\end{eqnarray*}
Defining a new noise
%\begin{equation}
$
\xi(t) =\eta(t) - \sum_{\alpha=1}^N \frac{c^2_\alpha}{m_\alpha\omega_\alpha^2} 
\cos(\omega_\alpha t) q(0) 
$,
%\end{equation}
it is a Gaussian random variable with 
\begin{eqnarray}
\langle \xi(t) \rangle &=&  
0 \;\;\;\;\;\;\; \mbox{for all times}
\, , 
\\
\langle \xi(t) \xi(t') 
\rangle &=&  
%k_B T \; \sum_{\alpha=1}^N \frac{c_\alpha^2}{m_\alpha \omega_\alpha^2}
%\cos(\omega_\alpha(t-t')) 
%= 
k_B T \; \gamma(t-t')
\; 
\end{eqnarray}
and the Langevin equation simplifies to 
\begin{equation}
\dot p(t) =
- \frac{\partial V(q)}{\partial q(t)} + \xi(t) - \int_0^t dt' \, 
\gamma(t-t') \dot q(t')
\; .
\label{eq:langevin_generalized2}
\end{equation}
A random force with non-vanishing correlations on a finite 
support is  usually called a {\it colored noise}.

Interestingly enough, $\gamma(t-t')$ and the 
noise-noise correlation are proportional, with a constant of
proportionality of value $k_B T$.  This is a generalized form of
the fluctuation-dissipation relation, and it applies to 
the environment. 
In this derivation it is clear that it is a consequence of having 
assumed the equilibration of the bath.

The third term on the {\sc rhs} of
Eq.~(\ref{eq:langevin_generalized2}) represents 
a rather complicated friction force. Its value at time $t$ depends
explicitly on the history of the particle at times $0\leq t'\leq t$. 
The  memory kernel $\gamma(t-t')$ plays the r{\^o}le of a 
retarded friction function. This term makes 
Eq.~(\ref{eq:langevin_generalized2}) non-Markovian.

 Different choices of 
the environment are possible by selecting different ensembles 
of harmonic oscillators. The simplest choice, that leads to an 
approximate Markovian equation, is to consider
%an ensemble of $N\to\infty$ oscillators all with the same mass $m$, 
identical oscillators coupled 
to the particle via the coupling constants $\pm c/\sqrt{N}$
but having a non-trivial distribution of frequencies, that in the limit
$N\to\infty$, can be treated as continuous. This allows one to 
introduce the spectral density
$I(\omega)$ and rewrite the kernel $\gamma$ as
\begin{equation}
\gamma(t-t') = \frac{c^2}{m} 
\int_0^\infty d\omega \, I(\omega) \, \frac{\cos(\omega(t-t'))}{\omega^2}
\; .
\end{equation} 
For a Debye distribution of frequencies 
\begin{equation}
I(\omega) = \frac{3\omega^2}{\omega^3_D} \theta(\omega_D-\omega)
\;\;\;\; \mbox{one has} \;\;\;\;
\gamma(t-t') = \frac{3 c^2}{m \omega^2_D} \; 
\frac{\sin(\omega_D (t-t'))}{\omega_D (t-t')} 
\; .
\end{equation}
If $\omega_D$ is sufficiently large,  $\gamma$ can be approximated by a 
delta function, $\gamma(t-t') = 2\gamma \delta(t-t')$
with $\gamma=3c^2\pi/(2m\omega_D^2)$, and  
Eq.~(\ref{eq:langevin_generalized}) becomes
Markovian. 

Different environments are characterized by different choices of the 
spectral density $I(\omega)$ at small $\omega$. 
For example, one has an Ohmic ($s=1$), sub-Ohmic ($s<1$)
or super-Ohmic ($s>1$) bath 
if $I(\omega)\sim\omega^s$ for $\omega\sim 0$.

\section{The Kubo formula}
\label{app:Kubo-formula}
\setcounter{equation}{0}
\renewcommand{\theequation}{B.\arabic{equation}}

The Kubo formula relates the linear response to the 
asymmetric correlation of a quantum process. It 
holds at the level if the linear response 
even of the system is out of equilibrium. 
The linear response reads
\begin{eqnarray*}
&& R_{AB}(t,t') = 
%\left. 
%\frac{\delta \langle A(t)\rangle_{h_B}}{\delta h_B(t')} 
%\right|_{h_B=0}
%=
\left.
\frac{\delta}{\delta h_B(t')} 
\left[ 
\frac{1}{Z(h_B)} \mbox{Tr} \left( \hat U_t(h_B) \hat A(0) 
\hat U^{-1}_t(h_B) \hat \rho(0) \right)
\right] \right|_{h_B=0}
\;\;\;\;\;\;\;\;\;\;\;
\nonumber\\
&&= 
\left.
-\frac{1}{Z^2(h_B)}
\frac{\delta Z(h_B)}{\delta h_B(t')} 
\; \mbox{Tr} \left[ \hat U_t(h_B) \hat A(0) 
\hat U^{-1}_t(h_B) \hat \rho(0)
\right] \right|_{h_B=0}
\nonumber\\
&& 
+
\frac{1}{Z(0)} 
\left.
\mbox{Tr} \left[
\left( 
\frac{\delta \hat U_t(h_B)}{\delta h_B} \hat A(0) \hat U^{-1}_t(h_B)
%\right.
%\right.
%\right.
%\nonumber\\
%&&
%\left.
%\left.
%\left. 
+ 
\hat U_t(h_B) \hat A(0) \frac{\delta \hat U^{-1}_t(h_B)}{\delta h_B} 
\right)
\hat \rho
\right]
\right|_{h_B=0}
\end{eqnarray*}
where $U_t(h_B)$ is the evolution  operator,
$U_t(h_B) \exp[i/\hbar \hat H \delta]$ for all infinitesimal 
time intervals
except from the one going from $t'-\delta/2$ to $t'+\delta/2$ where 
it takes the form
$U_t(h_B) \exp[i/\hbar (\hat H-h_B \hat B)]$. See the left-panel in 
Fig.~\ref{pert-resp}
for a graphical representation of a kick-like perturbation.
$t\geq t'$. 
Calculating the variations explicitly, and using $\langle \hat A(t) 
\rangle=0$ or $\langle \hat A(t) \rangle$,   we have 
\begin{eqnarray}
R_{AB}(t,t') &=&
\frac{1}{Z(0)} 
\left.
\mbox{Tr} \left[
\frac{i}{\hbar} 
\left(
-
B(t') U_t(0) A(0) U_t^{-1}(0) +
U_t(0) A(0) B(t') U_t^{-1}(0) 
\right)
\hat \rho
\right]
\right|_{h_B=0}
\nonumber\\
&=& 
\frac{i}{\hbar} \langle [ A(t), B(t') ]\rangle
\; .
\end{eqnarray}

\section{The response in a Langevin process}
\label{app:relations}
\setcounter{equation}{0}
\renewcommand{\theequation}{C.\arabic{equation}}

By the definition the linear response is given by 
\begin{eqnarray*}
\left. \frac{\delta \langle q(t)\rangle_h}{\delta h(t')} \right|_{h=0}
&=& 
\left. \frac{\delta}{\delta h(t')} 
\int {\cal D}q {\cal D}i\hat q {\cal D}\overline\psi {\cal D}\psi
\, q(t) \, \exp\left( -S_{\sc eff} + 
\int dt''\;  i\hat q(t'') h(t'') \right) \right|_{h=0} 
\end{eqnarray*}
with $S_{\sc eff}$ defined in Eq.~(\ref{Seff}) and evaluated at
vanishing sources. The {\sc rhs}  
immediately leads to $R(t,t') = \langle q(t) i\hat q(t')\rangle$,
{\it i.e.} Eq.~(\ref{Rqihatq}).

The proof of the relation (\ref{Rqxi}) is slightly more involved.  
The correlation between coordinate and noise can be obtained
from the variation with respect to $\lambda(t,t')$
of the generating functional~(\ref{generating1})
once the identities (\ref{identity1}) and (\ref{identity2})
have been used and the source
\begin{equation}
\int dt'' dt''' \; \lambda(t'',t''') \, q(t'') \xi(t''')
\end{equation}
 has been added.
Integrating over the noise and keeping only the linear 
terms in $\lambda$ in the effective action since all others 
will vanish when setting $\lambda=0$
\begin{eqnarray}
\mbox{Linear terms} &=& 
\frac{k_BT}{2} \int dt_1 dt_2 dt_3 dt_4 \;
\left[  
\lambda(t_1,t_2) q(t_1) \gamma(t_2,t_3) i\hat q(t_4) \delta(t_4-t_3) 
\right.
\nonumber\\
&& 
\;\;\;\;\;\;\;\;\;\;
\left.
+
i\hat q(t_1) \delta(t_1-t_2) \gamma(t_2,t_3) \lambda(t_4,t_3) q(t_4) 
\right]
\; .
\end{eqnarray}
The variation with respect to $\lambda(t,t')$ yields
$
(k_BT)/2 \int dt'' \; \left[ \gamma(t',t'') + \gamma(t'',t') \right] $$
\langle q(t) i\hat q(t'')\rangle = 
\langle q(t) \xi(t')\rangle
$.

\section{Grassmann variables and supersymmetry}
\label{app:Grassmann}
\setcounter{equation}{0}
\renewcommand{\theequation}{D.\arabic{equation}}

Grassmann variables anticommute
$
\theta^2={\overline \theta}^2=[\theta,\bar\theta]_+=0$. 
The integration rules are 
$\int d\theta = \int d\overline\theta =0$ and 
$\int d\theta \, \theta =\int d\overline \theta \, \overline \theta =1$
while the derivation is such that 
$\partial_\theta = \int d\theta$ and 
$\partial_{\overline\theta} =  \int d\overline \theta$.

In the supersymmetric formalism used in Section~\ref{generating-functionals} one 
enlarges the usual ``bosonic'' space to include two conjugate 
Grassmann variables $\theta$ and $\overline\theta$: $t \to
a=(t,\theta,\overline\theta)$. 
A ``superfield'' and its ``supercorrelator'' are then defined as 
\begin{equation}
\Phi(a) \equiv  q(t) + \psi(t) \overline \theta + \overline \psi(t) \theta  
+ i \hat q(t) \overline \theta \theta 
\; , \;\;\;\;\;\;\;\;\;\;\;
Q(a,b) \equiv \langle \Phi(a) \Phi(b) \rangle 
\; ,
\end{equation}
$b=(t',\theta,\overline\theta')$.
The latter encodes the usual correlations 
%\begin{equation}
$\langle q(t) q(t') \rangle$,
% \; , \;\;\;
$\langle q(t) i \hat q(t') \rangle$,
% \; , \;\;\;
$\langle i \hat q(t) q(t') \rangle$,
% \; , \;\;\;
$\langle i \hat q(t) i \hat q(t') \rangle$,
%  \; ,
%\label{eq-corr}
%\end{equation}
as well as ``fermionic'' correlators
%\begin{equation}
$\langle q(t) \psi(t') \rangle$,
% \; , \;\;\;
$\langle \overline \psi(t) i\hat q(t') \rangle$,
% \; , \;\;\;
$\langle \overline \psi(t) \psi(t') \rangle$,
%\; , \;\;\;
etc.
%\mbox{etc.}
%\end{equation}
The solutions we construct and study are such that 
all correlators that involve only one fermionic variable 
$\psi$ and $\overline\psi$ vanish. We are then left with the 
usual four correlators purely bosonic correlators and the 
fermion bilinears. One proves that the latter equal the linear 
response.  If, moreover, we 
only consider causal solutions, 
$\hat Q(t,t') \equiv \langle i \hat q(t) i \hat q(t') \rangle  =0$
and 
%in (\ref{eq-corr}), 
\begin{equation}
Q(a,b) = C(t,t') - (\overline\theta'-\overline\theta) 
\left( \theta' R(t,t') - \theta R(t',t) \right) 
\; .  
\label{causal}
\end{equation} 
Convolutions, or operational products, and Hadamard, or simple products,
are defined as 
\begin{eqnarray}
Q_1(a,b) \otimes Q_2(b,c) &=& \int db \,  Q_1(a,b) Q_2(b,c)
\; ,
\nonumber\\ 
Q_1(a,b) \bullet Q_2(a,b) &=& Q_1(a,b) Q_2(a,b)
\; ,
\end{eqnarray}
respectively, with $db \equiv dt d\overline\theta d\theta$. 

For correlators of the causal form (\ref{causal}), the convolution and 
the Hadamard product respect the structure of the correlator. Indeed, 
the result of the convolution is again of the form (\ref{causal})  with
\begin{eqnarray}
C_{\sc conv}(t,t'') &=& \int dt' \; 
\left[ C_1(t,t') R_2(t'',t') +  R_1(t,t') C_2(t',t'')  \right]
\; ,
\nonumber\\ 
R_{\sc conv}(t,t'') &=& \int dt' \; R_1(t,t') R_2(t',t'') 
\; ,
\end{eqnarray}
and the result of the Hadamard product is also  
of the form (\ref{causal})  with
\begin{eqnarray}
C_{\sc had}(t,t') &=& C_1(t,t') C_2(t,t')  
\; ,
\nonumber\\ 
R_{\sc had}(t,t') &=& C_1(t,t') R_2(t,t') + C_2(t,t') R_1(t,t')
\; .
\end{eqnarray}
The Dirac delta function is defined as 
$
\delta(a-b) = \delta(t - t') 
(\overline \theta- \overline \theta')
(\theta- \theta')
$.

\section{Integrals in the aging regime}
\label{integrals}
\setcounter{equation}{0}
\renewcommand{\theequation}{E.\arabic{equation}}

%\subsection{First type of integral}

Integrals of the form
\begin{equation}
I_1(t) \equiv \int_0^t dt'' A(t,t'') B(t,t'')
\; 
\end{equation}
appear, for example, in the equation for $\mu(t)$. 
We separate the integration time-interval as in Eq.~(\ref{interval1}).
If $\delta$ is chosen to be a finite time,
$A(t,t'')$ and $B(t,t'')$ in the first interval 
can be approximated by $A(t,0)$ that vanishes
when $t\to \infty$. Since  the integration interval is finite, 
this term can be neglected.
In the second interval the functions vary in the aging regime and 
in the third interval they vary in the stationary regime. Thus
\begin{eqnarray}
I_1(t) 
&\sim&
\int_\delta^{\Delta_t} dt'' A_{\sc ag}(t,t'') B_{\sc ag}(t,t'')
+
\int_{\Delta_t}^t dt'' \, 
\left[ 
\left(
A_{\sc st}(t-t'')
+
\lim_{t-t''\to\infty} \lim_{t''\to\infty}  A(t,t'')  
\right)
\right.
\nonumber\\
& & 
\;\;\;\;\;\;\;\;\;
\times
\left.
\left(
B_{\sc st}(t-t'')
+
\lim_{t-t''\to\infty} \lim_{t''\to\infty}  B(t,t'') 
\right)
\right]
\; .
\nonumber
\end{eqnarray}
We assume that this separation is sharp and that we can 
neglect the corrections associated to mixing of the three
regimes. In the third term we replaced $A$ and $B$ 
in terms of $A_{\sc st}$, $B_{\sc st}$.
We can now replace the lower limit of the first integral by $0$
and its upper limit by $t$. 
In addition, assuming that $B$ is proportional to 
the response, 
\begin{equation}
\lim_{t-t''\to\infty} \lim_{t''\to\infty}  B(t,t'') =0
\; ,
\end{equation}
and that $A$ is a function of the correlation such that 
\begin{equation}
\lim_{t-t''\to\infty} \lim_{t''\to\infty}  A(t,t'')
= A_{q_{\sc ea}}
\end{equation}
we have
\begin{eqnarray*}
 I_1(t)
&\sim&
\int_0^t dt'' A_{\sc ag}(t,t'') B_{\sc ag}(t,t'')
+
A_{q_{\sc ea}}
\int_0^{\infty} d\tau' \; B_{\sc st}(\tau') 
+
\int_0^{\infty} d\tau' \; 
 A_{\sc st}(\tau') B_{\sc st}(\tau') 
\end{eqnarray*}
where the upper limit tending to infinity is 
$t-\Delta_t\to\infty$.

%\subsection{Second type of integral}

Another type of integrals is:
%\begin{equation}
$I_2(t,t') \equiv \int_{t'}^t dt'' A(t,t'') B(t'',t')$.
%\end{equation}
In particular, if $B=1$, $A(t,t'')=R(t,t'')$, $t'=0$ and $t\to\infty$,
this integrals yields the static susceptibility. If instead, $t'$ is 
long and $t$ too we have the type of integral appearing in the equation
for the response. 
Let us assume that $t$ and $t'$ are 
far apart; we start by dividing the 
time interval in three subintervals
\begin{equation}
\int_{t'}^t = \int_{t'}^{\Delta_{t'}} + \int_{\Delta_{t'}}^{\Delta_t} +
\int_{\Delta_{t}}^t 
\end{equation}
and by approximating the functions $A$ and $B$ assuming that they
have a two-step decay as the one in Sections \ref{subsec:web} and 
\ref{subsec:wltm}: 
\begin{eqnarray}
I_2(t,t') &\sim& 
\int_{t'}^{\Delta_{t'}} dt''  A_{\sc ag}(t,t'') B(t''-t')
+ 
\int_{\Delta_{t'}}^{\Delta_t} dt''  A_{\sc ag}(t,t'') B_{\sc ag}(t'',t') 
\nonumber\\
& & 
+ 
\int_{\Delta_{t}}^t  dt'' A(t-t'') B_{\sc ag}(t'',t')\; .
\end{eqnarray}
In the first term $B(t-t'')$ can
be replaced by $B(t-t'')= 
\lim_{t-t''\to\infty} \lim_{t''\to\infty}B(t,t'')$ 
$+B_{\sc st}(t-t'')$.
The same applies to $A$ in the last term.
All functions vary fast in the stationary regime but very slowly in the 
aging regime. The next assumption is that functions
in the aging regime that are convoluted with functions
in the stationary regime, can be considered to be 
constant and taken out of 
the integrals. That is to say
\begin{eqnarray}
I_2(t,t') 
&\sim&
A_{\sc ag}(t,t') \int_{t'}^{\Delta_{t'}} dt'' \, B(t''-t') 
+
\int_{\Delta_{t'}}^{\Delta_t} dt''  A_{\sc ag}(t,t'') B_{\sc ag}(t'',t')
\nonumber\\
& & 
+
B_{\sc ag}(t,t')
\int_{\Delta_t}^t dt'' \, A(t-t'')
\nonumber\\
&\sim&
A_{\sc ag}(t,t') \int_0^\infty d\tau' \, B(\tau') 
+
\int_{t'}^t dt''  A_{\sc ag}(t,t'') B_{\sc ag}(t'',t')
\nonumber\\
& & 
+
B_{\sc ag}(t,t')
\int_0^\infty d\tau' \, A(\tau')
\; ,
\end{eqnarray}
where we used $t-\Delta_t \to\infty$ and $\Delta_{t'}-t'\to\infty$.
When this integral appears in the equation for the response, 
$A(\tau')$ and $B(\tau')$ are proportional to the response
function since one is the response itself and the other is the 
self-energy. Using {\sc fdt} the integrals in the first and 
third term can be computed in the classical limit or they can 
be expressed as functions of the correlation in the quantum case.  
The second term instead depends exclusively on the 
aging dynamic sector.

All other integrals can be evaluated, in the large-time limit,
in a similar way.

\vspace{1cm}
\noindent{\underline{Acknowledgements}}
The author  specially thanks J. Kurchan for his collaboration on this 
and other subjects, and G. Semerjian for very useful discussions and 
the careful correction of the manuscript. 
LFC is ICTP research scientist,
acknowledges financial support from 
the Guggenheim Foundation and the ACI 
``Algorithmes d'optimisation et syst\`emes d\'esordonn\'es
quantiques'' and thanks the Universities of Buenos Aires and 
La Plata (Argentina) and Harvard University for hospitality 
during the preparation of these notes.


\begin{thebibliography}{99}
\bibitem{metastable} One can think as the super-cooled liquid as 
a metastable state in the usual sense: its properties can be 
computed using a restricted partition function and, in particular, 
the dynamics is stationary and satisfies {\sc fdt} even if the system
is not equilibrated in its true most favorable state -- the crystal.
The glass, instead, is also metastable but of a different kind:
one cannot describe all its properties with a stationary probablility 
distribution.
\bibitem{comment-renorm} Note that the sign of the mass term generated 
by the coupling to the environment is negative. If strong enough, 
it may render unstable an initially  stable potential well. 
One way to avoid this 
problem is to introduce the counterterm.
One could also think that the coupling to the bath must be very weak 
and that the last term must vanish by itself (say, all $c_a\to 0$). 
However, this might be problematic when dealing with the stochastic 
dynamics since a very weak coupling to the bath implies also a very slow
relaxation.  
It is then conventional to include the counterterm to cancel the mass
renormalization.
\bibitem{comment-self} Note that the nonequilibrium correlation and response
functions are self-averaging.
\bibitem{eqglass} Note that many glass formers have a 
crystalline structure that corresponds to most convenient 
equilibrium configuration. How much does the crystal 
influence the behaviour or the glass is an interesting 
still unanswered question, see the discussion in ~\cite{Grigera}
\end{thebibliography}
\end{document}